\documentstyle[12pt, epsf]{report}

\def \a{\alpha}
\def \b{\beta}
\def \g{\gamma}
\def \d{\delta}
\def \ep{\epsilon}

\def \l{\lambda}
\def \m{\mu}
\def \n{\nu}
\def \x{\xi}
\def \vp{\varpi}
\def \r{\rho}

\def \s{\sigma}

\def \t{\tau}
\def \u{\upsilon}
\def \ph{\phi}
\def \vp{\varphi}
\def \c{\chi}
\def \ps{\psi}

\def \L{\Lambda}
\def \X{\Xi}
\def \S{\Sigma}

\def \Ph{\Phi}
\def \Ps{\Psi}

\def \re{{\rm e}}
\def \im{{\rm i}}

\def \Tr{{\rm Tr}}
\def \ma{{\rm matrix}}
\def \hatleftix{{\hat{\mathit L}_{\Lambda}}}
\def \leftix{{{\mathit L}_{\Lambda}}}
\def \hatrightix{{\hat{\mathit R}_{\Lambda}}}
\def \rightix{{{\mathit R}_{\Lambda}}}
\def \hatmultix{{\hat{\mathit M}_{\Lambda}}}
\def \multix{{{\mathit M}_{\Lambda}}}
\def \hatcentrix{{\hat{\mathit \Sigma}_{\Lambda}}}
\def \centrix{{{\mathit \Sigma}_{\Lambda}}}
\def \vectrix{{\cal V}_{\Lambda}}

\def \hatheterix{{\hat{\Gamma}_{\L | \L}}}
\def \supersalt{{\mathit F}_{\Lambda|\Lambda}}

\def \salt{{\mathit F}_{\Lambda}}
\def \tsalt{{\tilde{\mathit F}}_{\Lambda}}
\def \saltone{{{\mathit F}_1'}}
\def \tsaltone{{\tilde{\mathit F}}_1'}
\def \ignore#1{}
\def \gl{gl_{+\infty}}

\def \la#1{\label{#1}}
\def \ift{\infty}
\def \le{\left}
\def \ri{\right}
\def \da{\dagger}
\def \ti#1{\tilde{#1}}
\def \lb{\lbrack}
\def \rb{\rbrack}

\def \rar{\rightarrow}
\def \uar{\uparrow}
\def \dar{\downarrow}
\def \lrar{\leftrightarrow}
\def \ld{\ldots}
\def \cd{\cdots}
\def \nn{\nonumber}
\def \pa{\partial}

\newcommand \beq{\begin{eqnarray}}
\newcommand \eeq{\end{eqnarray}}
\newcommand \beqs{\begin{eqnarray}}
\newcommand \eeqs{\begin{eqnarray}}
\newcommand \ba{\begin{array}}
\newcommand \ea{\end{array}}

\newtheorem{lemma}{Lemma}

\begin{document}
\pagenumbering{roman}

\thispagestyle{empty}
\begin{flushright}
Ph. D. thesis (internet version)\\
hep-th/9907130 \\
\end{flushright}

\begin{center}
   {\LARGE\bf Symmetry Algebras of Quantum Matrix Models in the Large-$N$ Limit} \\
   \vspace{1.5cm}
   {\large\bf C.-W. H. Lee} \\
   {\it Department of Physics and Astronomy, P.O. Box 270171, University of Rochester, Rochester, New York 14627} \\
   \vspace{.5cm}
   {July 15, 1999} \\
   \vspace{1.5cm}
   {\Large\bf Abstract}
\addcontentsline{toc}{chapter}{Abstract}
\end{center}

Quantum matrix models in the large-$N$ limit arise in many physical systems like Yang--Mills theory with or without
supersymmetry, quantum gravity, string-bit models, various low energy effective models of string theory, M(atrix) 
theory, quantum spin chain models, and strongly correlated electron systems like the Hubbard model.  We introduce, 
in a unifying fashion, a hierachy of infinite-dimensional Lie superalgebras of quantum matrix models.  One of these 
superalgebras pertains to the open string sector and another one the closed string sector.  Physical observables of 
quantum matrix models like the Hamiltonian can be expressed as elements of these Lie superalgebras.  This indicates 
the Lie superalgebras describe the symmetry of quantum matrix models.  We present the structure of these Lie 
superalgebras like their Cartan subalgebras, root vectors, ideals and subalgebras.  They are generalizations of 
well-known algebras like the Cuntz algebra, the Virasoro algebra, the Toeplitz algebra, the Witt algebra and the 
Onsager algebra.  Just like we learnt a lot about critical phenomena and string theory through their conformal 
symmetry described by the Virasoro algebra, we may learn a lot about quantum chromodynamics, quantum gravity and 
condensed matter physics through this symmetry of quantum matrix models described by these Lie superalgebras.

\pagebreak

\begin{center}
   {\Large\bf Foreword}
\end{center}
\addcontentsline{toc}{chapter}{Foreword}

In the past year or so, my advisor, S. G. Rajeev, and I published a number of articles \cite{prl, opstal, clstal,
plb} on algebraic approaches to Yang--Mills theory, D-branes, M-theory and quantum spin chain systems.  Each of 
these papers focuses on a particular aspect.  For example, in one paper we discussed exclusively the case when the 
physical system consists of open singlet states only, and in another one we confined ourselves to closed singlet 
states.  Moreover, we omitted a number of proofs and intermediate steps of calculations which led to the 
conclusions presented in those articles in order to keep them of reasonable sizes.  Taking the opportunity to write 
this Ph.D. thesis, I now present our results in a more coherent fashion and a more detailed manner so that it is 
more accessible to those who want to acquire a deeper understanding.  We hope that the following pages have 
achieved these purposes.  We had in mind junior graduate students when we wrote this article, and hope that this 
article is accessible to them.

The following table shows whether and where the material of a section has been published:
\begin{verse}
Section~\ref{s2.2}: Refs.\cite{prl}, \cite{opstal}, \cite{clstal} and \cite{plb} and some yet to be published 
material. \\
Section~\ref{s2.3}: Ref.\cite{clstal}. \\
Section~\ref{s2.4}: Ref.\cite{opstal}. \\
Sections~\ref{s2.5} and \ref{s2.6}: to be published. \\
Section~\ref{s2.7}: Refs.\cite{prl}, \cite{clstal} and \cite{plb}. \\
Section~\ref{s2.8}: Refs.\cite{opstal} and \cite{plb}. \\
Chapter~\ref{c4}: Ref.\cite{opstal}. \\
Section~\ref{s3.2}: to be published. \\
Section~\ref{s3.3}: Ref.\cite{clstal}. \\
Section~\ref{s5.2}: to be published. \\
Section~\ref{s5.3}: Ref.\cite{opstal} and some yet to be published material. \\
Section~\ref{s5.4}: Refs.\cite{prl} and \cite{clstal} and some yet to be published material. \\
Section~\ref{s5.5}: Refs.\cite{prl} and \cite{clstal}.
\end{verse}

Ref.\cite{ijmp} is a very recent review article which is a simplified version of this thesis.  It introduces to 
readers only the bosonic part of the theory.  Readers who prefer a simpler and more basic exposition of the subject 
matter are referred to it.

\pagebreak

\begin{center}
{\Large \bf Acknowledgement} 
\end{center}
\addcontentsline{toc}{chapter}{Acknowledgement}

First of all, I would like to thank my advisor, S. G. Rajeev, for all the years of his patient guidance and 
instructions.  Besides teaching me a lot of physics, mathematics and sometimes even English, he distilled in me
valuable wisdom in putting a physics problem in a proper perspective and understanding its elegance.  For example, 
he demonstrated to me that if it is possible to think of a physics problem geometrically or diagrammatically, a lot 
of times the geometry or the diagrams will unveil simple but profound properties of the physics, and help you solve 
the problem elegantly.  I hope that the many diagrams in this whole thesis illustrate this point well.  More 
importantly, he demonstrates to us what a proper attitude towards scientific research should be.  One should pick a 
physics problem that is truly important to the progress of science, and be persistent and patient in looking for 
its solution.  No doubt understanding Yang--Mills theory is an important problem of science in this generation.  I 
hope that the approach described in these pages provide us a clue to the ultimate solution to Yang--Mills theory.

I would also like to thank to Y. Shapir and E. Wolf for their guidance of my research on roughening transition and 
coherence theory, respectively and thus gave me some valuable personal lessons in conducting scientific research.  
My special thanks is also due to my very first mentor, H. M. Lai, who introduced scientific research to me and 
taught some fundamental lessons of it.  For example, he taught me that if a problem looks too difficult, one may 
try to tackle a simpler version of it, and go back to the full problem later.  This simple philosophy proves to be 
very useful in this work.

Many other people have offered help to this research.  I would like to express my special gratitude to O. T. Turgut
for introducing to me an algebraic formulation of gluon dynamics which provides the basis for this whole work.  
Moreover, V. John, S. Okubo and T. D. Palev give us good suggestion on certain parts of the thesis.  Many other
people have taken the time to listen to our results in seminars, talks or informal discussions.  Regrettably, there
are too many of them that I am unable to list all their names here.

I would also like to thank T. Guptill and C. Macesanu for their technical help in the computer part of this thesis,
and L. Orr and S. Gitler for their careful proofreading.

Of course, without the instructions of basic knowledge in language, science and others from my former teachers, and 
without the encouragement from my former classmates at all the tertiary, secondary and primary institutions I 
attended, it is impossible to finish this research.  They laid a good foundation for me to pursue this work.

Last, I would like to express my greatest gratitude to my parents, Lee Kun Hung and Cheng Yau Ying, for all their 
years of care, patience and sacrifice.

This work was supported in part by funds provided by the U.S. Department of Energy under grant DE-FG02-91ER40685.

\pagebreak

\addcontentsline{toc}{chapter}{Table of Contents}
\tableofcontents
\pagebreak

\pagenumbering{arabic}

\chapter{Introduction}
\la{c1}

\section{Hadron Structure, Quantum Gravity and Condensed Matter Phenomena}
\la{s1.0.1}

As physicists, there are a number of phenomena in strong interaction, quantum gravity and condensed matter physics 
which we want to understand.

Quantum chromodynamics (QCD) is the widely accepted theory of strong interaction.  It postulates that the basic 
entities participating strong interaction are quarks, antiquarks and gluons.  In the high-energy regime, the theory
displays asymptotic freedom.  The coupling among these entities becomes so weak that we can use perturbative means
to calculate experimentally measurable quantities like differential cross sections in particle reactions.  Indeed, 
the excellent agreement between perturbative QCD and high energy particle phenomena form the experimental basis of 
the theory.

Nevertheless, strong interaction manifests itself not only in the high-energy regime but also in the low-energy one.
Here, the strong coupling constant becomes large.  Quarks, anti-quarks and gluons are permanently confined to form 
bound states called hadrons, like protons and neutrons.  Hadrons can be observed in laboratories.  We can measure 
their charges, spins, masses and other physical quantities.  One challenging but important problem in physics is to 
understand the structures of hadrons within the framework of QCD; this serves as an experimental verification of
QCD in the low-energy regime.  Hadronic structure can be described by something called a structure function which 
tells us the (fractional) numbers of constituent quarks, antiquarks or gluons carrying a certain fraction of the 
total momentum of the hadron.  The structure function of a proton has been measured carefully \cite{cteq}.  There 
has been no systematic theoretical attempt to explain the structure function until very recently \cite{krra}.  In 
this work, the number of colors $N$ is taken to be infinitely large as an approximation.  The resulting model can 
be treated as a classical mechanics \cite{berezin, yaffe, rajeev94}; i.e., the space of observables form a phase 
space of position and momentum, and the dynamics of a point on this phase space is governed by the Hamiltonian of 
this classical system and a Poisson bracket\footnote{Ref.\cite{arnold} provides an excellent discussion for such a 
geometric formulation of classical mechanics.}.  This is because quantum fluctuations abate in the large-$N$ limit 
--- the Green function of a product of color singlets is dominated by the product of the Green functions of these 
color singlets, and other terms are of subleading order\footnote{For an introductory discussion on this point, see 
Refs.\cite{witten78} and \cite{coleman}.}.  It is possible to derive a Poisson bracket for Yang--Mills theory in 
the large-$N$ limit.  This Poisson bracket can be incorporated into a commutative algebra of dynamical variables to 
form something called a Poisson algebra \cite{chpr}.  

As an initial attempt, only quarks and anti-quarks in the QCD model in Ref.\cite{krra} are dynamical.  A more 
realistic model should have dynamical gluons in addition to quarks.  Gluons carry a sizeable portion of the total
momentum of a proton \cite{cteq} and are thus significant entities.  One notable feature of a gluon field is that it
is in the adjoint representation of the gauge group and carries two color indices.  These can be treated as row and
column indices of a {\em matrix}.  This suggests gluon dynamics can be described by an abstract model of matrices.  

Besides strong interaction phenomena, another fundamental question in physics is how one quantizes gravity.  The 
most promising solution to this problem is superstring theory \cite{polchinski}.  Here we postulate that the basic 
dynamical entities are one-dimensional objects called strings.  Quarks, gluons, gravitons and photons all arise as 
excitations of string states.  If the theory consists of bosonic strings only, the ground states will be tachyons.
To remove tachyons, fermions are introduced into the theory in such a way that there exists a symmetry between 
bosons and fermions.  This boson--fermion symmetry is called supersymmetry\footnote{Supersymmetry could also be
viewed as a symmetry which unifies, in a non-trivial manner, the space-time symmetry described by the Poincar\'{e}
group, and the local gauge symmetry at each point of space-time.  The symmetry between bosons and fermions then 
come as a corollary.  See Refs.\cite{weba} or \cite{buku} for further details.}  A superstring theory which is free 
of quantum anomaly must be ten-dimensional.  Since we see only four macroscopic dimensions, the extra ones have to 
be compactified.  

A partially non-perturbative treatment of superstring theory is through entities called D$p$-branes 
\cite{polchinski}.  They are extended objects spanning $p$ dimensions.  Open strings stretch between D$p$-branes in 
the remaining dimensions, which are all compactified.  There is a non-trivial background gauge field permeating the 
whole ten-dimensional space-time.  The dynamics of the end points of open strings can be regarded as the dynamics 
of D$p$-branes themselves, each of which behaves like space-time of $p$ dimensions.  

Different versions of string theory were put forward in the 80's.  Lately, evidence suggests that there exist 
duality relationship among these different versions of string theory and so there is actually only one theory for
strings.  The most fundamental formulation of string theory is called M(atrix)-theory \cite{bfss}.  Currently, there
is a widely-believed M-theory conjecture which states that in the infinite momentum frame (a frame in which the 
momentum of a physical entity in one dimension is very large), M-theory \cite{bfss} can be described by the quantum 
mechanics of an infinite number $N$ of point-like D0-branes, the dynamics of which is in turn described by a {\em 
matrix model with supersymmetry}.

Besides the M-theory conjecture, there are a number of different ways of formulating the low-energy dynamics of
superstring theory as supersymmetric matrix models \cite{polchinski}.

We can use matrix models to describe condensed matter phenomena, too.  One major approach condensed matter 
physicists use to understand high-$T_c$ superconductivity, quantum Hall effect and superfluidity is to mimic them 
by integrable models like the Hubbard model \cite{hubbard}.  (This is a model for strongly correlated electron 
systems.  Its Hamiltonian consists of some terms describing electron hopping from site to site, and a term which 
suppresses the tendency of two electrons to occupy the same site.  We will write down the one-dimensional version 
of this model in a later chapter.)  It turns out that the Hubbard model and many other integrable models can 
actually be formulated as matrix models with or without supersymmetry.

Thus matrix models provide us a unifying formalism for a vast variety of physical phenomena.  Now, we would like to 
propose an {\em algebraic} approach to matrix models.  The centerpiece of the classical mechanical model of QCD in 
Refs.\cite{rajeev94} and \cite{krra} is a Poisson algebra.  We can write the Hamiltonian as an element of this 
Poisson algebra, and can describe the dynamics of hadrons and do calculations through it.  In string theory, the 
string is a one-dimensional object and so it sweeps out a two-dimensional surface called a worldsheet as time goes 
by.  Worldsheet dynamics possesses a remarkable symmetry called conformal symmetry --- the Lagrangian is invariant 
under an invertible mapping of worldsheet coordinates $x \rar x'$ which leaves the worldsheet metric tensor 
$g_{\m\n}(x)$ invariant up to a scale, i.e., $g'_{\m\n}(x') = \L (x) g_{\m\n}(x)$, where $\L (x)$ is a non-zero 
function of worldsheet coordinates.  This invertible mapping is called the conformal transformation.  It turns out 
that the conformal charges, i.e., the conserved charges associated with conformal symmetry, in particular the 
Hamiltonian of bosonic string theory, can be written as elements of a Lie algebra called the Virasoro algebra 
\cite{virasoro, gool}.  Through the Virasoro algebra, we learn a lot about string theory like the mass spectrum and 
the S-matrix elements.  

The relationship between conformal symmetry and the Virasoro algebra illustrates one powerful approach to physics
--- identify the symmetry of a physical system, express the symmetry in terms of an algebra, and use the properties
of the algebra to work out the physical behavior of the system.  Sometimes, the symmetry of the physical system is
so perfect that it completely determines the key properties of the system.

The above argument suggests that we may get fruitful discovery in gluon dynamics, M-theory and superconductivity if 
there is a Lie algebra for a generic matrix model, and we are able to write its Hamiltonian in terms of this Lie 
algebra, which expresses a new symmetry in physics.

\section{Classical Matrix Models in the Large-$N$ Limit}
\la{s1.2}

It helps to give a more precise definition of matrix models.  There are two broad families, namely classical matrix 
models and quantum matrix models.

Let us consider classical matrix models first.  They are models in which the classical dynamical variables form 
Hermitian matrices $Q(1)$, $Q(2)$, \ld, and $Q({\L})$ with the same dimension, say, $N$.  A matrix entry of, say, 
$Q(1)$ is denoted by $Q^{\m_1}_{\m_2}(1)$, where $\m_1$ and $\m_2$ are row and column indices and take any integer 
value between 1 and $N$ inclusive.  Let $\dot{Q}(1)$, $\dot{Q}(2)$, \ld, and $\dot{Q}({\L})$ be their time 
derivatives, respectively.  The Lagrangian  
\[ L(Q(1), Q(2), \ld, Q({\L}); \dot{Q}(1), \dot{Q}(2), \ld, \dot{Q}({\L})) \]
is the trace of a polynomial of these matrices.  As usual, the conjugate momenta $P^{m_2}_{m_1}(k)$ for $k = 1$, 2, 
\ld, and $\L$ and the Hamiltonian 
\[ H(Q(1), Q(2), \ld, Q(\L); P(1), P(2), \ld, P(\L)) \]
are defined by the formulae
\[ P^{m_2}_{m_1}(k) \equiv \frac{\pa L}{\pa\dot{Q}^{m_1}_{m_2}(k)} \]
and
\[ H \equiv \Tr \sum_{k=1}^{\L} P(k) \dot{Q}(k) - L. \]
The partition function can then be written as
\beq
   \int dQ(1) dQ(2) \cd dQ({\L}) dP(1) dP(2) \cd dP({\L}) \re^{- \b H},  
\la{1.2.1}
\eeq
where $1/\b$ is the product of the Boltzmann constant and the temperature of the system.  There are systematic 
approaches to evaluate this kind of integrals \cite{britpazu, chmame, mehta, douglas, sesz}.

Interest in matrix models in the large-$N$ limit was stimulated by a pioneering work of 't Hooft \cite{thooft74a}.  
He showed that if there are an infinite number of colors in Yang--Mills theory, then the Feynman diagrams can be 
dramatically simplified.  A gauge boson carries two color indices, and instead of drawing only one line for a gauge 
boson propagator, we draw two lines juxtaposed with each other.  A quark or an anti-quark carries one color index 
only and so we still draw just one line for its propagator.  The vertices are modified in such a way that 
propagation lines carrying the same color are joined together.  It turns out that in the large-$N$ limit, only 
planar diagrams --- diagrams in which no two lines cross each other --- with the least possible number of quark 
loops survive.  Feynman diagrams with higher genus, i.e., those with `handles', are of subleading order.  As pointed
out by 't Hooft, so long as a theory with a global $U(N)$ symmetry contains fields with two $U(N)$ indices, i.e.,
so long as the theory is a matrix model, the above argument will apply.

This observation led to an immediate triumph.  't Hooft was able to show that in two dimensions (one spatial and 
one temporal), the meson spectrum displays a Regge trajectory, i.e., the spectrum consists of discrete points lying 
roughly on a straight line with no upper bound \cite{thooft74b}.  In other words, quarks are permanently confined 
in the large-$N$ limit in two dimensions.

An alternative approach to describe the dynamics of gauge theory is to solve for the Schwinger--Dyson equation for
Wilson loops.  Since a Wilson loop consists of a series of the traces of path-ordered products of classical gauge 
boson fields, it is natural for us to think of these gauge boson fields as matrix fields.  One famous example is the
Eguchi--Kawai model \cite{egka}.  In this model, space-time is approximated by a discrete set of points (lattice), 
and parallel transport operators between adjacent points are assumed to be the same as long as the parallel 
transports are in the same directions.  Then the theory becomes a classical matrix model with a small number of 
different matrices.  This model and its variations provide an interesting way to understand phase transition 
between confinement and deconfinement in large-$N$ gauge theory.

't Hooft's idea can be adapted to quantum gravity, too. The dual of a planar Feynman diagram can be taken as a
triangulation of a planar surface, which serves as a lattice approximation of a geometrical surface.  The partition
function for a quantum gravitational theory can then be approximated by a classical matrix model \cite{david, 
kazakov85}.  This approach, called lattice quantum gravity, has drawn a lot of attention.  Ref.\cite{frgizi} gives
an interesting review on the subject.

Classical matrix models have also found application in the study of spin systems.  Kazakov showed that if Ising
spins are placed on a random lattice, the resulting Ising model is equivalent to a classical one-matrix model
\cite{kazakov86}.  Such kind of random spin models can be useful to mimic some condensed matter phenomena like
spin glass.  Moreover, there are some common properties of random spin models and quantum gravity; 
cross-pollination of the two disciplines can bring fruitful progress to both.

\section{Quantum Matrix Models in the Large-$N$ Limit}
\la{s1.3}

The major focus of this whole article is not classical matrix models, but quantum matrix models instead.  A quantum
matrix model is a matrix model whose matrix entries are not dynamical variables but quantum operators instead. More 
specifically, again consider the set of $N \times N$ time-dependent matrices $Q(1)$, $Q(2)$, \ld, and $Q({\L})$, 
and again let
\[ L(Q(1), Q(2), \ld, Q({\L}); \dot{Q}(1), \dot{Q}(2), \ld, \dot{Q}({\L})) \]
be the Lagrangian.  Instead of treating the $Q$'s and $P$'s as dynamical variables, here we impose the canonical 
commutation relation
\beq
   \le[ Q^{\m_1}_{\m_2}(k_1), P^{\m_3}_{\m_4}(k_2) \ri] = \im \hbar \d^{\m_1}_{\m_4} \d^{\m_3}_{\m_2}
   \d(k_1, k_2).
\la{1.3.1}
\eeq
We now have a quantum system.  The partition function becomes\footnote{Readers who are not familiar with the way 
the partition function of a quantum system is expressed as a path integral are referred to Ref.\cite{fehi}.}
\beq
   & & \int^{\ift}_{-\ift} du_1 \int^{\ift}_{-\ift} du_2 \cd \int^{\ift}_{-\ift} du_{\L} \nn \\ 
   & & \cdot \int_{Q(1,0) = u_1}^{Q(1,\b) = u_1} {\cal D}Q(1, t_1) 
       \int_{Q(2,0) = u_2}^{Q(2,\b) = u_2} {\cal D}Q(2, t_2) \cd 
       \int_{Q(\L, 0) = u_{\L}}^{Q(\L, \b) = u_{\L}} {\cal D}Q(\L, t_{\L}) \nn \\
   & & \cdot \re^{-\int_0^{\b} dt H(Q(1,t), Q(2,t), \ld, Q(\L,t); \frac{\pa L}{\pa Q(1,t)}, 
   \frac{\pa L}{\pa Q(2,t)} \frac{\pa L}{\pa Q(\L,t)})}.
\la{1.3.2}
\eeq
Comparing Eqs.(\ref{1.2.1}) and (\ref{1.3.2}), we see that ordinary integrals over matrix entries are replaced with 
path integrals over them.

Define the annihilation operators 
\beq
   a^{\mu_1}_{\mu_2}(k) \equiv \frac{1}{\sqrt{2}} \le( Q^{\mu_1}_{\mu_2}(k) + \im P^{\mu_1}_{\mu_2}(k) \ri)
\la{1.3.3}
\eeq
and creation operators
\beq
   a^{\da\mu_1}_{\mu_2}(k) \equiv \frac{1}{\sqrt{2}} \le( Q^{\mu_1}_{\mu_2}(k) - \im P^{\mu_1}_{\mu_2}(k) \ri).
\la{1.3.4}
\eeq
Then these operators satisfy the canonical commutation relation
\beq 
   \le[ a^{\m_1}_{\m_2}(k_1), a^{\da\m_3}_{\m_4}(k_2) \ri] = \hbar \d^{\m_1}_{\m_4} \d^{\m_3}_{\m_2} 
   \d(k_1, k_2).
\la{1.3.5}
\eeq
The Hamiltonian $H$ can be written as the trace of a linear combination of products of these operators.  In
practical applications, the Hamiltonian is usually written in this canonical form.

Like classical matrix models, quantum matrix models arise in a diversity of physical systems.  The models we 
discussed in Section~\ref{s1.0.1} are all quantum matrix models.  We will now briefly how these models fit in this 
formalsim of quantum matrix models, and introduce a number of other systems which can also be expressed as quantum
matrix models.  A more systematic account of how these systems are formulated as quantum matrix models will be 
provided in the next chapter.  

Consider Yang--Mills theory as a quantum theory.  Then Yang--Mills matrix fields, with colors as row and column 
indices, should be treated as quantum fields.  If we choose a certain gauge called the light-cone gauge 
\cite{brpapi} (to be explained in the next chapter), the Yang--Mills Hamiltonian will take the form described in 
the previous paragraph.  The physical states are formulated as linear combinations of the traces of products of 
matrix fields only, or products of matrix and vector fields.  Physically speaking, the vector fields represent 
quarks and anti-quarks, and the matrix fields represent gluons.  The traces are color singlet states.  A remarkable 
feature of this formulation of Yang--Mills theory is that if we treat two matrix fields sharing a common color 
index, which is being summed over, as adjacent segments of a `string', these color singlet states can be envisaged 
as closed strings or open strings.  A glueball, which has gluons only, is a closed string.  A meson, which has a 
quark, an anti-quark and gluons, is an open string with fermionic fields attached to the two ends of it.

This formulation of Yang--Mills theory leads us naturally to the idea of the string-bit model \cite{beth}.  The key
idea is that a string can be approximated as a collection of particle-like entities called string bits.  Each
string bit carries its own momentum, and the behavior of the whole string is manifested as a collective behavior
of these string bits.  The physical states of a quantum matrix model now represent strings, and the Hamiltonian 
acts as a linear operator, replacing a segment of the string by another segment of the same or a different length.

We can extend Yang--Mills theory to incorporate supersymmetry.  The resulting supersymmetric Yang--Mills (SYM) 
theory can be formulated as a quantum matrix model in the same manner.  The only complication is to add fermionic 
annihilation and creation operators to the model.  The fact that SYM theories with different space--time dimensions 
and numbers of supersymmetries can be used to mimic a large number of superstring models further broadens the 
applications of quantum matrix models.  For example, the low-energy dynamics of D$p$-branes in an essentially flat 
space-time can be approximated very well by an ${\cal N} = 1$ SYM theory\footnote{The number ${\cal N}$ counts the 
number of supersymmetries in a physical model \cite{sohnius, weba}.} dimensionally reduced from 10 down to $p + 1$ 
\cite{leigh, witten96}.  (In other words, the fields are independent of $9 - p$ transverse dimensions.)
 
M-theory \cite{polchinski} is conjectured to be the same as the $N \rar \ift$ limit of 0-brane quantum mechanics.  
It cannot be described by a quantum matrix model in the form introduced at the beginning of this section.  However, 
a natural corollary of this conjecture is that light-front type-IIA superstring theory can be described by an 
${\cal N} = 1$ SYM theory dimensionally reduced from 10 to 2 in the large-$N$ limit \cite{diveve}.  This so-called 
matrix string theory can be formulated as a quantum matrix model described above.

Quantum matrix models can be used to study spin systems, too.  This time the spins do not lie on a random lattice
as in classical matrix models.  The systems are one-dimensional quantum spin systems, many of which are well known
to be equivalent to two-dimensional classical ones \cite{gorusi}.  The Hamiltonian of a quantum spin system 
involves nearest neighbour interaction only and is translationally invariant.  Its action on the spin chain is to 
change the quantum states of adjacent spins.  The key observation is that a spin chain can be viewed as a string, 
and each spin can be regarded as a string bit.  Then the action of the Hamiltonian is to replace a segment of the 
string with another segment with an equal number of string bits.  This is typical of the action of the Hamiltonian 
of a string model.  Through the above construction of a string model as a quantum matrix model, we can formulate a 
one-dimensional quantum spin system as a quantum matrix model, too.  As we have noted in Section~\ref{s1.0.1}, many 
condensed matter phenomena can be mimicked by integrable models in the form of one-dimensional quantum spin system. 
This provides us a way to study condensed matter physics through quantum matrix models.

\section{Symmetry Algebra}
\la{s1.1}
  
Recall that we proposed to study matrix models by an algebraic approach in Section~\ref{s1.0.1}.  Let us look at a 
number of examples to see how this works.

We will start with the following pedagogical model.  Consider a $(3 + 1)$-dimensional quantum system whose 
Hamiltonian is nothing but the square of the angular momentum operator $L^2$.  Clearly the Hamiltonian $H$ is 
invariant under a rotation in the 3 spatial dimensions.  This rotational symmetry is described by the Lie group 
$SO(3)$.  The Hamiltonian is the generator of temporal translation, and it is natural for us to put it on the same 
footing as the generators of $SO(3)$.  Indeed, $H$ can be written in terms of the elements of the associated Lie 
algebra, $so(3)$\footnote{Those readers who are familiar with the general theory of Lie algebras may take note that 
the Hamiltonian is an element of the enveloping algebra \cite{humphreys} of $so(3)$.}.  This viewpoint of relating 
physical variables with a Lie algebra can be pushed further.  As we know, $H$ acts as an operator on a quantum 
state to produce another quantum state.  This is akin to treating the elements of a Lie algebra as linear 
transformations of vector spaces.  These vector spaces are called representation spaces.  The Hilbert space of 
quantum states can be regarded as a particular finite- or infinite-dimensional representation space, which is 
always a direct sum of irreducible representations.  $H$ can be seen as a matrix acting on this representation.  
Now, the representation of $so(3)$ is very well developed.  We know all the irreducible representations of it.  An 
irreducible representation is characterized by a positive integer $l$.  Each irreducible representation space is 
$(2l + 1)$-dimensional.  The highest weight vector is an eigenvector of $L_z$, the component of the angular 
momentum in the $z$-direction, with the eigenvector $l$.  The lowest weight vector is also an eigenvector of $L_z$ 
with the eigenvalue $-l$.  There is a whole set of eigenvectors of $L_z$ with integer eigenvalues between $-l$ and 
$l$.  In addition, every vector in this representation space is an eigenvector of $H$ with the eigenvalue 
$l (l + 1)$.  Thus, rotational symmetry alone dictates the spectrum of this physical system completely, and we can 
even determine the degeneracy of each eigenenergy up to internal symmetries not shown up in the Hamiltonian.

A number of more realistic models can be solved along the same line.  Take the hydrogenic atom as an example.  
Again the Hamiltonian respects rotational symmetry and so it commutes with the generators of spatial rotations.  
However, the central potential of the hydrogenic atom is so unique that it possesses an extra symmetry, described 
by the Runge--Lenz vector \cite{goldstein}.  Hence, the Hamiltonian commutes with a properly normalized Runge--Lenz 
vector which is dependent on the eigenvalue of the Hamiltonian.  The angular momentum operator, whose components
form the generators of the rotational symmetry, together with the Runge--Lenz vector form a bigger Lie algebra ---
$so(4)$.  We can form a certain function of the angular momentum and Runge--Lenz vector which is not dependent on 
the eigenvalues of the angular momentum and is dependent on the Runge--Lenz vector through the eigenvalue of the 
Hamiltonian only.  In the language of Lie algebra, this function can be written in terms of the elements of 
$so(4)$\footnote{More precisely speaking, the Hamiltonian is an element of the enveloping algebra of $so(4)$.}.  
Since we know the representation of $so(4)$ very well, too, we know all the irreducible representations of $so(4)$ 
and from them the eigenvalues of this function.  Consequently, we can obtain all the eigenvalues of the 
Hamiltonian, or in other words, the spectrum of the hydrogenic atom\footnote{Miller \cite{miller} has given a 
concise but thorough account of this argument.}.

Other physical systems possess other symmetries described by other Lie algebras.  For instance, the Hamiltonian of 
the simple harmonic oscillator can be written in terms of the symplectic algebra $sp(2)$.  We can use our 
knowledge of the representation theory of the symplectic algebra to deduce the spectrum of the simple harmonic 
oscillator.  Another example is the 2-dimensional Ising model, a well-known exactly integrable model.  Its 
Hamiltonian is an element of a more sophisticated algebra called the Onsager algebra \cite{onsager, davies90}, 
whose definition will be provided a later chapter.  Briefly speaking, the Onsager algebra is isomorphic to the 
fixed-point subalgebra of the $sl(2)$ loop algebra ${\cal L}(sl(2))$ with respect to the action of a certain 
involution \cite{roan}.  The properties of finite-dimensional versions of the Onsager algebra then help us solve 
for the spectrum of the Ising model.  Nowadays, many exactly integrable models are derived by solving for the
irreducible representations of Lie algebras or even quantum groups \cite{chpr}\footnote{G\'{o}mez, Ruiz-Altaba and 
Sierra \cite{gorusi} have provided a systematic account on the relationship between exactly integrable models and 
quantum groups.}.

Another notable example is conformal field theory \cite{bepoza}\footnote{Francesco, Mathieu and S\'{e}n\'{e}chal 
\cite{frmase} have provided a detailed introductory exposition and a comprehensive list of relevant literature on 
the subject.}, a field theory which possesses conformal symmetry which we discussed in Section~\ref{s1.0.1}.  
Besides string theory, critical phenomena manifests conformal symmetry also.  Our knowledge of the Virasoro algebra 
thus finds application in critical phenomena.  Indeed, it helps us determine the correlation functions at the 
critical points of a physical system displaying critical phenomena.

To understand models with both conformal symmetry and a classical-group symmetry like the Wess--Zumino--Witten
model \cite{witten84}, we need an algebra which is an apt amalgamation of the Virasoro algebra and a classical Lie 
algebra.  Such an amalgamation yields the Kac--Moody algebra, or the affine Lie algebra \cite{kac, hakiobcl}, 
another algebra which plays a vital role in physics.

All the above examples illustrate the following truth.  A fruitful approach to study a physical system is to 
identify the underlying symmetry of the system, express the symmetry in terms of a Lie algebra (or even a quantum
group), elucidate the structure of this Lie algebra, and use the irreducible representations of it to understand 
key properties of the system.

\section{An Algebraic Approach to Quantum Matrix Models in the Large-$N$ Limit}
\la{s1.4}

Superstring theory and its latest incarnation, M-theory, is a promising candidate of a theory of everything; strong
interaction is described by Yang--Mills theory; it is conjectured that the essential properties of high-$T_c$ 
superconductivity are captured by the Hubbard model.  The previous section shows that all these theories can be 
written as (supersymmetric) quantum matrix models.  Therefore it is of interest to develop methods to tackle them, 
something we have done for their classical counterparts.

Is there an underlying symmetry for quantum matrix models?  Can we use an algebra to characterize the symmetry?  
What are the mathematical properties and physical implications of this symmetry algebra?  These are the questions 
we are going to address in this whole article.

The first thing we need to do is to set up a formalism, with supersymmetry from the very outset, for quantum matrix 
models.  We will show that in the large-$N$ limit, planarity dramatically simplifies the formalism \cite{thooft74a, 
thorn79}.  In the language of string bit models, what will happen is that a physical observable will send a single 
closed string state to a linear combination of closed string states, and an open string state to a linear 
combination of open string states.  A string will not split into several strings, and several strings will not 
combine to form a string.  Equipped with this crucial observation, we will then be able to give a systematic 
account on how quantum matrix models arise in different physical system, as mentioned in the previous section.  We 
will do all these in Chapter~\ref{c2}.

In subsequent chapters, we will develop an argument for the existence of a Lie superalgebra\footnote{Those readers
who are not familiar with the notion of a Lie superalgebra are referred to Ref.\cite{buku} for its definition.}, 
which we will call the `grand string superalgebra', for quantum matrix models with both open and closed strings in 
the large-$N$ limit.  Because of the complexity of the full formalism, we will study an easier special case in 
which there is only one degree of freedom for the fundamental and conjugate matter fields, and there is no 
fermionic adjoint matter first in Chapter~\ref{c4}.  There is a Lie algebra of physical observables in this special 
case, and we will call it the `centrix algebra'.  It has a number of subalgebras.  Interestingly, they are all 
related to the Cuntz algebra \cite{cuntz}.  (Roughly speaking, the Cuntz algebra is an algebra of isometries acting 
on an infinite-dimensional Hilbert space.  Its versatility lies in the fact that as long as these isometric 
operators satisfy certain algebraic relations, it does not matter which Hilbert space and isometries we choose.  We 
will give a preliminary but more systematic account of the formalism in Chapter~\ref{c4}.  Some knowledge in 
${\rm C}^*$-algebras \cite{dixmier} and K-theory \cite{blackadar} is very helpful in understanding the Cuntz 
algebra.)  This implies that a well-developed representation theory of the Cuntz algebra should be very useful in 
elucidating the spectra of various quantum matrix models for open strings.  We remark in passing that the Cuntz 
algebra emerges in a number of other physical contexts, too \cite{szve, gogr, arvo}.

Then we will use the techniques we will have learnt in Chapter~\ref{c4} to study the most general case in 
subsequent chapters.  There indeed exists a superalgebra which we will call the `grand string superalgebra', for 
quantum matrix models with both open and closed strings in the large-$N$ limit.  This will be done by showing that 
this superalgebra is a subalgebra of a precursor superalgebra which we will call the `heterix superalgebra', and 
which we will derive in Chapter~\ref{c3}.  We will not give a rigorous proof of the existence of precursor 
superalgebra, as the proof will be too long to be written down.  However, we will point out some important 
graphical features of quantum matrix models to convince the reader of the result.  In the rest of the chapter, we 
will study the mathematical structure, e.g., the Cartan subalgebra and its root vectors, of the corresponding 
heterix algebra, which can be otained from the heterix superalgebra by removing all fermionic fields.

In Chapter~\ref{c5}, we will obtain the grand string superalgebra as a subalgebra of the heterix algebra.  In 
practical applications, open and closed strings essentially form distinct physical sectors.  This means there will 
be one superalgebra for each sector.  We will call them the open string superalgebra and `cyclix superalgebra'.  
These two superalgebras will both be quotient algebras of the grand string superalgebra.  The open string algebra 
associated with the corresponding superalgebra will be a direct product of the algebra for infinite-dimensional
matrices and the centrix algebra.  On the other hand, the cyclix algebra associated with the corresponding cyclix
superalgebra is closely related to the Onsager algebra.  This provides a glimpse of how the cyclic algebra helps
us determine the integrability of certain spin systems.

A major field of application of quantum matrix models in the large-$N$ limit is Yang--Mills theory.  As mentioned in
Section~\ref{s1.0.1}, this theory can be formulated as a classical mechanical system.  There is a Poisson algebra 
of dynamical variables in this `classical' Yang--Mills theory \cite{ratu, loop}.  We conjecture that the Lie 
algebras described in the previous paragraphs are approximations of this Poisson algebra in some geometric sense.

\chapter{Formulation and Examples of Quantum Matrix Models in the Large-$N$ Limit}
\la{c2}

\section{Introduction}
\la{s2.1}

In this chapter, we will demonstrate systematically how a number of physical systems are formulated as quantum
matrix models in the large-$N$ limit.  These systems fall into three broad classes: Yang--Mills theory with or 
without supersymmetry, string theory and one-dimensional quantum spin systems.  They were concisely introduced in 
Section~\ref{s1.3}.  

It should be obvious why Yang--Mills theory can be expressed as a quantum matrix model.  Identify the trivial 
vacuum state $|0\rangle$ with respect to the annihilation operators characterized by Eq.(\ref{1.3.5}), i.e., the 
action of any of these annihilation operators on this vacuum state yields 0.  A typical {\em physical state} is a 
linear combination of the traces of products of creation operators acting on this vacuum state.  (In the context of
Yang--Mills theory, this is a color-singlet state of a single meson or glueball.)  A typical {\em observable} is a 
linear combination of the trace of a product of creation and annihilation operators.  Naively, we may think that 
the ordering of these operators is arbitrary, and which ordering is used in an actual physical application depends 
on the details of that physical model only.  Actually, the planar property of the large-$N$ limit \cite{thooft74a} 
comes into the picture here, and only those traces in which all creation operators come to the left of all 
annihilation operators survive the large-$N$ limit.  Moreover, the large-$N$ limit brings about a dramatic 
simplification in the action of this collective operator on a physical state --- the unique trace of a product of 
creation operators acting on the vacuum will not be broken up into a product of more than one trace of products of 
creation operators acting on the vacuum, and it is not possible for a product of more than one trace of products of 
creation operators to merge into just one trace of products of creation operators.  In other words, {\em an 
observable propagates a color-singlet state into a linear combination of color-singlet states} \cite{thorn79}.

This explains why we can mimic a string as a collection of string bits.  In this context, a physical state is a
string, open or closed, and the concluding statement of the previous paragraph implies that an observable replaces
a segment of the string with another segment, without breaking it and without joining several strings together.  
This is a desirable simplifying feature for some string models.  An important feature of this string-bit model is
that wherever this segment lies on the string, it will be replaced with the new segment at exactly the same 
location.  In other words, the action of an observable involves neighbouring string bits only, and is 
translationally invariant with respect to the string.

This brings us to quantum spin chain models.  Since the two most important properties of a quantum spin chain model 
is that a spin interacts with neighbouring spins (not necessarily nearest neighbouring spins though) only, and
the interaction is translationally invariant, we can write down a typical quantum spin chain model as a quantum
matrix model in the large-$N$ limit, so long as the boundary conditions match.

We are going to present these ideas systematically in this chapter.

In Section~\ref{s2.2}, we will present an abstract formalism of quantum matrix models which is a generalization of
the one discussed in Section~\ref{s1.3}.  There the models possessed only bosonic degrees of freedom.  Here we will
include fermionic degrees of freedom as well.  This is needed in SYM theory, and in fermionic spin chains like
strongly correlated electron systems.

Sections~\ref{s2.3} to \ref{s2.5} will be devoted to two-dimensional Yang--Mills theories.  We will show how to
formulate these theories with different matter contents as quantum matrix models.  To be more specific, we will
consider Yang--Mills theory with only bosonic adjoint matter fields in Section~\ref{s2.3}, with bosonic adjoint
matter together with fundamental and conjugate matter fields in Section~\ref{s2.4}, and with fermionic and bosonic
adjoint matter fields in Section~\ref{s2.5}.  These models are suitable for studying glueballs, mesons and SYM 
theory, respectively.

In Section~\ref{s2.6}, we will focus on the relationship between string theories and quantum matrix models.  We will
see that there are many ways to approximate a string theory as a quantum matrix model like treating the string as a
collection of discrete entities called string bits \cite{beth}, taking the low-energy limit of superstrings, 
D-branes or even M-theory, or associating a string model with a quantum gravity model and then approximating the
quantum gravity model by a quantum matrix model.

In Sections~\ref{s2.7} and \ref{s2.8}, we will present a formal way of transcribing a quantum spin chain system 
into a quantum matrix model in the large-$N$ limit, and give examples from bosonic spin systems to strongly 
correlated electron systems.  Since many of them are known to be exactly integrable, and have even been exactly 
solved, we thus give here some examples of quantum matrix models, with or without supersymmetry, which are exactly 
integrable or even exactly solved.  The models in Section~\ref{s2.7} obey cyclicity and the periodic boundary 
condition.  The models in Section~\ref{s2.8} obey open boundary conditions.

\section{Formulation of Quantum Matrix Models in the Large-$N$ Limit}
\la{s2.2}

We are going to generalize the bosonic formalism in Section~\ref{s1.3} to include fermions.  We advise the reader
to take a look at Appendices~\ref{sa1.1} and \ref{sa1.2} first to learn the notations we will use throughout this 
article.

Think of the row and column indices of those annihilation and creation operators as the row and column indices of 
an element in $U(N)$.  We will call them {\em color indices}, as this is the case in Yang--Mills theory 
(Section~\ref{s2.3}).  Let $a^{\mu_1}_{\mu_2}(k)$ be an annihilation operator of a boson in the adjoint 
representation if $1 \leq k \leq \L$, or a fermion in the adjoint representation if $\L + 1 \leq k \leq 2 \L$.  
(Here $\L$ is a positive integer.)  Let $\c^{\mu}(\l)$ be an annihilation operator of a boson in the fundamental 
representation if $1 \leq \l \leq \L_F$, or a fermion in the fundamental representation if $\L_F + 1 \leq \l \leq 2 
\L_F$.  (Here $\L_F$ is also a positive integer.)  Lastly, let $\bar{\c}_{\mu}(\l)$ be an annihilation operator of 
a boson in the conjugate representation if $1 \leq \l \leq \L_F$, or a fermion in the conjugate representation if 
$\L_F + 1 \leq \l \leq 2 \L_F$.  We will call $k$ and $\l$ {\em quantum states other than color}.  The 
corresponding creation operators are $a^{\da\mu_1}_{\mu_2}(k)$, $\c^{\da}_{\mu}(\l)$ and $\bar{\c}^{\da\mu}(\l)$ 
with appropriate values for $k$ and $\l$.  We will say that these operators create an {\em adjoint parton}, a 
{\em fundamental parton} and a {\em conjugate parton}, respectively.  Most of these operators commute with each 
another except the following non-trivial cases:
\beq
   \le[ a^{\mu_1}_{\mu_2}(k_1), a^{\da\mu_3}_{\mu_4}(k_2) \ri] = 
   \d_{k_1 k_2} \d^{\mu_3}_{\mu_2} \d^{\mu_1}_{\mu_4}
\la{2.2.1}
\eeq
for $1 \leq k_1, k_2 \leq \L$;
\beq
   \le[ a^{\mu_1}_{\mu_2}(k_1), a^{\da\mu_3}_{\mu_4}(k_2) \ri]_+ =
   \d_{k_1 k_2} \d^{\mu_3}_{\mu_2} \d^{\mu_1}_{\mu_4} 
\la{2.2.2}
\eeq
for $\L + 1 \leq k_1, k_2 \leq 2 \L$;
\beq
   \le[ \bar{\c}_{\mu_1}(\l_1), \bar{\c}^{\da\mu_2}(\l_2) \ri] = \d_{\l_1 \l_2} \d^{\mu_2}_{\mu_1} 
\la{2.2.3}
\eeq
for $1 \leq \l_1, \l_2 \leq \L_F$; 
\beq
   \le[ \bar{\c}_{\mu_1}(\l_1), \bar{\c}^{\da\mu_2}(\l_2) \ri]_+ = \d_{\l_1 \l_2} \d^{\mu_2}_{\mu_1} 
\la{2.2.3.1}
\eeq
for $\L_F + 1 \leq \l_1, \l_2 \leq 2 \L_F$;
\beq
   \le[ \c^{\mu_1}(\l_1), \c^{\da}_{\mu_2}(\l_2) \ri] = \d_{\l_1 \l_2} \d^{\mu_1}_{\mu_2}
\la{2.2.3.2}
\eeq
for $1 \leq \l_1, \l_2 \leq \L_F$; and
\beq
   \le[ \c^{\mu_1}(\l_1), \c^{\da}_{\mu_2}(\l_2) \ri]_+ = \d_{\l_1 \l_2} \d^{\mu_1}_{\mu_2}
\la{2.2.4}
\eeq
for $\L_F + 1 \leq \l_1, \l_2 \leq 2 \L_F$.

There are two families of physical states (or color-singlet states in the context of gauge theory).  One family
consists of linear combinations of states of the form
\beq
   \bar{\ph}^{\l_1} \otimes s^{\dot{K}} \otimes \ph^{\l_2} & \equiv & N^{-(\dot{c}+1)/2} \bar{\c}^{\da\u_1}(\l_1) 
   a^{\da\u_2}_{\u_1}(k_1) a^{\da\u_3}_{\u_2}(k_2) \cd \nn \\
   & & \cdot a^{\da\u_{\dot{c}+1}}_{\u_{\dot{c}}}(k_{\dot{c}}) \c^{\da}_{\u_{\dot{c}+1}}(\l_2) |0 \rangle.
\la{2.2.5}
\eeq
Here we use the capital letter $K$ to denote the integer sequence $k_1$, $k_2$, \ld, $k_c$.  {\em Unless otherwise
specified, the summation convention applies to all repeated color indices throughout this whole article.}  The 
justification of the use of the notation $\otimes$ will be given later.  This term carries a factor of $N$ to make
its norm finite in the large-$N$ limit.  (The proof of this is similar to that given in Appendix~\ref{sa1.3} in 
which we will prove a closely related statement.)  We will call these {\em open singlet states}.  They are single
meson states in QCD, and open superstring states in the string-bit model.  The other family consists of linear 
combinations of states of the form
\beq
   \Ps^K \equiv N^{-c/2} a^{\da\u_2}_{\u_1}(k_1) a^{\da\u_3}_{\u_2}(k_2) \cd
   a^{\da\u_1}_{\u_c}(k_c) |0 \rangle.
\la{2.2.6}
\eeq
This will be called a {\em closed singlet state}.  They are single glueball states in QCD, and closed superstring
states in the string-bit model.

Fig.\ref{f2.1}(a) shows a typical open singlet state in detail.  The solid square at the top is the creation 
operator of a conjugate parton of the quantum state $\l_1$.  Following it is a series of 4 adjoint partons of the 
quantum states $k_1$, $k_2$, and $k_3$, respectively.  The creation operators of these gluons are represented by 
solid circles.  The solid square at the bottom is the creation operator of a fundamental parton of the quantum 
state $\l_2$.  Note that all the creation operators carry color indices.  A solid line, no matter how thick it is, 
connecting two circles, or a circle and a square, is used to mean that the two corresponding creation operators 
share a color index, and this color index is being summed over.  The arrow indicates the direction of the integer 
sequence $\dot{K}$.  Fig.\ref{f2.1}(b) is a simplified diagram of an open singlet state.  Conjugate and fundamental 
partons are neglected.  They will be consistently ignored in all brief diagrammatic representations.  The series of 
adjoint partons in between are represented by the integer sequence $\dot{K}$.  Fig.\ref{f2.1}(c) shows a typical 
closed singlet state $\Ps^K$ with 5 indices in detail.  There is a series of adjoint partons of the quantum states 
$k_1$, $k_2$, \ldots, and $k_5$.  This state is cyclically symmetric up to a sign.  Fig.\ref{f2.1}(d) is a 
simplified diagram of a closed singlet state.  The state is represented by the integer sequence $K$.  We are 
ignoring the sign carried by this state, and thus it is legitimate to use a cyclically symmetric diagram to 
represent it.  The size of this big circle does not indicate the number of indices the circle carries.  

\begin{figure}
\epsfxsize=5in
\centerline{\epsfbox{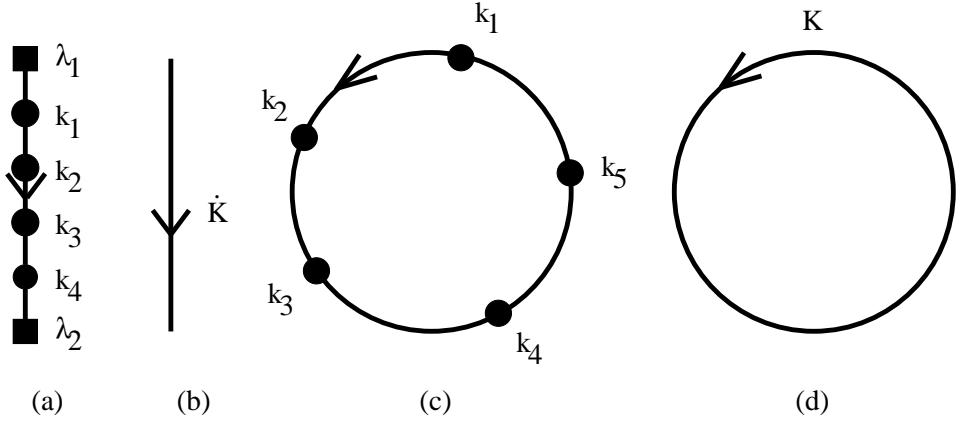}}
\caption{\em Color-invariant states.}     
\label{f2.1}
\end{figure}

To move on, we need the notion of a grade.  The {\em grade} of the integer sequence $\dot{I}$, which we will denote 
by $\ep(\dot{I})$, is defined to be 0 if the number of integers within $\dot{I}$ which are larger than $\L$ but 
smaller than $2\L + 1$ is even, and 1 if the number is odd.  Moreover, we define the {\em grades} of the open 
singlet state $s^K$ and the closed singlet state $\Ps^L$ to be the grades of $K$ and $L$ respectively.  
We call a state a {\em pure state} if it is a linear combination of states of the same grade.  Clearly, the grade 
of a pure singlet state is 0 if the number of adjoint fermions in each term of the linear combination forming 
the state is even, and 1 if the number is odd.  In these cases, we say that the singlet states are {\em even} 
and {\em odd}, respectively.

We now remark that
\beq
   \Ps^K = (-1)^{\ep(K_1) \ep(K_2)} \Ps^{K_2 K_1}.
\la{2.2.7}
\eeq
Thus a closed singlet state is cyclic up a sign.  The special case in which there is no fermionic
creation operator in $\Ps^K$ will be of considerable interest.  Then Eq.(\ref{2.2.7}) reduces to
\beq
   \Ps^{K_1 K_2} = \Ps^{K_2 K_1}.
\la{2.2.8}
\eeq
This state is thus manifestly cyclic.  To emphasize this cyclicity, we may denote such $\Ps^K$'s as $\Ps^{(K)}$'s. 

Now let us construct physical operators acting on these singlet states.  It turns out that there are five 
families of them.  The first family consists of {\em finite} linear combinations of operators of the form
\beq
   \bar{\X}^{\l_1}_{\l_2} \otimes f^{\dot{I}}_{\dot{J}} \otimes \X^{\l_3}_{\l_4} & \equiv &
   N^{-(\dot{a} + \dot{b} + 2)/2} \bar{\c}^{\da\mu_1}(\l_1) a^{\da\mu_2}_{\mu_1}(i_1) \cd 
   a^{\da\mu_{\dot{a}+1}}_{\mu_{\dot{a}}}(i_{\dot{a}}) \c^{\da}_{\mu_{\dot{a}+1}}(\l_3) \nn \\
   & & \cdot \bar{\c}_{\n_1}(\l_2) a^{\n_1}_{\n_2}(j_1) \cd a^{\n_{\dot{b}}}_{\n_{\dot{b}+1}}(j_{\dot{b}}) 
   \c^{\n_{\dot{b}+1}}(\l_4),
\la{2.2.9}
\eeq
where $\dot{a} = \#(\dot{I})$ and $\dot{b} = \#(\dot{J})$.  We say that this is an {\em operator of the first 
kind}.  Fig.\ref{f2.2}(a) shows a typical operator of the first kind.  The solid squares and circles are creation 
operators of conjugate, fundamental and adjoint partons.  The hollow squares are annihilation operators of a 
conjugate parton of the quantum state $\lambda_2$ and a fundamental parton of the quantum state $\lambda_4$, and 
the hollow circles are annihilation operators of adjoint partons.  In this particular example, there are 2 creation 
and 4 annihilation operators of adjoint partons.  The creation operators are joined by thick lines, whereas the 
annihilation operators are joined by thin lines.  Note that the sequence $\dot{J}$ is in reverse.  Fig.\ref{f2.2}(b)
is a simplified diagram of an operator of the first kind.  The thick line represents a sequence of creation 
operators of adjoint partons, whereas the thin line represents a sequence of annihilation operators of them.  
$\dot{J}$ carries an asterisk to signify the fact that $\dot{J}$ is put in reverse.  Note that the lengths of the 
two lines have no bearing on the numbers of creation or annihilation operators they represent.

\begin{figure}
\epsfxsize=5.5in
\centerline{\epsfbox{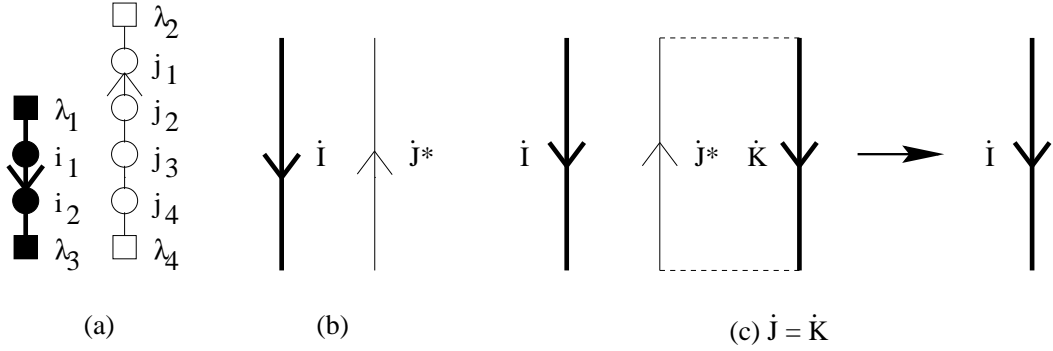}}
\caption{\em Operators of the first kind.}
\la{f2.2}
\end{figure}

In the planar large-$N$ limit, a typical term of an operator of the first kind propagates an open singlet state to 
another open singlet state:
\beq
   \bar{\X}^{\l_1}_{\l_2} \otimes f^{\dot{I}}_{\dot{J}} \otimes \X^{\l_3}_{\l_4} 
   \le( \bar{\ph}^{\l_5} \otimes s^{\dot{K}} \otimes \ph^{\l_6} \ri) = 
   \d^{\l_5}_{\l_2} \d^{\dot{K}}_{\dot{J}} \d^{\l_6}_{\l_4} 
   \bar{\ph}^{\l_1} \otimes s^{\dot{I}} \otimes \ph^{\l_3}.
\la{2.2.10}
\eeq
We can visualize Eq.(\ref{2.2.10}) in Fig.~\ref{f2.2}(c).  In this figure, the dotted lines connect the line 
segments to be `annihilated' together.  The figure on the right of the arrow is the resultant open singlet state.  
On the other hand, an operator of the first kind annihilates a closed singlet state:
\beq
   \bar{\X}^{\l_1}_{\l_2} \otimes f^{\dot{I}}_{\dot{J}} \otimes \X^{\l_3}_{\l_4} \le( \Ps^K \ri) = 0.
\la{2.2.11}
\eeq
The fact that this operator does not split a single singlet state into a product of states, and it does not combine 
a product of singlet states together to form a single singlet state has been known for a long time \cite{thorn79}, 
and is ultimately related to the planarity of the large-$N$ limit \cite{thooft74a}.  We will provide a non-rigorous 
diagrammatic proof of this fact in Appendix~\ref{sa1.3}, where we work on the action of an operator of the second 
kind (to be defined below) on an open singlet state.  The reader can easily work out the actions of operators of 
other kinds by the same reasoning.

It is now clear why we are using the direct product symbol $\otimes$ ---  the open singlet state can be regarded as 
a direct product of $\bar{\ph}^{\l_5}$, $s^{\dot{K}}$ and $\ph^{\l_6}$.  The set of all $\bar{\ph}^{\l_5}$'s, where 
$\l_1 = 1, 2$, \ldots, and $2\L_F$, form a basis of a $2\L_F$-dimensional vector space.  The set of all 
$\ph^{\l_6}$'s, where again $\l_2 = 1, 2$, \ldots, and $2\L_F$, form a basis of another $2\L_F$-dimensional vector 
space.  The operator of the first kind can be regarded as a direct product of the operators 
$\bar{\X}^{\l_1}_{\l_2}$, $f^I_J$ and $\X^{\l_3}_{\l_4}$.  The first operator acts as a $2\L_F \times 2\L_F$ matrix 
on $\bar{\ph}^{\l_5}$, the second one acts on $s^{\dot{K}}$, whereas the last one acts as another $2\L_F \times 
2\L_F$ matrix on $\ph^{\l_6}$.  It is therefore clear that that an open singlet state lies within a direct product 
of two $2\L_F$-dimensional vector spaces (labelling the fundamental and conjugate parton states)  and a countably 
infinite-dimensional vector space spanned by all $s^{\dot{K}}$'s labelling the adjoint parton states (including the 
state containing no adjoint partons).  An operator of the first kind lies within the direct product $gl(2\L_F) 
\otimes \supersalt \otimes gl(2\L_F)$. Here, $gl(2\L_F)$ is the Lie algebra of the general linear group 
$GL(2\L_F)$.  Also the infinite-dimensional Lie superalgebra $\supersalt$ is spanned by $f^{\dot{I}}_{\dot{J}}$. We 
will see that the Lie algebra $\salt$, which is the even part of $\supersalt$, is isomorphic to the inductive limit 
$gl_{+\infty}$ of the $gl(n)$'s as $n\to \infty$.

\begin{figure}
\epsfxsize=3.8in
\epsfysize=3.8in
\centerline{\epsfbox{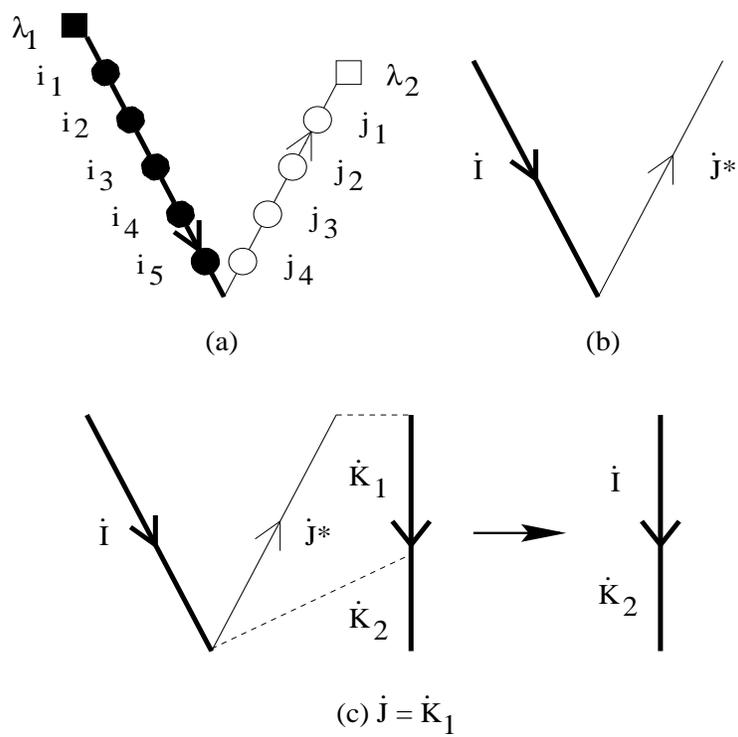}}
\caption{\em  Operators of the second kind.}
\label{f2.3}
\end{figure}

{\em Operators of the second kind} are {\em finite} linear combinations of operators of the form
\beq
   \bar{\X}^{\l_1}_{\l_2} \otimes l^{\dot{I}}_{\dot{J}} & \equiv & N^{-(\dot{a} + \dot{b})/2}
   \bar{\c}^{\da\mu_1}(\l_1) a^{\da\mu_2}_{\mu_1}(i_1) a^{\da\mu_3}_{\mu_2}(i_2) \cd
   a^{\da\mu_{\dot{a} + 1}}_{\mu_{\dot{a}}}(i_{\dot{a}}) \nn \\
   & & \cdot a^{\n_{\dot{b}}}_{\mu_{\dot{a} + 1}}(j_{\dot{b}}) a^{\n_{\dot{b} - 1}}_{\n_{\dot{b}}}(j_{\dot{b} - 1}) 
   \cd a^{\n_1}_{\n_2}(j_1) \bar{\c}_{\n_1}(\l_2).
\la{2.2.12}
\eeq
A typical operator of this kind is depicted in Fig.\ref{f2.3}(a).  There are 5 creation and 4 annihilation 
operators of adjoint partons in this example.  The solid square is a creation operator of a conjugate parton, 
whereas the hollow square is an annihilation operator of it.  Fig.\ref{f2.3}(b) is an abbreviated diagram of
Fig.\ref{f2.3}(a).  An operator of the second kind acts on the end with a conjugate parton and propagates an open 
singlet state to a linear combination of open singlet states:
\beq
   \bar{\X}^{\l_1}_{\l_2} \otimes l^{\dot{I}}_{\dot{J}} 
   \le( \bar{\ph}^{\l_3} \otimes s^{\dot{K}} \otimes \ph^{\l_4} \ri) = 
   \d^{\l_3}_{\l_2} \sum_{\dot{K_1} \dot{K_2} = \dot{K}} \d^{\dot{K}_1}_{\dot{J}} 
   \bar{\ph}^{\l_1} \otimes s^{\dot{I} \dot{K}_2} \otimes \ph^{\l_4}.
\la{2.2.13}
\eeq
In words, the action of $l^{\dot{I}}_{\dot{J}}$ is such that it checks if the beginning segment of $\dot{K}$ is 
identical to $\dot{J}$.  If this is the case, then the beginning segment is replaced with $\dot{I}$ whereas the 
rest remains unchanged; otherwise, the action yields 0.  On the R.H.S. of this equation, there will only be a 
finite number of non-zero terms in the sum (bounded by the number of ways of splitting $\dot{K}$ into subsequence), 
so there is no problem of convergence.  For example, if $\dot{K}$ is shorter than $\dot{J}$, the right hand side 
will vanish.  We can visualize this equation in Fig.~\ref{f2.3}(c).  This equation shows why we can treat this 
operator as a direct product of the operators $\bar{\Xi}^{\lambda_1}_{\lambda_2}$, $l^{\dot{I}}_{\dot{J}}$ and the 
identity operator.  The first operator acts as a $2\Lambda_F \times 2\Lambda_F$ matrix on $\bar{\phi}^{\l_3}$, the 
second one acts on $s^{\dot{K}}$, whereas the last one acts trivially on $\phi^{\l_4}$.  This operator annihilates 
a closed singlet state:
\beq
   \bar{\X}^{\l_1}_{\l_2} \otimes l^{\dot{I}}_{\dot{J}} \le( \Ps^K \ri) = 0.
\la{2.2.14}
\eeq

\begin{figure}
\epsfxsize=3.8in
\epsfysize=3.8in
\centerline{\epsfbox{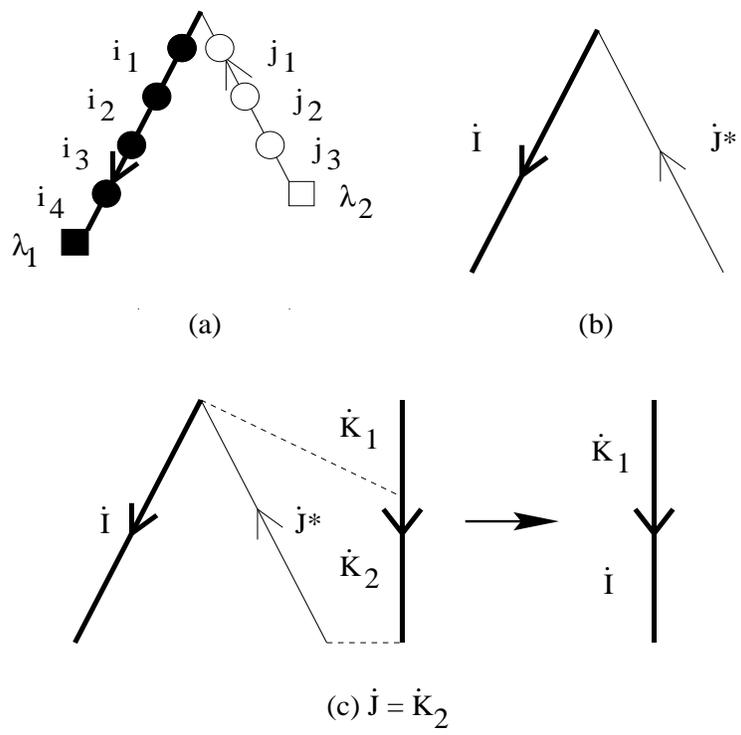}}
\caption{\em  Operators of the third kind.}
\label{f2.4}
\end{figure}

{\em Operators of the third kind} are very similar to those of the second kind.  They are linear combinations of
operators of the form
\beq
   r^{\dot{I}}_{\dot{J}} \otimes \X^{\l_1}_{\l_2} & \equiv & N^{-(\dot{a} + \dot{b})/2}
   a^{\da\mu_2}_{\mu_1}(i_1) a^{\da\mu_3}_{\mu_2}(i_2) \cd a^{\da\mu_{\dot{a} + 1}}_{\mu_{\dot{a}}}(i_{\dot{a}})
   \c^{\da}_{\mu_{\dot{a} + 1}}(\l_1) \nn \\
   & & \cdot \c^{\n_{\dot{b}}}(\l_2) a^{\n_{\dot{b} - 1}}_{\n_{\dot{b}}}(j_{\dot{b}})
   a^{\n_{\dot{b} - 2}}_{\n_{\dot{b} - 1}}(j_{\dot{b} - 1}) \cd a^{\mu_1}_{\n_1}(j_1).
\la{2.2.15}
\eeq
A typical operator of this kind is depicted in Fig.\ref{f2.4}(a).  There are 4 creation and 3 annihilation 
operators of adjoint partons in this example.  This time the solid square is a creation operator of a fundamental 
parton, whereas the hollow square is an annihilation operator of it.  Compare the orientations of the arrows with 
those in Fig.~\ref{f2.3}(a).  The choices of the orientations reflect the fact that the color indices are 
contracted differently in Eqs.(\ref{2.2.12}) and (\ref{2.2.15}).  Fig.\ref{f2.4}(b) is a simplified diagram of an
operator of the second kind.  They act on the end with a fundamental parton instead of a conjugate parton as shown 
below:
\beq
   r^{\dot{I}}_{\dot{J}} \otimes \X^{\l_1}_{\l_2} \le( \bar{\ph}^{\l_3} \otimes s^{\dot{K}} \otimes \ph^{\l_4} \ri) 
   & = & \d^{\l_4}_{\l_2} \sum_{\dot{K}_1 \dot{K}_2 = \dot{K}} 
   (-1)^{\ep(\dot{K}_1) \lb \ep(\dot{I}) + \ep(\dot{J}) \rb} \d^{\dot{K}_2}_{\dot{J}} \nn \\
   & & \bar{\ph}^{\l_3} \otimes s^{\dot{K}_1 \dot{I}} \otimes \ph^{\l_1}.
\la{2.2.16}
\eeq
Fig.~\ref{f2.4}(c) shows this action diagrammatically. This equation shows that a term of an operator of the third
kind is a direct product of the identity operator, the operator $r^{\dot{I}}_{\dot{J}}$ and the operator 
$\Xi^{\lambda_1}_{\lambda_2}$.  Like the previous two kinds of operators, an operator of the third kind also 
annihilates a closed singlet state:
\beq
   r^{\dot{I}}_{\dot{J}} \otimes \X^{\l_3}_{\l_4} \le( \Ps^K \ri) = 0.
\la{2.2.17}
\eeq

\begin{figure}
\epsfxsize=5.5in
\centerline{\epsfbox{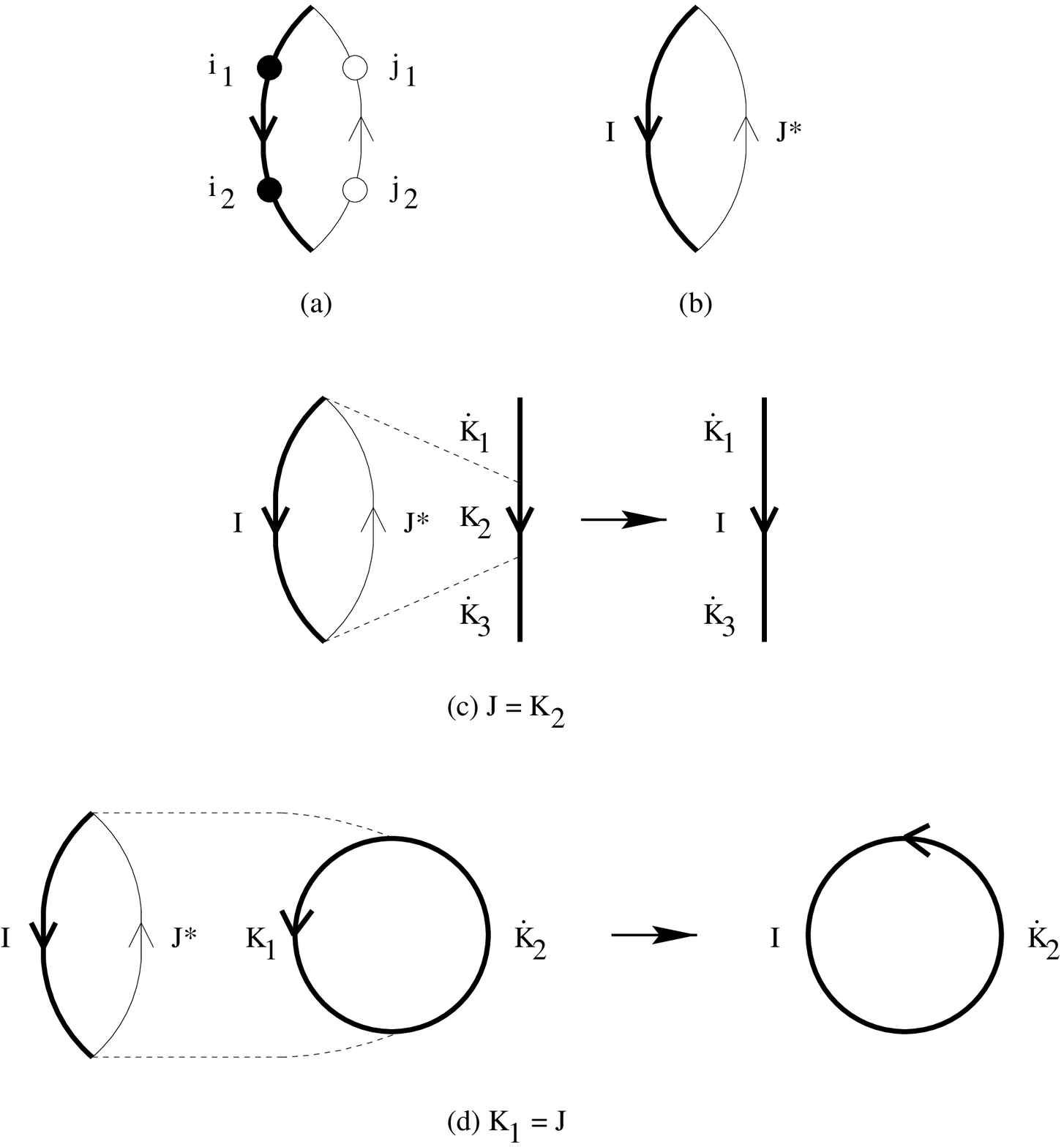}}
\caption{\em Operators of the fourth kind.}
\label{f2.5}
\end{figure}

{\em Operators of the fourth kind} are the most non-trivial among all physical operators.  They are finite linear 
combinations of operators of the form
\beq
   \g^I_J & \equiv & N^{-(a + b - 2)/2} a^{\da\mu_2}_{\mu_1}(i_1) a^{\da\mu_3}_{\mu_2}(i_2) \cd 
   a^{\da\n_b}_{\mu_a}(i_a) \nn \\
   & & \cdot a^{\n_{b-1}}_{\n_b}(j_b) a^{\n_{b-2}}_{\n_{b-1}}(j_{b-1}) \cd a^{\mu_1}_{\n_1}(j_1).
\la{2.2.18}
\eeq
{\em Sometimes we write $\g^I_J$ as $\s^I_J$}.  Unlike the operators of the first three kinds, {\em both $I$ and 
$J$ in the operators of the fourth kind must be non-empty sequences.}  Fig.\ref{f2.5}(a) shows a typical operator 
of the fourth kind in detail.  There are 2 creation and 2 annihilation operators of adjoint partons in this 
example.  There are no operators acting on a conjugate or fundamental parton.  Fig.\ref{f2.5}(b) is an abbreviated
version of Fig.\ref{f2.5}(a).  In the planar large-$N$ limit, it replaces some sequences of adjoint partons on an 
open singlet state with some other sequences, producing a linear combination of this kind of states:
\beq
   \lefteqn{ \g^I_J \le( \bar{\ph}^{\l_1} \otimes s^{\dot{K}} \otimes \ph^{\l_2} \ri) = } \nn \\
   & & \bar{\ph}^{\l_1} \otimes \le( \sum_{\dot{K}_1 K_2 \dot{K}_3 = \dot{K}} 
   (-1)^{\dot{K}_1 \lb \ep(I) + \ep(J) \rb} \d^{K_2}_J s^{\dot{K}_1 I \dot{K}_3} \ri) \otimes \ph^{\l_2}.
\la{2.2.19}
\eeq
In words, the action of $\s^I_J$ is such that if it detects a segment of $\dot{K}$, no matter where this segment 
lies within $\dot{K}$, to be identical to $J$, then this particular segment is replaced with $I$.  If there are $n$ 
segments within $\dot{K}$ identical to $J$, then the action returns a sum of $n$ sequences in each of which one of 
the segments is replaced by $I$.  This action is depicted in Fig.\ref{f2.5}(c).  When it acts on a closed 
singlet state, it also replaces some sequences of adjoint partons with others:
\beq
   \g^I_J \Ps^K & = & \d^K_J \Ps^I + \sum_{K_1 K_2 = K} (-1)^{\ep(K_1) \ep(K_2)} \d^{K_2 K_1}_J \Ps^I
   + \sum_{K_1 K_2 = K} \d^{K_1}_J \Ps^{I K_2} \nn \\
   & & + \sum_{K_1 K_2 K_3 = K} (-1)^{\ep(K_1) \lb \ep(K_2) + \ep(K_3) \rb} \d^{K_2}_J \Ps^{I K_3 K_1} \nn \\
   & & + \sum_{K_1 K_2 = K} (-1)^{\ep(K_1) \ep(K_2)} \d^{K_2}_J \Ps^{I K_1} \nn \\
   & & + \sum_{J_1 J_2 = J} \sum_{K_1 K_2 K_3 = K} (-1)^{\ep(K_3) \lb \ep(K_1) + \ep(K_2) \rb} \d^{K_3}_{J_1}
   \d^{K_1}_{J_2} \Ps^{I K_2};
\la{2.2.20}
\eeq
This action is depicted in Fig.~\ref{f2.5}(d).

\begin{figure}
\epsfxsize=4.4in
\centerline{\epsfbox{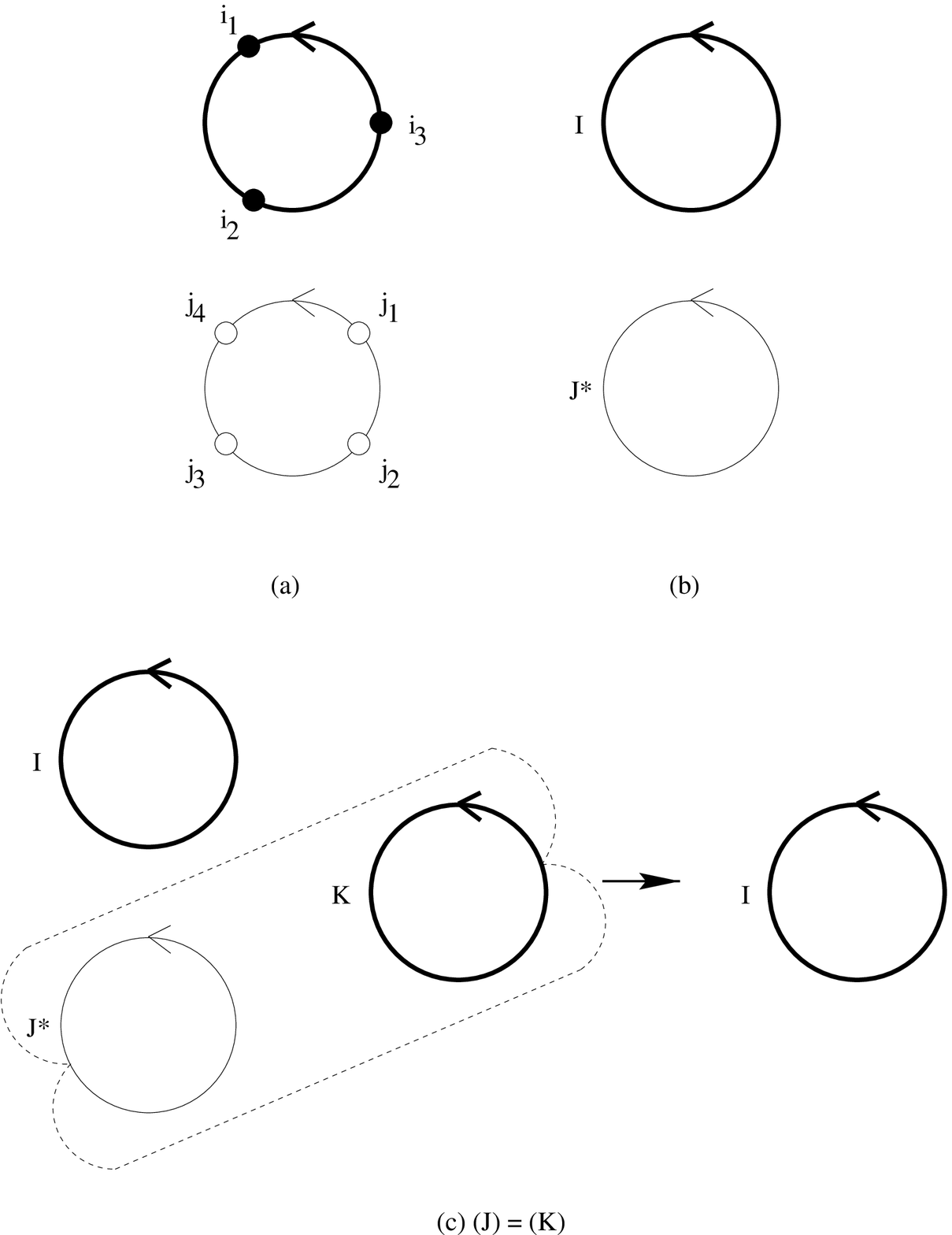}}
\caption{\em Operators of the fifth kind.}
\label{f2.6}
\end{figure}

{\em Operators of the fifth kind} form the last family of physical operators.  They are linear combinations of 
operators of the form
\beq
   \ti{f}^I_J & \equiv & N^{-(a+b)/2} a^{\da\mu_2}_{\mu_1}(i_1) a^{\da\mu_3}_{\mu_2}(i_2) \cd
   a^{\da\mu_1}_{\mu_a}(i_a) \nn \\
   & & \cdot a^{\n_b}_{\n_1}(j_b) a^{\n_{b-1}}_{\n_b}(j_{b-1}) \cd a^{\n_1}_{\n_2}(j_1).
\la{2.2.21}
\eeq
A typical operator of this kind is shown in detail in Fig.\ref{f2.6}(a).  In this diagram, the upper circle 
represents $I$, whereas the lower circle represents $J$.  There are 3 indices in $I$ and 4 indices in $J$.  Note 
the sequence $J$ is put in the clockwise instead of the anti-clockwise direction.  Fig.\ref{f2.6}(b) is a simplified
diagram of an operator of the fifth kind.  $J$ carries an asterisk to signify the fact that $J$ is in the 
anti-clockwise direction.  This operator possesses the following interesting properties:
\beq
   \ti{f}^{I_2 I_1}_J & = & (-1)^{\ep(I_1) \ep(I_2)} \ti{f}^{I_1 I_2}_J \mbox{; and} \nn \\
   \ti{f}^I_{J_2 J_1} & = & (-1)^{\ep(J_1) \ep(J_2)} \ti{f}^I_{J_1 J_2}.
\la{2.2.21a}
\eeq
It annihilates an open singlet state:
\beq
   \ti{f}^I_J \le( \bar{\ph}^{\l_1} \otimes s^{\dot{K}} \otimes \ph^{\l_2} \ri) = 0.
\la{2.2.22}
\eeq
It may replace a closed singlet state with another one, though:
\beq
   \ti{f}^I_J \Ps^K & = & \d^K_J \Ps^I + \sum_{J_1 J_2 = J} (-1)^{\ep(J_1) \ep(J_2)} \d^K_{J_2 J_1} \Ps^I.
\la{2.2.23}
\eeq
This action is depicted in Fig.\ref{f2.6}(c).

The {\em grades} of $f^{\dot{I}}_{\dot{J}}$, $l^{\dot{I}}_{\dot{J}}$, $r^{\dot{I}}_{\dot{J}}$, $\g^I_J$ and
$\ti{f}^I_J$ are defined to be the grade of the concatenated sequence $\dot{I} \dot{J}$ or $IJ$.  An observable is 
said to be {\em even} or {\em odd} if the grades of all terms in the linear combination are all 0 or all 1, 
respectively.  Clearly, an even operator sends an even singlet state to an even singlet state, and an odd state to 
an odd state.  On the contrary, an odd operator sends an even singlet state to an odd singlet state, and an odd 
state to an even state.  A {\em pure} operator is either an even or odd operator. 

It is easy for the reader to derive from the above formulae the actions of these five families of operators on the 
physical states when there is no adjoint fermion in the theory.  Alternatively, the reader can find the formulae in 
Refs.\cite{opstal} and \cite{clstal}.  We will say more about the mathematical properties of these physical 
operators in later chapters.

\section{Examples: Yang--Mills Theory with Adjoint Matter Fields}
\la{s2.3}

Let us see how the physical states and operators show up in actual physical systems.  Our first example is 
2-dimensional Yang--Mills theory with adjoint matter fields.  Dalley and Klebanov have provided a detailed 
introduction to the formulation of this model \cite{dakl} and the following presentation is closely parallel to
theirs.  As will be shown below, there is no dynamics for the gauge bosons in 2 dimensions.  The quantum matrices 
in the above abstract formalism are realized as canonically quantized adjoint matter fields.  Different matrices 
carry different momenta.  The linear momentum and Hamiltonian of the model can be expressed as linear combinations 
of the five families of operators introduced above.

We need a number of preliminary definitions.  Let $g$ be the Yang--Mills coupling constant, $\a$ and $\b$ ordinary 
space-time indices, $A_{\a}$ a Yang--Mills potential and $\ph$ a scalar field in the adjoint representation of the 
gauge group $U(N)$, and $m$ the mass of this scalar field.  Both $\ph$ and $A_{\a}$ are $N \times N$ Hermitian 
matrix fields.  If we treat the Yang--Mills potential as the Lie-algebraically valued connection form and $\ph$ a 
cross section in the language of differential geometry\footnote{An authoritative text on differential geometry 
which explains the notion of a fiber bundle is Ref.\cite{kono}.  However, those readers who are not interested in 
differential geometry may well take the covariant derivative to be given shortly for granted.}, then the covariant 
derivative is 
\[ D_{\a} \phi = \pa_{\a} \ph + {\rm i} \le[ A_{\a},\ph \ri]. \]

The Minkowski space action of this model is
\beq
   S = \int d^2 x {\rm Tr} \left[ \frac{1}{2} D_{\alpha}\phi D^{\alpha}\phi - \frac{1}{2} m^2 \phi^2 
   - \frac{1}{4g^2} F_{\alpha\beta} F^{\alpha\beta} \right].
\la{2.3.1}
\eeq
Introduce the light-cone coordinates
\[ x_+ = \frac{1}{\sqrt{2}} \le( x_0 + x_1 \ri) \]
and
\[ x_- = \frac{1}{\sqrt{2}} \le( x_0 - x_1 \ri). \]
Take $x^+$ as the time variable.  Choose the {\em light-cone gauge}
\[ A_- = 0. \]
The action is now simplified to
\beq
   S = \int dx^+ dx^- \Tr \le[ \pa_+ \ph \pa_- \ph - \frac{1}{2} m^2 \ph^2 + \frac{1}{2g^2} \le( \pa_- A_+ \ri)^2
       + A_+ J^+ \ri],
\la{2.3.2}
\eeq
where
\[ J^{+\m_1}_{\m_2} = \im \le[ \ph, \pa_- \ph \ri]^{\m_1}_{\m_2} \]
is the longitudinal momentum current.  Note that $A_+$ is no time dependence in Eq.(\ref{2.3.2}) and so the gauge 
field is not dynamical at all in the light-cone gauge.  Let us split $A_+$ into its zero mode $A_{+,0}$ (this is a
mode which is independent of $x_-$.) and non-zero mode $A_{+,n}$, i.e.,
\[ A_+ = A_{+,0} + A_{+,n}. \]
The Lagrange constraints for them are
\beq
   \int dx^- J^+ & = & 0 \mbox{, and}
\la{2.3.3} \\
   \pa_-^2 A_{+,n} - g^2 J^+ & = & 0,
\la{2.3.4}
\eeq
respectively.  Now we can use these two equations to eliminate $A_+$ in Eq.(\ref{2.3.2}) and get
\beq
   S = \int dx^+ dx^- \Tr \le( \pa_+ \ph \pa_- \ph - \frac{1}{2} m^2 \ph^2 + \frac{g^2}{2} J^+ \frac{1}{\pa_-^2} J^+
       \ri).
\la{2.3.5}
\eeq
The light-cone momentum and Hamiltonian are 
\[ P^{\pm} \equiv \int dx^- T^{+\pm}. \]
It follows from Eq.(\ref{2.3.5}) that their expressions in this model are 
\beq
   P^+ & = & \int dx^- \Tr \le( \pa_- \ph \ri)^2 \mbox{, and}
\la{2.3.6} \\
   P^- & = & \int dx^- \Tr \le( \frac{1}{2} m^2 \ph^2 - \frac{1}{2} g^2 J^+ \frac{1}{\pa_-^2} J^+ \ri).
\la{2.3.7}
\eeq

We now quantize the system.  Eq.(\ref{2.3.5}) implies that the canonical quantization condition is
\beq
   \le[ \ph^{\m_1}_{\m_2}(x^-), \pa_- \ph^{\m_3}_{\m_4}(\ti{x}^-) \ri] = 
   \frac{\im}{2} \d^{\m_1}_{\m_4} \d^{\m_3}_{\m_2} \d(x^- - \ti{x}^-). 
\la{2.3.8}
\eeq
Then a convenient field decomposition is
\beq
   \ph^{\m_1}_{\m_2}(x^+ = 0) = \frac{1}{\sqrt{2 \pi}} \int_0^{\ift} \frac{dk^+}{\sqrt{2k^+}} 
   \le[ a^{\m_1}_{\m_2}(k^+) \re^{- \im k^+ x^-} + a^{\da\m_1}_{\m_2}(k^+) \re^{\im k^+ x^-} \ri].
\la{2.3.9}
\eeq
The annihilation and creation operators here satisfy the canonical commutation relation Eq.(\ref{2.2.1}), except 
that the Kronecker delta function $\d_{k_1 k_2}$ in Eq.(\ref{2.2.1}) is replaced with the Dirac delta function 
$\d(k_1 - k_2)$ here, the adjoint matter field will satisfy Eq.(\ref{2.3.8}).  A state can be built out of a series 
of creation operators acting on the trivial vacuum, and it takes the form
\[ a^{\da\m_1}_{\n_1}(k^+_1) a^{\da\m_2}_{\n_2}(k^+_2) \cd a^{\da\m_c}_{\n_c}(k^+_c) |0\rangle. \]
However, in general this state does not satisfy the Lagrangian constraint Eq.(\ref{2.3.3}).  Only those of the form 
given in Eq.(\ref{2.2.6}) satisfy this constraint.  We thus obtain a model whose physical states are the physical 
states introduced in the previous section.

Using Eqs.(\ref{2.3.6}) and (\ref{2.3.8}), we can quantize the light-cone momentum and get
\beq
   P^+ = \int_0^{\ift} dk k \g^k_k,
\la{2.3.10}
\eeq
where $g^k_k$ is defined in Eq.(\ref{2.2.18}).  We have dropped the superscript $+$ in $k$ to simplify the 
notation.  This is a concrete example in which a physical observable is expressed as a linear combination of the 
operators introduced in the previous section.  (Here $k$ can take on an infinite number of values, and the 
regulator $\L \rar \ift$.  Nonetheless, taking the regulator to infinity has an influence on technical details 
only, since we are talking about a field theory without divergences in this limit.)  To obtain a similar formula 
for the light-cone Hamiltonian, we need to express, in momentum space, the longitudinal momentum current as a 
quantum operator first.  Define the longitudinal momentum current in momentum space to be
\beq
   \ti{J}^+ (k) & = & \frac{1}{\sqrt{2 \pi}} \int dx^- J^+ (x^-) \re^{- \im k^+ x^-}.
\la{2.3.11}
\eeq
Then, for $q > 0$,
\beq
   \ti{J}^{+\m_1}_{\m_2}(-q) & = & \frac{1}{2 \sqrt{2 \pi}} \int_0^{\ift} dp \frac{2p + q}{\sqrt{p(p + q)}}
   \le[ a^{\da\n}_{\m_2}(p) a^{\m_1}_{\n}(p + q) \ri. \nn \\
   & & \le. - a^{\da\m_1}_{\n}(p) a^{\n}_{\m_2}(p + q) \ri] \nn \\
   & & + \int_0^q dp \frac{q - 2p}{\sqrt{p(q - p)}} a^{\n}_{\m_2}(p) a^{\m_1}_{\n}(q - p),
\la{2.3.12}
\eeq
and
\beq
   \ti{J}^{+\m_1}_{\m_2}(q) = \le[ \ti{J}^{+\m_1}_{\m_2}(-q) \ri]^{\da}.
\la{2.3.13}
\eeq
We can now use Eqs.(\ref{2.3.7}), (\ref{2.3.8}), (\ref{2.3.11}), (\ref{2.3.12}) and (\ref{2.3.13}) to obtain
\begin{eqnarray}
   P^- & = & \frac{1}{2} m^2 \int_0^{\infty} \frac{dk}{k} g^k_k 
   + \frac{g^2 N}{4\pi} \int_0^{\infty} \frac{dk}{k} C g^k_k \nonumber \\
   & & + \frac{g^2 N}{8\pi} \int_0^{\infty} \frac{dk_1 dk_2 dk_3 dk_4}
   {\sqrt{k_1 k_2 k_3 k_4}}  
   \left\{ A \delta (k_1 + k_2 - k_3 - k_4) g^{k_3 k_4}_{k_1 k_2} \right. \nonumber \\  
   & & + \left. B \delta (k_1 - k_2 - k_3 - k_4) \le( g^{k_2 k_3 k_4}_{k_1} + g^{k_1}_{k_2 k_3 k_4} \ri) \ri\}, 
\label{2.3.14}
\end{eqnarray}     
where 
\begin{eqnarray*}
   A & = & \frac{(k_2 - k_1)(k_4 - k_3)}{(k_1 + k_2)^2} - \frac{(k_3 + k_1)(k_4 + k_2)}{(k_4 - k_2)^2}; \\
   B & = & \frac{(k_2 + k_1) (k_4 - k_3)}{(k_4 + k_3)^2} - \frac{(k_4 + k_1) (k_3 - k_2)}{(k_3 + k_2)^2} 
           \mbox{; and} \\
   C & = & \int_0^k dp \frac{(k+p)^2}{p(k-p)^2}.
\end{eqnarray*}
Eq.(\ref{2.3.14}) shows that the light-cone Hamiltonian is a linear combination of operators of the fourth kind.

The above formalism can be easily generalized to the case in which there are more than one adjoint matter fields.
This kind of models is useful for studying gluons in quantum chromodynamics \cite{anda96a} --- take the large-$N$ 
limit as an approximation for pure Yang--Mills theory.  If we assume that the gluon field, i.e., the gauge field, 
is not dependent on the transverse dimensions (in other words, we take {\em dimensional reduction} as a further 
approximation), the gluon field will precisely be the adjoint matter fields.  The number of adjoint matter fields is
the same as the number of transverse dimensions in the system.  The prototype model in the next section can serve as
an example; if we remove the fundamental and conjugate matter fields, we will obtain an approximate model for 
gluons in three dimensions 

\section{Examples: Yang--Mills Theory with Fundamental and Adjoint Matter Fields}
\la{s2.4}
 
We can incorporate fundamental and conjugate matter fields in the above model.  This is best understood in the 
context of large-$N$ Yang--Mills theory with quark and antiquark matter fields \cite{anda96b}.  Consider the 
simplest non-trivial example, an $SU(N)$ gauge theory in (2+1)-dimensions with one flavor of quarks of mass $m$.  
Let $\ti{g}$ be the strong coupling constant, and $\Ps$ a quark field in the fundamental representation of the 
gauge group $U(N)$.  Then $\Ps$ is a column vector of $N$ Grassman fields.  The covariant derivative for $\Ps$ is
\[ D_{\alpha} \Psi = \partial_{\alpha} \Psi + {\rm i} A_{\alpha} \Psi \]
The action of this model is
\begin{equation}
   S = \int d^4 x \left[ - \frac{1}{4\tilde{g}^2} {\rm Tr} F_{\alpha\beta} F^{\alpha\beta}
   + {\rm i} \bar{\Psi} \gamma^{\alpha} D_{\alpha} \Psi - m \bar{\Psi} \Psi \right]
\la{2.4.1}
\end{equation}
in the Weyl representation
\begin{eqnarray*}
   \gamma^0 = \left( \begin{array}{cc} 0 & - {\bf 1} \\ - {\bf 1} & 0 \end{array} \right)
   \; \mbox{and} \;
   \gamma^i = \left( \begin{array}{cc} 0 & \tau^i \\ - \tau^i & 0 \end{array} \right)
\end{eqnarray*}
for $i$ = 1 or 3.  Imposing the conditions for dimensional reduction
\[ \partial_1 A_{\alpha} = \partial_1 \Psi = 0, \]
we obtain an effectively two-dimensional Yang--Mills theory with adjoint matter fields and fundamental Dirac 
spinors. 

To express this model as a quantum matrix model, we can follow the same procedure described in detail in the 
previous section (and so here we will be brief).  Take the light-front gauge $A_- = 0$.  Let
\begin{equation}
   2^{1/4} \Psi \sqrt{\int dx^1} = \left( \begin{array}{c} u(+) \\ u_L(+) \\ u_L(-)
   \\ u(-) \end{array} \right).
\la{2.4.2}
\end{equation}
Then the fields $A_+$ and $u_L(\pm)$ are not dynamical.  Eliminating the constrained fields yields the light-front 
energy and momentum
\begin{eqnarray}
   \lefteqn{ P^- = \int dx^- \left[ - \frac{1}{2 g^2} J^{+\mu_1}_{\mu_2} \frac{1}{\partial^2_-} J^{+\mu_2}_{\mu_1} 
   \right. } \nonumber \\
   & & - \frac{\rm i}{2} \left( u^{\ast}_{\mu_1}(+) A^{\mu_1}_{1\mu_2} - m u^{\ast}_{\mu_2}(-) \right)
   \frac{1}{\partial_-} \left( A^{\mu_2}_{1\mu_3} u^{\mu_3}(+) - m u^{\mu_2}(-) \right) \nonumber \\
   & & \left. - \frac{\rm i}{2} \left( u^{\ast}_{\mu_1}(-) A^{\mu_1}_{1\mu_2} + m u^{\ast}_{\mu_2}(+) \right)
   \frac{1}{\partial_-} \left( A^{\mu_2}_{1\mu_3} u^{\mu_3}(-) - m u^{\mu_2}(+) \right) \right]
\label{2.4.3} \\
   \lefteqn{ P^+ = \int dx^- \left[ \partial_- A^{\mu_1}_{1\mu_2}(+) \partial_- A^{\mu_2}_{1\mu_1}(-) 
   + {\rm i} \left( u^{\ast}_{\mu}(+) \partial_- u^{\mu}(+) \right. \right. } \nonumber \\
   & & \left. \left. + u^{\ast}_{\mu}(-) \partial_- u^{\mu}(-) \right) \right] 
\label{2.4.4}
\end{eqnarray}
In Eqs.(\ref{2.4.3}) and (\ref{2.4.4}), $g^2 = \tilde{g}^2/ \int dx^1$.  The longitudinal momentum current is given 
by
\beq
   J^{+\mu_1}_{\mu_2} =  {\rm i} \lbrack A_1, \partial_- A_1 \rbrack^{\mu_1}_{\mu_2} +
   u^{\mu_1}(+) u^{\ast}_{\mu_2}(+) + u^{\mu_1}(-) u^{\ast}_{\mu_2}(-)
\la{2.4.5}
\eeq
We now quantize the dynamical fields as follows:
\begin{equation}
   u^{\mu}(\pm) = \frac{1}{\sqrt{2\pi}} \int_0^{\infty} dk^+ \left(
   \c^{\mu}(k^+, \pm) {\rm e}^{-ik^+ x^-} + \bar{\c}^{\dagger\mu}(k^+, \mp)
   {\rm e}^{-ik^+ x^-} \right)
\la{2.4.6}
\end{equation}
and
\begin{equation}
   A^{\mu_1}_{1\mu_2} = \frac{g}{\sqrt{2\pi}} \int_0^{\infty} \frac{dk^+}{\sqrt{2k^+}}
   \left(a^{\mu_1}_{\mu_2}(k^+) {\rm e}^{-{\rm i} k^+ x^-} +
   a^{\dagger\mu_1}_{\mu_2}(k^+) {\rm e}^{{\rm i} k^+ x^-} \right).
\la{2.4.7}
\end{equation}
$\c$ and $\c^{\da}$ satisfy Eq.(\ref{2.2.4}), and so are $\bar{\c}$ and $\bar{\c}^{\da}$.  Physically speaking,
$\c$ and $\bar{\c}$ are annihilation operators of a quark and an anti-quark, respectively.  $a$ and $a^{\da}$ 
satisfy Eq.(\ref{2.2.1}), as usual.  Again we will simply write $k^+$ as $k$ in the rest of this section.

There are two families of physical states in this model.  One family is of the form shown in Eq.(\ref{2.2.6}).  This
is a single glueball state.  Another family is of the form shown in Eq.(\ref{2.2.5}).  This is a single meson state.
The physical observables can be expressed as linear combinations of operators of the first four kinds.  For 
instance, the light-cone momentum and Hamiltonian read
\begin{eqnarray}
   \lefteqn{ P^+ = \int_0^{\infty} dk \, k \left[ \s^k_k + \sum_{j = +, -} \left(
   \bar{\X}^{kj}_{kj} \otimes l^{\phi}_{\phi} + r^{\phi}_{\phi} \otimes \Xi^{kj}_{kj} \right) \right] \mbox{; and} }
\label{2.4.8} \\
   \lefteqn{ P^- = \int_0^{\infty} dk \, h_{IV}(k) \s^k_k + 
   \int_0^{\infty} dk_1 dk_2 dk_3 dk_4 \cdot } \nonumber \\
   & & \left[ h_{IV}(k_1, k_2; k_3, k_4) \delta(k_1 + k_2 - k_3 - k_4) \s^{k_3 k_4}_{k_1 k_2}  \right. \nn \\
   & & \left. + h_{IV}(k_1; k_2, k_3, k_4) \delta(k_1 - k_2 - k_3 - k_4) \left( \s^{k_2 k_3 k_4}_{k_1}  
   + \s^{k_1}_{k_2 k_3 k_4} \right) \right]
   \nonumber \\
   & & + \sum_{j=+,-} \int_0^{\infty} dk h_{II}(k) \left( \bar{\Xi}^{kj}_{kj} \otimes l^{\emptyset}_{\emptyset} 
   + 1 r^{\emptyset}_{\emptyset} \otimes \Xi^{kj}_{kj} \right) \nonumber \\
   & & + \int_0^{\infty} dk_1 dk_2 dk_3 h_{II}(k_1; k_2, k_3) \delta(k_1 - k_2 - k_3)
   \left( - \bar{\Xi}^{k_1,+}_{k_3,-} \otimes l^{\emptyset}_{k_2} \right. \nonumber \\
   & & + \bar{\Xi}^{k_1,-}_{k_3,+} \otimes l^{\emptyset}_{k_2} +
   \bar{\Xi}^{k_3,+}_{k_1,-} \otimes l^{k_2}_{\emptyset} -
   \bar{\Xi}^{k_3,-}_{k_1,+} \otimes l^{k_2}_{\emptyset} +
   r^{\emptyset}_{k_2} \otimes \Xi_{k_3,-}^{k_1,+} \nn \\ 
   & & \left. - r^{\emptyset}_{k_2} \otimes \Xi_{k_3,+}^{k_1,-} -
   r^{k_2}_{\emptyset} \otimes \Xi_{k_1,-}^{k_3,+} +
   r^{k_2}_{\emptyset} \otimes \Xi_{k_1,+}^{k_3,-} \right) \nonumber \\
   & & + \sum_{j=+,-} \int_0^{\infty} dk_1 dk_2 dk_3 dk_4 \left[ h_{II}(k_1, k_2; k_3, k_4) 
   \delta(k_1 + k_2 - k_3 - k_4) \cdot \right. \nonumber \\
   & & \left( \bar{\Xi}^{k_1,j}_{k_4,j} \otimes l^{k_2}_{k_3} +
   r^{k_2}_{k_3} \otimes \Xi_{k_4, j}^{k_1, j} \right) \nonumber \\
   & & + h_{II}(k_1; k_2, k_3, k_4) \delta(k_1 - k_2 - k_3 - k_4) \left(
   \bar{\Xi}^{k_1,j}_{k_4,j} \otimes l^{\emptyset}_{k_3 k_2} + \right.\nonumber\\
   & & \left. \left. + \bar{\Xi}^{k_4,j}_{k_1,j} \otimes l^{k_3 k_2}_{\emptyset} +
   r^{\emptyset}_{k_2 k_3} \otimes \Xi_{k_4, j}^{k_1, j} +
   r^{k_2 k_3}_{\emptyset} \otimes \Xi_{k_1, j}^{k_4, j} \right)  \right] \nonumber \\
   & & + \int_0^{\infty} dk_1 dk_2 dk_3 dk_4 h_I(k_1, k_2; k_3, k_4)
   \delta(k_1 + k_2 - k_3 - k_4) \cdot \nonumber \\ 
   & & \left( \bar{\Xi}^{k_1, +}_{k_3, +} \otimes f^{\emptyset}_{\emptyset} \otimes \Xi_{k_4, -}^{k_2, -} +
   \bar{\Xi}^{k_1, -}_{k_3, -} \otimes f^{\emptyset}_{\emptyset} \otimes \Xi_{k_4, +}^{k_2, +} \right),
\label{2.4.9}
\end{eqnarray}
where
\begin{eqnarray*}
   h_{IV}(k) & = & \frac{g^2 N}{4 \pi k} \int_0^k dp \frac{(k+p)^2}{p(k-p)^2}; \\
   h_{IV}(k_1, k_2; k_3, k_4) & = & \frac{g^2 N}{8 \pi} \frac{1}{\sqrt{k_1 k_2 k_3 k_4}} \cdot \\
   & & \left[ \frac{(k_2 - k_1) (k_4 - k_3)}{(k_2 + k_1)^2}
   - \frac{(k_3 + k_1) (k_4 + k_2)}{(k_4 - k_2)^2} \right]; \\
   h_{IV}(k_1; k_2, k_3, k_4) & = & \frac{g^2 N}{8 \pi} \frac{1}{\sqrt{k_1 k_2 k_3 k_4}} \cdot \\
   & & \left[ \frac{(k_2 + k_1) (k_4 - k_3)}{(k_4 + k_3)^2}
   - \frac{(k_4 + k_1) (k_3 - k_2)}{(k_3 + k_2)^2} \right]; \\
   h_{II}(k) & = & \frac{m^2}{2k} + \frac{g^2 N}{4 \pi} \int_0^k dp \left[ \frac{1}{p^2} + 
   \frac{1}{2 p (k - p)} + \frac{1}{(k-p)^2} \right]; \\
   h_{II}(k_1; k_2, k_3) & = & \frac{g}{4} \sqrt{\frac{N}{\pi}} \frac{m}{\sqrt{k_2}} 
   (\frac{1}{k_1} - \frac{1}{k_3}); \\
   h_{II}(k_1, k_2; k_3, k_4) & = & \frac{g^2 N}{8 \pi} \frac{1}{\sqrt{k_2 k_3}}
   \left[ - \frac{2(k_2 + k_3)}{(k_2 - k_3)^2} + \frac{1}{k_1 + k_2} \right]; \\
   h_{II}(k_1; k_2, k_3, k_4) & = & \frac{g^2 N}{8 \pi} \frac{1}{\sqrt{k_2 k_3}}
   \left[ \frac{2(k_2 - k_3)}{(k_2 + k_3)^2} + \frac{1}{k_3 + k_4} \right] \mbox{; and} \\
   h_I(k_1, k_2; k_3, k_4) & = & - \frac{g^2 N}{2 \pi} \frac{1}{(k_1 - k_3)^2}.
\end{eqnarray*}
(remark: $h_{IV}(k)$, $h_{IV}(k_1, k_2; k_3, k_4)$ and $h_{IV}(k_1; k_2, k_3, k_4)$ are closely related to the 
coefficients $A$, $B$ and $C$ in the previous section.)  The Roman numerals $I$ and $IV$ carried by some $h$'s 
refer to the fact that these are coefficients of operators of the first and fourth kinds respectively, whereas the 
Roman number $II$ carried by other $h$'s signify that these are coefficients of operators of the second and third 
kinds\footnote{Eq.(\ref{2.4.9}) corresponds to Eq.(14) in Ref.~\cite{anda96b}, which shows terms corresponding to 
operators of the second kind only.  In fact, these two equations do not agree, but since these self-energy terms 
will anyway be absorbed into a redefinition of the mass of the scalar, they do not affect the final answer.}. 

This is a model suitable for studying meson spectrum.

\section{Examples: Supersymmetric Yang--Mills Theory}
\la{s2.5}

The adjoint matter fields in all the models discussed so far are bosonic.  In Wess--Zumino matrix model 
\cite{hakl} and supersymmetric Yang--Mills theory, there are both fermionic and bosonic adjoint matter fields.  
Because of the usefulness of SYM theory in studying the low-energy dynamics of D-branes and M-theory, it should be
of interest to see if these models can be expressed in terms of the physical states and observables introduced in 
Section~\ref{s2.2}, too.  Indeed, the answer is affirmative, and we will explicitly demonstrate this using the 
simplest SYM theory, ${\cal N} = 1$ 2-dimensional supersymmetric Yang-Mills theory \cite{ferrara, masasa}.  We will 
follow the notations in Ref.\cite{masasa} in this section.

In the Wess--Zumino gauge, the action is given by
\beq
   S = \int d^2x \Tr \le[ - \frac{1}{4g^2} F_{\a\b} F^{\a\b} + \frac{1}{2} D_{\a} \ph D^{\a} \ph +
       \im \bar{\Ps} \g^{\a} D_{\a} \Ps - 2 \im g \ph \bar{\Ps} \g_5  \Ps \ri].
\la{2.5.1}  
\eeq
In this equation, $A_{\a}$, $F_{\a\b}$ and $D_{\a}$ are Yang--Mills potential, Yang--Mills field and the covariant
derivative,  as usual.  $\ph$ and $\Ps$ are bosonic and fermionic adjoint matter fields, respectively.  
$\bar{\Ps} = \Ps^T \g^0$.  This Lagrangian density is invariant under the following transformation, which is a 
mixture of supersymmetric and gauge transformations:
\beq
   \d A_{\a} & = & \im g \bar{\ep} \g_5 \g_{\a} \sqrt{2} \Ps, \nn \\
   \d \ph & = & - \bar{\ep} \sqrt{2} \Ps \mbox{, and} \nn \\
   \d \Ps & = & - \frac{1}{2 \sqrt{2} g} \ep \ep^{\a\b} F_{\a\b} + \frac{\im}{\sqrt{2}} \g^{\a} \ep D_{\a} \ph,
\la{2.5.2}
\eeq
where $\ep$ is an infinitesimal spinor matrix, and $\ep^{01} = - \ep_{01} = 1$.  Noether's Theorem tells us that the
associated spinor supercurrent $j^{\a}$ is
\beq
   \bar{\ep} j^{\a} = \Tr \le[ - \sqrt{2} \bar{\ep} \Ps D^{\a} \ph + 
   \frac{\im}{\sqrt{2} g} \ep^{\b\l} F_{\b\l} \bar{\ep} \g^{\a} \Ps +
   \sqrt{2} \bar{\ep} \g_5 \Ps \ep^{\a\b} D_{\b} \ph \ri].
\la{2.5.3}
\eeq
Choose the following conventions for the spinor and gamma matrices:
\beq
   \Ps^{\m_1}_{\m_2} & = & 2^{-1/4} ( \ps^{\m_1}_{\m_2}, \c^{\m_1}_{\m_2} )^T, \nn \\
   \g^0 & = & \s_2, \nn \\
   \g^1 & = & \im \s_1 \mbox{, and} \nn \\
   \g_5 & = & \g^0 \g^1 = \s_3.
\la{2.5.4}
\eeq
If we now rewrite the action in terms of light-cone coordinates and take the usual light-cone gauge $A_- = 0$, we
will find that $A_+$ and $\c$ are not dynamical.  Quantize the dynamical fields as usual:
\beq
   \ph^{\m_1}_{\m_2} (x^-, 0) = \frac{1}{\sqrt{2 \pi}} \int_0^{\ift} \frac{dk^+}{\sqrt{2 k^+}} 
   \le[ a^{\m_1}_{\m_2}(k^+) \re^{- \im k^+ x^-} + a^{\da\m_1}_{\m_2}(k^+) \re^{\im k^+ x^-} \ri]
\la{2.5.5}
\eeq
and
\beq
   \ps^{\m_1}_{\m_2} (x^-, 0) = \frac{1}{2 \sqrt{\pi}} \int_0^{\ift} dk^+ 
   \le[ b^{\m_1}_{\m_2}(k^+) \re^{- \im k^+ x^-} + b^{\da\m_1}_{\m_2}(k^+) \re^{\im k^+ x^-} \ri].
\la{2.5.6}
\eeq
The annihilation and creation operators in Eqs.(\ref{2.5.5}) and (\ref{2.5.6}) satisfy
\beq
   \le[ a^{\m_1}_{\m_2}(k^+), a^{\da\m_3}_{\m_4}(\ti{k}^+) \ri] = 
   \le[ b^{\m_1}_{\m_2}(k^+), b^{\da\m_3}_{\m_4}(\ti{k}^+) \ri]_+ =
   \d(k^+ - \ti{k}^+) \d^{\m_1}_{\m_4} \d^{\m_3}_{\m_2}.
\la{2.5.7}
\eeq
Up to some notational difference, Eq.(\ref{2.5.7}) is the same as Eqs.(\ref{2.2.1}) and (\ref{2.2.2}).  This is why
the formalism in Section~\ref{s2.2} is applicable here.  We can now write the supercharge operators as operators of
the fourth kind, as follows:
\beq
   Q_1 \equiv \int dx^- j_1^+ = \im 2^{1/4} \int_0^{\ift} dk \sqrt{k} \le[ b^{\da\m_1}_{\m_2}(k) a^{\m_2}_{\m_1}(k)
   - a^{\da\m_1}_{m_2}(k) b^{\m_2}_{\m_1}(k) \ri]
\la{2.5.8}
\eeq
and
\beq
   Q_2 \equiv \int dx^- j_2^+ = - \im 2^{1/4} g \int_0^{\ift} \frac{dk}{k} 
   \le[ b^{\da\m_1}_{\m_2}(k) \ti{J}^{\m_2}_{\m_1}(-k) - \ti{J}^{\da\m_1}_{\m_2}(-k) b^{\m_2}_{\m_1}(k) \ri],
\la{2.5.9}
\eeq
where $\ti{J}$ is again the longitudinal momentum current, and again we have dropped the superscript $+$ carried by
the longitudinal momenta.  In this model, the expression of the longitudinal momentum current is
\beq
   \ti{J}^{\m_2}_{\m_1}(-k) & = & \frac{1}{2 \sqrt{2 \pi}} \int_0^{\ift} dp \frac{2p + k}{\sqrt{p (p + k)}}
   \le[ a^{\da \m_3}_{\m_1}(p) a^{\m_2}_{\m_3}(k + p) \ri. \nn \\
   & & \le. - a^{\da\m_2}_{\m_3}(p) a^{\m_3}_{\m_1}(k + p) \ri] \nn \\
   & & + \frac{1}{2 \sqrt{2 \pi}} \int_0^k dp \frac{k - 2p}{\sqrt{p (k - p)}} 
   a^{\m_3}_{\m_1}(p) a^{\m_2}_{\m_3}(k - p) \nn \\
   & & + \frac{1}{\sqrt{2 \pi}} \int_0^{\ift} dp \le[ b^{\da\m_3}_{\m_1}(p) b^{\m_2}_{\m_3}(k + p) -
   b^{\da\m_2}_{\m_3}(p) b^{\m_3}_{\m_1}(k + p) \ri] \nn \\
   & & + \frac{1}{\sqrt{2 \pi}} \int_0^k dp b^{\m_3}_{\m_1}(p) b^{\m_2}_{\m_3}(k - p)
\la{2.5.10}
\eeq
and $\ti{J}^{\m_2}_{\m_1}(k) = \ti{J}^{\da\m_2}_{\m_1}(-k)$ for $k > 0$.

The above trick can be applied to other supersymmetric matrix models.

\section{Examples: Quantum Gravity and String Theory}
\la{s2.6}

There are several ways to study quantum gravity and its most promising candidate, string theory, in terms of quantum
matrix models.  One idea is based on the crucial observation that the dual of a Feynman diagram can be taken as a 
triangulation of a planar surface, which serves as a lattice approximation to a geometrical surface.  The partition 
function for two-dimensional quantum gravity can then be approximated by a matrix model in the large-$N$ limit 
\cite{david, kazakov85}.  Whether the matrix model is classical or quantum, and the exact form of its action 
depends on what the conformal matter field coupled to quantum gravity is.  This, in turn, is equivalent to a string
theory with certain dimensions \cite{frgizi}.  This approach has the virtue that it reveals some non-perturbative 
behavior of string theory.  For example, a three-dimensional non-critical string theory is equivalent to a model of 
two-dimensional quantum gravity coupled with conformal matter with the conformal charge $c = 2$.  Then this model 
of quantum gravity is further mapped to a one-matrix model in the large-$N$ limit with $\phi^3$ interaction, i.e.,
the action of this matrix model is
\beq
   S = \int d^2 x \Tr \le( \frac{1}{2} \pa_{\a} \Ph \pa^{\a} \Ph + \frac{1}{2} \m \Ph^2 - 
       \frac{1}{3 \sqrt{N}} \l \Ph^3 \ri),
\la{2.6.1}
\eeq
where $\Ph$ is an $N \times N$ matrix, and $\m$ and $\l$ are constants.

Another idea is to approximate a string by a collection of string bits \cite{beth}.  A closed singlet state then 
represents a closed string, and an open singlet state an open string.

Another approach is to consider the low-energy dynamics of string theory \cite{polchinski}.  For example, if we 
exclude all Feynman diagrams with more than one loop in a bosonic open string theory with $N$ Chan--Paton factors, 
we will retain only a tachyonic matrix field $\vp$ with three-tachyon coupling, and Yang--Mills gauge bosons 
minimally coupled to tachyons.  The action is
\beq
   S & = & \frac{1}{g^2} \int d^{26} x \le[ - \frac{1}{2} \Tr ( D_{\a} \vp D^{\a} \vp ) +
   \frac{1}{2 \a'} \Tr ( \vp^2 ) + \frac{1}{3} \sqrt{\frac{2}{\a'}} \Tr ( \vp^3 ) \ri. \nn \\
   & & \le. - \frac{1}{4} \Tr ( F_{\a\b} F^{\a\b} ) \ri],
\la{2.6.2}
\eeq
where $\a'$ is the Regge slope, and $g$ is the gauge boson coupling constant.

The low-energy dynamics of bosonic D$p$-branes \cite{polchinski} can likewise be described by a quantum matrix 
model.  Let $\x^1$, $\x^2$, \ld, and $\x^p$ be the coordinates inside a $p$-brane.  The action turns out to be
\beq
   S & = & - T_p \int d^{p+1} \x \re^{-\Ph} \le[ - {\rm det} ( G_{ab} + B_{ab} + 2 \pi \a' F_{ab} ) \ri]^{1/2},
\la{2.6.3}
\eeq
where $T_p$ is the brane tension, 
\beq
   G_{ab}(\x) = \frac{\pa X^{\a}}{\pa \x^a} \frac{\pa X^{\b}}{\pa \x^b} G_{\a\b}(X(\x))
\la{2.6.4}
\eeq
and
\beq
   B_{ab}(\x) = \frac{\pa X^{\a}}{\pa \x^a} \frac{\pa X^{\b}}{\pa \x^b} B_{\a\b}(X(\x))
\la{2.6.5}
\eeq
are the induced metric and antisymmetric tensor on the brane, and $F_{ab}$ is the background Yang--Mills field.  

If supersymmetry is incorporated into the theory, and if we further assume that space--time is essentially flat, 
the low-energy dynamics of D$p$-branes can be approximated very well by an ${\cal N} = 1$ SYM theory dimensionally 
reduced from 10 down to $p + 1$ \cite{leigh, witten96}.  Here we have a $(p + 1)$-dimensional gauge field theory 
with $9 - p$ adjoint matter fields of bosons and a number of adjoint matter fields of fermions.  $N$, the dimension 
of the matrices, is the number of D-branes in the system,  the `color index' labels a particular D-brane, and the
`quantum state other than color', ranging from $p+1$ to 9, labels the transverse dimensions.  Classically speaking, 
diagonal elements of the matrices give the coordinates of the D-branes in the $k$-th dimension, and off-diagonal 
elements tell us the distances between the corresponding D-branes in that transverse dimension.

Closely associated with D-branes is the M-theory conjecture \cite{bfss}, namely that in the infinite momentum 
frame, M-theory is exactly described by the $N \rightarrow \infty$ limit of 0-brane quantum mechanics.  A natural 
corollary of this conjecture is that light-front type-IIA superstring theory can be described by an ${\cal N} = 1$ 
SYM theory dimensionally reduced from 10 to 2 \cite{diveve} in the large-$N$ limit.  The Hamiltonian thus obtained 
is essntially the Green-Schwarz light-front string Hamiltonian, except that the fields are matrices.  (A 
comprehensive review on SYM theory, D-branes and M-theory can be found in Ref.~\cite{taylor}.)

The existence of a multitude of methods to transcribe string theory to matrix models shows their intimate 
relationship.

\section{Examples: Quantum Spin Chains (1)}
\la{s2.7}

Quantum matrix models have a deep relationship with condensed matter systems, too.  The essence of many condensed 
matter systems can be captured by bosonic or fermionic quantum spin chain systems.  For example, we can use the
Ising model to study ferromagnetism, lattice gas and order-disorder phase transition \cite{domb}.  Another 
paradigmatic model is the Hubbard model in which the spins are fermions (electrons).  The Hubbard model is believed
to describe salient features of high-$T_c$ superconductivity, metal-insulator transition, fractional quantum Hall 
effect, superfluidity, etc. \cite{baeriswyl}.

Another factor which motivates us to study quantum spin chain systems is their integrability.  The Bethe ansatz
\cite{bethe} and the closely related Yang--Baxter equations \cite{yang, baxter} are well known to be powerful tools 
in studying and exactly solving many of these spin chain systems.  These tools provide us a way to determine the 
integrability of and solve the associated quantum matrix models.  A better knowledge in quantum matrix models
should thus help us understand exactly integrable models and condensed matter systems better, and vice versa.

There is a one-to-one correspondence between spin chain systems, bosonic and/or fermionic, satisfying periodic or 
open boundary conditions and quantum matrix systems in the large-$N$ limit.  (A connection between spin systems and 
matrix models was observed previously in Ref.\cite{klsu}.)  In this section, we will focus on spin chains that 
satisfy the periodic boundary condition.  Let $c$ be the number of sites in a spin chain.  Each site can be 
occupied either by a boson or a fermion.  Assume that there are $\L_B$ possible bosonic states, and $\L_F$ possible 
fermionic ones.  Let $A_p(1)$, $A_p(2)$, \ld, $A_p(\L_B)$ be the annihilation operators of these bosonic states at
the $p$-th site, and $A_p(\L_B + 1)$, $A_p(\L_B + 2)$, \ld, $A_p(\L_B + \L_F)$ be the annihilation operators of the 
fermionic ones.  Attaching daggers as superscripts to these operators turns them to the corresponding creation 
operators.  Then a typical state of a spin chain can be written as
\beq
    \ps^K \equiv A^{\da}_1(k_1) A^{\da}_2(k_2) \cd A^{\da}_c(k_c) |0 \rangle. 
\la{2.7.1}
\eeq
Define the {\em Hubbard operator} \cite{bablog} as follows:
\beq
   X^{ij}_p \equiv A^{\da}_p(i) A_p(j).
\la{2.7.2}
\eeq
Now, consider the action of a product of Hubbard operators on $\ps^K$:
\beq
   \lefteqn{X^{i_1 j_1}_p X^{i_2 j_2}_{p+1} \cd X^{i_a j_a}_{p+a-1} \ps^K = 
   (-1)^{\lb \ep(k_1) + \ep(k_2) + \cd + \ep(k_{p-1}) \rb \lb \ep(I) + \ep(J) \rb }} \nn \\
   & & \cdot (-1)^{\lb \ep(i_a) + \ep(j_a) \rb \lb \ep(k_p) + \ep(k_{p+1}) + \cd + \ep(k_{p+a-2}) \rb} 
   \d^{k_{p+a-1}}_{j_a} \nn \\
   & & \cdot (-1)^{\lb \ep(i_{a-1}) + \ep(j_{a-1}) \rb \lb \ep(k_p) + \ep(k_{p+1}) + \cd + \ep(k_{p+a-3}) \rb}
   \d^{k_{p+a-2}}_{j_{a-1}} \cd \nn \\
   & & \cdot (-1)^{\lb \ep(i_2) + \ep(j_2) \rb \ep(k_p)} \d^{k_{p+1}}_{j_2} \d^{k_p}_{j_1}
   \ps^{k_1 k_2 \ld k_{p-1} i_1 i_2 \ld i_a k_{p+a} k_{p+a+1} \ld k_n}
\la{2.7.3}
\eeq
for $p+a-1 \leq n$.  We can manipulate Eq.(\ref{2.7.3}) to be
\beq
   \lefteqn{(-1)^{\lb \ep(i_a) + \ep(j_a) \rb \lb \ep(j_1) + \ep(j_2) + \cd + \ep(j_{a-1}) \rb }} \nn \\
   \lefteqn{\cdot (-1)^{\lb \ep(i_{a-1}) + \ep(j_{a-1}) \rb \lb \ep(j_1) + \ep(j_2) + \cd + \ep(j_{a-2}) \rb} \cd} 
   \nn \\
   \lefteqn{\cdot (-1)^{\lb \ep(i_2) + \ep(j_2) \rb \ep(j_1)} X^{i_1 j_1}_p X^{i_2 j_2}_{p+1} \cd 
   X^{i_a j_a}_{p+a-1} \ps^K = } \nn \\
   & & (-1)^{\lb \ep(k_1) + \ep(k_2) + \cd + \ep(k_{p-1}) \rb \lb \ep(J) + \ep(k_{p+a}) + \ep(k_{p+a+1}) + \cd
   + \ep(k_n) \rb} \nn \\
   & & \cdot (-1)^{\lb \ep(k_1) + \ep(k_2) + \cd + \ep(k_{p-1}) \rb \lb \ep(I) + \ep(k_{p+a}) + \ep(k_{p+a+1}) + \cd
   + \ep(k_n) \rb} \nn \\
   & & \cdot \d^{k_p}_{j_1} \d^{k_{p+1}}_{j_2} \cd \d^{k_{p+a-1}}_{j_a} 
   \ps^{k_1 k_2 \ld k_{p-1} i_1 i_2 \ld i_a k_{p+a} k_{p+a+1} \ld k_n}.
\la{2.7.4}
\eeq
If $p + a - 1 > c = p + a_1 - 1$, then the action of this product of Hubbard operators on $\ps^K$ becomes instead
\beq
   \lefteqn{X^{i_1 j_1}_p X^{i_2 j_2}_{p+1} \cd X^{i_a j_a}_{p+a-1} \Ps^K =
   (-1)^{\lb \ep(i_a) + \ep(j_a) \rb \lb \ep(k_1) + \ep(k_2) + \cd + \ep(k_{a_2 - 1}) \rb} \d^{k_{a_2}}_{j_a} } 
   \nn \\
   & & \cdot (-1)^{\lb \ep(i_{a-1}) + \ep(j_{a-1}) \rb \lb \ep(k_1) + \ep(k_2) + \cd + \ep(k_{a_2 - 2}) \rb} 
   \d^{k_{a_2 - 1}}_{j_{a-1}} \cd \nn \\
   & & \cdot (-1)^{\lb \ep(i_{a_1 + 2}) + \ep(j_{a_1 + 2}) \rb \ep(k_1)} \d^{k_2}_{j_{a_1 + 2}} 
   \d^{k_1}_{j_{a_1 + 1}} \nn \\
   & & \cdot (-1)^{\lb \ep(i_{a_1 + 1}) + \cd + \ep(i_a) + \ep(k_{a_2+1}) + \cd + \ep(k_{p-1}) \rb 
   \lb \ep(i_1) + \cd + \ep(i_{a_1}) + \ep(j_1) + \cd + \ep(j_{a_1}) \rb} \nn \\
   & & \cdot (-1)^{\lb \ep(i_{a_1}) + \ep(j_{a_1}) \rb \lb \ep(k_p) + \ep(k_{p+1}) + \cd + \ep(k_{c-1}) \rb}
   \d^{k_c}_{j_{a_1}} \nn \\
   & & \cdot (-1)^{\lb \ep(i_{a_1 - 1}) + \ep(j_{a_1 - 1}) \rb \lb \ep(k_p) + \ep(k_{p+1}) + \cd + \ep(k_{c-2}) \rb}
   \d^{k_{c-1}}_{j_{a_1 - 1}} \cd \nn \\
   & & \cdot (-1)^{\lb \ep(i_2) + \ep(j_2) \rb \ep(k_p)} \d^{k_{p+1}}_{j_2} \d^{k_p}_{j_1} 
   \ps^{i_{a_1 + 1} i_{a_1 + 2} \ld i_a k_{a_2 + 1} k_{a_2 + 2} \ld k_{p - 1} i_1 i_2 \ld i_{a_1}},
\la{2.7.5}
\eeq
where $a_1 + a_2 = a$.  Again we can turn Eq.(\ref{2.7.5}) into
\beq
   \lefteqn{(-1)^{\lb \ep(i_a) + \ep(j_a) \rb \lb \ep(j_1) + \ep(j_2) + \cd + \ep(j_{a-1}) \rb }
   (-1)^{\lb \ep(i_{a-1}) + \ep(j_{a-1}) \rb \lb \ep(j_1) + \ep(j_2) + \cd + \ep(j_{a-2}) \rb } \cd } \nn \\
   \lefteqn{\cdot (-1)^{\lb \ep(i_2) + \ep(j_2) \rb \ep(j_1)} X^{i_1 j_1}_p X^{i_2 j_2}_{p+1} \cd 
   X^{i_a j_a}_{p+a-1} \ps^K = } \nn \\
   & & (-1)^{\lb \ep(j_{a_1 + 1}) + \ep(j_{a_1 + 2}) + \cd + \ep(j_a) + \ep(k_{a_2 + 1}) + \ep(k_{a_2 + 2}) +
   \cd + \ep(k_{p-1}) \rb \lb \ep(j_1) + \ep(j_2) + \cd + \ep(j_{a_1}) \rb } \nn \\
   & & \cdot (-1)^{\lb \ep(i_{a_1 + 1}) + \ep(i_{a_1 + 2}) + \cd + \ep(i_a) + \ep(k_{a_2 + 1}) + \ep(k_{a_2 + 2}) 
   + \cd + \ep(k_{p-1}) \rb \lb \ep(i_1) + \ep(i_2) + \cd + \ep(i_{a_1}) \rb } \nn \\
   & & \cdot \d^{k_p}_{j_1} \d^{k_{p+1}}_{j_2} \cd \d^{k_c}_{j_{a_1}} \d^{k_1}_{j_{a_1 + 1}} \d^{k_2}_{j_{a_1 + 2}}
   \cd \d^{k_{a_2}}_{j_a} \ps^{i_{a_1 + 1} i_{a_1 + 2} \ld i_a k_{a_2 + 1} k_{a_2 + 2} \ld k_{p-1} i_1 i_2 \ld 
   i_{a_1}}.
\la{2.7.6}
\eeq

We are now ready to identify quantum matrix models with quantum spin chains.  Consider the following weighted sum 
of all cyclic permutations of the states of a quantum spin chain in the following way:
\beq
   \Ps^K = \sum_{K_1 \dot{K}_2 = K} (-1)^{\ep(K_1) \ep(\dot{K}_2)} \ps^{\dot{K}_2 K_1}.
\la{2.7.7}
\eeq
This $\Ps^K$ obeys Eq.(\ref{2.2.7}).  Eqs.(\ref{2.7.4}), (\ref{2.7.6}) and (\ref{2.7.7}) together then imply that
\beq
   \lefteqn{(-1)^{\lb \ep(i_a) + \ep(j_a) \rb \lb \ep(j_1) + \ep(j_2) + \cd + \ep(j_{a-1}) \rb}} \nn \\ 
   \lefteqn{\cdot (-1)^{\lb \ep(i_{a-1}) + \ep(j_{a-1}) \rb \lb \ep(j_1) + \ep(j_2) + \cd + \ep(j_{a-2}) \rb} \cd} 
   \nn \\
   \lefteqn{\cdot (-1)^{\lb \ep(i_2) + \ep(j_2) \rb \ep(j_1) } \sum_{p=1}^n X^{i_1 j_1}_p X^{i_2 j_2}_{p+1} \cd 
   X^{i_a j_a}_{p+a-1} \Ps^K = } \nn \\
   & & \d^K_J \Ps^I + \sum_{K_1 K_2 = K} (-1)^{\ep(K_1) \ep(K_2)} \d^{K_2 K_1}_J \Ps^I
   + \sum_{K_1 K_2 = K} \d^{K_1}_J \Ps^{I K_2} \nn \\
   & & + \sum_{K_1 K_2 K_3 = K} (-1)^{\ep(K_1) \lb \ep(K_2) + \ep(K_3) \rb} \d^{K_2}_J \Ps^{I K_3 K_1} \nn \\
   & & + \sum_{K_1 K_2 = K} (-1)^{\ep(K_1) \ep(K_2)} \d^{K_2}_J \Ps^{I K_1} \nn \\
   & & + \sum_{J_1 J_2 = J} \sum_{K_1 K_2 K_3 = K} (-1)^{\ep(K_3) \lb \ep(K_1) + \ep(K_2) \rb} \d^{K_3}_{J_1}
   \d^{K_1}_{J_2} \Ps^{I K_2}.
\la{2.7.8}
\eeq
Comparing Eq.(\ref{2.7.8}) with Eq.(\ref{2.2.20}) then finally yields
\beq
   \g^I_J & = & (-1)^{\lb \ep(i_a) + \ep(j_a) \rb \lb \ep(j_1) + \ep(j_2) + \cd + \ep(j_{a-1}) \rb } \nn \\
   & & \cdot (-1)^{\lb \ep(i_{a-1}) + \ep(j_{a-1}) \rb \lb \ep(j_1) + \ep(j_2) + \cd + \ep(j_{a-2}) \rb } \cd
   (-1)^{\lb \ep(i_2) + \ep(j_2) \rb \ep(j_1)} \nn \\
   & & \cdot \sum_{p=1}^c X^{i_1 j_1}_p X^{i_2 j_2}_{p+1} \cd X^{i_a j_a}_{p+a-1}.
\la{2.7.9}
\eeq
As the Hamiltonian of a typical matrix model is of the form $\sum_{I, J} h^J_I \g^I_J$, where only a finite number 
of $h^J_I$'s are non-zero, Eq.(\ref{2.7.9}) provides us a way to transcribe the matrix model into a quantum spin 
chain with fermions and/or bosons, if for each non-zero $h^J_I$ the numbers of integers in $I$ and $J$ are the same.

Let us see how the above abstract formalism is put into practice in representative models.
\begin{itemize}
\item{\em the Ising model} \cite{onsager, frsu, kogut}. \\
The one-dimensional quantum Ising model is perhaps the simplest exactly integrable model.  Its Hamiltonian 
$H^{\rm spin}_{\rm Ising}$ is
\beq
   H^{\rm spin}_{\rm Ising}(\tau, \lambda) = \sum_{p=1}^c \tau^z_p + \lambda \sum_{p=1}^c \tau^x_p \tau^x_{p+1}.
\label{2.7.10}
\eeq
Here $\lambda$ is a constant, and $\tau^{x,y,z}_p$ are Pauli matrices at the $p$-th site.  Moreover, 
$\t^{x,y,z}_{c+1} = \t^{x,y,z}_1$.  Two Pauli matrices at different sites (i.e., with different subscripts) commute 
with each other.  This Ising spin chain is equivalent to a bosonic two-matrix model, as follows.  We set the 
bosonic quantum state 1 in the matrix model to correspond to the spin-up state in the Ising spin chain, and the 
bosonic quantum state 2 to correspond to the spin-down state.  Since
\begin{equation}
   \tau^z_p = X^{11}_p - X^{22}_p, \; \tau^x_p + {\rm i} \tau^y_p = 2 X^{12}_p \; \mbox{and} \;
   \tau^x_p - {\rm i} \tau^y_p = 2 X^{21}_p,
\la{2.7.11}
\end{equation}
The Hamiltonian can be rewritten as
\beq
	H^{\rm matrix}_{\rm Ising} = H_0 + \lambda V
\eeq
where
\beq
   H_0 & = & \g^1_1 - \g^2_2 \mbox{; and} \nonumber \\
     V & = & \left[ \g^{22}_{11} + \g^{21}_{12} + \g^{12}_{21} + \g^{11}_{22} \right].
\label{2.7.12}
\eeq
This is a two-matrix model with the Hamiltonian
\beq
   H^{\rm matrix}_{\rm Ising} & = & \Tr \le[ a^{\da}(1) a(1) - a^{\da}(2) a(2) \ri]
   + \frac{\l}{N} \Tr \le[ a^{\da}(2) a^{\da}(2) a(1) a(1) \right. \nonumber \\
   & & + a^{\da}(2) a^{\da}(1) a(2) a(1) + a^{\da}(1) a^{\da}(2) a(1) a(2) \nonumber \\
   & & \left. + a^{\da}(1) a^{\da}(1) a(2) a(2) \right] .
\label{2.7.13}
\eeq
Our results, along with known results of the Ising spin chain \cite{kogut}, give the
spectrum of this matrix model in the large-$N$ limit:
\beq
	E(n_p, c, \lambda) = -2\sum_{p=-c}^c \le( 1 + 2 \lambda \cos \le[ \frac{2 \pi p}{2c +1} \ri] + 
			     \lambda^2 \ri)^{1/2} n_p,
\label{2.7.14}
\eeq
where $n_p = 0$ or 1.  Also, we must impose the condition $\sum_{p=-c}^c n_p p=0$ to get cyclically symmetric 
states.  Let us underscore that {\em the matrix model defined by the Hamiltonian $H^{\ma}_{\rm Ising}$ in 
Eqs.(\ref{2.7.12}) or (\ref{2.7.13}) is an integrable matrix model in the large-$N$ limit}.  We will come back to
this model again in Chapter~\ref{c5}.
\item{\em the chiral Potts model} \cite{hokani, geri, aumcpetaya, bapeau}. \\
This is a bosonic model in which the number of quantum states available for a site is not restricted to 2 but is 
any finite positive integer.  The Hamiltonian is
\beq
	H^{{\rm spin}}_{{\rm CP}}=\sum_{p=1}^c \sum_{k=1}^{\L -1}[\tilde \alpha_k Q_p^k + 
        \lambda \alpha_k P_p^k P^{\L-k}_{p+1}],
\la{2.7.15}
\eeq
where
\begin{eqnarray*}
   & \alpha_k = \frac{{\rm e}^{{\rm i} \phi (\frac{2k}{n} - 1)}}{\sin \frac{\pi k}{n}}, \;
   \tilde{\alpha}_k = \frac{{\rm e}^{{\rm i} \varphi (\frac{2k}{n} - 1)}}{\sin \frac{\pi k}{n}}, \;
   \cos \varphi = \lambda \cos \phi, \; \omega = {\rm e}^{\frac{2 \pi {\rm i}}{n}}, & \nonumber \\
   & \sigma_j = \left( \begin{array}{ccccc}
   			1 & 0 & 0 & \ldots & 0 \\
   			0 & \omega & 0 & \ldots & 0 \\
   			0 & 0 & \omega^2 & \ldots & 0 \\
   			& & \ldots & & \\
   			0 & 0 & 0 & \ldots & \omega^{n-1}
   		     \end{array} \right), &
\end{eqnarray*}
and
\[ \Gamma_j = \left( \begin{array}{cccccc}
   			0 & 0 & 0 & \ldots & 0 & 1 \\
   			1 & 0 & 0 & \ldots & 0 & 0 \\
   			0 & 1 & 0 & \ldots & 0 & 0 \\
   			  &   &   & \ldots &   &   \\
   			0 & 0 & 0 & \ldots & 1 & 0
   		     \end{array} \right). \]
This model is exactly solvable by the Yang-Baxter method.  The Hamiltonian of the associated solvable multi-matrix 
model is
\beq
   H^{\rm matrix}_{\rm CP} = \sum_{k=1}^{\L-1} \left[ \tilde\alpha_k \sum_{j=1}^{\L} \omega^{k(j-1)} \g^j_j +
   \lambda \alpha_k \sum_{j_1, j_2 = 1}^{\L} \g^{j_1 + k, j_2 - k}_{j_1,j_2} \right]
\la{2.7.16}
\eeq
where $j_1 + k$ should be replaced with $j_1 + k - \L$ if $j_1 + k > \L$ and $j_2 - k$ should be replaced with
$j_2 + \L - k$ if $j_2 - k \leq 0$ in $g^{j_1 + k, j_2 - k}_{j_1, j_2}$ in the above equation.
\item{the Hubbard Model} \cite{hubbard}. \\
We will see that the matrix model associated with the Hubbard model is a model with two fermionic and two bosonic 
states.  Let $c_{p, \uar}$ and $c_{p, \dar}$ be the annihilation operators of spin-up and spin-down electrons at 
the $p$-th site respectively.  The Hamiltonian of the Hubbard model is
\beq
   H_{\rm Hubbard} = - \sum_{p=1}^n \sum_{\s = \uar, \dar} \le( c^{\da}_{p\s} c_{p+1, \s} + 
   c^{\da}_{p+1, \s} c_{p\s} \ri) + U \sum_{p=1}^n n_{p\uar} n_{p\dar},
\la{2.7.17}
\eeq
where $n_{p\uar} = c^{\da}_{p\uar} c_{p\uar}$ and $n_{p\dar} = c^{\da}_{p\dar} c_{p\dar}$ are the number operators 
for spin-up and spin-down states at the $p$-th site.  Rewrite the Hamiltonian in terms of Hubbard operators as 
follows.  Identify the states 1, 2, 3, 4 to be the vacuum state, the state with one spin-up and one spin-down 
electrons, the state with a spin-up electron only, and the state with a spin-down electron only, respectively.  
Then the following correspondence holds true:
\beq
   c_{p, \uar} \lrar A_p(3); \; c_{p, \dar} \lrar A_p(4) \mbox{; and} \; c_{p, \dar} c_{p, \uar} \lrar A_p(2).
\la{2.7.18}
\eeq
We can then use Eq.(\ref{2.7.18}) to rewrite Eq.(\ref{2.7.17}) as
\beq
   H_{\rm Hubbard} & = & - \sum_{p=1}^n \sum_{\s = 3, 4} \le( X^{\s 1}_p X^{1 \s}_{p+1} - 
   X^{1 \s}_p X^{\s 1}_{p+1} \ri) \nn \\
   & & - \sum_{p=1}^n \le( X^{31}_p X^{42}_{p+1} + X^{24}_p X^{13}_{p+1}
   + X^{14}_p X^{23}_{p+1} + X^{32}_p X^{41}_{p+1} \ri. \nn \\
   & & \le. - X^{42}_p X^{31}_{p+1} - X^{13}_p X^{24}_{p+1} - X^{23}_p X^{14}_{p+1} - X^{41}_p X^{32}_{p+1} \ri) 
   \nn \\
   & & - \sum_{p=1}^n \sum_{\s = 3, 4} \le( X^{2 \s}_p X^{\s 2}_{p+1} - X^{\s 2}_p X^{2 \s}_{p+1} \ri) \nn \\
   & & + U \sum_{p=1}^n X^{22}_p.
\la{2.7.19}
\eeq
Then Eq.(\ref{2.7.9}) tells us that the corresponding quantum matrix model is
\beq
   \lefteqn{ H^{\ma}_{\rm Hubbard} = - \le( \g^{34}_{12} + \g^{34}_{21} - \g^{43}_{12} - \g^{43}_{21} 
   + \g^{12}_{34} + \g^{21}_{34} - \g^{12}_{43} - \g^{21}_{43} \ri) } \nn \\
   & & - \le( \g^{13}_{31} + \g^{31}_{13} + \g^{14}_{41} + \g^{41}_{14} \ri) +
   \le( \g^{23}_{32} + \g^{32}_{23} + \g^{24}_{42} + \g^{42}_{24} \ri) \nn \\
   & & + U \g^2_2.
\la{2.7.20}
\eeq
Using the Bethe ansatz, Lieb and Wu showed that this model is exactly integrable \cite{liwu}.  
\item {\em t-J model} \cite{sutherland, schlottmann87, schlottmann97, bablog}. \\
It is not necessary for the number of bosonic and fermionic quantum states in a model to be the same in order for 
the model to be exactly integrable.  Consider, for instance, the Hamiltonian of the t-J model:
\beq
   H_{\rm tJ} & = & -t \sum_{p=1}^n \sum_{\s = \uar, \dar} P \le( c^{\da}_{i\s} c_{p+1, \s} +
   c^{\da}_{p+1, \s} c_{p\s} \ri) P \nn \\
   & & + \frac{J}{2} \sum_{p=1}^n \le[ {\bf S}_p \cdot {\bf S}_{p+1} - 
   \frac{1}{4} \le( n_{p\uar} + n_{p\dar} \ri) \le( n_{p+1, \uar} n_{p+1, \dar} \ri) \ri].
\la{2.7.21}
\eeq
In this equation,
\[ P = \prod_{p=1}^n (1 - n_{p\uar} n_{p\dar}) \]
is the projection operator to the collective state in which no site has 2 electrons.  ${\bf S}_i$ is the usual spin
operator:
\begin{eqnarray*}
   S^x_p & = & \frac{1}{2} \le( S_p + S^{\da}_p \ri); \\
   S^y_p & = & \frac{1}{2 {\rm i}} \le( S_p - S^{\da}_p \ri); \\
   S^z_p & = & \frac{1}{2} \le( n_{p\uar} - n_{p\dar} \ri); \\
   S^{\da}_p & = & c^{\da}_{p\dar} c_{p\uar}; \; \mbox{and} \\
   S_p & = & c^{\da}_{p\uar} c_{p\dar}.  
\end{eqnarray*}
If $J \ll t$, then this model is equivalent to the Hubbard model in the large-$U$ limit.  However, this model is in 
general different from the Hubbard model.  When the values of $J$ and $t$ are such that $J = 2t$, we say that the 
model is at the supersymmetric point, and it is integrable.  The corresponding integrable matrix model in the 
large-$N$ limit is
\beq
   H^{\ma}_{\rm tJ} = -t \le( \g^{0\uar}_{\uar 0} + \g^{\uar 0}_{0\uar} + \g^{0\dar}_{\dar 0} + 
   \g^{\dar 0}_{0 \dar} \ri) - \frac{J}{2} \le( \g^{\uar\dar}_{\dar\uar} + \g^{\dar\uar}_{\uar\dar} 
   - \g^{\uar\dar}_{\uar\dar} - \g^{\dar\uar}_{\dar\uar} \ri).
\la{2.7.22}
\eeq
Note that there are 2 fermionic states ($\uar$ and $\dar$) but only 1 bosonic state (0) in this integrable model.
\end{itemize}

Some other examples of exactly integrable quantum spin chain models and the corresponding quantum matrix models in 
the large-$N$ limit can be found in Refs.\cite{clstal} and \cite{plb}.

\section{Examples: Quantum Spin Chains (2)}
\la{s2.8}

So far, we have been discussing quantum spin chains satisfying the periodic boundary condition.  Quantum spin chains
satisfying open boundary conditions are also of considerable interest.  Recent years have seen significant progress 
in the understanding of the integrability of open spin chains.  Consider vertex models with open boundary 
conditions.  Sklyanin \cite{sklyanin} found that if a set of equations, now know as the Sklyanin equations 
\cite{gorusi}, are satisfied in a vertex model, then this model is integrable.  We can further derive an expression
for the associated one-dimensional quantum spin chain model. 

There is also a one-to-one correspondence between the open spin chains and quantum matrix models in the large-$N$ 
limit.  The relation between a $\g$ and Hubbard operators is almost the same as that in Eq.(\ref{2.7.9}), except 
that the summation index $p$ runs from 1 to $c - a + 1$ this time.  It can be easily seen that the corresponding 
Hubbard operators for an $l^I_J$ and an $r^I_J$ are
\beq
   l^I_J & = & (-1)^{\lb \ep(i_a) + \ep(j_a) \rb \lb \ep(j_1) + \ep(j_2) + \cd + \ep(j_{a-1}) \rb } \cdot \nn \\
   & & (-1)^{\lb \ep(i_{a-1}) + \ep(j_{a-1}) \rb \lb \ep(j_1) + \ep(j_2) + \cd + \ep(j_{a-2}) \rb } \cd
   (-1)^{\lb \ep(i_2) + \ep(j_2) \rb \ep(j_1)} \cdot \nn \\
   & & X^{i_1 j_1}_1 X^{i_2 j_2}_2 \cd X^{i_a j_a}_a
\la{2.8.1}
\eeq
and
\beq
   r^I_J & = & (-1)^{\lb \ep(i_a) + \ep(j_a) \rb \lb \ep(j_1) + \ep(j_2) + \cd + \ep(j_{a-1}) \rb } \cdot \nn \\
   & & (-1)^{\lb \ep(i_{a-1}) + \ep(j_{a-1}) \rb \lb \ep(j_1) + \ep(j_2) + \cd + \ep(j_{a-2}) \rb } \cd
   (-1)^{\lb \ep(i_2) + \ep(j_2) \rb \ep(j_1)} \cdot \nn \\
   & & X^{i_1 j_1}_{c-a+1} X^{i_2 j_2}_{c-a+2} \cd X^{i_a j_a}_c
\la{2.8.2}
\eeq
respectively.

Let us give representative examples to see how the transcription is put into practice.
\begin{itemize}
\item{integrable spin-$\frac{1}{2}$ XXZ model} \cite{albaba}. \\
The Hamiltonian $H^{\rm spin}_{\rm XXZ}$ of this spin chain model corresponding to the six-vertex model which 
satisfies the Sklyanin equations is
\begin{eqnarray}
   H^{\rm spin}_{\rm XXZ} & = & \frac{1}{2 \sin \gamma} \left[ \sum_{j=1}^{c-1} 
   (\tau_j^x \tau_{j+1}^x + \tau_j^y \tau_{j+1}^y + \cos \gamma \tau_j^z
   \tau_{j+1}^z)
   \right. \nonumber \\
   & & + \left. i \sin \gamma ( \coth \xi_- \tau_1^z + \coth \xi_+ \tau_c^z )
   \right],
\label{2.8.3}
\end{eqnarray}
where $\gamma \in (0, \pi)$ and both $\xi_-$ and $\xi_+$ are arbitrary constants.  The Hamiltonian 
$H^{\ma}_{\rm XXZ}$ of the associated integrable matrix model is
\begin{eqnarray}
   H^{\ma}_{\rm XXZ} & = & \frac{1}{2 \sin \gamma} \left\{ 2 ( \s_{12}^{21} + \s_{21}^{12} 
   + \cos \gamma (\s_{11}^{11} - \s_{12}^{12} - \s_{21}^{21} + \s_{22}^{22} ) \right. \nonumber \\
   & & + \im \left. \sin \gamma \lbrack \coth \xi_- (l_1^1 - l_2^2) + \coth \xi_+ (r_1^1 - r_2^2) \rbrack \right\}.
\label{2.8.4}
\end{eqnarray}
We can further rewrite this formula in terms of the creation and annihilation operators $a^{\dagger}$ and $a$:
\begin{eqnarray}
   \lefteqn{ H^{\ma}_{\rm XXZ} = \frac{1}{2N\sin\gamma} \left\{ 2 \left( 
   a^{\dagger\mu_2}_{\mu_1}(1) a^{\dagger\nu_2}_{\mu_2}(2) a^{\nu_1}_{\nu_2}(1) a^{\mu_1}_{\nu_1}(2) 
   \ri. \ri. } \nn \\
   & & \le. + a^{\dagger\mu_2}_{\mu_1}(2) a^{\dagger\nu_2}_{\mu_2}(1) a^{\nu_1}_{\nu_2}(2) a^{\mu_1}_{\nu_1}(1) 
   \right) \nonumber \\
   & & \cos\gamma \left( 
   a^{\dagger\mu_2}_{\mu_1}(1) a^{\dagger\nu_2}_{\mu_2}(1) a^{\nu_1}_{\nu_2}(1) a^{\mu_1}_{\nu_1}(1) - 
   a^{\dagger\mu_2}_{\mu_1}(1) a^{\dagger\nu_2}_{\mu_2}(2) a^{\nu_1}_{\nu_2}(2) a^{\mu_1}_{\nu_1}(1)
   \right. \nonumber \\
   & & - \left. 
   a^{\dagger\mu_2}_{\mu_1}(2) a^{\dagger\nu_2}_{\mu_2}(1) a^{\nu_1}_{\nu_2}(1) a^{\mu_1}_{\nu_1}(2) + 
   a^{\dagger\mu_2}_{\mu_1}(2) a^{\dagger\nu_2}_{\mu_2}(2) a^{\nu_1}_{\nu_2}(2) a^{\mu_1}_{\nu_1}(2)
   \right) \nonumber \\
   & & + \im \sin\gamma \left[ \coth\xi_- ( 
   \bar{q}^{\dagger\mu_1} a^{\dagger\mu_2}_{\mu_1}(1) a^{\mu_3}_{\mu_2}(1) \bar{q}_{\mu_3} - 
   \bar{q}^{\dagger\mu_1} a^{\dagger\mu_2}_{\mu_1}(2) a^{\mu_3}_{\mu_2}(2) \bar{q}_{\mu_3}   
   ) \right. \nonumber \\
   & & \left. \left. \coth\xi_+ (
   q^{\dagger}_{\mu_1} a^{\dagger\mu_1}_{\mu_2}(1) a^{\mu_2}_{\mu_3}(1) q^{\mu_3} - 
   q^{\dagger}_{\mu_1} a^{\dagger\mu_1}_{\mu_2}(2) a^{\mu_2}_{\mu_3}(2) q^{\mu_3}
   \right] \right\}.
\label{2.8.5}
\end{eqnarray}
\item{\em the Hubbard model} \cite{schultz, assu, zhou}.  Many Hubbard models with different open boundary 
conditions have been found to be exactly integrable.  One of them \cite{assu} has the Hamiltonian
\begin{eqnarray}
   \lefteqn{ H_{\rm Hubbard(o)} = - \sum_{i=1}^{n-1} \sum_{\s = \uar, \dar} 
   \le( c^{\da}_{i\s} c_{i+1, \s} + c^{\da}_{i+1, \s} c_{i\s} \ri) + U \sum_{i=1}^n n_{i\uar} n_{i\dar} } \nn \\
   & & + \mu \sum_{i=1}^n \le( n_{i\uar} + n_{i\dar} \ri) - p_{\uar} \le( n_{1\uar} + n_{n\uar} \ri)
   - p_{\dar} \le( n_{1\dar} + n_{n\dar} \ri).
\la{2.8.6}
\end{eqnarray}
Using Eqs.(\ref{2.7.9}) (subject to the remark in the second paragraph of this section), (\ref{2.8.1}) and 
(\ref{2.8.2}), we can derive the Hamiltonian of the associated matrix model:
\begin{eqnarray}
   \lefteqn{H^{\ma}_{\rm Hubbard(o)} = - \le( \g^{34}_{12} + \g^{34}_{21} - \g^{43}_{12} - \g^{43}_{21} 
   + \g^{12}_{34} + \g^{21}_{34} - \g^{12}_{43} - \g^{21}_{43} \ri) } \nn \\
   & & - \le( \g^{13}_{31} + \g^{31}_{13} + \g^{14}_{41} + \g^{41}_{14} \ri) +
   \le( \g^{23}_{32} + \g^{32}_{23} + \g^{24}_{42} + \g^{42}_{24} \ri) \nn \\
   & & - U \g^2_2 + \mu \le( \g^3_3 + \g^4_4 + 2 \g^2_2 \ri) \nn \\
   & & - p_{\uar} \le( l^3_3 + r^3_3 + l^2_2 + r^2_2 \ri) - p_{\dar} \le( l^4_4 + r^4_4 + l^2_2 + r^2_2 \ri).
\la{2.8.7}
\end{eqnarray}
\end{itemize}

Other examples of exactly integrable spin chain models satisfying open boundary conditions and the corresponding
quantum matrix models in the large-$N$ limit can be found in Refs.\cite{opstal} and \cite{plb}.

\chapter{A Lie Algebra for Open Strings}
\la{c4}

\section{Introduction}
\la{s4.1}

As we have just been seen in the previous chapter, quantum matrix models in the large-$N$ limit are of very wide 
applicability in physics.  This leads us naturally to the following question: can we do something more in addition
to writing down the physical observables of these models as linear combinations of the operators introduced in 
Section~\ref{s2.2}?  One major milestone in understanding the physics of these models would be to obtain the 
spectra of some of these observables, or at least to obtain their key qualitative properties.

As we have discussed in Section~\ref{s1.2}, a number of systematic analytic approaches have been developed to solve 
classical matrix models in the large-$N$ limit, which are also widely used in physics.  On the other hand, the 
method employed for studying the spectra of the Hamiltonians or supercharges of various versions of Yang--Mills 
theory in recent years is numerical in nature.  A lot of effort has been spent to develop more accurate and
sophisticated numerical analysis, and there is now a fairly large literature in the numerical solutions of these
quantum matrix models.  (Refs.\cite{dakl, hakl, masasa, anda96a, anda96b} are just a random sample of the 
literature in this area.)  It should be nice to see how much this approach can tell us about the various interesting
physical systems.  Nonetheless, it should also be nice if we have several other approaches to the same problem; 
some results which are difficult to obtain in one approach may be pretty trivial in another.  A combination of the 
knowledge gained from different approaches can lead us to much deeper understandings of the physics.

In Section~\ref{s1.1}, we have underscored the advantage of understanding the symmetry of a physical system in 
elucidating important features of it.  This suggests another approach: what is the symmetry of quantum matrix 
models?  A continuous symmetry in physics is usually expressed as a Lie group, or its infinitesimal form, a Lie 
algebra.  Is there a Lie algebra for quantum matrix models?  More specifically speaking, are the physical 
observables of quantum matrix models elements of a Lie algebra, or a Lie superalgebra?  

Yes, such a Lie superalgebra exists.

The derivation of this superalgebra is, however, not that easy to understand.  For pedagogical purpose, we will 
consider a special case in this chapter --- we will study the open strings with adjoint bosons, one degree of 
freedom for the fermion in the fundamental representation, and one degree of freedom for the anti-fermion in the
conjugate representation only.  The physical operators acting on these open strings form a Lie algebra, which we 
will call the {\em centrix algebra} $\hatcentrix$.  We will see that the centrix algebra has very rich mathematical 
properties.  There are a hierachy of subalgebras.  The most notable ones are the {\em extended Cuntz--Lie algebras} 
\cite{cuntz, evans} formed by all operators of the second and third kinds.  The centrix algebra itself is an 
extension of the Lie algebra of a set of outer derivations of the Cuntz algebra $\vectrix$ by a Lie algebra which 
we will call the {\em multix algebra} $\hatmultix$ and which is formed out of the extended Cuntz--Lie algebra.  In 
the simplest case when there is only one degree of freedom for the adjoint bosons, the centrix algebra reduces to 
an extension of the Witt algebra \cite{zassenhaus, chang, seligman}, the central extension of which is the Virasoro 
algebra \cite{virasoro}, by the commutative algebra of polynomials on a unit circle after the proper ideal formed 
by all finite-rank operators, i.e., all operators of the first kind, has been quotiented out.  Thus this set of 
outer derivations of the Cuntz algebra form a generalization of the Witt algebra, and plays an important role in 
governing the dynamics of many physical systems.

We will study $\hatcentrix$ progressively from its most easily understood subalgebra to its full version.  In 
Section~\ref{s4.2}, we will study the Lie algebra $\salt$ formed by the operators of the first kind.  We will see
that $\salt$ is equivalent to $\gl$ \cite{kac} and is well understood.  In Section~\ref{s4.3}, we will move on to 
study the Lie algebra $\hatleftix$ formed by the operators of the second kind.  We will see that this $\hatleftix$ 
is precisely the extended Cuntz--Lie algebra.  We will study its structure like its Cartan subalgebra and root 
vectors, and will see that $\salt$ form a proper ideal of $\hatleftix$.  If there is only one degree of freedom, 
i.e., $\L = 1$, the quotient algebra $\hat{L}_1 / \saltone$ will be the Toeplitz algebra \cite{murphy}.  In 
Section~\ref{s4.4}, we will study the Lie algebra $\hatrightix$ formed by the operators of the third kind.  This 
algebra is isomorphic to $\hatleftix$.  We will also consider the Lie algebra $\hatmultix$ formed by the operators 
of the first three kinds.  

Finally, in Section~\ref{s4.5}, we will study the centrix algebra $\hatcentrix$.  We will show in an accompanying
appendix that this is indeed a Lie algebra.  We will study its Cartan subalgebra, root vectors and ideals.  We will
see that $\hatleftix$, $\hatrightix$ and $\hatmultix$ are proper ideals of $\hatcentrix$.  In Section~\ref{s4.6}, 
we will consider the special case when there is only one degree of freedom.  We will see why the smallest quotient 
algebra of $\hat{\S}_1$ we know so far is equivalent to the Witt algebra. 

After acquiring some expertise in this kind of Lie algebraic arguments, we will study the full superalgebra, the 
{\em grand string superalgebra}, for all five kinds of operators, and some more subalgebras of it in subsequent 
chapters.

\section{An Algebra for Finite-Rank Operators}
\la{s4.2}

In this section, we are going to study operators of the first kind with no adjoint fermions, and restrict our 
attention to the action of the middle of these operators on adjoint matter.  We will see that these operators form
an associative algebra equivalent to $\gl$; it is thus not an essentially new object.

Consider the action of an operator of the first kind on an open string state.  We will ignore the action on the 
fundamental and conjugate matter fields; this corresponds to setting $\L_F=1$ temporarily. Then Eq.(\ref{2.2.10})
tells us that
\beq
   f^{\dot{I}}_{\dot{J}} s^{\dot{K}} = \d^{\dot{K}}_{\dot{J}} s^{\dot{I}}
\la{4.2.1}
\eeq
(c.f., Fig.\ref{f2.2}(c)).  In particular, $f^{\emptyset}_{\emptyset}$, where $\emptyset$ is the null sequence, is 
nothing but the projection operator to the state with no adjoint matter.  Define the composite operator 
$f^{\dot{I}}_{\dot{J}} f^{\dot{K}}_{\dot{L}}$ by the requirement that its action on $s^{\dot{M}}$ give 
$f^{\dot{I}}_{\dot{J}} (f^{\dot{K}}_{\dot{L}} s^{\dot{M}})$.   Then it follows that
\beq
   f^{\dot{I}}_{\dot{J}} f^{\dot{K}}_{\dot{L}} = \d^{\dot{K}}_{\dot{J}} f^{\dot{I}}_{\dot{L}}.
\la{4.2.2}
\eeq
Thus the $f$'s form an algebra.  Moreover, it is obvious that this algebra is the associative algebra of 
finite-rank matrices on the infinite-dimensional space spanned by all $s^{\dot{I}}$'s.  $f^{\dot{I}}_{\dot{J}}$ is
a Weyl matrix in this orthogonal basis.

Let us define a Lie algebra $\salt$ spanned by the $f^{\dot{I}}_{\dot{J}}$'s with the usual commutator 
\beq
   \lbrack f^{\dot{I}}_{\dot{J}}, f^{\dot{K}}_{\dot{L}} \rbrack \equiv 
   f^{\dot{I}}_{\dot{J}} f^{\dot{K}}_{\dot{L}} - f^{\dot{K}}_{\dot{L}} f^{\dot{I}}_{\dot{J}}.
\la{4.2.3}
\eeq
Then 
\beq
   \left[ f^{\dot{I}}_{\dot{J}}, f^{\dot{K}}_{\dot{L}} \right] = 
   \delta^{\dot{K}}_{\dot{J}} f^{\dot{I}}_{\dot{L}} - \delta^{\dot{I}}_{\dot{L}} f^{\dot{K}}_{\dot{J}}.
\la{4.2.4}
\eeq
The set of all `diagonal' elements $f^{\dot{I}}_{\dot{I}}$'s form a Cartan subalgebra of $\salt$, with each 
$f^{\dot{K}}_{\dot{L}}$, where $\dot{K} \neq \dot{L}$, forming a one-dimensional root vector space:
\beq
   \left[ f^{\dot{I}}_{\dot{I}}, f^{\dot{K}}_{\dot{L}} \right] = 
   \left( \delta^{\dot{K}}_{\dot{I}} - \delta^{\dot{I}}_{\dot{L}} \right) f^{\dot{K}}_{\dot{L}}.
\la{4.2.5}
\eeq

Eq.(\ref{4.2.5}) reveals clearly that $\salt$ is isomorphic to $\gl$, the Lie algebra, with the usual bracket, of 
all complex matrices $(a_{ij})_{i, j} \in Z_+$ such that the number of nonzero $a_{ij}$ is finite.  The isomorphism 
is given by a one-one correspondence between the multi-indices $I$ and the set of natural numbers $Z_+$,  for 
example by the lexicographic ordering introduced in Appendix~\ref{sa1.1}.  Thus each of the algebras $\salt$ with 
any value of $\L$ is isomorphic to $\gl$, the inductive limit of the general linear algebras.  $\gl$ has already 
been thoroughly studied \cite{kac, kape}.

\section{Operators of the Second Kind and the Cuntz Algebra}
\label{s4.3}

We now study the algebra formed by the adjoint portion of the operators of the second kind with no fermions.  We 
will see that this associative algebra is an infinite-dimensional associative algebra closely related to the Cuntz 
algebra.  The corresponding Lie algebra is a kind of current algebra in our theory.  In a special case, we will 
relate this current algebra with another well-known algebra, the Toeplitz algebra \cite{murphy}.  

Ignoring the action on the two ends of an open string, we can simplify Eq.(\ref{2.2.13}) to
\beq
   l^{\dot{I}}_{\dot{J}} s^{\dot{K}} =  
   \sum_{\dot{K}_1 \dot{K}_2 = \dot{K}} \d^{\dot{K}_1}_{\dot{J}} s^{\dot{I} \dot{K}_2}.  
\la{4.3.1}
\eeq
This action was illustrated in Fig.\ref{f2.3}(c).  Note that $l^{\emptyset}_{\emptyset}=1$, the identity operator.  

Let us consider the multiplication rule for the $l^{\dot{I}}_{\dot{J}}$'s.  Define the composite operator 
$l^{\dot{I}}_{\dot{J}} l^{\dot{K}}_{\dot{L}}$ by the requirement that its action on $s^{\dot{M}}$ give 
$l^{\dot{I}}_{\dot{J}} (l^{\dot{K}}_{\dot{L}} s^{\dot{M}})$.  Then it follows that
\beq
   l^{\dot{I}}_{\dot{J}} l^{\dot{K}}_{\dot{L}} = \delta^{\dot{K}}_{\dot{J}} l^{\dot{I}}_{\dot{L}} +
   \sum_{\dot{J}_1 J_2 = \dot{J}} \delta^{\dot{K}}_{\dot{J}_1} l^{\dot{I}}_{\dot{L} J_2} + 
   \sum_{\dot{K}_1 K_2 = \dot{K}} \delta^{\dot{K}_1}_{\dot{J}} l^{\dot{I} K_2}_{\dot{L}}.
\la{4.3.2}
\eeq
Here we sum over all possible sequences except that both $J_2$ and $K_2$ are required to be non-empty.  If we do 
not impose this non-empty restriction, there will be an overcounting problem.  $\dot{J_2}$ being empty and 
$\dot{K_2}$ being empty describe really the first term on the R.H.S. and we want to make sure that this term is 
counted only once in the sums. 

Thus the $l$'s form an algebra.  Moreover, this algebra is associative.  The associativity is proved in 
Appendix~\ref{sa4.1}.  We can now define the corresponding Lie algebra of the $l$'s as follows:
\beq
   \lbrack l^{\dot{I}}_{\dot{J}}, l^{\dot{K}}_{\dot{L}} \rbrack & = & 
   \delta^{\dot{K}}_{\dot{J}} l^{\dot{I}}_{\dot{L}} +
   \sum_{\dot{J}_1 J_2 = \dot{J}} \delta^{\dot{K}}_{\dot{J}_1} l^{\dot{I}}_{\dot{L} J_2} + 
   \sum_{\dot{K}_1 K_2 = \dot{K}} \delta^{\dot{K}_1}_{\dot{J}} l^{\dot{I} K_2}_{\dot{L}} \nonumber \\
   & & - \delta^{\dot{I}}_{\dot{L}} l^{\dot{K}}_{\dot{J}} -
   \sum_{\dot{L}_1 L_2 = \dot{L}} \delta^{\dot{I}}_{\dot{L}_1} l^{\dot{K}}_{\dot{J} L_2} - 
   \sum_{\dot{I}_1 I_2 = \dot{I}} \delta^{\dot{I}_1}_{\dot{L}} l^{\dot{K} I_2}_{\dot{J}}.
\la{4.3.3}
\eeq
The second and third terms on the R.H.S. of Eq.(\ref{4.3.3}) are shown in Fig.\ref{f4.1}(a) and (b), respectively.
We will call the Lie algebra defined by Eq.(\ref{4.3.3}) the {\em leftix algebra} or $\hatleftix$. (We justify this 
name as an abbreviation of `left multi-matrix algebra'.) 

\begin{figure}
\epsfxsize=4in
\centerline{\epsfbox{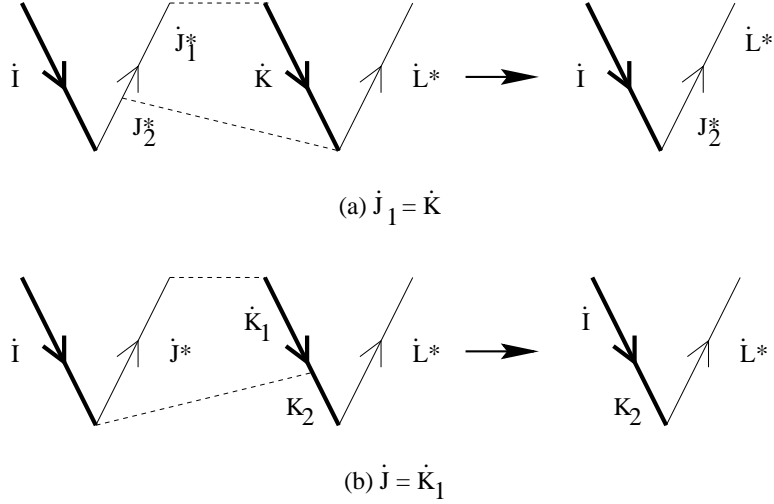}}
\caption{\em The leftix algebra.}
\label{f4.1}
\end{figure}

To understand the properties of this Lie algebra better, let us determine its Cartan subalgebra and the root spaces 
with respect to this Cartan subalgebra.  We will follow the definition of a Cartan subalgebra given by Humphreys 
\cite{humphreys}.  To understand this definition, we need a number of preliminary notions.  Let ${\cal L}$ be a Lie 
algebra.  Define the {\em descending central series} by ${\cal L}^0 = {\cal L}$, ${\cal L}^1 = \lb {\cal L}, 
{\cal L} \rb$, ${\cal L}^2 = \lb {\cal L}, {\cal L}^1 \rb$, \ld, ${\cal L}^i = \lb {\cal L}, {\cal L}^{i-1} \rb$.  
${\cal L}$ is {\em nilpotent} if ${\cal L}^n = 0$ for some $n$.  The {\em normalizer} of a subalgebra ${\cal K}$ of 
${\cal L}$ is defined by $N_{\cal L} ({\cal K}) = \{ x \in {\cal L} | \lb x, {\cal K} \rb \subset {\cal K} \}$.  A 
{\em Cartan subalgebra} of a Lie algebra ${\cal L}$ is a nilpotent subalgebra which is equal to its normalizer in 
${\cal L}$.  It turns out that all vectors of the form $l^{\dot{I}}_{\dot{I}}$, where $\dot{I}$ is either empty or 
an arbitrary finite integer sequence of integers between $1$ and $\L$ inclusive, span a Cartan subalgebra of the 
algebra $\hatleftix$.  The proof of this proposition can be found in Appendix~\ref{sa4.2}. 

Now we note the following identity:
\beq
    \left( l^{\dot{K}}_{\dot{L}} - \sum_{j=1}^{\L} l^{\dot{K}j}_{\dot{L}j} \right) s^{\dot{M}} = 
    \d^{\dot{M}}_{\dot{L}} s^{\dot{K}}.
\la{4.3.4}
\eeq
This equation tells us that the operator on the L.H.S. is nothing but $f^{\dot{K}}_{\dot{L}}$ we saw earlier:
\beq
   f^{\dot{K}}_{\dot{L}} = l^{\dot{K}}_{\dot{L}} - \sum_{j=1}^{\L} l^{\dot{K}j}_{\dot{L}j}.
\la{4.3.5}
\eeq
In particular, the projection operator to the state with no adjoint partons can be written as
\beq
   f^{\emptyset}_{\emptyset} = 1 - \sum_{j=1}^{\L} l^j_j.
\la{4.3.6}
\eeq
We can obtain from Eqs.(\ref{4.2.1}) and (\ref{4.3.1}) the following relation:
\beq
   \le[ l^{\dot{I}}_{\dot{J}}, f^{\dot{K}}_{\dot{L}} \ri] = 
   \sum_{\dot{K}_1 \dot{K}_2 = \dot{K}} \delta^{\dot{K}_1}_{\dot{J}} f^{\dot{I}\dot{K}_2}_{\dot{L}}
   - \sum_{\dot{L}_1 \dot{L}_2 = \dot{L}} \delta^{\dot{I}}_{\dot{L}_1} f^{\dot{K}}_{\dot{J} \dot{L}_2}.
\la{4.3.7}
\eeq
The first term on the R.H.S. of this formula is depicted in Fig.\ref{f4.2}.

\begin{figure}
\epsfxsize=4in
\centerline{\epsfbox{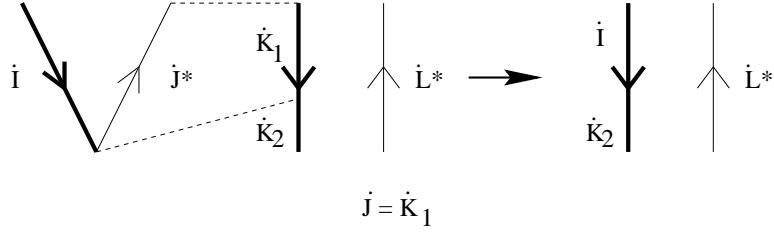}}
\caption{\em A root vector of the leftix algebra.}
\la{f4.2}
\end{figure}

It follows from Eq.(\ref{4.3.7}) that $\salt$ forms an ideal of the algebra $\hatleftix$.  This is a proper ideal 
because obviously {\em finite} linear combinations of $f^{\dot{I}}_{\dot{J}}$ do not span the whole leftix algebra.
The quotient $\leftix = \hatleftix / \salt$ is thus also a Lie algebra.  Put it another way, $\hatleftix$ is the 
extension of $\leftix$ by $\salt$: we have the exact sequence of Lie algebras
\[ 0 \to \salt \to \hatleftix \to \leftix \to 0. \]

$\hatleftix$, $\leftix$ and the above exact sequence have an intimate relationship with the Cuntz algebra.  To 
understand this, let us recall some basic properties of the Cuntz algebra now, following Evans \cite{evans}.  Let 
${\cal H}$ be a $\L$-dimensional Hilbert space spanned by $v^1$, $v^2$, \ld, and $v^{\L}$, and 
${\cal F} ({\cal H})$ be the full Fock space $\oplus_{r=0}^{\infty} (\otimes^r {\cal H})$, where 
$(\otimes^0 {\cal H})$ is a one-dimensional Hilbert space spanned by a unit vector $\Omega$, the 
`vacuum'\footnote{For us the state $\Omega$ is not exactly the vacuum, but rather has the physical meaning of the 
state with no adjoint matter.  This state still contains a fundamental and a conjugate matter field.}.  Then there 
is a bijective mapping between the Fock space of physical states spanned by open string states and the space 
${\cal F} ({\cal H})$.  The state $s^{\emptyset}$ is mapped to $\Omega$, and the state $s^K$ is mapped to 
$v^{k_1} \otimes v^{k_2} \otimes \cdots \otimes v^{k_c}$.  Define the operators
$a^{1\dagger}$, $a^{2\dagger}$, \ldots, $a^{{\Lambda} \dagger}$ as follows:
\begin{eqnarray}
   a^{i\dagger} v^{k_1} \otimes v^{k_2} \otimes \cdots \otimes v^{k_c} & = &
   v^i \otimes v^{k_1} \otimes v^{k_2} \otimes \cdots \otimes v^{k_c} \mbox{; and}
   \nonumber \\
   a^{i\dagger} \Omega & = & v^i
\la{4.3.8}
\end{eqnarray}
where $i$ is an integer between $1$ and $\Lambda$ inclusive.  The corresponding adjoint operators $a_1$, $a_2$, 
\ldots, $a_{{\Lambda}}$ have the following properties:
\begin{eqnarray}
   a_i v^{k_1} \otimes v^{k_2} \otimes \cdots \otimes v^{k_c} & = &
   \delta^{k_1}_i v^{k_2} \otimes \cdots \otimes v^{k_c} \mbox{; and} 
   \nonumber \\
   a_i \Omega & = & 0.
\la{4.3.9}
\end{eqnarray}
As a result, the $a$'s and $a^{\dagger}$'s satisfy the following properties:
\begin{eqnarray}
   a_i a^{j\dagger} & = & \delta^j_i \mbox{; and} 
\la{4.3.10} \\
   \sum_{i=1}^{{\Lambda}} a^{i\dagger} a_i & = & 1 - P_\Omega ,
\la{4.3.11}
\end{eqnarray}
where $P_{\Omega}$ is the projection operator to the vacuum $\Omega$.  Eqs.(\ref{4.3.10}) and (\ref{4.3.11}) give
the defining relations for the annihilation operators $a$ and creation operators $a^{\da}$ of the {\em extended 
Cuntz algebra}.  Furthermore,
\begin{equation}
   v^{k_1} \otimes v^{k_2} \otimes \cdots \otimes v^{k_c} =
   a^{k_1 \dagger} a^{k_2 \dagger} \cdots a^{k_c \dagger} \Omega .
\la{4.3.12}
\end{equation}
It is now straightforward to see that there is a one-to-one correspondence among the operators characterized by
Eq.(\ref{4.3.1}) and the operators acting on ${\cal F} ({\cal H})$.  $l^I_J$ corresponds to $a^{i_1 \dagger} 
a^{i_2 \dagger} \cdots a^{i_a \dagger} a_{j_b} a_{j_{b-1}} \cdots a_1$, $l^I_{\emptyset}$ to $a^{i_1 \dagger} 
a^{i_2 \dagger} \cdots a^{i_a \dagger}$, $l^{\emptyset}_J$ to $a_{j_b} a_{j_{b-1}} \cdots a_1$, and 
$l^\emptyset_\emptyset$ to the identity operator.

There is a technical difference between our point of view and that of Evans or Cuntz. They regard all algebras as 
Banach algebras\footnote{A Banach algebra is an algebra among elements in a complete normed space.  Its elementary
properties are discussed in Ref.\cite{kreyszig}.}, while we regard our algebras as {\em finite} linear combinations 
of basis elements.  For example, the analogue of our algebra $\salt$ will be the algebra ${\cal K}({\cal H})$ of 
{\em compact} operators\footnote{A compact linear operator is a linear operator $T$ from a normed space $X$ to a 
normed space $Y$ such that for every bounded subset $M$ of $X$, the closure of the image $T(M)$ is compact 
\cite{kreyszig}.} on ${\cal H} = C^{\L}$.  This is because $\cal K({\cal H})$ is just the completion of $\salt$ in 
the topology defined by the operator norm.  It follows from our previous discussion that ${\cal K} ({\cal H})$ is 
an ideal of the Banach algebra generated by $a$'s and $a^{\da}$'s.  If we quotient the extended Cuntz algebra by 
${\cal K} ({\cal H})$, then we get the {\em Cuntz algebra}, with the $a$'s and $a^{\da}$'s satisfying the following 
relations:
\beq
   a_i a^{j\dagger} & = & \delta^j_i \mbox{; and} 
\la{4.3.13} \\
   \sum_{i=1}^{{\Lambda}} a^{i\dagger} a_i & = & 1.
\la{4.3.14}
\eeq
The role of the last relation is to set all $f^{\dot{I}}_{\dot{J}}$'s to zero; i.e., to quotient by $\salt$.
Consequently, up to the issue of completeness in the operator norm, we can make these identifications: $\salt$ is 
the Lie algebra associated with the Banach algebra of compact operators on ${\cal H}$, $\hatleftix$ is the Lie 
algebra associated with the extended Cuntz algebra, or the {\em extended Cuntz--Lie algebra}, and $\leftix$ is the 
Lie algebra associated with the Cuntz algebra, or the {\em Cuntz--Lie algebra}.

We will see that in fact the presence of $\salt$ as a proper ideal is a generic feature of most of the algebras we 
will study. Quotienting by this ideal will get us the  essentially new algebras we are interested in studying. But 
it is only the extension that will have interesting representations.  It appears that the extension by 
$gl_{+\infty}$ plays a role in our theory that central extensions play in the theory of Kac--Moody and
Virasoro algebras. Similar extensions have appeared in previous approaches to current algebras \cite{mira}.

As a special case, we get from Eq.(\ref{4.3.7}) that:
\beq
   \left[ l^{\dot{I}}_{\dot{I}}, f^{\dot{K}}_{\dot{L}} \right] = \left(  
   \sum_{\dot{K}_1 \dot{K}_2 = \dot{K}} \delta^{\dot{K}_1}_{\dot{I}} 
   - \sum_{\dot{L}_1 \dot{L}_2 = \dot{L}} \delta^{\dot{I}}_{\dot{L}_1} \right) f^{\dot{K}}_{\dot{L}}.
\la{4.3.15}
\eeq
As a result, every $f^{\dot{K}}_{\dot{L}}$ is a root vector of $\leftix$.  Moreover, these are the only root 
vectors of the algebra $\hatleftix$. A proof that there are no root vectors other than the 
$f^{\dot{K}}_{\dot{L}}$'s will be provided in Appendix~\ref{sa4.3}.  We therefore conclude that every root space is 
one-dimensional.

We can get further insight of $\hatleftix$ and $\leftix$ by considering the special case $\L = 1$.  Then all 
integer sequences are repetitions of the number 1 a number of times.  We can then simplify the notations and write 
$s^{\dot{K}}$ as $s^{\#(\dot{K})}$ and $l^{\dot{I}}_{\dot{J}}$ as $l^{\#(\dot{I})}_{\#(\dot{J})}$.  We can deduce 
from Eq.(\ref{4.3.1}) that the action of $l^{\dot{a}}_{\dot{b}}$, where $\dot{a}$ and $\dot{b}$ are the numbers of 
integers in the various sequences, on $s^{\dot{c}}$, where $\dot{c}$ is also a non-negative integer, is given by
\begin{equation}
   l^{\dot{a}}_{\dot{b}} s^{\dot{c}} = \theta (\dot{b} \leq \dot{c}) s^{\dot{a} + \dot{c} - \dot{b}},
\la{4.3.16}
\end{equation}
where $\theta( \mbox{condition} )$ is 1 if the condition holds, and 0 otherwise.  The Lie bracket Eq.(\ref{4.3.3}) 
is simplified to
\begin{equation}
   \left[ l^{\dot{a}}_{\dot{b}}, l^{\dot{c}}_{\dot{d}} \right] = 
   \theta (\dot{c} \leq \dot{b}) l^{\dot{a}}_{\dot{b} + \dot{d} - \dot{c}} +
   \theta (\dot{b} < \dot{c}) l^{\dot{a} + \dot{c} - \dot{b}}_{\dot{d}} - 
   \theta (\dot{a} \leq \dot{d}) l^{\dot{c}}_{\dot{b} + \dot{d} - \dot{a}} -
   \theta (\dot{d} < \dot{a}) l^{\dot{a} + \dot{c} - \dot{d}}_{\dot{b}}.
\la{4.3.17}
\end{equation}
The set of all vectors of the form $l^{\dot{a}}_{\dot{a}}$ span a Cartan subalgebra.  The root vectors are given by 
$f^{\dot{c}}_{\dot{d}} = l^{\dot{c}}_{\dot{d}} - l^{\dot{c} + 1}_{\dot{d} + 1}$.  This can be deduced from 
Eq.(\ref{4.3.5}).  The action of $f^{\dot{a}}_{\dot{b}}$ on $s^{\dot{c}}$, which can be derived from 
Eq.(\ref{4.2.1}), is
\begin{equation}
   f^{\dot{a}}_{\dot{b}} s^{\dot{c}} = \delta^{\dot{c}}_{\dot{b}} s^{\dot{a}}.
\label{4.3.18}
\end{equation}
The corresponding eigenequation, which can be deduced from Eq.(\ref{4.3.15}), is
\begin{equation}
   \left[ l^{\dot{a}}_{\dot{a}}, f^{\dot{c}}_{\dot{d}} \right] = 
   \left( \theta (\dot{a} \leq \dot{c}) - \theta (\dot{a} \leq \dot{d}) \right) f^{\dot{c}}_{\dot{d}}.
\la{4.3.19}
\end{equation}

As in an earlier discussion, the subspace ${{\mathit F}_1}$ spanned by all the vectors of the form 
$f^{\dot{c}}_{\dot{d}}$ form a proper ideal of this $\Lambda=1$ algebra $\hat{\mathit L}_1$.  We can deduce from 
Eq.(\ref{4.3.7}) that
\begin{equation}
   \left[ l^{\dot{a}}_{\dot{b}}, f^{\dot{c}}_{\dot{d}} \right] = 
   \theta (\dot{b} \leq \dot{c}) f^{\dot{a} + \dot{c} - \dot{b}}_{\dot{d}} - 
   \theta (\dot{a} \leq \dot{d}) f^{\dot{c}}_{\dot{b} + \dot{d} - \dot{a}}. 
\la{4.3.20}
\end{equation}
We can now form the quotient algebra of cosets of the form $v + \salt$, where $v$ is an arbitrary vector of the 
algebra $\hat{\mathit L}_1$.  This quotient algebra is spanned by the cosets $l^{\dot{a}}_0 + {\salt}$ and 
$l^0_{\dot{b}} + {{\mathit F}_1}$, where $a$ and $b$ run over all $0, 1, \cdots, \infty$.  It is straightforward to 
show that the following Lie brackets hold:
\begin{eqnarray}
   \left[ l^{\dot{a}}_0 + {{\mathit F}_1}, l^{\dot{c}}_0 + {{\mathit F}_1} \right] & = & {{\mathit F}_1} ;  
   \nonumber \\
   \left[ l^{\dot{a}}_0 + {{\mathit F}_1}, l^0_{\dot{d}} + {{\mathit F}_1} \right] & = & {{\mathit F}_1} 
   \mbox{; and} \nonumber \\
   \left[ l^0_{\dot{b}} + {{\mathit F}_1}, l^0_{\dot{d}} + {{\mathit F}_1} \right] & = & {{\mathit F}_1}.
\label{4.3.21}
\end{eqnarray}
Therefore, this quotient algebra is an Abelian algebra.  Furthermore, consider Eq.(\ref{4.3.11}), one of the 
relations for the extended Cuntz algebra .  When $\Lambda = 1$, the superscript in $a^{\dagger}$ and the subscript 
in $a$ have to be 1 and so there is no danger of ambiguity to omit them.  This relation is now simplified to: 
\beq
   a a^{\dagger} = 1 .
\label{4.3.22}
\eeq
This is the defining relation for the {\em Toeplitz algebra} \cite{murphy}.  Note that the commutator between $a$ 
and $a^{\dagger}$ is a finite-rank operator, the projection operator to the vacuum state.  Thus if we quotient the 
Toeplitz algebra by ${\mathit F}_1$, we will get an Abelian algebra generated by the operators satisfying
\beq
   aa^{\dag}=1, \quad \; \hbox{\rm and} \; a^{\dag}a=1.
\la{4.3.23}
\eeq
This is just the algebra of functions on a circle, and is consistent with the fact that the quotient algebra 
characterized by Eq.(\ref{4.3.21}) is Abelian.  Thus we can regard the Cuntz algebra as a non-commutative
multi-dimensional generalization of the algebra of functions on the circle.

\section{Current Algebra in the Large-$N$ Limit}
\la{s4.4}

In this section, we will study the Lie Algebra for all operators of the third kind, and the Lie algebra spanned by
the operators of the first three kinds.

The theory of the Lie algebra $\hatrightix$ spanned by the operators of the third kind is closely parallel to that 
of $\hatleftix$.  Considering only the action of an operator of the third kind on the adjoint matter portion of an
open string state in Eq.(\ref{2.2.16}) gives
\beq
   r^{\dot{I}}_{\dot{J}} s^{\dot{K}} =  
   \sum_{\dot{K}_1 \dot{K}_2 = \dot{K}} \delta^{K_2}_{\dot{J}} s^{\dot{K}_1 \dot{I}}.   
\label{4.4.1}
\eeq

The Lie bracket between two $r$'s is:
\begin{eqnarray}
   \lbrack r^{\dot{I}}_{\dot{J}}, r^{\dot{K}}_{\dot{L}} \rbrack & = & 
   \delta^{\dot{K}}_{\dot{J}} r^{\dot{I}}_{\dot{L}} +
   \sum_{J_1 \dot{J}_2 = \dot{J}} \delta^{\dot{K}}_{\dot{J}_2} l^{\dot{I}}_{J_1 \dot{L}} + 
   \sum_{K_1 \dot{K}_2 = \dot{K}} \delta^{\dot{K}_2}_{\dot{J}} l^{K_1 \dot{I}}_{\dot{L}} \nonumber \\
   & & - \delta^{\dot{I}}_{\dot{L}} l^{\dot{K}}_{\dot{J}} -
   \sum_{L_1 \dot{L}_2 = \dot{L}} \delta^{\dot{I}}_{\dot{L}_2} l^{\dot{K}}_{L_1 \dot{J}} - 
   \sum_{I_1 \dot{I}_2 = \dot{I}} \delta^{\dot{I}_2}_{\dot{L}} l^{I_1 \dot{K}}_{\dot{J}}.
\label{4.4.2}
\end{eqnarray}
A vector in $\salt$ can be expressed as an element in $\hatrightix$ as well:
\begin{equation}
   f^{\dot{K}}_{\dot{L}} = r^{\dot{K}}_{\dot{L}} - \sum_{i=1}^{\Lambda} r^{i\dot{K}}_{i\dot{L}}.
\label{4.4.3}
\end{equation}
The Lie bracket between an $r$ and an $f$ is
\begin{equation}
   \left[ r^{\dot{I}}_{\dot{J}}, f^{\dot{K}}_{\dot{L}} \right] = 
   \sum_{\dot{K}_1 \dot{K}_2 = \dot{K}} \delta^{\dot{K}_2}_{\dot{J}} f^{\dot{K}_1 \dot{I}}_{\dot{L}}
   - \sum_{\dot{L}_1 \dot{L}_2 = \dot{L}} \delta^{\dot{I}}_{\dot{L}_2} f^{\dot{K}}_{\dot{L}_1 \dot{J}}.
\label{4.4.4}
\end{equation}

The algebra $\hatmultix$ spanned by all operators of the forms $f^{\dot{I}}_{\dot{J}}$, $l^{\dot{I}}_{\dot{J}}$, and
$r^{\dot{I}}_{\dot{J}}$ together is also interesting.  $\hatleftix$ and $\hatrightix$ are subalgebras of 
$\hatmultix$.  In addition, the Lie bracket between an element of $\hatleftix$ and an element of $\hatrightix$ is
\begin{equation}
   \left[ l^{\dot{I}}_{\dot{J}}, r^{\dot{K}}_{\dot{L}} \right] = 
   \sum_{\begin{array}{l}
   	    \dot{J}_1 \dot{J}_2 = \dot{J} \\
   	    \dot{K}_1 \dot{K}_2 = \dot{K}
   	 \end{array}} 
   \delta^{\dot{K}_1}_{\dot{J}_2} f^{\dot{I} \dot{K}_2}_{\dot{J}_1 \dot{L}} -
   \sum_{\begin{array}{l}
   	    \dot{I}_1 \dot{I}_2 = \dot{I} \\
   	    \dot{L}_1 \dot{L}_2 = \dot{L}
   	 \end{array}}
   \delta^{\dot{I}_2}_{\dot{L}_1} f^{\dot{I}_1 \dot{K}}_{\dot{J} \dot{L}_2}.
\label{4.4.5}
\end{equation}
A heuristic way to see that Eq.(\ref{4.4.5}) is true is by verifying that the action of the R.H.S. on an arbitrary 
state $s^{\dot{M}}$ gives $l^{\dot{I}}_{\dot{J}} (r^{\dot{K}}_{\dot{L}} s^{\dot{M}}) - r^{\dot{K}}_{\dot{L}} 
(l^{\dot{I}}_{\dot{J}} s^{\dot{M}})$.  This, however, does not automatically imply that this definition of the Lie 
bracket between an $l$ and an $r$ satisfies the Jacobi identity because Eqs.(\ref{4.3.5}) and (\ref{4.4.3}) 
together imply that the set of all finite linear combinations of the operators in $\hatleftix$ and $\hatrightix$ is 
not linearly independent.  We will properly justify Eq.(\ref{4.4.5}) in Chapter~\ref{c5} where we will see that we 
can treat $\multix$ as a subalgebra of yet another larger Lie algebra.  The first term on the R.H.S. of the above 
equation is depicted in Fig.\ref{f4.3}.

\begin{figure}
\epsfxsize=4in
\centerline{\epsfbox{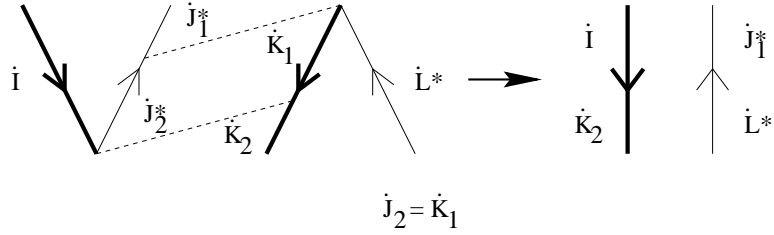}}
\caption{\em The multix algebra.}
\label{f4.3}
\end{figure}

Eq.(\ref{4.4.5}) tells us that $\hatleftix$ and $\hatrightix$ are proper ideals of $\hatmultix$.  In particular, 
$\salt$ is a proper ideal of $\hatmultix$, so there exists the quotient algebra $\multix = \hatmultix / \salt$. 
Since the commutator between an $l$ and an $r$ is an element in $\salt$, $l$ and $r$ commute when they are viewed 
as elements in $\multix$.  In fact  $\multix$ is almost the direct sum of $\leftix$ and $\rightix$. The only 
subtlety is that the identity operator can be identified with either $l^\emptyset_\emptyset$ or 
$r^\emptyset_\emptyset$. In order not to count it twice, we must require $l^\emptyset_\emptyset - 
r^\emptyset_\emptyset = 0$ in $\leftix \oplus \rightix$ to get $\multix$. 

These operators are regularized versions of the left- and right-handed current operators of Yang--Mills theory. We 
see that the chiral structure of the current algebra is reproduced beautifully in the multix algebra. The fact that 
only the extension of the current algebra by $\gl$ has a representation on the space of open string states 
is also reminiscent of what happens in lower dimensional theories. There must be a way to extend this
representation on the space of open string states to a sort of Fock space of multi-open string states. 

We have summarized the relationship among the various algebras discussed in this section in Table~\ref{t4.1}.

\begin{table}
\centerline{
\begin{tabular}[t]{|| c | c | l | rcl ||} \hline
operators    & extended 	        & \multicolumn{1}{c|}{comment}
& \multicolumn{3}{c||}	{quotient algebra}			   	      \\
             & algebra                  &                                       & 		      &   &                 			       \\ \hline \hline
$f^{\dot{I}}_{\dot{J}}$      & ${\mathit F}_{\Lambda}$      &${\mathit F}_{\Lambda}\equiv  gl_{+\infty}$&  & &  \\ \hline
$l^{\dot{I}}_{\dot{J}}$      & $\hat{\mathit L}_{\Lambda}$  & ${\mathit F}_{\Lambda}$ is a proper
             						    & ${\mathit L}_{\Lambda}$ & $\equiv$ & $\hat{\mathit L}_{\Lambda} / {\mathit F}_{\Lambda}$  \\
$f^{\dot{K}}_{\dot{L}} = l^{\dot{K}}_{\dot{L}} -  l^{\dot{K}j}_{\dot{L}j}$& & ideal of $\hat{\mathit L}_{\Lambda}$. 	        & 		      &	  &					  	      \\ \hline
$r^{\dot{I}}_{\dot{J}}$      & $\hat{\mathit R}_{\Lambda}$  & ${\mathit F}_{\Lambda}$ is a proper & ${\mathit R}_{\Lambda}$ & $\equiv$ & $\hat{\mathit R}_{\Lambda} / {\mathit F}_{\Lambda}$  \\
$f^{\dot{K}}_{\dot{L}} = r^{\dot{K}}_{\dot{L}} - r^{i\dot{K}}_{i\dot{L}}$ & & ideal of $\hat{\mathit R}_{\Lambda}$. 	        & 		      &	  &					  	      \\ \hline
$l^{\dot{I}}_{\dot{J}}$ and $r^{\dot{I}}_{\dot{J}}$ & $\hat{\mathit M}_{\Lambda}$ & $\hat{\mathit M} \neq \hat{\mathit L}_{\Lambda} \oplus \hat{\mathit R}_{\Lambda}$ & 
  ${\mathit M}_{\Lambda}$ & $\equiv$ & $\hat{\mathit M}_{\Lambda} / {\mathit F}_{\Lambda}$ \\ \hline \hline
\end{tabular}}
\caption{\em Relationship among $\salt$, $\hatleftix$, $\hatrightix$ and $\hatmultix$.  The summation convention is 
adopted for repeated indices in this table.}
\label{t4.1}
\end{table}

\section{A Lie Algebra for the Operators of the Fourth Kind}
\la{s4.5}

In previous sections we studied the actions of the operators of the first three kinds, and obtained associative 
algebras closely related to known algebras such as the Cuntz algebra.  It turns out that {\em products} of 
operators of the fourth kind cannot be written as {\em finite} linear combinations of these operators; they do not 
span an algebra under multiplication.  However, the {\em commutator} of two operators of the fourth kind can be 
written as a finite linear combination of these operators. Thus operators of the fourth kind acting on open singlet 
states form a Lie algebra, which we will call the {\em centrix} algebra $\hatcentrix$.  This new Lie algebra we 
discover reduces to the Witt algebra \cite{zassenhaus, chang, seligman} in the special case $\Lambda=1$.  Thus the 
centrix algebra is a sort of generalization of the Lie algebra of vector fields on the circle in noncommutative 
geometry \cite{connes}. {\it A priori} many such generalizations are possible; we have identified the one that is 
relevant to the planar large-$N$ limit of matrix models.

We will see that there is a proper ideal $\salt'$ of the Lie algebra $\hatcentrix$ which is equivalent to $\gl$; 
hence, there is a quotient Lie algebra $\centrix = \hatcentrix / \salt'$.  Furthermore, we will see that 
$\hatleftix$ and $\hatrightix$ share some elements with $\hatcentrix$. This is perhaps a bit surprising since they
were originally introduced using operators acting on the ends of an open string state, and $\hatcentrix$ has no
effect on the ends. More precisely, the generators $l^I_J$ and $r^I_J$ with non-empty $I$ and $J$ are in fact some
linear combinations of $\s^I_J$.  We can form a quotient $\hatcentrix$ by the algebra among all $l^I_J$'s and 
$r^I_J$'s.  This is the essentially new object we have discovered.

Once again we will focus on the action of $\s^I_J$ on the adjoint matter portion of an open string state.  We can 
capture the essence of this action given in Eq.(\ref{2.2.19}) by the following equation\footnote{Of course, 
$\sigma^I_J$ acting on a state with no adjoint matter gives zero.}:
\begin{equation}
   \s^I_J s^{\dot{K}} = \sum_{\dot{K}_1 K_2 \dot{K}_3 = \dot{K}} \delta^{K_2}_{J} s^{\dot{K_1} I \dot{K_3}}.
\label{4.5.1}    
\end{equation}
This action was shown diagrammatically in Fig.\ref{f2.5}(c).  Using the identities in Appendix~\ref{sa1.1}, we can
rewrite the action of $\sigma$ as
\begin{equation}
   \s^I_J s^{\dot{K}} = \sum_{\dot{A}, \dot{B}} \delta^{\dot{K}}_{\dot{A} J \dot{B}} s^{\dot{A} I \dot{B}}.
\label{4.5.2}   
\end{equation}
This form is more convenient in some calculations.  From Eq.(\ref{4.5.1}), we can see that the set of $\s^I_J$'s 
with all possible non-empty sequences $I$ and $J$'s is linearly independent. 

\begin{figure}
\epsfxsize=5in
\centerline{\epsfbox{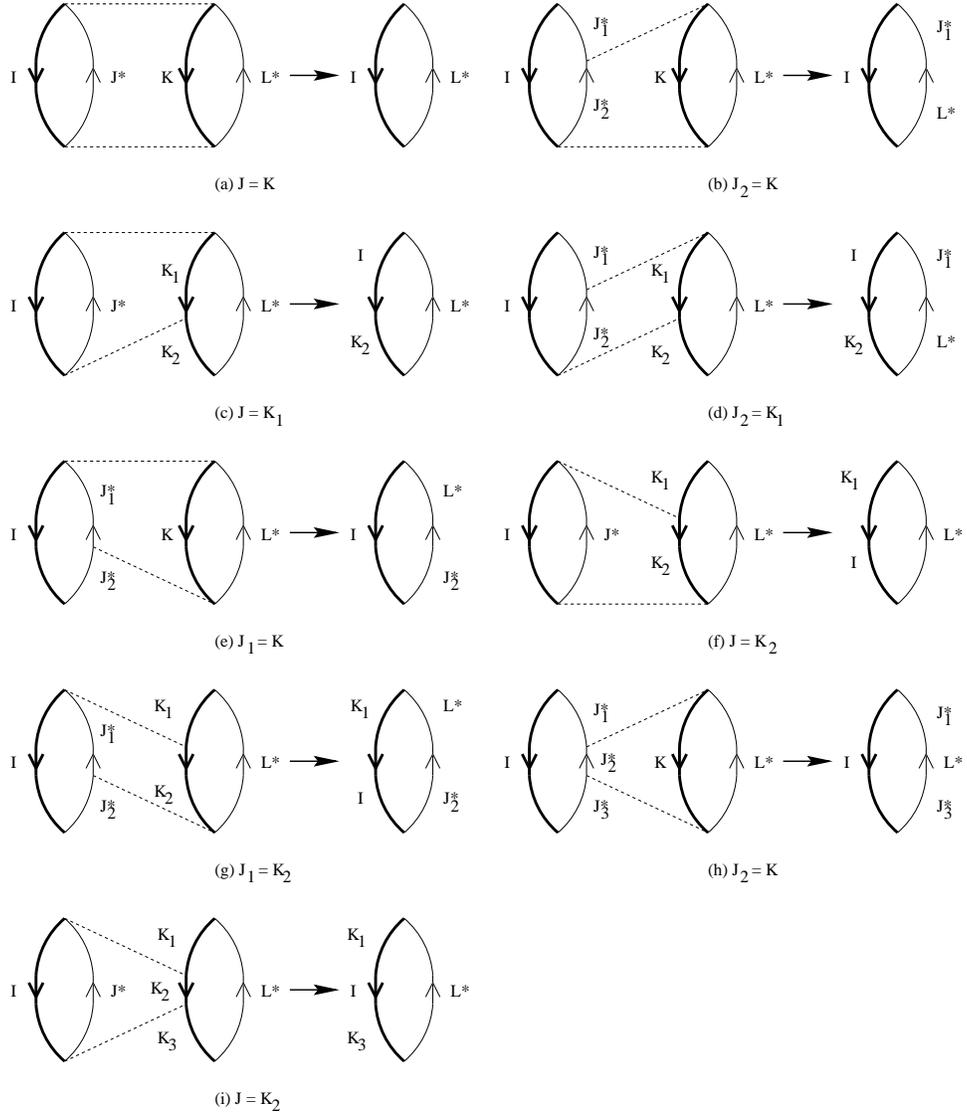}}
\caption{\em  The centrix or cyclix algebra.}
\label{f4.4}
\end{figure}

Unlike the case of the algebra $\hatleftix$ considered in Section~\ref{s4.3}, we cannot define the composite 
operator $\s^I_J \s^K_L$ by the requirement that its action on $s^{\dot{M}}$ give $\s^I_J (\s^K_L s^{\dot{M}})$ 
because in general this action cannot be written as a finite linear combination of the $s$'s.  (We will put off 
proving this assertion because we are going to prove a similar but more general statement in the next chapter in 
Appendix~\ref{sa3.2}.)  Nonetheless, the Lie bracket between two $\s$'s is well defined by the requirement that
\begin{equation}
   \left( \lbrack \s^I_J, \s^K_L\rbrack \right) s^{\dot{P}} \equiv \s^I_J \left( \s^K_L s^{\dot{P}} \right) -
   \s^K_L \left( \s^I_J s^{\dot{P}} \right)
\label{4.5.3}
\end{equation}
for any arbitrary sequence $\dot{P}$.  Then it can be shown that the expression of the Lie bracket is
\begin{eqnarray}
   \lefteqn{ \left[ \s^I_J, \s^K_L \right] =
   \delta^K_J \s^I_L + \sum_{J_1 J_2 = J} \delta^K_{J_2} 
   \s^I_{J_1 L} + \sum_{K_1 K_2 = K} \delta^{K_1}_J \s^{I K_2}_L } \nonumber \\
   & & + \sum_{\begin{array}{l}
		  J_1 J_2 = J \\
		  K_1 K_2 = K
	       \end{array}}
   \delta^{K_1}_{J_2} \s^{I K_2}_{J_1 L} 
   + \sum_{J_1 J_2 = J} \delta^K_{J_1} \s^I_{L J_2}
   + \sum_{K_1 K_2 = J} \delta^{K_2}_J \s^{K_1 I}_L \nonumber \\
   & & + \sum_{\begin{array}{l}
		  J_1 J_2 = J \\
		  K_1 K_2 = K
	       \end{array}}
   \delta^{K_2}_{J_1} \s^{K_1 I}_{L J_2}
   + \sum_{J_1 J_2 J_3 = J} \delta^K_{J_2} \s^I_{J_1 L J_3} 
   + \sum_{K_1 K_2 K_3 = K} \delta^{K_2}_J \s^{K_1 I K_3}_L \nonumber \\
   & & - (I \leftrightarrow K, J \leftrightarrow L). 
\label{4.5.4}
\end{eqnarray}
The proof of this equation can be found in Appendix~\ref{sa4.4}.  The diagrammatic representations of the first 
nine terms are given in Fig.\ref{f4.4}).  We will call the Lie algebra defined by Eq.(\ref{4.5.4}) the {\em centrix 
algebra} $\hatcentrix$.

As in the discussion of the algebra $\hatleftix$ (Section~\ref{s4.3}), we are going to determine a Cartan 
subalgebra of the centrix algebra and the root spaces with respect to this Cartan subalgebra.  It turns out that 
all vectors of the form $\s^I_I$, where $I$ is an arbitrary finite integer sequence of integers between $1$ and 
$\L$ inclusive, span a Cartan subalgebra $\hat{\Sigma}^0_{{\Lambda}}$ of the centrix algebra.  The proof is pretty 
much the same as the one shown in Appendix~\ref{sa4.2}. 

Now let us note the following identities that follow easily from the action of $\sigma^I_J$:
\beq
   \left( \s^I_J - \sum_{i=1}^{\Lambda} \s^{iI}_{iJ} \right) s^{\dot{K}} 
   = \sum_{K_1 \dot{K}_2 = \dot{K}} \delta^{K_1}_J s^{I\dot{K}_2}
\la{4.5.5}
\eeq
and
\beq
   \left( \s^I_J - \sum_{j=1}^{\Lambda} \s^{Ij}_{Jj} \right) s^{\dot{K}} 
   = \sum_{\dot{K}_1 K_2 = \dot{K}} \delta^{K_2}_J s^{\dot{K}_1 I}.
\la{4.5.6}
\eeq
These are exactly the action of the opeartors $l^I_J$ and $r^I_J$. Thus,
\begin{eqnarray}
   l^I_J & = & \s^I_J - \sum_{i=1}^{{\Lambda}} \s^{iI}_{iJ}\mbox{; and}
\label{4.5.7} \\
   r^I_J & = & \s^I_J - \sum_{j=1}^{{\Lambda}} \s^{Ij}_{Jj}.
\label{4.5.8}
\end{eqnarray}
Of course, we must require that the sequences $I$ and $J$ be non-empty for this to be true. 

The operators $l^I_J$'s with non-empty sequences $I$ and $J$ form a subalgebra $\hatleftix'$ of $\hatleftix$, the 
operators $r^I_J$'s form another subalgebra $\hatrightix'$ of $\hatrightix$, and the operators $l^I_J$'s and 
$r^I_J$'s together form yet another subalgebra $\hatmultix'$ of $\hatmultix$.  What we have just seen is that 
$\hatleftix'$, $\hatrightix'$ and $\hatmultix'$ can also be viewed as subalgebras of $\hatcentrix$; in fact they 
are even ideals of $\hatcentrix$.  This will be evident from the (more general) commutation relations between
$\sigma^I_J$ and $l^{\dot{I}}_{\dot{J}}$, etc. given below.

Analogously, we can define a vector space $\salt'$ spanned by $f^I_J$ with non-empty $I$ and $J$.  By the same 
counting argument as before, this is also isomorphic to $\gl$. We have already seen that 
\[ f^K_L = l^K_L - \sum_{j=1}^{\Lambda} l^{Kj}_{Lj} \]       
and 
\[ f^K_L = r^K_L - \sum_{i=1}^{\Lambda} r^{iK}_{iL}. \]
Thus we have
\beq
  f^I_J = \sigma^I_J - \sum_{j=1}^{\Lambda} \sigma^{Ij}_{Jj} - \sum_{i=1}^{\Lambda} \sigma^{iI}_{iJ}
  + \sum_{i,j=1}^{\Lambda} \sigma^{iIj}_{iJj}.
\label{4.5.9}
\eeq
$\salt'$ is an ideal of $\hatcentrix$ as well. 

Now we give the commutation relations between centrix and multix operators:
\begin{eqnarray}
   \lefteqn{\left[ \s^I_J, l^{\dot{K}}_{\dot{L}} \right] = \delta^{\dot{K}}_J l^I_{\dot{L}} + 
   \sum_{J_1 J_2 = J} \delta^{\dot{K}}_{J_1} l^I_{\dot{L} J_2} +
   \sum_{K_1 K_2 = \dot{K}} \delta^{K_2}_J l^{K_1 I}_{\dot{L}} } \nonumber \\
   & & + \sum_{K_1 K_2 = \dot{K}} \delta^{K_1}_J l^{I K_2}_{\dot{L}} +
   \sum_{\begin{array}{l}
	    J_1 J_2 = J \\
	    K_1 K_2 = \dot{K}
	 \end{array}}
   \delta^{K_2}_{J_1} l^{K_1 I}_{\dot{L} J_2}
   + \sum_{K_1 K_2 K_3 = \dot{K}} \delta^{K_2}_J l^{K_1 I K_3}_{\dot{L}} \nonumber \\
   & & - \delta^I_{\dot{L}} l^{\dot{K}}_J -
   \sum_{I_1 I_2 = I} \delta^{I_1}_{\dot{L}} l^{\dot{K} I_2}_J -
   \sum_{L_1 L_2 = \dot{L}} \delta^I_{L_2} l^{\dot{K}}_{L_1 J} \nonumber \\
   & & - \sum_{L_1 L_2 = \dot{L}} \delta^I_{L_1} l^{\dot{K}}_{J L_2} -
   \sum_{\begin{array}{l}
   	    L_1 L_2 = \dot{L} \\
   	    I_1 I_2 = I
   	 \end{array}}
   \delta^{I_1}_{L_2} l^{\dot{K} I_2}_{L_1 J} 
   - \sum_{L_1 L_2 L_3 = \dot{L}} \delta^I_{L_2} l^{\dot{K}}_{L_1 J L_3};
\label{4.5.10} \\
   \lefteqn{\left[ \s^I_J, r^{\dot{K}}_{\dot{L}} \right] = \delta^{\dot{K}}_J r^I_{\dot{L}} + 
   \sum_{J_1 J_2 = J} \delta^{\dot{K}}_{J_2} r^I_{J_1 \dot{L}} +
   \sum_{K_1 K_2 = \dot{K}} \delta^{K_2}_J r^{K_1 I}_{\dot{L}} } \nonumber \\
   & & + \sum_{K_1 K_2 = \dot{K}} \delta^{K_1}_J r^{I K_2}_{\dot{L}} +
   \sum_{\begin{array}{l}
	    J_1 J_2 = J \\
	    K_1 K_2 = \dot{K}
	 \end{array}}
   \delta^{K_1}_{J_2} r^{I K_2}_{J_1 \dot{L}}
   + \sum_{K_1 K_2 K_3 = \dot{K}} \delta^{K_2}_J r^{K_1 I K_3}_{\dot{L}} \nonumber \\
   & & - \delta^I_{\dot{L}} r^{\dot{K}}_J -
   \sum_{I_1 I_2 = I} \delta^{I_2}_{\dot{L}} r^{I_1 \dot{K}}_J -
   \sum_{L_1 L_2 = \dot{L}} \delta^I_{L_2} r^{\dot{K}}_{L_1 J} \nonumber \\
   & & - \sum_{L_1 L_2 = \dot{L}} \delta^I_{L_1} r^{\dot{K}}_{J L_2} -
   \sum_{\begin{array}{l}
   	    L_1 L_2 = \dot{L} \\
   	    I_1 I_2 = I
   	 \end{array}}
   \delta^{I_2}_{L_1} r^{I_1 \dot{K}}_{J L_2} 
   - \sum_{L_1 L_2 L_3 = \dot{L}} \delta^I_{L_2} r^{\dot{K}}_{L_1 J L_3};
\label{4.5.11}
\end{eqnarray}
and
\beq
   \left[ \s^I_J, f^{\dot{K}}_{\dot{L}} \right] = 
   \sum_{\dot{K}_1 K_2 \dot{K}_3 = \dot{K}} \delta^{K_2}_J f^{\dot{K}_1 I \dot{K}_3}_{\dot{L}}
   - \sum_{\dot{L}_1 L_2 \dot{L}_3 = \dot{L}} \delta^I_{L_2} f^{\dot{K}}_{\dot{L}_1 J \dot{L}_3}.
\label{4.5.12}
\eeq
The first six terms on the R.H.S. of Eq.(\ref{4.5.10}), the first six terms on the R.H.S. of Eq.(\ref{4.5.11}) and 
the first term on the R.H.S. of Eq.(\ref{4.5.12}) are illustrated in Figs.\ref{f4.5}, \ref{f4.6} and \ref{f4.7}, 
respectively.

\begin{figure}
\epsfxsize=5in
\centerline{\epsfbox{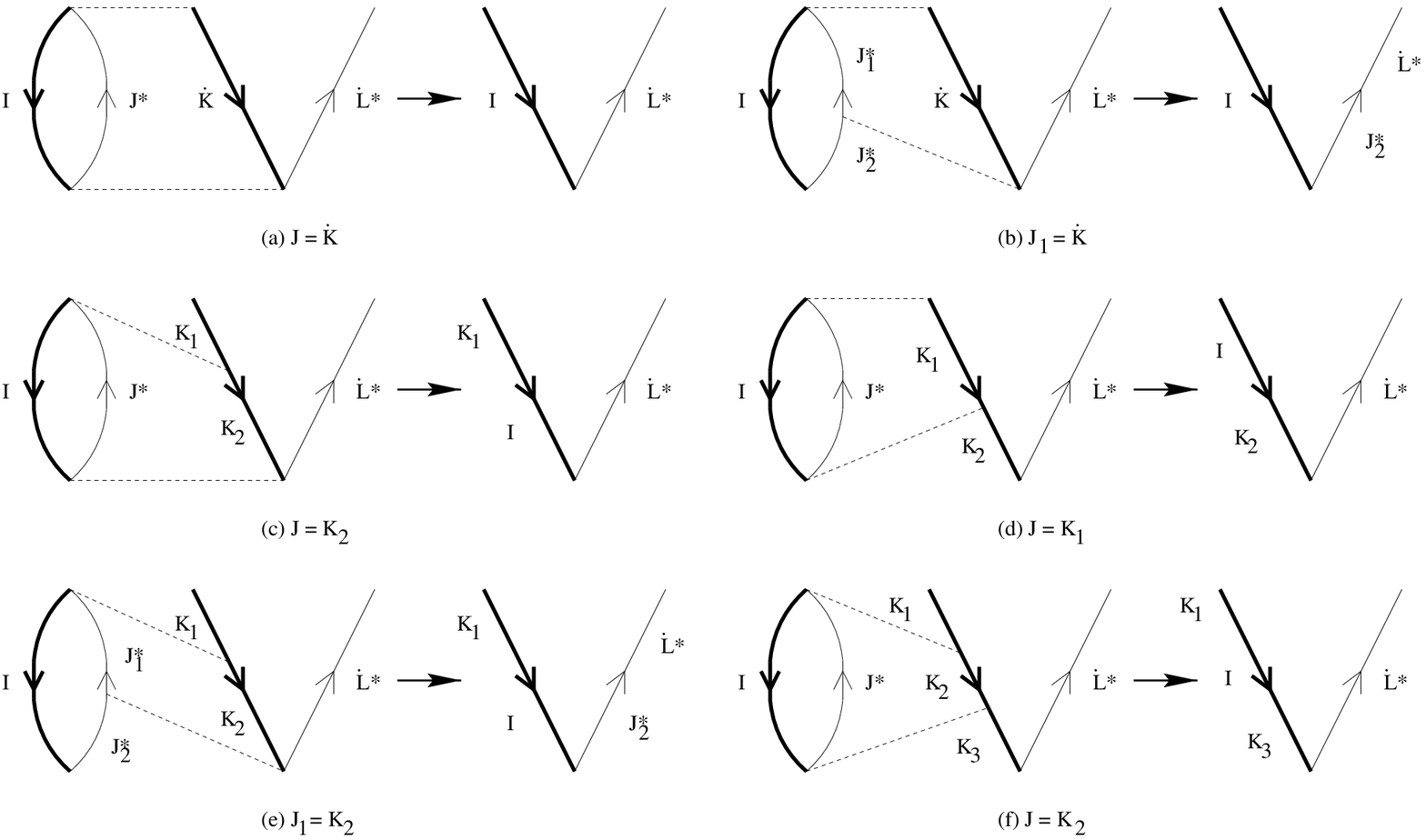}}
\caption{\em  The Lie bracket between an opertor of the fourth kind and an operator of the second kind.}
\label{f4.5}
\end{figure}

\begin{figure}
\epsfxsize=5in
\centerline{\epsfbox{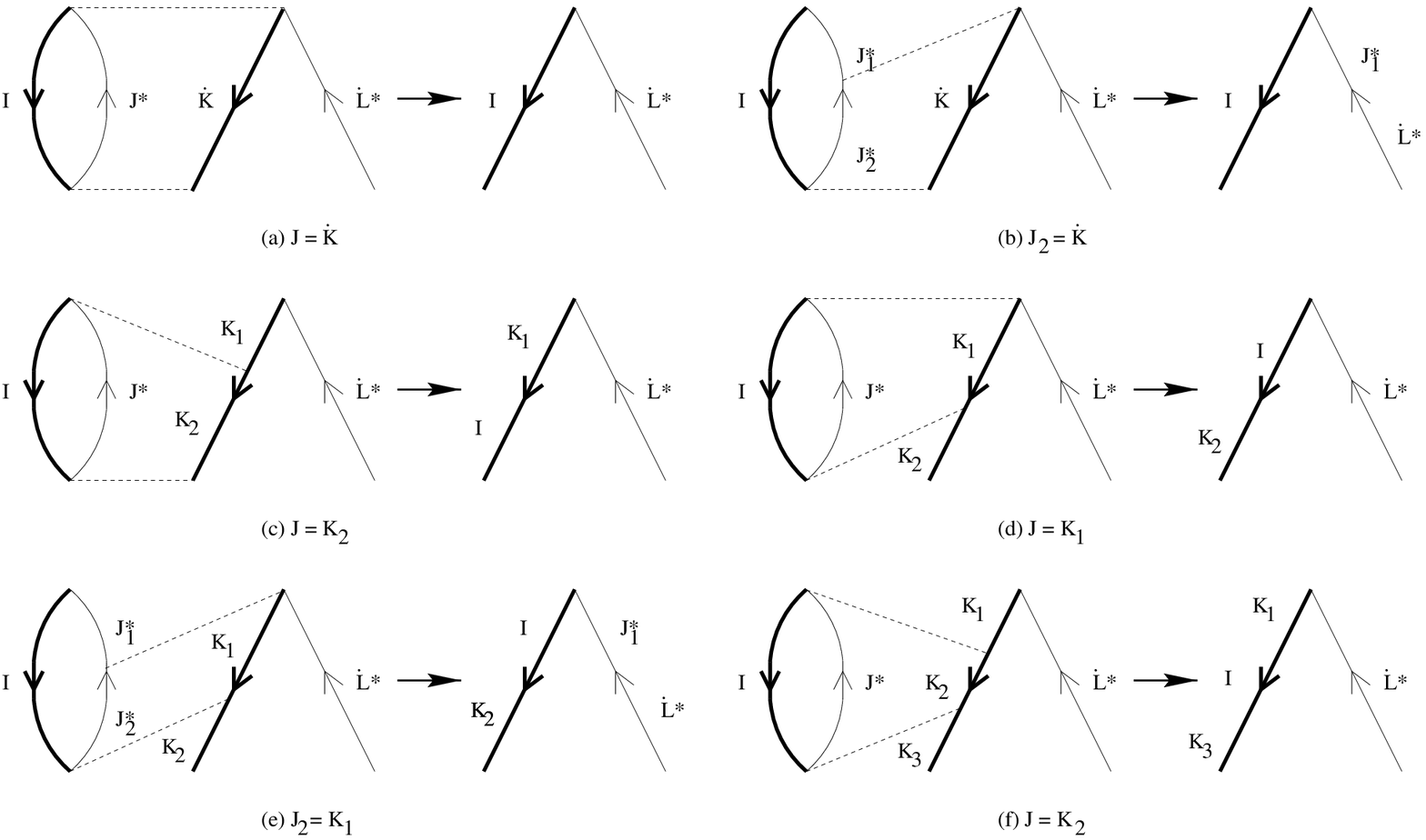}}
\caption{\em  The Lie bracket between an operator of the fourth kind and an operator of the third kind.}
\label{f4.6}
\end{figure}

\begin{figure}
\epsfxsize=2.5in
\centerline{\epsfbox{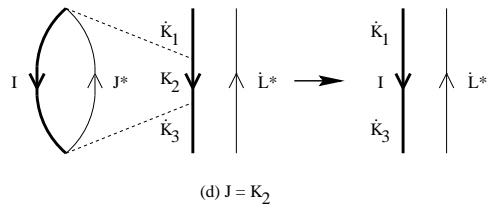}}
\caption{\em  The Lie bracket between an operator of the fourth kind and an operator of the first kind.}
\label{f4.7}
\end{figure}

These prove, in particular, that the algebras $\hatleftix'$, $\hatrightix'$ and $\salt'$ are proper ideals of the 
centrix algebra.  In addition, Eq.(\ref{4.5.10}) shows that an element which does not belong to $\hatleftix'$ in 
the centrix algebra is an outer derivation of the algebra $\hatleftix$, which is the extended Cuntz--Lie algebra.  
Let us summarize the various relationships among the Lie algebras in Table~\ref{t4.2}.

\begin{table}
\centerline{
\begin{tabular}[t]{|| c | c | l | rcl ||} \hline
operators    & extended 	        & \multicolumn{1}{c|}{comment}
& \multicolumn{3}{c||}	{quotient algebra(s)}			   	      \\
             & algebra                  &                                       & 		      &   &                 			       \\ \hline \hline
$f^I_J$      & ${\mathit F'}_{\Lambda}$      &${\mathit F'}_{\Lambda}\equiv  gl_{+\infty}$&  & &  \\ \hline
$l^I_J$      & $\hat{\mathit L'}_{\Lambda}$  & ${\mathit F'}_{\Lambda}$ is a
             proper ideal
                                                                                & ${\mathit L'}_{\Lambda}$ & $\equiv$ & $\hat{\mathit L'}_{\Lambda} / {\mathit F'}_{\Lambda}$  \\ 
${f}^K_L = l^K_L -  l^{Kj}_{Lj}$& & of $\hat{\mathit L'}_{\Lambda}$. 	        & 		      &	  &					  	      \\ \hline
$r^I_J$      & $\hat{\mathit R'}_{\Lambda}$  & ${\mathit F'}_{\Lambda}$ is a proper ideal & ${\mathit R'}_{\Lambda}$ & $\equiv$ & $\hat{\mathit R'}_{\Lambda} / {\mathit F'}_{\Lambda}$  \\
${f}^K_L = r^K_L - r^{iK}_{iL}$ & & of $\hat{\mathit R'}_{\Lambda}$. 	        & 		      &	  &					  	      \\ \hline
$l^I_J$ and $r^I_J$ & $\hat{\mathit M'}_{\Lambda}$ & $\hat{\mathit M'} \neq \hat{\mathit L'}_{\Lambda} \oplus \hat{\mathit R'}_{\Lambda}$ & 
  ${\mathit M'}_{\Lambda}$ & $\equiv$ & $\hat{\mathit M'}_{\Lambda} / {\mathit F'}_{\Lambda}$ \\ \hline
$\sigma^I_J$ & $\hat{\Sigma}_{\Lambda}$ & ${\mathit F'}_{\Lambda}$, $\hat{\mathit L'}_{\Lambda}$ and $\hat{\mathit R'}_{\Lambda}$ are & 
  $\Sigma_{\Lambda}$ & $\equiv$ & $\hat{\Sigma}_{\Lambda} / {\mathit F'}_{\Lambda}$
\\
  $l^I_J =\sigma^I_J-\sigma^{iI}_{iJ}$& & proper ideals of $\hat{\Sigma}_{\Lambda}$. & 
${\cal V}_{\Lambda}$ & $\equiv$ & $\hat{\Sigma}_{\Lambda} / \hat{M}_{\Lambda}'$ \\
  $r^I_J =\sigma^I_J-\sigma^{Ij}_{Jj}$& &  &
& = & $\Sigma_{\Lambda} / M_{\Lambda}'$ \\	
$f^I_J = \sigma^I_J - \sigma^{Ij}_{Jj}$ & &  & 		      &   & \\					       
$- \sigma^{iI}_{iJ} + \sigma^{iIj}_{iIj}$ & & & & & \\ \hline \hline
\end{tabular}}
\caption{\em $\hatcentrix$, its subalgebras and quotient algebras.  The summation convention is adopted
for repeated indices in this table.}
\label{t4.2}
\end{table}

Now, recall that outer derivations of the algebra of functions on a manifold are just vector fields.  We have seen 
that the Cuntz algebra is a noncommutative generalization of the algebra of functions on a circle. Thus our centrix 
Lie algebra should be thought of as the corresponding noncommutative generalization of the Lie algebra of vector 
fields on a circle. Indeed, in the special case $\L=1$ we will see that our algebra reduces exactly to the Witt 
algebra, the central extension of which is the Virasoro algebra; the Witt algebra is just the algebra of vector 
fields on a circle.

Let us make some remarks on root vectors parallelling our discussion in Section~\ref{s4.3}.  We obtain from 
Eq.(\ref{4.5.12}) that 
\begin{equation}
   \left[ \s^I_I, f^K_L \right] = \left( \sum_{\dot{K}_1 K_2 \dot{K}_3 = K} \delta^{K_2}_I 
   - \sum_{\dot{L}_1 L_2 \dot{L}_3 = L} \delta^I_{L_2} \right) f^K_L	 
\label{4.5.13}
\end{equation}
As a result, every $f^K_L$ is a root vector.  Moreover, there are no root vectors other than $f^K_L$'s, and a
proof of this statement will be given in Appendix~\ref{sa4.5}.  Hence every root space is one-dimensional.   

There are various ways to construct different quotient algebras from the centrix algebra.  For instance, we can 
make the set of all cosets $\s^I_J + \salt'$ into a quotient algebra $\centrix$.  Furthermore, within this quotient 
algebra, all cosets of the form $l^I_J + \salt$ span a proper ideal $\leftix'$, and all cosets of the form 
$r^I_J + \salt$ span another ideal $\rightix$.  We can further extract the quotient algebra
\begin{equation}
   {\cal V}_{\L} \equiv \hat{\S}_{\L} / \hat{M}_{\L}' = \S_{\L} / M_{\L}'
\la{4.5.14}
\end{equation}
This Lie algebra $\vectrix$ seems not to have been discussed before; we will see in the next section that 
${\cal V}_1$ is just the algebra of vector fields on the circle. More generally, the previous discussion shows that 
the generators of $\hatcentrix$ act as derivations on the Cuntz algebra.  When we quotient out the above ideal, we 
are basically extracting the part which corresponds to `inner derivations' (up to some extensions by standard 
algebras such as $\gl$).  Thus $\vectrix$ should be viewed as a generalization of the algebra of vector fields 
of a circle. It is the Lie algebra of vector fields on a sort of noncommutative generalization of the circle, which 
has the Cuntz algebra as its algebra of functions.  It would be interesting to investigate this further.

There is a way of splitting $\hatcentrix$ into `raising' operators, `diagonal' operators and `lowering' operators.
Let $\hat{\Sigma}_{{\Lambda}}^+$ be the vector space spanned by all elements of the form $\s^I_J$ such that $I > J$ 
(Appendix~\ref{sa1.1}).  Then it can be proved by checking term by term on the R.H.S. of Eq.(\ref{4.5.4}) that 
$\hat{\Sigma}_{{\Lambda}}^+$ is indeed a subalgebra of the centrix algebra\footnote{Take the fourth term as an 
example.  If both $\s^I_J$ and $\s^K_L \in \hat{\Sigma}_{{\Lambda}}^+$, then $IK_2 > JK_2 = J_1 J_2 K_2 = 
J_1 K_1 K_2 = J_1 K > J_1 L$.  Hence $\s^{I K_2}_{J_1 L} \in \hat{\Sigma}_{{\Lambda}}^+$ also.}.  Likewise, let 
$\hat{\Sigma}_{{\Lambda}}^-$ be the vector space spanned by all elements of the form $\s^I_J$ such that $I < J$.  
Then it also follows from Eq.(\ref{4.5.4}) that $\hat{\Sigma}_{{\Lambda}}^-$ is also a subalgebra.  Moreover, we 
have
\[ \hat{\Sigma}_{{\Lambda}} = \hat{\Sigma}^+_{{\Lambda}} \oplus 
\hat{\Sigma}^0_{{\Lambda}} \oplus \hat{\Sigma}^-_{{\Lambda}}. \] 

\section{Operators of the Fourth Kind and the Witt Algebra} 
\label{s4.6}

We are going to consider the simplest version of the centrix algebra --- the $\Lambda = 1$ centrix algebra.  We 
will see that it could be obtained by first forming a semi-direct product of the Virasoro algebra and the algebra 
of functions on a circle, and then extending the semi-direct product by $gl_{+\infty}$.  This will suggest a deep 
relationship between the Witt algebra and an algebra of a set of outer derivations of the Cuntz algebra.

Consider the centrix algebra for the case $\Lambda = 1$.  Again we write $s^{\dot{K}}$ as $s^{\#(\dot{K})}$ and 
$\s^I_J$ as $\s^{\#(I)}_{\#(J)}$.  We can deduce from Eq.(\ref{4.5.1}) that the action of $\s^a_b$, where $a$ and 
$b>0$ are the number of integers in the various sequences, on $s^{\dot{c}}$, where $\dot{c}$ is a non-negative 
integer, is
\begin{equation}
   \s^a_b s^{\dot{c}} = (\dot{c} - b + 1) \theta (b \leq \dot{c}) s^{a + \dot{c} - b}.
\la{4.6.1}
\end{equation}
The Lie bracket given in Eq.(\ref{4.5.4}) is simplified to
\begin{eqnarray}
   \left[ \s^a_b, \s^c_d \right] & = & \theta (c \leq b) \left[  
   2 \s^{a+c-1}_{b+d-1} + 2 \s^{a+c-2}_{b+d-2} + \cdots + 2 \s^{a+1}_{b+d-c+1}
   \right. \nonumber \\
   & & \left. + (b-c+1) \s^a_{d+b-c} \right] + \theta (b < c) \left[ 2 
   \s^{a+c-1}_{b+d-1} + 2 \s^{a+c-2}_{b+d-2} \right. \nonumber \\
   & & \left. + \cdots + 2 \s^{a+c-b+1}_{d+1} + (c-b+1) \s^{a+c-b}_d \right] \nonumber \\
   & & - (a \leftrightarrow c, b \leftrightarrow d).
\la{4.6.2}
\end{eqnarray}
All the vectors of the form $\s^a_a$ span a Cartan subalgebra.  A vector in $\hat{\mathit M}_1'$ is a linear 
combination of vectors of the form $l^c_d = r^c_d = \s^c_d - \s^{c+1}_{d+1}$.  The root vectors are given by 
$f^c_d = \s^c_d - 2 \s^{c+1}_{d+1} + \s^{c+2}_{d+2}$.  This can be deduced from Eq.(\ref{4.5.9}). The action of 
$f^a_b$ on $s^{\dot{c}}$ is also given by Eq.(\ref{4.3.18}).  The corresponding eigenequation, which can be deduced 
from Eq.(\ref{4.5.13}), is
\begin{equation}
   \left[ \s^a_a, f^c_d \right] = \left( \theta (a \leq c) (c - a + 1)
   - \theta (a \leq d) (d - a + 1) \right) f^c_d.
\la{4.6.3}
\end{equation}

As in the previous section, the subspaces $\saltone$ and ${\mathit M}_1'$ form proper ideals of this $\Lambda=1$
centrix algebra.  Let us consider the quotient algebra of cosets of the form $v + \salt'$, where $v$ is an 
arbitrary vector of the algebra $\hat{\mathit \Sigma}_1$, in detail.  This quotient algebra is spanned by the 
cosets $l^a_1 + \saltone$, $l^1_b + \saltone$, $\s^a_1 + \saltone$ and $\s^1_b + \saltone$, where $a$ and $b$ run 
over all positive integers.  It is a straightforward matter to show that the following Lie brackets are true:
\begin{eqnarray}
   \left[ l^a_1 + \saltone, l^c_1 + \saltone \right] & = & \saltone ; \nonumber \\
   \left[ l^a_1 + \saltone, l^1_d + \saltone \right] & = & \saltone ; \nonumber \\
   \left[ l^1_b + \saltone, l^1_d + \saltone \right] & = & \saltone ; \nonumber \\
   \left[ \s^a_1 + \saltone, \s^c_1 + \saltone \right] & = & 
   (c - a) (\s^{a+c-1} + \saltone) ; \nonumber \\
   \left[ \s^a_1 + \saltone, \s^1_d + \saltone \right] & = &
   \left\{ \begin{array}{ll}
   (2 - a - d) ( \s^{a-d+1}_1 + \saltone ) & \mbox{if $d \leq a$, or} \\
   (2 - a - d) ( \s^1_{d-a+1} + \saltone ) & \mbox{if $a \leq d$;}
   \end{array} \right. \nonumber \\
   \left[ \s^1_b + \saltone, \s^1_d + \saltone \right] & = & 
   (b - d) (\s^1_{b+d-1} + \saltone); \nonumber \\
   \left[ \s^a_1 + \saltone, l^c_1 + \saltone \right] & = & 
   (c - 1) (l^{a+c-1}_1 + \saltone); \nonumber \\
   \left[ \s^1_b + \saltone, l^c_1 + \saltone \right] & = & \left\{
   \begin{array}{ll}
      (c - 1) (l^{c-b+1}_1 + \saltone)& \mbox{if $b \leq c$, or} \\
      (c - 1) (l^1_{b-c+1} + \saltone)& \mbox{if $c \leq b$;}
   \end{array} \right. \nonumber \\
   \left[ \s^1_b + \saltone, l^1_d + \saltone \right] & = & 
   - (d - 1) (\s^1_{b+d-1} + \saltone) \mbox{; and} \nonumber \\
   \left[ l^a_1 + \saltone, l^1_d + \saltone \right] & = & \left\{
   \begin{array}{ll}
      - (d - 1) (l^1_{d-a+1} + \saltone) & \mbox{if $a \leq d$, or} \\
      - (d - 1) (l^{a-d+1}_1 + \saltone) & \mbox{if $d \leq a$.}
   \end{array} \right. .
\label{4.6.4}
\end{eqnarray}

Let us make the following identifications:
\begin{eqnarray}
   l^a_1 + \saltone & \rightarrow & z^{a-1} ; \nonumber \\
   l^1_b + \saltone & \rightarrow & z^{-(b-1)} ; \nonumber \\
   \s^a_1 + \saltone & \rightarrow & z^{a-1} \frac{d}{dz} = - L_{a-1} 
   \mbox{; and} \nonumber \\
   \s^1_b + \saltone & \rightarrow & z^{-(b-1)} \frac{d}{dz} = - L_{1-b}.
\label{4.6.5}
\end{eqnarray}
Here $z$ is a complex number with $|z| = 1$.  Note that the $L$'s here are {\em not} integer sequences.  Then 
Eq.(\ref{4.6.1}) becomes
\begin{eqnarray}
   \left[ z^p, z^q \right] & = & 0 ; \nonumber \\
   \left[ L_p, L_q \right] & = & (p - q) L_{p+q} \mbox{; and} \nonumber \\
   \left[ L_p, z^q \right] & = & -q z^{p+q-1},
\la{4.6.6}
\end{eqnarray}    
where $p$ and $q$ are integers which may be positive, negative or zero.  Thus this is the extension of the Witt 
algebra by the functions on a unit circle on the complex plane.  Now we also see that the algebra ${\cal V}_1$ is 
just the Lie algebra of vector fields on the circle, spanned by all $L_p$'s with $p$ being all integers.

We can also infer from the results in Sections~\ref{s4.3}, \ref{s4.5} and this one that the Cuntz--Lie algebra 
corresponds to a generalization of the algebra of functions on the unit circle which is the leftix algebra with an 
arbitrary positive integer $\L$, and the quotient Lie algebra of a set of outer derivations of the Cuntz algebra by
the elements of the Cuntz algebra corresponds to a generalization of the Witt algebra which is the centrix algebra, 
also with an arbitrary positive integer $\L$.

\chapter{A Lie Superalgebra}
\la{c3}

\section{Introduction}
\la{s3.1}

In the previous chapter, we studied a Lie algebra for bosonic open strings, and gained some expertise in how to
study its structure.  We will use this skill to fulfill our promise in the preceding chapter, which is to study the 
full Lie superalgebra for open and closed string-bit states, or in other words, the Lie superalgebra for the five 
kinds of operators discussed in Section~\ref{s2.2}.  

To achieve this purpose, we need to define a precursor Lie superalgebra which we will call the {\em heterix 
superalgebra} among some physical operators.  These physical operators are nothing but the ones defined in 
Section~\ref{s2.2}, except that the ranges of values of the quantum states other than color (see the second
paragraph in Section~\ref{s2.2}) are changed.  Then the grand string superalgebra will be seen as a subalgebra of 
the heterix superalgebra.  Readers who are only interested in the definition of this superalgebra may skip this 
chapter. occasionally returning to Section~\ref{s3.2} to look up the relevant definitions.  In Section~\ref{s3.3}, 
we will study the mathematical structure of the heterix algebra, which can be obtained from the heterix superalgebra
by considering only bosonic adjoint matter fields.  Again we will derive a Cartan subalgebra and the associated
root vectors of the heterix algebra. 

\section{Derivation of a Precursor Superalgebra}
\la{s3.2}

We are going to derive a precursor superalgebra.  The grand string superalgebra will be identified as a subalgebra
of this superalgebra in the next chapter.  (Those readers who do not know what a superalgebra is are referred to
Ref.\cite{buku} for its definition.)

Let $\a^{\mu}_{\n}(k)$ be an annihilation operator for a boson in the adjoint representation for $1 \leq k \leq \L$ 
and $2 \L + 1 \leq k \leq 2 \L + 4 \L_F$, or a fermion in the adjoint representation for $\L + 1 \leq k \leq 2 \L$ 
(the apparently weird choices of the ranges of $k$ are for later convenience), and let $\bar{\c}_{\mu}$ and 
$\c^{\mu}$ be annihilation operators for an antifermion in the conjugate representation and a fermion in the 
fundamental representation, respectively.  Moreover, let $\a^{\da\mu}_{\n}(k)$, $\bar{\c}^{\da\mu}$ and 
$\c^{\da}_{\mu}$ be the corresponding creation operators.  The annihilation and creation operators satisfy the 
usual canonical (anti)-commutation relations, the non-vanishing ones being
\beq
   \le[ \a^{\mu_1}_{\mu_2}(k_1), \a^{\da\mu_3}_{\mu_4}(k_2) \ri] =  
   \d_{k_1 k_2} \d^{\mu_3}_{\mu_2} \d^{\mu_1}_{\mu_4} 
\la{3.2.1}
\eeq
for both $k_1$ and $k_2 \in \{ 1, 2, \ld, \L, 2\L+1, 2\L+2, \ldots 2\L+4\L_F \}$;
\beq
   \le[ \a^{\mu_1}_{\mu_2}(k_1), \a^{\da\mu_3}_{\mu_4}(k_2) \ri]_+ = 
   \d_{k_1 k_2} \d^{\mu_3}_{\mu_2} \d^{\mu_1}_{\mu_4} 
\la{3.2.2}
\eeq
for both $k_1$ and $k_2 \in \{ \L+1, \L+2, \ldots, 2\L \}$;
\beq
   \le[ \bar{\c}_{\mu_1}, \bar{\c}^{\da\mu_2} \ri]_+ = \d^{\mu_2}_{\mu_1}; 
\la{3.2.3}
\eeq
and
\beq
   \le[ \c^{\mu_1}, \c^{\da}_{\mu_2} \ri]_+ = \d^{\mu_1}_{\mu_2}.
\la{3.2.4}
\eeq

Again we introduce two families of color singlet states.  A typical open singlet state is a linear combination of 
the states of the form
\beq
   s'^K \equiv N^{-(c+1)/2} \bar{\c}^{\da\u_1} \a^{\da\u_2}_{\u_1}(k_1) \a^{\da\u_3}_{\u_2}(k_2) \cd 
   \a^{\da\u_{c+1}}_{\u_c}(k_c) \c^{\da}_{\u_{c+1}} |0 \rangle.
\la{3.2.5}
\eeq
We denote by ${\cal T}'_o$ the Hilbert space of all these open singlet states.  A typical closed singlet state is a 
linear combination of the states of the form
\beq
   \Ps'^K \equiv N^{-c/2} \a^{\da\u_2}_{\u_1}(k_1) \a^{\da\u_3}_{\u_2}(k_2) \cd \a^{\da\u_1}_{\u_c}(k_c) 
   |0 \rangle.
\la{3.2.6}
\eeq
The Hilbert space of all closed singlet states will be denoted by ${\cal T}'_c$.

We need only two families of color-invariant operators acting on ${\cal T}'_o$ and ${\cal T}'_c$ to establish our
main results.  One family consists of operators of the form 
\beq 
   \ti{f}'^I_J & \equiv & N^{-(a+b)/2} \a^{\da\mu_2}_{\mu_1} (i_1) \a^{\da\mu_3}_{\mu_2} (i_2) \cd 
   \a^{\da\mu_1}_{\mu_a} (i_a, \ep(i_a)) \nn \\
   & & \a^{\n_b}_{\n_1} (j_b) \a^{\n_{b-1}}_{\n_b} (j_{b-1}) \cd \a^{\n_1}_{\n_2} (j_1).
\la{3.2.7}
\eeq
In the large-$N$ limit, the actions of this operator on singlet states read (c.f. Ref.\cite{thorn79} or 
Appendix~\ref{sa1.3} of this article)
\beq
   \ti{f}'^I_J s'^K & = & 0 \mbox{; and}
\la{3.2.8} \\
   \ti{f}'^I_J \Ps'^K & = & \d^K_J \Ps'^I + \sum_{J_1 J_2 = J} (-1)^{\ep(J_1) \ep(J_2)} \d^K_{J_2 J_1} \Ps'^I,
\la{3.2.9}
\eeq
It is clear that the sum on the right hand side of Eq.(\ref{3.2.9}) is a finite one.  The other family consists of 
operators of the form
\beq
   \g'^I_J & \equiv & N^{-(a+b-2)/2} \a^{\da\mu_2}_{\mu_1} (i_1) \a^{\da\mu_3}_{\mu_2} (i_2) \cd 
   \a^{\da\n_b}_{\mu_a} (i_a) \nn \\
   & & \a^{\n_{b-1}}_{\n_b} (j_b) \a^{\n_{b-2}}_{\n_{b-1}} (j_{b-1}) \cd \a^{\mu_1}_{\n_1} (j_1).
\la{3.2.10}
\eeq
In the large-$N$ limit, this operator propagates singlet states in the following manner:
\beq
   \lefteqn{\g'^I_J s'^K \equiv \sum_{\dot{K}_1 K_2 \dot{K}_3 = K} \d^{K_2}_J 
   (-1)^{\ep(\dot{K}_1) \lb \ep(I) + \ep(J) \rb} s'^{\dot{K_1} I \dot{K_3}} \mbox{; and} } 
\la{3.2.11} \\
   \lefteqn{\g'^I_J \Ps'^K \equiv \d^K_J \Ps'^I + \sum_{K_1 K_2 = K} (-1)^{\ep(K_1) \ep(K_2)} \d^{K_2 K_1}_J \Ps'^I
   } \nn \\
   & & + \sum_{K_1 K_2 = K} \d^{K_1}_J \Ps'^{I K_2} + \sum_{K_1 K_2 K_3 = K} (-1)^{\ep(K_1) \lb \ep(K_2) 
       + \ep(K_3) \rb} \d^{K_2}_J \Ps'^{I K_3 K_1} \nn \\
   & & + \sum_{K_1 K_2 = K} (-1)^{\ep(K_1) \ep(K_2)} \d^{K_2}_J \Ps'^{I K_1} \nn \\
   & & + \sum_{J_1 J_2 = J} \sum_{K_1 K_2 K_3 = K} (-1)^{\ep(K_3) \lb \ep(K_1) + \ep(K_2) \rb} \d^{K_3}_{J_1}
   \d^{K_1}_{J_2} \Ps'^{I K_2}.
\la{3.2.12}
\eeq

The set of all $\ti{f}'^I_J$'s and $\g'^I_J$'s is linearly independent.  This fact, the proof of which can be found 
in Appendix~\ref{sa3.1}, is of some importance, as we will define a Lie superalgebra with this set as a basis soon.
On the contrary, the set of all $g'^I_J$'s acting on closed string states alone is not linearly independent.

The next thing we are going to do is to construct a precursor Lie superalgebra out of the two families of 
color-invariant operators.  If the product of any two of these operators were well defined, i.e., if we could write
down the product as a finite linear combination of color-invariant operators, the easiest way to obtain a 
superalgebra would certainly be to define the Lie superbracket of two operators as a sum or difference of their 
products in different orders of the operators.  However, such a product is actually not well defined; this is shown
in Appendix~\ref{sa3.2}.  Now consider the commutator of two pure operators, at least one of which being even, and 
the anticommutator of two odd operators.  {\em It can be shown that the actions of these (anti-)commutators on 
singlet states are identical to the actions of some observables}.  If we define the color-invariant-operator-valued 
binary operation $\lb \cdot, \cdot \rb_{\pm}$ on two color-invariant operators to be the color-invariant operator 
whose action on ${\cal T}'_o \oplus {\cal T}'_c$ is identical to that of the (anti)-commutator of the pair of 
operators, then we have
\beq
   \lefteqn{ \le[ \g'^I_J, \g'^K_L \ri]_{\pm} = \d^K_J \g'^I_L + 
   \sum_{J_1 J_2 = J} (-1)^{\ep(J_1) \lb \ep(K) + \ep(L) \rb} \d^K_{J_2} \g'^I_{J_1 L} } \nn \\
   & & + \sum_{K_1 K_2 = K} \delta^{K_1}_J \g'^{I K_2}_L
   + \sum_{\ba{l}
	      J_1 J_2 = J \\
	      K_1 K_2 = K
	   \ea}
   (-1)^{\ep(J_1) \lb \ep(K) + \ep(L) \rb} \d^{K_1}_{J_2} \g'^{I K_2}_{J_1 L} \nn \\ 
   & & + \sum_{J_1 J_2 = J} \d^K_{J_1} \g'^I_{L J_2}
   + \sum_{K_1 K_2 = J} (-1)^{\ep(K_1) \lb \ep(I) + \ep(J) \rb} \d^{K_2}_J \g'^{K_1 I}_L \nn \\
   & & + \sum_{\ba{l}
		  J_1 J_2 = J \\
	    	  K_1 K_2 = K
	       \ea}
   (-1)^{\ep(K_1) \lb \ep(I) + \ep(J) \rb} \d^{K_2}_{J_1} \g'^{K_1 I}_{L J_2} \nn \\
   & & + \sum_{J_1 J_2 J_3 = J} (-1)^{\ep(J_1) \lb \ep(K) + \ep(L) \rb} \d^K_{J_2} \g'^I_{J_1 L J_3} \nn \\ 
   & & + \sum_{K_1 K_2 K_3 = K} (-1)^{\ep(K_1) \lb \ep(I) + \ep(J) \rb} \d^{K_2}_J \g'^{K_1 I K_3}_L \nn \\
   & & + \sum_{\ba{l}
   		  J_1 J_2 = J \\
   		  K_1 K_2 = K
   	       \ea}
   (-1)^{\ep(J_1) \ep(K_1)} \d^{K_1}_{J_2} \d^{K_2}_{J_1} \ti{f}'^I_L \nn \\
   & & + \sum_{\ba{l}
   		 J_1 J_2 J_3 = J \\
   		 K_1 K_2 = K
   	       \ea}
   (-1)^{\lb \ep(J_1) + \ep(J_2) \rb \ep(K_1)} \d^{K_1}_{J_3} \d^{K_2}_{J_1} \ti{f}'^I_{L J_2} \nn \\
   & & + \sum_{\ba{l}
   		  J_1 J_2 = J \\
   		  K_1 K_2 K_3 = K
   	       \ea}
   (-1)^{\ep(J_1) \lb \ep(K_1) + \ep(K_2) \rb} \d^{K_1}_{J_2} \d^{K_3}_{J_1} \ti{f}'^{I K_2}_L \nn \\
   & & + \sum_{\ba{l}
		  J_1 J_2 J_3 = J \\
		  K_1 K_2 K_3 = K
	       \ea}
   (-1)^{\lb \ep(J_1) + \ep(J_2) \rb \lb \ep(K_1) + \ep(K_2) \rb} 
   \d^{K_1}_{J_3} \d^{K_3}_{J_1} \ti{f}'^{I K_2}_{L J_2} \nn \\
   & & - (-1)^{\lb \ep(I) + \ep(J) \rb \lb \ep(K) + \ep(L) \rb} (I \lrar K, J \lrar L), 
\la{3.2.13}
\eeq
\beq
   \lefteqn{ \le[ \g'^I_J, \ti{f}'^K_L \ri]_{\pm} = \le\{ \d^K_J \ti{f}'^I_L +
   \sum_{K_1 K_2 = K} (-1)^{\ep(K_1) \ep(K_2)} \d^{K_2 K_1}_J \ti{f}'^I_L +
   \sum_{K_1 K_2 = K} \d^{K_1}_J \ti{f}'^{I K_2}_L \ri. } \nn \\
   & & + \sum_{K_1 K_2 K_3 = K} \d^{K_2}_J (-1)^{\ep(K_1) \lb \ep(I) + \ep(J) \rb} \ti{f}'^{K_1 I K_3}_L \nn \\ 
   & & + \sum_{K_1 K_2 = K} \d^{K_2}_J (-1)^{\ep(K_1) \lb \ep(I) + \ep(J) \rb} \ti{f}'^{K_1 I}_L \nn \\
   & & \le. + \sum_{K_1 K_2 K_3 = K} \d^{K_3 K_1}_J (-1)^{\lb \ep(K_1) + \ep(K_2) \rb \ep(K_3)} \ti{f}'^{I K_2}_L
   \ri\} \nn \\
   & & - (-1)^{\lb \ep(I) + \ep(J) \rb \lb \ep(K) + \ep(L) \rb} \le\{ \d^I_L \ti{f}'^K_J 
   - \sum_{L_1 L_2 = L} (-1)^{\ep(L_1) \ep(L_2)} \d^I_{L_2 L_1} \ti{f}'^K_J \ri. \nn \\
   & & - \sum_{L_1 L_2 = L} (-1)^{\ep(L_1) \lb \ep(I) + \ep(J) \rb} \d^I_{L_2} \ti{f}'^K_{L_1 J} \nn \\
   & & - \sum_{L_1 L_2 L_3 = L} (-1)^{\ep(L_1) \lb \ep(I) + \ep(J) \rb} \d^I_{L_2} \ti{f}'^K_{L_1 J L_3}
   - \sum_{L_1 L_2 = L} \d^I_{L_1} \ti{f}'^K_{J L_2} \nn \\
   & & \le. - \sum_{L_1 L_2 L_3 = L} (-1)^{\ep(L_3) \lb \ep(L_1) + \ep(L_2) \rb} \d^I_{L_3 L_1} \ti{f}'^K_{J L_2}
   \ri\}
\la{3.2.14}
\eeq
and
\beq
   \le[ \ti{f}'^I_J, \ti{f}'^K_L \ri]_{\pm} & = & \d^K_J \ti{f}'^I_L + 
   \sum_{K_1 K_2 = K} (-1)^{\ep(K_1) \ep(K_2)} \d^{K_2 K_1}_J \ti{f}'^I_L \nn \\
   & & - (-1)^{\lb \ep(I) + \ep(J) \rb \lb \ep(K) + \ep(L) \rb} (I \lrar K, J \lrar L).
\la{3.2.15}
\eeq

To see that Eqs.(\ref{3.2.13}), (\ref{3.2.14}) and (\ref{3.2.15}) indeed hold, we can verify that the actions of 
the left hand sides of these equations on an arbitrary singlet state are the same of those on the right hand sides 
of the same equations.  We have already done part of the job in Appendix~\ref{sa4.4}, which accompanies 
Section~\ref{s4.5}.  We will do another part of the verification in Appendix~\ref{sa3.3}, which accompanies
Section~\ref{s3.3}.  These two long appendices together will complete the proof for the case in which there are no 
fermions.  We will not present the full proof here because it would take up too many pages.  To enlighten the 
reader, we would like to discuss the following special cases in Eq.(\ref{3.2.13}) to see why it makes sense for the 
right hand sides of these equations to appear this way.

Consider $\g'^I_J (\g'^K_L \Ps'^M)$, and consider a term $t_1$ in $\g'^I_J (\g'^K_L \Ps'^M)$ produced by the 
operations of $\g'^I_J$ and $\g'^K_L$ on disjoint sequences of creation operators in $\Ps'^M$ 
(Fig.~\ref{f3.2.1}(a)), i.e., no annihilation operator in $\g'^I_J$ acts on any creation operator in $\g'^K_L$.  
This term is identical, up to a minus sign, to the term $t_2$ in $\g'^K_L (\g'^I_J \Ps'^M)$ produced by the 
operations of $\g'^I_J$ and $\g'^K_L$ on the same disjoint sequences in $\Ps'^M$.  Since during the process of 
producing $t_1$ and $t_2$, there is no contraction at all between any operator in $\g'^I_J$ and any in $\g'^K_L$ 
(any contraction between $\g'^I_J$ and $\g'^K_L$ will produce terms which we are not considering right now), $t_2$ 
can be obtained from $\g'^I_J (\g'^K_L \Ps'^M)$ as well by interchanging $\g'^I_J$ and $\g'^K_L$ first before these 
two color-invariant operators act on $\Ps'^M$.  This interchange produces a factor of 
$(-1)^{\lb \ep(I) + \ep(J) \rb \lb \ep(K) + \ep(L) \rb}$.  As a result, $t_1$ is cancelled by 
$(-1)^{\lb \ep(I) + \ep(J) \rb \lb \ep(K) + \ep(L) \rb} t_2$ in the (anti)-commutator.

\begin{figure}
\epsfxsize=5in
\centerline{\epsfbox{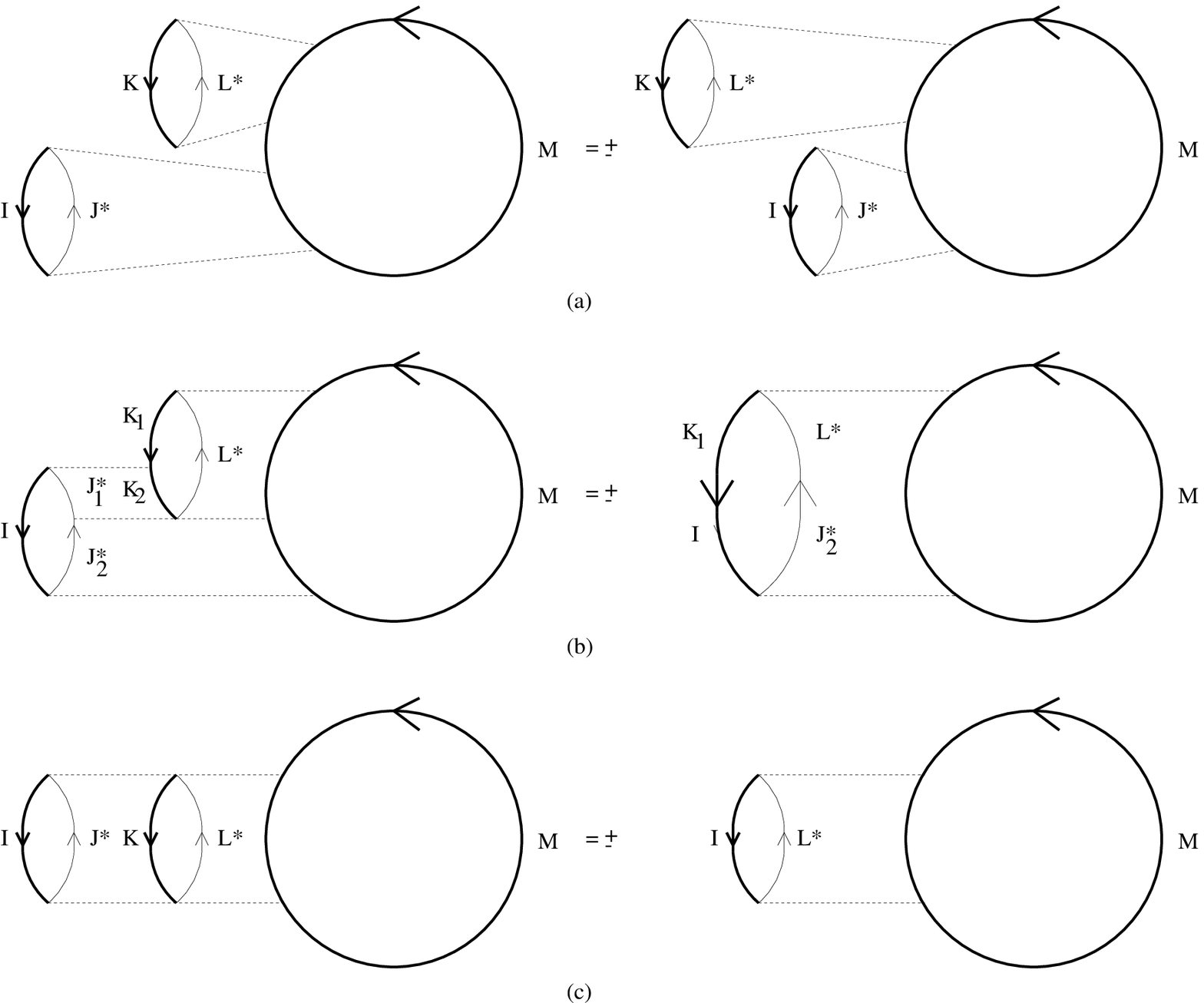}}
\caption{\em Possible terms in $\g'^I_J (\g'^K_L \Ps'^M)$.}
\la{f3.2.1}
\end{figure}

\begin{figure}
\epsfxsize=5in
\centerline{\epsfbox{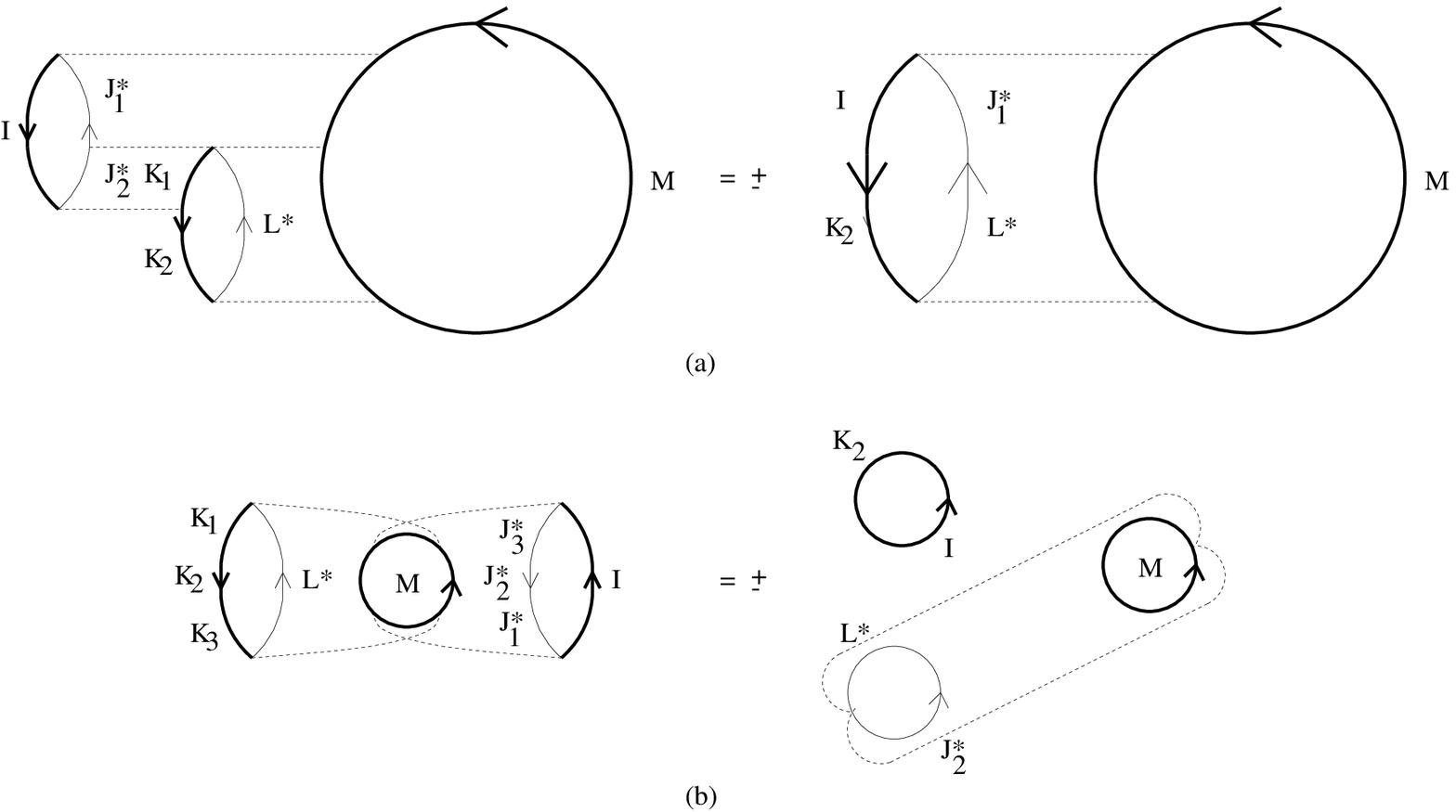}}
\caption{\em Other possible terms in $\g'^I_J (\g'^K_L \Ps'^M)$.}
\la{f3.2.2}
\end{figure}

Therefore, during the process of producing terms that are not killed by the (anti)-commutator, there must be
contraction(s) among some operators in $\g'^I_J$ and some in $\g'^K_L$.  If we perform such contractions first
before contracting the annihilation operators in $\g'^I_J$ and $\g'^K_L$ with the creation operators in $\Ps'^M$,
in general we will obtain a color-invariant operator of the form ${\rm Tr} ( \a^{\da} \cd \a^{\da} \a \cd \a 
\a^{\da} \cd \a^{\da} \a \cd \a \cd) \cd {\rm Tr} ( \a^{\da} \cd \a^{\da} \a \cd \a \a^{\da} \cd)$ multiplied by a 
factor of $N$ raised to some power.  Figs.\ref{f3.2.1} and \ref{f3.2.2} tell us there are only five different
possibilities.  Fig.\ref{f3.2.1}(a) shows the case when $\g'^I_J$ and $g'^K_L$ act on disjoint segments in 
$\Ps'^M$.  Obviously this is the same as the corresponding term in $\g'^K_L (\g'^I_J \Ps'^M)$ up to a sign.  This 
term is cancelled in the Lie superbracket.  Fig.\ref{f3.2.1}(b) shows a partial overlap between $J$ and $K$.  This 
corresponds to the case $J_1 = K_2$.  This term is equivalent to the corresponding one in $\g'^{K_1 I}_{L J_2} 
\Ps'^M$.  Fig.\ref{f3.2.1}(c) shows a complete overlap between $J$ and $K$.  This time $J = K$.  This term is 
equivalent to a term in $\g'^I_L \Ps'^M$.  Fig.\ref{f3.2.2}(a) shows another partial overlap between $J$ and $K$.
This time $J_2 = K_1$.  This term is equivalent to a term in $\g'^{I K_2}_{J_1 L} \Ps'^M$.  Lastly, 
Fig.\ref{f3.2.2}(b) shows a partial overlap in which $J_1 = K_3$ and $J_3 = K_1$.  Note that we have turned 
$\g'^I_J$ by 180 degrees.  This term is equivalent to $\ti{f}'^{I K_2}_{J_2 L} \Ps'^M$.  These five figures 
together clearly reveal that only color-invariant operators of the forms ${\rm Tr} (\a^{\da} \cd \a^{\da} \a \cd 
\a)$ and ${\rm Tr} (\a^{\da} \cd \a^{\da}) {\rm Tr} (\a \cd \a)$ survive the large-$N$ limit.  All the 
possibilities of producing color-invariant operators in these two forms are contained in Eqs.(\ref{3.2.13}) to 
(\ref{3.2.15}).  To understand the factors of -1, consider, say, the term 
\[ \sum_{\ba{l} J_1 J_2 = J \\ K_1 K_2 = K \ea}
   (-1)^{\ep(J_1) \lb \ep(K) + \ep(L) \rb} \d^{K_1}_{J_2} \g'^{I K_2}_{J_1 L} \]
in Eq.(\ref{3.2.13}).  Here the annihilation operators in the subsequence $J_2$ are contracted with the creation
operators in $K_1$.  Thus this term is derived by the following procedure (in the following expressions, 
$b = \#(J)$, $b_1 = \#(J_1)$, $b_2 = \#(J_2)$, $c = \#(K)$, $c_1 = \#(K_1)$ and $c_2 = \#(K_2)$): 
\begin{eqnarray*}
   & & N^{-(a + b - 2) / 2} \a^{\da\mu_2}_{\mu_1} (i_1, \ep(i_1)) \a^{\da\mu_3}_{\mu_2} (i_2, \ep(i_2)) \cd \\
   & & \a^{\da\mu_{a+1}}_{\mu_a} (i_a, \ep(i_a)) \a^{\mu_{a+2}}_{\mu_{a+1}} (j_b, \ep(j_b)) 
   \a^{\mu_{a+3}}_{\mu_{a+2}} (j_{b - 1}, \ep(j_{b - 1})) \cd \\
   & & \a^{\mu_1}_{\mu_{a + b}} (j_1, \ep(j_1)) N^{-(c + d - 2) / 2} \a^{\da\n_2}_{\n_1} (k_1, \ep(k_1)) 
   \a^{\da\n_3}_{\n_2} (k_2, \ep(k_2)) \cd \\
   & & \a^{\da\n_{c + 1}}_{\n_c} (k_c, \ep(k_c)) \a^{\n_{c + 2}}_{\n_{c + 1}} (l_d, \ep(l_d)) 
   \a^{\n_{c + 3}}_{\n_{c + 2}} (l_{d-1}, \ep(l_{d-1})) \cd \\
   & & \a^{\n_1}_{\n_{c + d}} (l_1, \ep(l_1)) \\
   & = & (-1)^{\ep(J_1) \lb \ep(K) + \ep(L) \rb} N^{-(a + b + c + d - 4)/2} \a^{\da\mu_2}_{\mu_1} (i_1, \ep(i_1))
   \a^{\da\mu_3}_{\mu_2} (i_2, \ep(i_2)) \cd \\
   & & \a^{\da\mu_{a+1}}_{\mu_a} (i_a, \ep(i_a)) \a^{\mu_{a+2}}_{\mu_{a+1}} (j_b, \ep(j_b)) 
   \a^{\mu_{a+3}}_{\mu_{a+2}} (j_{b-1}, \ep(j_{b-1})) \cd \\
   & & \a^{\mu_{a + b_2 + 1}}_{\mu_{a + b_2}} (j_{b_1 + 1}, \ep(j_{b_1 + 1})) \a^{\da\n_2}_{\n_1} (k_1, \ep(k_1)) 
   \a^{\da\n_3}_{\n_2} (k_2, \ep(k_2)) \cd \\ 
   & & \a^{\da\n_{c + 1}}_{\n_c} (k_c, \ep(k_c)) \a^{\n_{c + 2}}_{\n_{c + 1}} (l_d, \ep(l_d)) 
   \a^{\n_{c + 3}}_{\n_{c + 2}} (l_{d-1}, \ep(l_{d-1})) \cd \\
   & & \a^{\n_1}_{\n_{c + d}} (l_1, \ep(l_1)) \a^{\mu_{a + b_2 + 2}}_{\mu_{a + b_2 + 1}} (j_{b_1}, \ep(j_{b_1}))
   \a^{\mu_{a + b_2 + 3}}_{\mu_{a + b_2 + 2}} (j_{b_1 - 1}, \ep'(j_{b_1 - 1})) \cd \\
   & & \a^{\mu_1}_{\mu_{a + b}} (j_1, \ep(j_1)) + \cd \\
   & = & (-1)^{\ep(J_1) \lb \ep(K) + \ep(L) \rb} N^{-(a + b_1 + c_2 + d - 2)/2} 
   \a^{\da\mu_2}_{\mu_1} (i_1, \ep(i_1)) \a^{\da\mu_3}_{\mu_2} (i_2, \ep(i_2)) \cd \\
   & & \a^{\da\mu_{a+1}}_{\mu_a} (i_a, \ep(i_a)) \a^{\da\n_{c_1 + 2}}_{\mu_{a + 1}} (k_{c_1 + 1}, \ep(k_{c_1 + 1})) 
   \a^{\da\n_{c_1 + 3}}_{\n_{c_1 + 2}} (k_{c_1 + 2}, \ep(k_{c_1 + 2})) \cd \\
   & & \a^{\da\n_{c + 1}}_{\n_c} (k_c, \ep(k_c)) \a^{\n_{c + 2}}_{\n_{c + 1}} (l_d, \ep(l_d)) 
   \a^{\n_{c + 3}}_{\n_{c + 2}} (l_{d-1}, \ep(l_{d-1})) \cd \\
   & & \a^{\n_1}_{\n_{c + d}} (l_1, \ep(l_1)) \a^{\mu_{a + b_2 + 2}}_{\n_1} (j_{b_1}, \ep(j_{b_1}))
   \a^{\mu_{a + b_2 + 3}}_{\mu_{a + b_2 + 2}} (j_{b_1 - 1}, \ep(j_{b_1 - 1})) \cd \\
   & & \a^{\mu_1}_{\mu_{a + b}} (j_1, \ep(j_1)) + \cd.
\end{eqnarray*}
Other terms in Eqs.(\ref{3.2.13}), (\ref{3.2.14}) and (\ref{3.2.15}) can be understood in a similar fashion.

The {\em heterix superalgebra} is the Lie superalgebra defined by the Lie superbrackets in Eqs.(\ref{3.2.13}) to 
(\ref{3.2.15}).  If we consider only the subspace of the heterix superalgebra spanned by all color-invariant 
operators with no fermions in the adjoint representation, we will obtain the {\em heterix algebra} $\hatheterix$.  

\section{Structure of the Precursor Algebra}
\la{s3.3}

{\em A remark on the notations in the remaining sections of this chapter and their accompanying appendices: $s'^K$ 
and $\g'^I_J$ introduced in the previous section will be written as $s^K$ and $\g^I_J$, respectively.  $\Ps'^K$ and 
$\ti{f}'^I_J$ will be written as $\Ps^{(K)}$ and $\ti{f}^{(I)}_{(J)}$, respectively because with the absence of 
adjoint fermions, any cyclic permutations of $K$, $I$ and $J$ leave them unchanged.  Moreover, we will take $\L$ to 
be the number of degrees of freedom in this algebra.  These notational changes are not valid in other chapters.}

\begin{figure}
\epsfxsize=3.3in
\epsfysize=6.7in
\centerline{\epsfbox{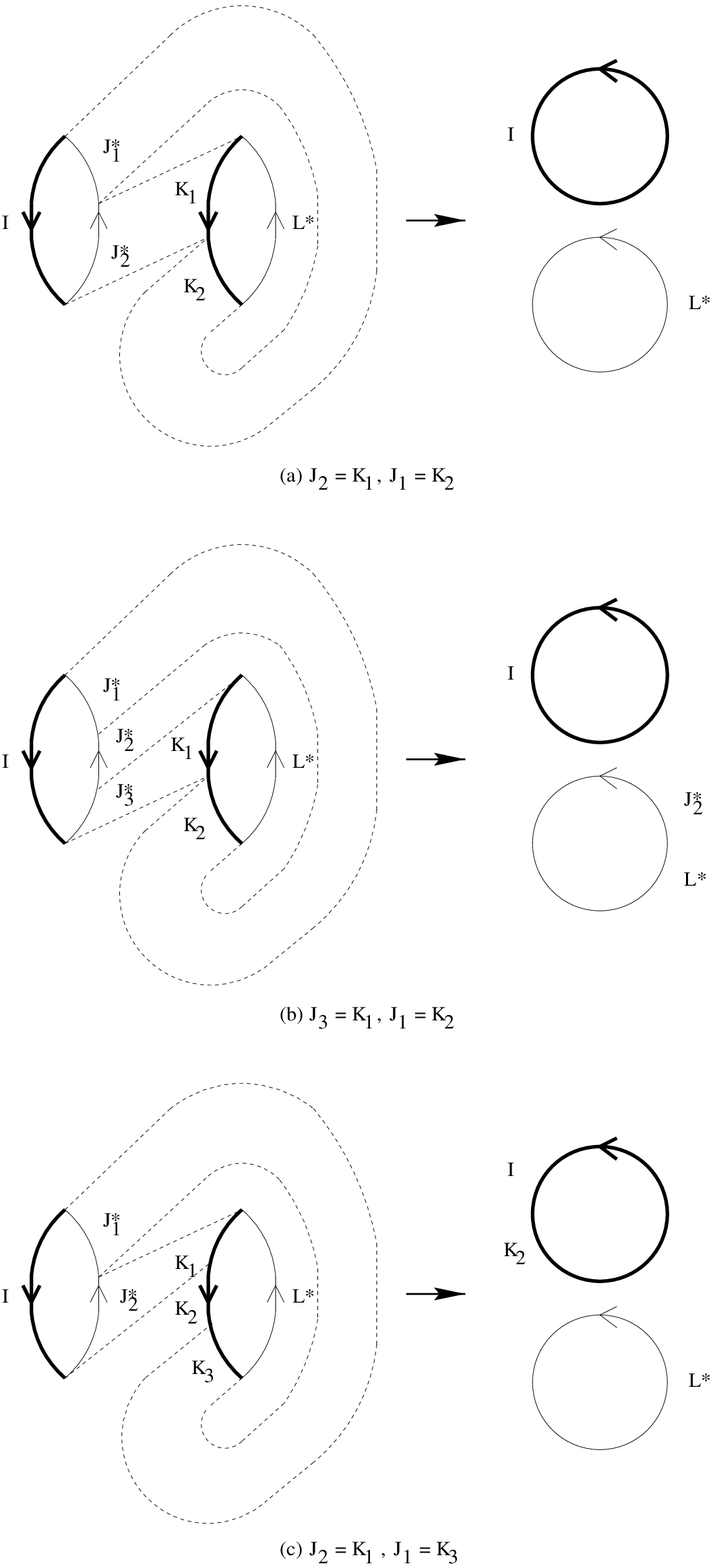}}
\caption{\em Some terms on the R.H.S. of Eq.(\ref{3.3.1}).}
\label{f3.3.1}
\end{figure}

\begin{figure}
\epsfxsize=3.5in
\centerline{\epsfbox{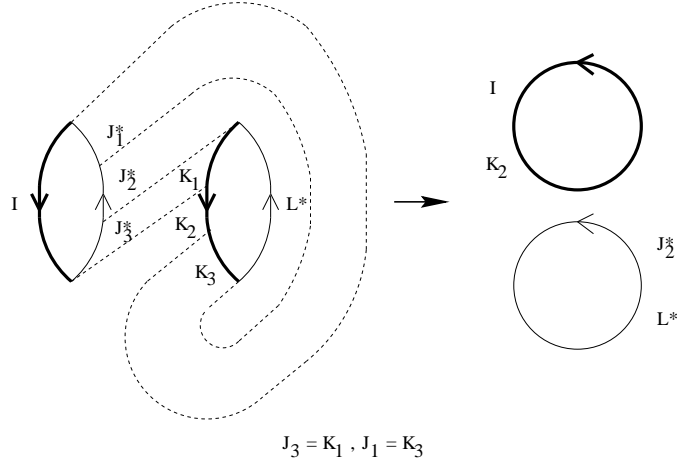}}
\caption{\em Another term on the R.H.S. of Eq.(\ref{3.3.1}).}
\label{f3.3.2}
\end{figure}

We are going to find a Cartan subalgebra of the heterix algebra and its associated root vectors.  We will also 
consider the special case when $\L = 1$ to see what the algebra looks like.  Readers who are only interested in the
grand string superalgebra and its subalgebras can skip this section altogether.

From Eqs.(\ref{3.2.13}), (\ref{3.2.14}) and (\ref{3.2.15}), we can obtain the following Lie brackets of the heterix
algebra:
\begin{eqnarray}
   \lefteqn{ \left[ \g^I_J, \g^K_L \right] = 
   \delta^K_J \g^I_L + \sum_{J_1 J_2 = J} \delta^K_{J_2} \g^I_{J_1 L} + 
   \sum_{K_1 K_2 = K} \delta^{K_1}_J \g^{I K_2}_L } \nonumber \\
   & & + \sum_{\begin{array}{l}
		  J_1 J_2 = J \\
		  K_1 K_2 = K
	       \end{array}}
   \delta^{K_1}_{J_2} \g^{I K_2}_{J_1 L} 
   + \sum_{J_1 J_2 = J} \delta^K_{J_1} \g^I_{L J_2}
   + \sum_{K_1 K_2 = J} \delta^{K_2}_J \g^{K_1 I}_L \nonumber \\
   & & + \sum_{\begin{array}{l}
		  J_1 J_2 = J \\
		  K_1 K_2 = K
	       \end{array}}
   \delta^{K_2}_{J_1} \g'^{K_1 I}_{L J_2}
   + \sum_{J_1 J_2 J_3 = J} \delta^K_{J_2} \g^I_{J_1 L J_3} 
   + \sum_{K_1 K_2 K_3 = K} \delta^{K_2}_J \g^{K_1 I K_3}_L \nonumber \\
   & & + \sum_{\begin{array}{l}
   		  J_1 J_2 = J \\
   		  K_1 K_2 = K
   	       \end{array}}
   \delta^{K_1}_{J_2} \delta^{K_2}_{J_1} \ti{f}^{(I)}_{(L)}
   + \sum_{\begin{array}{l}
   	      J_1 J_2 J_3 = J \\
   	      K_1 K_2 = K
   	   \end{array}}
   \delta^{K_1}_{J_3} \delta^{K_2}_{J_1} \ti{f}^{(I)}_{(J_2 L)}
   \nonumber \\
   & & + \sum_{\begin{array}{l}
   		  J_1 J_2 = J \\
   		  K_1 K_2 K_3 = K
   	       \end{array}}
   \delta^{K_1}_{J_2} \delta^{K_3}_{J_1} \ti{f}^{(I K_2)}_{(L)} \nonumber \\
   & & + \sum_{\begin{array}{l}
		  J_1 J_2 J_3 = J \\
		  K_1 K_2 K_3 = K
	       \end{array}}
   \delta^{K_1}_{J_3} \delta^{K_3}_{J_1} \ti{f}^{(I K_2)}_{(J_2 L)} 
    - (I \leftrightarrow K, J \leftrightarrow L). 
\label{3.3.1}
\end{eqnarray}
Moreover, the commutators between a $\g$ and an $\ti{f}$, and between two $\ti{f}$'s, are
\begin{equation}
   \left[ \g^I_J, \ti{f}^{(K)}_{(L)} \right] =
   \delta^K_{(J)} \ti{f}^{(I)}_{(L)} + \sum_{K_1 K_2 = (K)} \delta^{K_1}_J      
   \ti{f}^{(I K_2)}_{(L)} - \delta^I_{(L)} \ti{f}^{(K)}_{(J)} -
   \sum_{L_1 L_2 = (L)} \delta^I_{L_2} \ti{f}^{(K)}_{(L_1 J)}  
\label{3.3.2}
\end{equation}
and
\begin{equation} 
   \left[ \ti{f}^{(I)}_{(J)}, \ti{f}^{(K)}_{(L)} \right] = 
   \delta^K_{(J)} \ti{f}^{(I)}_{(L)} - \delta^I_{(L)} \ti{f}^{(K)}_{(J)},
\label{3.3.3}
\end{equation}
respectively.  In the above equations, the summation $\sum_{K_1 K_2 = (K)}$ and the delta function $\d^I_{(J)}$ are 
introduced in Appendix~\ref{sa1.2}.  The proof of these equations will be relegated to Appendix~\ref{sa3.3}.  The 
first nine terms on the R.H.S. of Eq.(\ref{3.3.1}) are the same as those of Eq.(\ref{4.5.4}), and have been 
depicted in Fig.\ref{f4.4}.  The next four terms on the R.H.S. of Eq.(\ref{3.3.1}) are new terms which have not 
shown up before.  The tenth to the twelfth terms are depicted in Fig.\ref{f3.3.1}, and the thirteenth term is
depicted in Fig.\ref{f3.3.2}.  Illustrated in Fig.~\ref{f3.3.3} are the first two terms on the R.H.S. of
Eq.(\ref{3.3.2}), and the first term on the R.H.S. of Eq.(\ref{3.3.3}).  These two equations together clearly show 
that the subspace spanned by all $\ti{f}^{(I)}_{(J)}$'s, which we will call $\ti{F}_{\L}$, forms an ideal of 
$\hat{F}_{\L}$.

\begin{figure}
\epsfxsize=4in
\epsfysize=6.5in
\centerline{\epsfbox{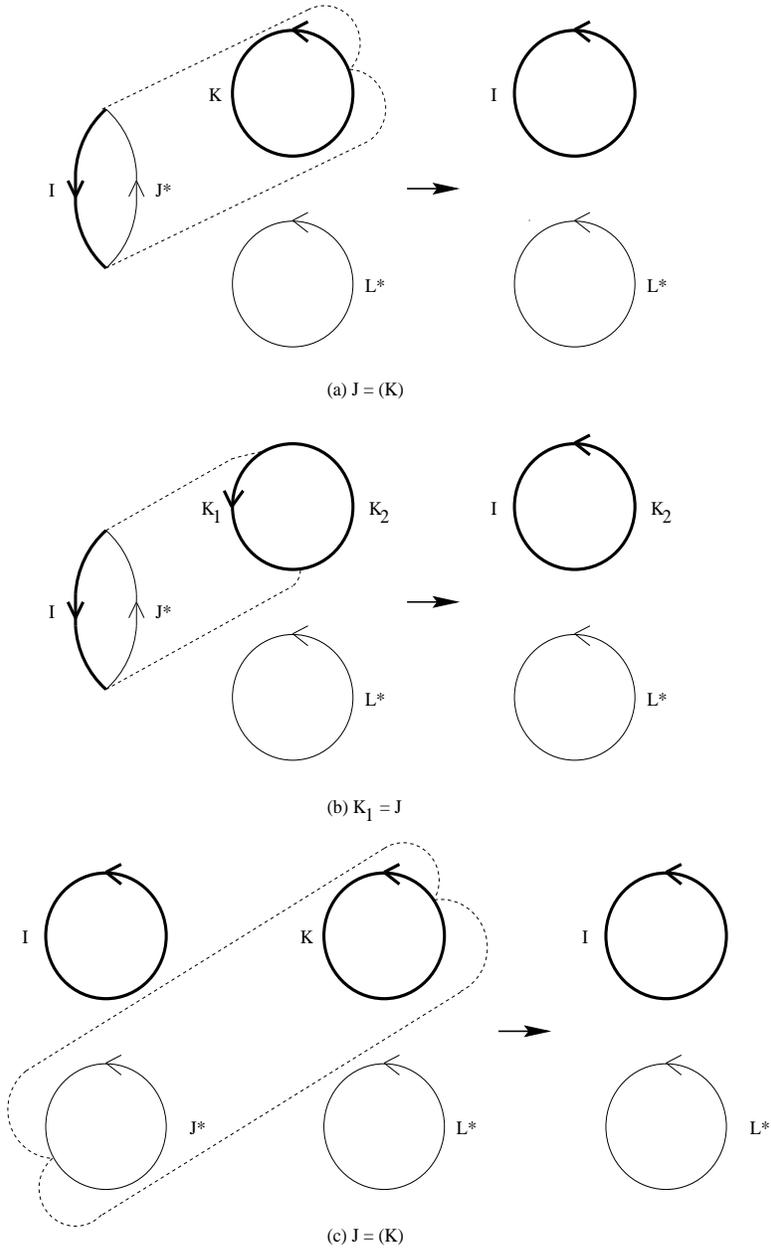}}
\caption{\em The Lie brackets involving an operator of the fifth kind.}
\label{f3.3.3}
\end{figure}

Let us explore the structure of the heterix algebra.  Consider the subspace ${\cal H}$ spanned by all vectors of the
forms $\g^I_I$ and $\ti{f}^{(I)}_{(I)}$, where $I$ is an arbitrary finite integer sequence of integers between 1 and
$\Lambda$ inclusive.  It turns out that ${\cal H}$ is a Cartan subalgebra of the heterix algebra.  (The proof is 
given in Appendix~E of Ref.\cite{clstal}.  This is essentially just a more sophisticated version of 
Appendix~\ref{sa4.2}.)  In addition, we have the following special cases of Eqs.(\ref{3.3.2}) and (\ref{3.3.3}):
\begin{equation}
   \left[ \g^I_I, \ti{f}^{(K)}_{(L)} \right] = \left( \delta^K_{(I)}
   + \sum_{K_1 K_2 = (K)} \delta^{K_1}_I - \delta^I_{(L)} - 
   \sum_{L_1 L_2 = (L)} \delta^I_{L_1} \right) \ti{f}^{(K)}_{(L)}
\label{3.3.4}
\end{equation}
and
\begin{equation}
   \left[ \ti{f}^{(I)}_{(I)}, \ti{f}^{(K)}_{(L)} \right] = \left(
   \delta^K_{(I)} - \delta^I_{(L)} \right) \ti{f}^{(K)}_{(L)}.
\label{3.3.5}
\end{equation}
Therefore, every $\ti{f}^{(K)}_{(L)}$ is a root vector of the heterix algebra with respect to the subalgebra 
${\cal H}$.

Suppose we define some operators as follows:
\begin{eqnarray}
   l^I_J & \equiv & \g^I_J - \sum_{i=1}^{\Lambda} \g^{iI}_{iJ} - \ti{f}^{(I)}_{(J)} \mbox{; and} \nonumber \\
   r^I_J & \equiv & \g^I_J - \sum_{j=1}^{\Lambda} \g^{Ij}_{Jj} - \ti{f}^{(I)}_{(J)}.
\label{3.3.6}
\end{eqnarray}
The actions of these operators are:
\begin{eqnarray}
   l^I_J s^K & = & \delta^K_J s^I + \sum_{K_1 K_2 = K} \delta^{K_1}_J s^{I K_2}; \nonumber \\
   l^I_J \Ps^{(K)} & = & 0; \nonumber \\
   r^I_J s^K & = & \delta^K_J s^I + \sum_{K_1 K_2 = K} \delta^{K_2}_J s^{K_1 I} \mbox{; and} \nonumber \\
   r^I_J \Ps^{(K)} & = & 0.
\label{3.3.7}
\end{eqnarray}
Hence they act only on open singlet states.  $l^I_J$ and $r^I_J$ are precisely the operators discussed in 
Sections~\ref{s4.3} and \ref{s4.4}.  Thus the algebra consisting of finite linear combinations of $l^I_J$'s, 
$\hat{L}'_{\Lambda}$, is a subalgebra of the heterix algebra.  Similarly, the algebra consisting of finite linear
combinations of $r^I_J$'s, $\hat{R}'_{\Lambda}$, is also a subalgebra of the heterix algebra.  

Now define
\begin{eqnarray}
   f^I_J & \equiv & \g^I_J - \sum_{i=1}^{\Lambda} \g^{iI}_{iJ} - \sum_{j=1}^{\Lambda} \g^{Ij}_{Jj}
		    \sum_{i,j=1}^{\Lambda} \g^{iIj}_{iJj} - \ti{f}^{(I)}_{(J)} + 
		    \sum_{i=1}^{\Lambda} \ti{f}^{(iI)}_{(iJ)}
\nonumber \\
   & = & l^I_J - \sum_{j=1}^{\Lambda} l^{Ij}_{Jj} \nonumber \\
   & = & r^I_J - \sum_{i=1}^{\Lambda} r^{iI}_{iJ}.
\label{3.3.8}
\end{eqnarray}
Its actions read
\begin{eqnarray}
   f^I_J s^K & = & \delta^K_J s^I \mbox{; and} \nonumber \\
   f^I_J \Ps^{(K)} & = & 0.
\label{3.3.9}
\end{eqnarray}
Hence this $f^I_J$ is the one we discussed in Section~\ref{s4.2}.  The algebra consisting of finite linear 
combinations of $f^I_J$'s, $F'_{\Lambda}$, is a subalgebra of $\hat{L}'_{\Lambda}$, $\hat{R}'_{\Lambda}$ and the 
heterix algebra.  We deduce from Eqs.(\ref{3.2.9}), (\ref{3.2.12}) and (\ref{3.3.9}) that
\begin{eqnarray}
   \lefteqn{ \left[ \g^I_J, f^K_L \right] = \delta^K_J f^I_L + \sum_{K_1 K_2 = K} \delta^{K_1}_J f^{I K_2}_L
   + \sum_{K_1 K_2 = K} \delta^{K_2}_J f^{K_1 I}_L } \nonumber \\
   & & + \sum_{K_1 K_2 K_3 = K} \delta^{K_2}_J f^{K_1 I K_3}_L 
   - \delta^I_L f^K_J - \sum_{L_1 L_2 = L} \delta^I_{L_1} f^K_{J L_2} \nonumber \\ 
   & & - \sum_{L_1 L_2 = L} \delta^I_{L_2} f^K_{L_1 J} - \sum_{L_1 L_2 L_3 = L} \delta^I_{L_2} f^K_{L_1 J L_3}
   \mbox{; and} \nonumber \\
   \lefteqn{ \left[ \ti{f}^{(I)}_{(J)}, f^K_L \right] = 0. }
\end{eqnarray}
In particular,
\begin{eqnarray}
   \left[ \g^I_I, f^K_L \right] & = & \left( \delta^K_I + \sum_{K_1 K_2 = K} \delta^{K_1}_I 
   + \sum_{K_1 K_2 = K} \delta^{K_2}_I + \sum_{K_1 K_2 K_3 = K} \delta^{K_2}_I \right. \nonumber \\
   & & \left. - \delta^I_L - \sum_{L_1 L_2 = L} \delta^I_{L_1} - \sum_{L_1 L_2 = L} \delta^I_{L_2}
   - \sum_{L_1 L_2 L_3 = L} \delta^I_{L_2} \right) f^K_L \mbox{; and} \nonumber \\
   \left[ \ti{f}^{(I)}_{(I)}, f^K_L \right] & = & 0.
\label{3.3.10}
\end{eqnarray}
Thus every $f^K_L$ is also a root vector with respect to the Cartan subalgebra ${\cal H}$.  In fact, a root vector
with respect to ${\cal H}$ is either of the form $\tilde{f}^{(K)}_{(L)}$ or $f^K_L$.  The proof of this assertion 
can be found in Appendix~F of Ref.\cite{clstal}.  (This is essentially a more sophisticated version of 
Appendix~\ref{sa4.5}.)  Hence every root space is one-dimensional.

Now let us consider various quotient algebras of the heterix algebra.  Since $\tsalt$ is a proper ideal, we
can make the set of all cosets $\g^I_J + \tsalt$ into a quotient algebra.  This quotient algebra is nothing but
the centrix algebra $\hat{\Sigma}_{\Lambda}$ defined in Section~\ref{s4.5}.  The multix algebra 
$\hat{M}'_{\Lambda}$ discussed in Section~\ref{s4.4} is another proper ideal.  Hence we can make the set of all 
cosets $\g^I_J + \hat{M}'_{\Lambda}$ and $\ti{f}^{(I)}_{(J)} + \hat{M}'_{\L}$ a quotient algebra as well.  Note 
that every operator in $\hat{M}'_{\Lambda}$ kills a closed singlet state.  We conjecture that the vector space 
spanned by all closed string states provides a faithful representation for the quotient algebra of all cosets 
$\g^I_J + \hat{M}'_{\Lambda}$ and $\tilde{f}^{(I)}_{(J)} + \hat{M}'_{\Lambda}$.  As we will see in 
Section~\ref{s5.4}, the corresponding $\Lambda = 1$ quotient algebra is again the Witt algebra.  Therefore the 
quotient algebra with $\Lambda > 1$ can be regarded as a generalization of the Witt algebra.  At present we know 
nothing about the relationship between this generalization of the Witt algebra, and the other generalization 
$\centrix$ discussed in Section~\ref{s4.5}.  We would like to investigate this in the future.

The Lie bracket of the subalgebra $\tsalt$ is given by Eq.(\ref{3.3.3}).  If we rescale each vector by defining
\begin{equation}
   f^{(I)}_{(J)} \equiv \frac{\ti{f}^{(I)}_{(J)}}{\sqrt{\delta^I_{(I)}\delta^J_{(J)}}},
\la{3.3.11}
\end{equation}
then the Lie bracket between two $f^{(I)}_{(J)}$'s is similar to that in Eq.(\ref{3.3.3}), except that the delta 
function there with the generic form $\delta^I_{(J)}$ should be replaced with another delta function such that this 
new delta function yields simply 1 if $(I) = (J)$, and remains 0 otherwise.  Then there is a one-to-one 
correspondence between each  $f^{(I)}_{(J)}$ and each complex matrix $(a_{ij})_{i,j} \in Z_+$ such that all but one 
$a_{ij}$ is nonzero, and that nonzero entry is 1.  Thus the Lie algebras $\tsalt$, for any $\Lambda$ are all 
isomorphic to $\gl$, the inductive limit of the general linear algebras.  

\chapter{Open String Superalgebra and Closed String Superalgebra}
\la{c5}

\section{Introduction}
\la{s5.1}

In the previous chapter, we justified the existence of a superalgebra, and studied its mathematical 
structure.  Building upon this result, we will finally fulfill our promise in this chapter by deriving a Lie
superalgebra for open superstrings, and one for closed superstrings, as subalgebras of this precursor superalgebra. 
This is done as follows.  In Section~\ref{s5.2}, we will give the definition of a superalgebra of operators acting
on open and closed singlet states.  We will call this the `grand string superalgebra'.  The difference between 
this superalgebra and the precursor superalgebra is that there are more than one degree of freedom at the ends of 
the open singlet states.  At a first glance, this superalgebra is somewhat `larger' than the precursor 
superalgebra.  Ironically, we will derive this grand string superalgebra as a {\em subalgebra} of the precursor 
superalgebra by manipulating the numbers of degrees of freedom in an accompanying appendix.

In Section~\ref{s5.3}, we will derive a superalgebra just for open superstrings.  This will be a quotient algebra
of the grand string superalgebra.  In Section~\ref{s5.4}, we will derive a superalgebra for closed superstrings as
another quotient algebra of the grand string superalgebra.  We will see that the corresponding Lie algebra, which we
will call the `cyclix algebra', for bosonic closed strings has a close relationship with the Witt algebra, too.

To have a glimpse of how to use these algebras in the study of physical systems, we will turn to the Ising model
again in Section~\ref{s5.5}.  It is well known that there are many ways of solving the quantum Ising model in one
dimension.  One method which is close to the spirit of the original way Onsager himself solved the model 
\cite{onsager} is via the `Onsager algebra' \cite{davies90}.  We will see that actually this Onsager algebra is
a subalgebra of the cyclix algebra, and we can use the cyclix algebra directly to obtain some conserved quantities
of the Ising matrix model.  This example may give us a clue of how to use these Lie algebras more effectively in the
future.

\section{Definition of the Grand String Superalgebra}
\la{s5.2}

We are going to give a precise definition of a Lie superalgebra of the operators defined in Section~\ref{s2.2} 
acting on closed or open singlet states.  Unlike the open singlet states in Section~\ref{s3.2}, the numbers of 
degrees of freedom of the fundamental and conjugate fields are arbitrary.  The proof that this is really a Lie
superalgebra will be given in Appendix~\ref{sa5.1}.

Consider a vector space of all finite linear combinations of vectors of the forms given in Table~\ref{t5.1}, where 
$1 \leq \l_1, \l_2, \l_3, \; \mbox{and} \; \l_4 \leq 2 \L_F$, $I$ and $J$ are non-empty sequences of integers
between 1 and $2 \L$ inclusive, and $\dot{I}$ and $\dot{J}$ are empty or non-empty sequences of integers also
between 1 and $2 \L$ inclusive.  In the above table, $\ep(I)$ is 0 if the number of integers between $\L + 1$ and 
$2\L$ inclusive in $I$ is even, and 1 if it is odd.  The definition is analogous for $\ep(\dot{I})$.  The reader 
can regard the five types of vectors as those operators defined in Eq.(\ref{a5.1.1}), or as axiomatic entities 
satisfying a set of Lie superbrackets to be described immediately.  We call a vector or an operator an {\em 
operator of the first, second, third, fourth or fifth kind} if it is a finite linear combination of operators, all 
of which are of the first, second, third, fourth or fifth form enlisted above, respectively.

\begin{table}
\begin{center}
\begin{tabular}{||c|c|c||}
\hline
operator of & expression & grade \\
which kind & & \\ \hline \hline
first & $\bar{\X}^{\l_1}_{\l_2} \otimes f^{\dot{I}}_{\dot{J}} \otimes \X^{\l_3}_{\l_4}$ & 
$\ep(\dot{I}) + \ep(\dot{J}) \pmod{2}$ \\
second & $\bar{\X}^{\l_1}_{\l_2} \otimes l^{\dot{I}}_{\dot{J}}$ & $\ep(\dot{I}) + \ep(\dot{J}) \pmod{2}$ \\
third & $r^{\dot{I}}_{\dot{J}} \otimes \X^{\l_3}_{\l_4}$ & $\ep(\dot{I}) + \ep(\dot{J}) \pmod{2}$ \\
fourth & $\g^I_J$ & $\ep(I) + \ep(J) \pmod{2}$ \\
fifth & $\ti{f}^I_J$ & $\ep(I) + \ep(J) \pmod{2}$ \\
\hline \hline
\end{tabular}
\caption{\em The grades of operators.}
\end{center}
\la{t5.1}
\end{table}

Let us describe the Lie superbrackets among different kinds of operators.  The Lie superbracket between an
operator of the fifth kind, and an operator of the first, second or third kind is trivial:
\beq
   \le[ \bar{\X}^{\l_1}_{\l_2} \otimes f^{\dot{I}}_{\dot{J}} \otimes \X^{\l_3}_{\l_4}, 
   \ti{f}^K_L \ri]_{\pm} & = & 0; \nn \\
   \le[ \bar{\X}^{\l_1}_{\l_2} \otimes l^{\dot{I}}_{\dot{J}}, \ti{f}^K_L \ri]_{\pm} & = & 0 \mbox{; and} \nn \\
   \le[ r^{\dot{I}}_{\dot{J}} \otimes \X^{\l_1}_{\l_2}, \ti{f}^K_L \ri]_{\pm} & = & 0.
\la{5.2.1}
\eeq
The operators of the first kind form a proper ideal of the Lie superalgebra:
\beq
   \lefteqn{\le[ \bar{\X}^{\l_1}_{\l_2} \otimes f^{\dot{I}}_{\dot{J}} \otimes \X^{\l_3}_{\l_4},
   \bar{\X}^{\l_5}_{\l_6} \otimes f^{\dot{K}}_{\dot{L}} \otimes \X^{\l_7}_{\l_8} \ri]_{\pm} = 
   \d^{\l_5}_{\l_2} \d^{\dot{K}}_{\dot{J}} \d^{\l_7}_{\l_4} \bar{\X}^{\l_1}_{\l_6} \otimes f^{\dot{I}}_{\dot{L}} 
   \otimes \X^{\l_3}_{\l_8} } \nn \\
   & & - (-1)^{\lb \ep(I) + \ep(J) \rb \lb \ep(K) + \ep(L) \rb} \d^{\l_1}_{\l_6} \d^{\dot{I}}_{\dot{L}} 
   \d^{\l_3}_{\l_8} \bar{\X}^{\l_5}_{\l_2} \otimes f^{\dot{K}}_{\dot{J}} \otimes \X^{\l_7}_{\l_4};
\nn \\
   \lefteqn{\le[ \bar{\X}^{\l_1}_{\l_2} \otimes f^{\dot{I}}_{\dot{J}} \otimes \X^{\l_3}_{\l_4},
   \bar{\X}^{\l_5}_{\l_6} \otimes l^{\dot{K}}_{\dot{L}} \ri]_{\pm} = 
   \d^{\l_5}_{\l_2} \bar{\X}^{\l_1}_{\l_6} \otimes \sum_{\dot{J}_1 \dot{J}_2 = \dot{J}} 
   \d^{\dot{K}}_{\dot{J}_1} f^{\dot{I}}_{\dot{L} \dot{J}_2} \otimes \X^{\l_3}_{\l_4} } \nn \\
   & & - (-1)^{\lb \ep(I) + \ep(J) \rb \lb \ep(K) + \ep(L) \rb} \d^{\l_1}_{\l_6} \bar{\X}^{\l_5}_{\l_2} \otimes
   \sum_{\dot{I}_1 \dot{I}_2 = \dot{I}} \d^{\dot{I}_1}_{\dot{L}} f^{\dot{K} \dot{I}_2}_{\dot{J}} \otimes
   \X^{\l_3}_{\l_4};
\nn \\
   \lefteqn{\le[ \bar{\X}^{\l_1}_{\l_2} \otimes f^{\dot{I}}_{\dot{J}} \otimes \X^{\l_3}_{\l_4},
   r^{\dot{K}}_{\dot{L}} \otimes \X^{\l_5}_{\l_6} \ri]_{\pm} = } \nn \\
   & & \d^{\l_5}_{\l_4} \bar{\X}^{\l_1}_{\l_2} \otimes \sum_{\dot{J}_1 \dot{J}_2 = \dot{J}} 
   (-1)^{\ep(\dot{J}_1) \lb \ep(\dot{K}) + \ep(\dot{L}) \rb} \d^{\dot{K}}_{\dot{J}_2} 
   f^{\dot{I}}_{\dot{J}_1 \dot{L}} \otimes \X^{\l_3}_{\l_6} \nn \\
   & & - (-1)^{\lb \ep(I) + \ep(J) \rb \lb \ep(K) + \ep(L) \rb}
   \d^{\l_3}_{\l_6} \bar{\X}^{\l_1}_{\l_2} \otimes \sum_{\dot{I}_1 \dot{I}_2 = \dot{I}}
   (-1)^{\ep(\dot{I}_1) \lb \ep(\dot{K}) + \ep(\dot{L}) \rb} \d^{\dot{I}_2}_{\dot{L}} 
   f^{\dot{I}_1 \dot{K}}_{\dot{J}} \otimes \X^{\l_5}_{\l_4};
\nn
\eeq
and
\beq 
   & & \le[ \bar{\X}^{\l_1}_{\l_2} \otimes f^{\dot{I}}_{\dot{J}} \otimes \X^{\l_3}_{\l_4}, 
   \g^K_L \ri]_{\pm} = \bar{\X}^{\l_1}_{\l_2} \otimes \le( \sum_{\dot{J}_1 J_2 \dot{J}_3 = \dot{J}} 
   (-1)^{\ep(\dot{J}_1) \lb \ep(K) + \ep(L) \rb} \d^K_{J_2} f^{\dot{I}}_{\dot{J}_3 L \dot{J}_1} \ri. \nn \\
   & & - \le. (-1)^{\lb \ep(I) + \ep(J) \rb \lb \ep(K) + \ep(L) \rb}
   \sum_{\dot{I}_1 I_2 \dot{I}_3 = \dot{I}} (-1)^{\ep(\dot{I}_1) \lb \ep(K) + \ep(L) \rb}
   \d^{I_2}_L f^{\dot{I}_1 K \dot{I}_3}_{\dot{J}} \ri) \otimes \X^{\l_3}_{\l_4}. 
\la{5.2.2}
\eeq
The operators of the second kind form a subalgebra.  So are the operators of the third kind:
\beq
   \lefteqn{\le[ \bar{\X}^{\l_1}_{\l_2} \otimes l^{\dot{I}}_{\dot{J}},
   \bar{\X}^{\l_3}_{\l_4} \otimes l^{\dot{K}}_{\dot{L}} \ri]_{\pm} = } \nn \\ 
   & & \d^{\l_3}_{\l_2} \bar{\X}^{\l_1}_{\l_4} \otimes \le( \d^{\dot{K}}_{\dot{J}} l^{\dot{I}}_{\dot{L}}
   + \sum_{\dot{J}_1 J_2 = \dot{J}} \d^{\dot{K}}_{\dot{J}_1} l^{\dot{I}}_{\dot{L} J_2} +
   \sum_{\dot{K}_1 K_2 = \dot{K}} \d^{\dot{K}_1}_{\dot{J}} l^{\dot{I} K_2}_{\dot{L}} \ri) \nn \\
   & & - (-1)^{\lb \ep(I) + \ep(J) \rb \lb \ep(K) + \ep(L) \rb}
   \d^{\l_1}_{\l_4} \bar{\X}^{\l_3}_{\l_2} \otimes \le( \d^{\dot{I}}_{\dot{L}} l^{\dot{K}}_{\dot{J}}
   + \sum_{\dot{L}_1 L_2 = \dot{L}} \d^{\dot{I}}_{\dot{L}_1} l^{\dot{K}}_{\dot{J} L_2} \ri. \nn \\
   & & \le. + \sum_{\dot{I}_1 I_2 = \dot{I}} \d^{\dot{I}_1}_{\dot{L}} l^{\dot{K} I_2}_{\dot{J}} \ri)
\nn
\eeq
and
\beq
   \lefteqn{\le[ r^{\dot{I}}_{\dot{J}} \otimes \X^{\l_1}_{\l_2},
   r^{\dot{K}}_{\dot{L}} \otimes \X^{\l_3}_{\l_4} \ri]_{\pm} =
   \d^{\l_3}_{\l_2} \le( \d^{\dot{K}}_{\dot{J}} r^{\dot{I}}_{\dot{L}} + \sum_{J_1 \dot{J}_2 = \dot{J}} 
   (-1)^{\ep(J_1) \lb \ep(\dot{K}) + \ep(\dot{L}) \rb} \d^{\dot{K}}_{\dot{J}_2} r^{\dot{I}}_{J_1 \dot{L}} 
   \ri. } \nn \\
   & & + \le. \sum_{K_1 \dot{K}_2 = \dot{K}} (-1)^{\lb \ep(K_1) \lb \ep(\dot{I}) + \ep(\dot{J}) \rb} 
   \d^{\dot{K}_2}_{\dot{J}} r^{K_1 \dot{I}}_{\dot{L}} \ri) \otimes \X^{\l_1}_{\l_4} \nn \\
   & & - (-1)^{\lb \ep(I) + \ep(J) \rb \lb \ep(K) + \ep(L) \rb}
   \d^{\l_1}_{\l_4} \le( \d^{\dot{I}}_{\dot{L}} r^{\dot{K}}_{\dot{J}} 
   + \sum_{L_1 \dot{L}_2 = \dot{L}} (-1)^{\ep(L_1) \lb \ep(\dot{I}) + \ep(\dot{J}) \rb} 
   \d^{\dot{I}}_{\dot{L}_2} r^{\dot{K}}_{L_1 \dot{J}} \ri. \nn \\
   & & + \le. \sum_{I_1 \dot{I}_2 = \dot{I}} (-1)^{\lb \ep(I_1) \lb \ep(\dot{K}) + \ep(\dot{L}) \rb} 
   \d^{\dot{I}_2}_{\dot{L}} r^{I_1 \dot{K}}_{\dot{J}} \ri) \otimes \X^{\l_3}_{\l_2}.
\la{5.2.3}
\eeq
Eqs.(\ref{5.2.1}) to (\ref{5.2.3}) together with the following relations show that operators of the first three 
kinds as a whole form another ideal:
\beq   
   \lefteqn{\le[ \bar{\X}^{\l_1}_{\l_2} \otimes l^{\dot{I}}_{\dot{J}},
   r^{\dot{K}}_{\dot{L}} \otimes \X^{\l_3}_{\l_4} \ri]_{\pm} = } \nn \\
   & & \bar{\X}^{\l_1}_{\l_2} \otimes \le( \sum_{\ba{l} \dot{J}_1 \dot{J}_2 = \dot{J} \\ 
   \dot{K}_1 \dot{K}_2 = \dot{K} \ea} (-1)^{\ep(\dot{J}_1) \lb \ep(\dot{K}) + \ep(\dot{L}) \rb} 
   \d^{\dot{K}_1}_{\dot{J}_2} f^{\dot{I} \dot{K}_2}_{\dot{J}_1 \dot{L}} \ri. \nn \\
   & & - \le. (-1)^{\lb \ep(I) + \ep(J) \rb \lb \ep(K) + \ep(L) \rb}
   \sum_{\ba{l} \dot{I}_1 \dot{I}_2 = \dot{I} \\ \dot{L}_1 \dot{L}_2 = \dot{L} \ea}
   (-1)^{\ep(\dot{I}_1) \lb \ep(\dot{K}) + \ep(\dot{L}) \rb} \d^{\dot{I}_2}_{\dot{L}_1}
   f^{\dot{I}_1 \dot{K}}_{\dot{J} \dot{L}_2} \ri) \otimes \X^{\l_3}_{\l_4};
\nn \\
   \lefteqn{\le[ \bar{\X}^{\l_1}_{\l_2} \otimes l^{\dot{I}}_{\dot{J}}, \g^K_L \ri]_{\pm} = } \nn \\
   & & \bar{\X}^{\l_1}_{\l_2} \otimes \le[ \le(
   \d^K_{\dot{J}} l^{\dot{I}}_L + \sum_{K_1 K_2 = K} \d^{K_1}_{\dot{J}} l^{\dot{I} K_2}_L
   + \sum_{J_1 J_2 = \dot{J}} (-1)^{\ep(J_1) \lb \ep(K) + \ep(L) \rb} \d^K_{J_2} l^{\dot{I}}_{J_1 L} \ri. \ri. 
   \nn \\ 
   & & + \sum_{J_1 J_2 = \dot{J}} \d^K_{J_1} l^{\dot{I}}_{L J_2} +
   \sum_{\ba{l} J_1 J_2 = \dot{J} \\ K_1 K_2 = K \ea} (-1)^{\ep(J_1) \lb \ep(K) + \ep(L) \rb} 
   \d^{K_1}_{J_2} l^{\dot{I} K_2}_{J_1 L} \nn \\
   & & \le. + \sum_{J_1 J_2 J_3 = \dot{J}} (-1)^{\ep(J_1) \lb \ep(K) + 
   \ep(L) \rb} \d^K_{J_2} l^{\dot{I}}_{J_1 L J_3} \ri) \nn \\
   & & - (-1)^{\lb \ep(I) + \ep(J) \rb \lb \ep(K) + \ep(L) \rb} \le(
   \d^{\dot{I}}_L l^K_{\dot{J}} + \sum_{L_1 L_2 = L} \d^{\dot{I}}_{L_1} l^K_{\dot{J} L_2} \ri. \nn \\ 
   & & + \sum_{I_1 I_2 = \dot{I}} (-1)^{\ep(I_1) \lb \ep(K) + \ep(L) \rb} \d^{I_2}_L l^{I_1 K}_{\dot{J}}
   + \sum_{I_1 I_2 = \dot{I}} \d^{I_1}_L l^{K I_2}_{\dot{J}} \nn \\
   & & + \sum_{\ba{l} L_1 L_2 = L \\ I_1 I_2 = \dot{I} \ea} (-1)^{\ep(I_1) \lb \ep(K) + \ep(L) \rb} 
   \d^{I_2}_{L_1} l^{I_1 K}_{\dot{J} L_2} \nn \\ 
   & & \le. \le. + \sum_{I_1 I_2 I_3 = \dot{I}} (-1)^{\ep(I_1) \lb \ep(K) + \ep(L) \rb} 
   \d^{I_2}_L l^{I_1 K I_3}_{\dot{J}} \ri) \ri] \mbox{; and}
\nn \\
   \lefteqn{\le[ r^{\dot{I}}_{\dot{J}} \otimes \X^{\l_1}_{\l_2}, \g^K_L \ri]_{\pm} = } \nn \\
   & & \le[ \le( \d^K_{\dot{J}} r^{\dot{I}}_L + 
   \sum_{K_1 K_2 = K} (-1)^{\ep(K_1) \lb \ep(\dot{I}) + \ep(\dot{J}) \rb} 
   \d^{K_2}_{\dot{J}} r^{K_1 \dot{I}}_L \ri. \ri. \nn \\
   & & + \sum_{J_1 J_2 = \dot{J}} (-1)^{\ep(J_1) \lb \ep(K) + \ep(L) \rb} \d^K_{J_2} r^{\dot{I}}_{J_1 L} 
   + \sum_{J_1 J_2 = \dot{J}} \d^K_{J_1} r^{\dot{I}}_{L J_2} \nn \\
   & & + \sum_{\ba{l} J_1 J_2 = \dot{J} \\ K_1 K_2 = K \ea} 
   (-1)^{\ep(K_1) \lb \ep(\dot{I}) + \ep(\dot{J}) \rb} \d^{K_2}_{J_1} r^{K_1 \dot{I}}_{L J_2} \nn \\
   & & \le. + \sum_{J_1 J_2 J_3 = \dot{J}} (-1)^{\ep(J_1) \lb \ep(K) + \ep(L) \rb} \d^K_{J_2} 
   r^{\dot{I}}_{J_1 L J_3} \ri) \nn \\
   & & - (-1)^{\lb \ep(I) + \ep(J) \rb \lb \ep(K) + \ep(L) \rb} \le( 
   \d^{\dot{I}}_L r^K_{\dot{J}} - \sum_{L_1 L_2 = L} (-1)^{\ep(L_1) \lb \ep(\dot{I}) + \ep(\dot{J}) \rb} 
   \d^{\dot{I}}_{L_2} r^K_{L_1 \dot{J}} \ri. \nn \\
   & & + \sum_{I_1 I_2 = \dot{I}} (-1)^{\ep(I_1) \lb \ep(K) + \ep(L) \rb} \d^{I_2}_L r^{I_1 K}_{\dot{J}}
   + \sum_{I_1 I_2 = \dot{I}} \d^{I_1}_L r^{K I_2}_{\dot{J}} \nn \\
   & & + \sum_{\ba{l} L_1 L_2 = L \\ I_1 I_2 = \dot{I} \ea} 
   (-1)^{\ep(L_1) \lb \ep(\dot{I}) + \ep(\dot{J}) \rb} \d^{I_1}_{L_2} r^{K I_2}_{L_1 \dot{J}} \nn \\
   & & + \le. \le. \sum_{I_1 I_2 I_3 = \dot{I}} (-1)^{\ep(I_1) \lb \ep(K) + \ep(L) \rb} 
   \d^{I_2}_L r^{I_1 K I_3}_{\dot{J}} \ri) \ri] \otimes \X^{\l_1}_{\l_2}.
\la{5.2.4}
\eeq
These equations also reveal that operators of the first kind form a proper ideal of the superalgebra spanned
by operators of the first three kinds.

The operators of the fifth kind form yet another proper ideal of this superalgebra:
\beq
   \le[ \ti{f}^I_J, \ti{f}^K_L \ri]_{\pm} & = & \d^K_J \ti{f}^I_L + 
   \sum_{K_1 K_2 = K} (-1)^{\ep(K_1) \ep(K_2)} \d^{K_2 K_1}_J \ti{f}^I_L \nn \\
   & & - (-1)^{\lb \ep(I) + \ep(J) \rb \lb \ep(K) + \ep(L) \rb} (I \lrar K, J \lrar L)
\la{5.2.5}
\eeq
and
\beq
   \lefteqn{ \le[ \g^I_J, \ti{f}^K_L \ri]_{\pm} = \le\{ \d^K_J \ti{f}^I_L +
   \sum_{K_1 K_2 = K} (-1)^{\ep(K_1) \ep(K_2)} \d^{K_2 K_2}_J \ti{f}^I_L +
   \sum_{K_1 K_2 = K} \d^{K_1}_J \ti{f}^{I K_2}_L \ri. } \nn \\
   & & + \sum_{K_1 K_2 K_3 = K} \d^{K_2}_J (-1)^{\ep(K_1) \lb \ep(I) + \ep(J) \rb} \ti{f}^{K_1 I K_3}_L \nn \\ 
   & & + \sum_{K_1 K_2 = K} \d^{K_2}_J (-1)^{\ep(K_1) \lb \ep(I) + \ep(J) \rb} \ti{f}^{K_1 I}_L \nn \\
   & & \le. + \sum_{K_1 K_2 K_3 = K} \d^{K_3 K_1}_J (-1)^{\lb \ep(K_1) + \ep(K_2) \rb \ep(K_3)} \ti{f}^{I K_2}_L
   \ri\} \nn \\
   & & - (-1)^{\lb \ep(I) + \ep(J) \rb \lb \ep(K) + \ep(L) \rb} \le\{ \d^I_L \ti{f}^K_J 
   - \sum_{L_1 L_2 = L} (-1)^{\ep(L_1) \ep(L_2)} \d^I_{L_2 L_1} \ti{f}^K_J \ri. \nn \\
   & & - \sum_{L_1 L_2 = L} (-1)^{\ep(L_1) \lb \ep(I) + \ep(J) \rb} \d^I_{L_2} \ti{f}^K_{L_1 J} \nn \\
   & & - \sum_{L_1 L_2 L_3 = L} (-1)^{\ep(L_1) \lb \ep(I) + \ep(J) \rb} \d^I_{L_2} \ti{f}^K_{L_1 J L_3}
   - \sum_{L_1 L_2 = L} \d^I_{L_1} \ti{f}^K_{J L_2} \nn \\
   & & \le. - \sum_{L_1 L_2 L_3 = L} (-1)^{\ep(L_3) \lb \ep(L_1) + \ep(L_2) \rb} \d^I_{L_3 L_1} \ti{f}^K_{J L_2}
   \ri\}.
\la{5.2.6}
\eeq
Finally, the Lie superbracket between two operators of the fourth kind is a linear combination of operators of the
fourth and fifth kinds:
\beq
   \lefteqn{ \le[ \g^I_J, \g^K_L \ri]_{\pm} = \d^K_J \g^I_L + 
   \sum_{J_1 J_2 = J} (-1)^{\ep(J_1) \lb \ep(K) + \ep(L) \rb} \d^K_{J_2} \g^I_{J_1 L} } \nn \\
   & & + \sum_{K_1 K_2 = K} \delta^{K_1}_J \g^{I K_2}_L 
   + \sum_{\ba{l}
	      J_1 J_2 = J \\
	      K_1 K_2 = K
	   \ea}
   (-1)^{\ep(J_1) \lb \ep(K) + \ep(L) \rb} \d^{K_1}_{J_2} \g^{I K_2}_{J_1 L} \nn \\ 
   & & + \sum_{J_1 J_2 = J} \d^K_{J_1} \g^I_{L J_2} 
   + \sum_{K_1 K_2 = J} (-1)^{\ep(K_1) \lb \ep(I) + \ep(J) \rb} \d^{K_2}_J \g^{K_1 I}_L \nn \\
   & & + \sum_{\ba{l}
		  J_1 J_2 = J \\
	    	  K_1 K_2 = K
	       \ea}
   (-1)^{\ep(K_1) \lb \ep(I) + \ep(J) \rb} \d^{K_2}_{J_1} \g^{K_1 I}_{L J_2} \nn \\
   & & + \sum_{J_1 J_2 J_3 = J} (-1)^{\ep(J_1) \lb \ep(K) + \ep(L) \rb} \d^K_{J_2} \g^I_{J_1 L J_3} \nn \\ 
   & & + \sum_{K_1 K_2 K_3 = K} (-1)^{\ep(K_1) \lb \ep(I) + \ep(J) \rb} \d^{K_2}_J \g^{K_1 I K_3}_L \nn \\
   & & + \sum_{\ba{l}
   		  J_1 J_2 = J \\
   		  K_1 K_2 = K
   	       \ea}
   (-1)^{\ep(J_1) \ep(K_1)} \d^{K_1}_{J_2} \d^{K_2}_{J_1} \ti{f}^I_L \nn \\
   & & + \sum_{\ba{l}
   		 J_1 J_2 J_3 = J \\
   		 K_1 K_2 = K
   	       \ea}
   (-1)^{\lb \ep(J_1) + \ep(J_2) \rb \ep(K_1)} \d^{K_1}_{J_3} \d^{K_2}_{J_1} \ti{f}^I_{L J_2} \nn \\
   & & + \sum_{\ba{l}
   		  J_1 J_2 = J \\
   		  K_1 K_2 K_3 = K
   	       \ea}
   (-1)^{\ep(J_1) \lb \ep(K_1) + \ep(K_2) \rb} \d^{K_1}_{J_2} \d^{K_3}_{J_1} \ti{f}^{I K_2}_L \nn \\
   & & + \sum_{\ba{l}
		  J_1 J_2 J_3 = J \\
		  K_1 K_2 K_3 = K
	       \ea}
   (-1)^{\lb \ep(J_1) + \ep(J_2) \rb \lb \ep(K_1) + \ep(K_2) \rb} 
   \d^{K_1}_{J_3} \d^{K_3}_{J_1} \ti{f}^{I K_2}_{L J_2} \nn \\
   & & - (-1)^{\lb \ep(I) + \ep(J) \rb \lb \ep(K) + \ep(L) \rb} (I \lrar K, J \lrar L).
\la{5.2.7}
\eeq
We call the Lie superalgebra defined by the Lie superbrackets from Eqs.(\ref{5.2.1}) to (\ref{5.2.7}) the {\em 
grand string superalgebra}.  We have given diagrammatic representations of these equations in the last two 
chapters with a more detailed exposition of the mathematical implications of these Lie superbrackets.

Let ${\cal T}_o$ be the vector space of all finite linear combinations of singlet states of the form 
$\bar{\ph}^{\r_1} \otimes s^{\dot{K}} \otimes \ph^{\r_2}$ where $1 \leq \r_1, \r_2 \leq 2 \L_F$ and all integers 
are between 1 and $2 \L$ inclusive in $\dot{K}$, which may be empty.  Also let ${\cal T}_c$ be the vector space of 
all finite linear combinations of singlet states of the form $\Ps^K$ such that all integers are again between 1 and 
$2 \L$ inclusive in $K$, which has to be non-empty, and that $\Ps^K$ satisfies Eq.(\ref{2.2.7}).  If we treat the 
operators ${\cal T}_o \oplus {\cal T}_c$ as the ones defined in the Section~\ref{s3.2}, we will find that the 
actions of an operator of the first kind are precisely given by Eqs.(\ref{2.2.10}) and (\ref{2.2.11}); those of the 
second kind by Eqs.(\ref{2.2.13}) and (\ref{2.2.14}); those of the third kind by Eqs.(\ref{2.2.16}) and 
(\ref{2.2.17}); those of the fourth kind by Eqs.(\ref{2.2.19}) and (\ref{2.2.20}); and those of the fifth kind by 
Eqs.(\ref{2.2.22}) and (\ref{2.2.23}).  Alternatively, we can {\em define} the actions of the operators of the five 
kinds on ${\cal T}_o \oplus {\cal T}_c$ by these ten equations, and show that ${\cal T}_o \oplus {\cal T}_c$
provides a representation for this superalgebra.

\section{Open String Superalgebra}
\la{s5.3}

Now that we have the grand string superalgebra at hand, we are going to derive the superalgebra of operators acting
on open singlet states only as a quotient algebra of it.  We will call this the open string superalgebra, and 
identify some subalgebras of it.  The superalgebra of operators acting on closed singlet states only will be 
considered in the next section.

Since every element of the open-closed string superalgebra maps a state in ${\cal T}_o$ to a state in ${\cal T}_o$,
and a state in ${\cal T}_c$ to a state in ${\cal T}_c$, ${\cal T}_o$ and ${\cal T}_c$ furnish two representation
spaces to the superalgebra.  The representations of the operators in the superalgebra form a Lie superalgebra on 
each of these two representation spaces.  However, since none of these two representation spaces provide faithful 
representations to the superalgebra, some operators vanish.

Consider the Lie superalgebra generated by the representation ${\cal T}_o$ first.  We are going to call this
superalgebra the {\em open string superalgebra}.  Since all operators of the fifth kind sends any state in 
${\cal T}_o$ to zero, there are only four kinds of operators in this superalgebra.  The Lie brackets of these 
operators are the same as those in the previous section, except that we set all $\ti{f}^{I}_{J}$'s to be zero, 
i.e., only Eq.(\ref{5.2.7}) needs to be modified.  Let us write $\g$ as $\s$ in this open string superalgebra.  
Note that the operators in this superalgebra are not linearly independent; there are many relations among them.  
For example,
\beq
   \sum_{\l=1}^{2\L_F} \bar{\X}^{\l}_{\l} \otimes l^I_J & = & \s^I_J - 
      \sum_{i=1}^{2\L} (-1)^{\ep(i) \lb \ep(I) + \ep(J) \rb} \s^{iI}_{iJ}; \nn \\
   \sum_{\l=1}^{2\L_F} r^I_J \otimes \X^{\l}_{\l} & = & \s^I_J - \sum_{j=1}^{2\L} \s^{Ij}_{Jj}; \nn \\
   \sum_{\l_3=1}^{2\L_F} \bar{\X}^{\l_1}_{\l_2} \otimes f^{\dot{I}}_{\dot{J}} \otimes \X^{\l_3}_{\l_3}
      & = & \bar{\X}^{\l_1}_{\l_2} \otimes l^{\dot{I}}_{\dot{J}} -
      \bar{\X}^{\l_1}_{\l_2} \otimes \sum_{j=1}^{2\L} l^{\dot{I}j}_{\dot{J}j} \mbox{; and} \nn \\
   \sum_{\l_1=1}^{2\L_F} \bar{\X}^{\l_1}_{\l_1} \otimes f^{\dot{I}}_{\dot{J}} \otimes \X^{\l_3}_{\l_4}
      & = & r^{\dot{I}}_{\dot{J}} \otimes \X^{\l_3}_{\l_4} -
      \sum_{i=1}^{2\L} (-1)^{\ep(i) \lb \ep(I) + \ep(J) \rb} r^{i \dot{I}}_{i \dot{J}} \otimes \X^{\l_3}_{\l_4}.
\la{5.3.1}
\eeq

There are many subalgebras and ideals in the open string superalgebra.  These are tabulated in Table~\ref{t5.2}.  
The results of this table can be easily deduced from the results of the previous section.

\begin{table}
\begin{tabular}{|| c | c | l ||} \hline
operators & algebra & \multicolumn{1}{c||}{comment} \\ \hline \hline
$\bar{\X}^{\l_1}_{\l_2} \otimes f^{\dot{I}}_{\dot{J}} \otimes \Xi^{\l_3}_{\l_4}$ & 
${\mathit F}_{\L | \L, \L_F | \L_F}$ & 
${\mathit F}_{\L | \L, \L_F | \L_F} \equiv$ \\
& & $gl(\L_F|\L_F) \otimes {\mathit F}_{\L|\L} \otimes gl(\L_F|\L_F)$ \\ \hline
$\bar{\X}^{\l_1}_{\l_2} \otimes l^{\dot{I}}_{\dot{J}} \otimes 1$ & 
$gl(\L_F|\L_F) \otimes \hat{\mathit L}_{\L|\L}$ & \\ \hline
$1 \otimes r^{\dot{I}}_{\dot{J}} \otimes \X^{\l_3}_{\l_4}$ & 
$\hat{\mathit R}_{\L|\L} \otimes gl(\L_F|\L_f)$ & \\ \hline
$\bar{\X}^{\l_1}_{\l_2} \otimes f^{\dot{I}}_{\dot{J}} \otimes \X^{\l_3}_{\l_4}$, & 
$\hat{\mathit M}_{\L|\L, \L_F|\L_F}$ & 
${\mathit F}_{\L|\L, \L_F|\L_F}$ is a proper ideal of \\
$\bar{\X}^{\l_1}_{\l_2} \otimes l^{\dot{I}}_{\dot{J}} \otimes 1$, and & & $\hat{\mathit M}_{\L|\L, \L_F|\L_F}$. \\
$1 \otimes r^{\dot{I}}_{\dot{J}} \otimes \Xi^{\l_3}_{\l_4}$ & & \\ \hline
$\s^I_J$ & $\hat{\S}_{\L|\L}$ & \\ \hline
all of the above & open string & ${\mathit F}_{\L, \L_F}$ and $\hat{\mathit M}_{ \L, \L_F}$ are proper \\
& superalgebra & ideals of the open string \\
& & superalgebra. \\ \hline \hline
\end{tabular}
\caption{\em Open string superalgebra, its ideals and subalgebras.}
\la{t5.2}
\end{table}

Like the centrix algebra $\hat{\S}_{\L}$ defined in Section~\ref{s4.5}, the `centrix superalgebra' 
$\hat{\S}_{\L|\L}$ also has a myriad of subalgebras and ideals.  The reader can obtain these subalgebras and ideals
easily by generalizing the results in Table~\ref{t4.2}.

\section{Closed String Superalgebra}
\la{s5.4}

Let us turn our attention to the Lie superalgebra of operators acting on closed singlet states only.  It is
generated by the representation space ${\cal T}_c$.  We will call this the {\em closed string superalgebra} or the 
{\em cyclix superalgebra} $\hat{C}_{\L|\L_F}$.  Since acting any operator of the first three kinds on ${\cal T}_c$ 
yields zero, this Lie superalgebra is obtained by considering the Lie superbracket among operators of the fourth 
and fifth kinds only.  Thus the closed string superalgebra is characterized by Eqs.(\ref{5.2.5}), (\ref{5.2.6}) and 
(\ref{5.2.7}).  We will write $\g$ as $g$ in the closed string superalgebra.  Again the operators are not linearly 
independent; in fact, any operator of the fifth kind can be written as a linear combination of operators of the 
fourth kind as
\beq
   \ti{f}^I_J = g^I_J - \sum_{k=1}^{2\L} g^{I k}_{J k}
\la{5.4.1}
\eeq
or
\beq
   \ti{f}^I_J = g^I_J - \sum_{k=1}^{2\L} (-1)^{\ep(k) \lb \ep(I) + \ep(J) \rb} g^{k I}_{k J}. 
\la{5.4.2}
\eeq
These two relations together with
\beq
   \ti{f}^{I_1 I_2}_J = (-1)^{\ep(I_1) \ep(I_2)} \ti{f}^{I_2 I_1}_J
\la{5.4.3}
\eeq
and
\beq
   \ti{f}^I_{J_1 J_2} = (-1)^{\ep(J_1) \ep(J_2)} \ti{f}^I_{J_2 J_1}
\la{5.4.4}
\eeq
can generate many other relations.  From Eq.(\ref{5.2.6}), we see that the set of all $\ti{f}^I_J$'s span a 
proper ideal $\ti{F}'_{\L|\L}$ for the closed string superalgebra.

We stated a conjecture in Section~\ref{s3.3} that the vector space spanned by all closed string states provides a 
faithful representation for the quotient algebra of all cosets $\g^I_J + \hat{M}'_{\Lambda}$ and 
$\tilde{f}^{(I)}_{(J)} + \hat{M}'_{\Lambda}$ of the heterix algebra.  In other words, we conjecture that this 
quotient algebra is precisely the cyclix algebra.  Likewise, we have a conjecture that the cyclix superalgebra can
be obtained as a quotient algebra of the superalgebra defined by Eqs.(\ref{5.2.5}), (\ref{5.2.6}) and (\ref{5.2.7}),
where $\ti{f}^I_J$ and $\g^I_J$ are independent, by $\hat{\mathit M}_{ \L, \L_F}$.

We know that this conjecture is true for $\L =1$ and $\L_F = 0$.  In the remaining paragraphs, we will show why 
this is so, and will see that the cyclix algebra, like the centrix algebra, has a close relationship with the Witt 
algebra.

Let $\L = 1$ and $\L_F = 0$.  Then all the sequences are repetitions of the number $1$ a number of times.  We can 
simplify the notations and write $\Ps^K$ as $\Ps^{(\#(K))}$, $s^K$ as $s^{\#(K)}$, $\g^I_J$ as $\g^{\#(I)}_{\#(J)}$ 
and $\ti{f}^{(I)}_{(J)}$ as $\ti{f}^{(\#(I))}_{(\#(J))}$.  We can deduce from Eqs.(\ref{3.2.8}), (\ref{3.2.9}), 
(\ref{3.2.11}) and (\ref{3.2.12}) that the actions of $\ti{f}^{(a)}_{(b)}$ and $\g^a_b$, where $a$ and $b$ are the 
number of integers in the various sequences, on $\Ps^{(c)}$ and $s^c$, where $c$ is also a positive integer, are 
given by
\begin{eqnarray}
   \ti{f}^{(a)}_{(b)} \Ps^{(c)} & = & c \delta^c_b \Ps^{(a)}; 
\label{5.4.5} \\
   \ti{f}^{(a)}_{(b)} s^c & = & 0; 
\label{5.4.6} \\
   \g^a_b \Ps^{(c)} & = & c \theta (b \leq c) \Ps^{(a+c-b)} \mbox{; and}
\label{5.4.7} \\
   \g^a_b s^c & = & (c - b + 1) \theta (b \leq c) s^{a+c-b}.
\label{5.4.8}
\end{eqnarray}
where $\theta( \mbox{condition} )$ is 1 if the condition holds, and 0 otherwise.  The Lie brackets for the 
$\Lambda = 1$ cyclix algebra are
\begin{eqnarray}
   \left[ \g^a_b, \g^c_d \right] & = & \theta (c \leq b) \left[ 2 \g^{a+c-1}_{b+d-1} + 2 \g^{a+c-2}_{b+d-2} +
   \cdots + 2 \g^{a+1}_{b+d-c+1} \right. \nonumber \\
   & & \left. + (b-c+1) \g^a_{b+d-c} \right] + \theta (2 \leq c \leq b) \left[ \ti{f}^{(a+c-2)}_{(b+d-2)} +
   2 \ti{f}^{(a+c-3)}_{(b+d-3)} \right. \nonumber \\
   & & \left. + \cdots + (c-1) \ti{f}^{(a)}_{(b+d-c)} \right] + \theta (b < c) \left[ 2 \g^{a+c-1}_{b+d-1} +
   2 \g^{a+c-2}_{b+d-2} \right. \nonumber \\
   & & \left. + \cdots + 2 \g^{a+c-b+1}_{d+1} + (c-b+1) \g^{a+c-b}_d \right] + \theta (2 \leq b < c)
   \left[ \ti{f}^{(a+c-2)}_{(b+d-2)} \right. \nonumber \\
   & & \left. + \ti{f}^{(a+c-3)}_{(b+d-3)} + \cdots + (b-1) \ti{f}^{(a+c-d)}_{(b)} \right] \nonumber \\
   & & - (a \leftrightarrow c, b \leftrightarrow d);
\label{5.4.9}
\end{eqnarray}
\begin{equation}
   \left[ \g^a_b, \ti{f}^{(c)}_{(d)} \right] = c \theta (b \leq c) \ti{f}^{(a+c-b)}_{(d)} -
   d \theta (a \leq d) \ti{f}^{(c)}_{(b+d-a)}
\label{5.4.10}
\end{equation}
and
\begin{equation}
   \left[ \ti{f}^{(a)}_{(b)}, \ti{f}^{(c)}_{(d)} \right] = 
   b \delta^c_b \ti{f}^{(a)}_{(d)} - a \delta^a_d \ti{f}^{(c)}_{(b)}.
\label{5.4.11}
\end{equation}
These three equations can be deduced from Eqs.(\ref{3.3.1}), (\ref{3.3.2}) and (\ref{3.3.3}), respectively.

The Cartan subalgebra ${\cal H}$ is now spanned by vectors of the forms $\g^a_a$ and $\ti{f}^{(a)}_{(a)}$.  A root
vector reads $\ti{f}^{(c)}_{(d)}$ or $f^c_d = \g^c_d - 2 \g^{c+1}_{d+1} + \g^{c+2}_{d+2} - \ti{f}^{(c)}_{(d)} +
\ti{f}^{(c+1)}_{(d+1)}$.  The expression for $f^c_d$ can be deduced from Eq.(\ref{3.3.8}).  The eigenequations are
\begin{eqnarray}
   \left[ \g^a_a, \ti{f}^{(c)}_{(d)} \right] & = & \left( c \theta (a \leq c) - d \theta (a \leq d) \right) 
   \ti{f}^{(c)}_{(d)}; \nonumber \\
   \left[ \ti{f}^{(a)}_{(a)}, \ti{f}^{(c)}_{(d)} \right] & = & 
   a \left( \delta^c_a - \delta^a_d \right) \ti{f}^{(c)}_{(d)}; 
\nonumber \\
   \left[ \g^a_a, f^c_d \right] & = & \left( \theta (a \leq c) (c - a + 1) - \theta (a \leq d) (d - a + 1) \right)
   f^c_d \mbox{; and} \nn \\
   \left[ \ti{f}^{(a)}_{(a)}, f^c_d \right] & = & 0.
\label{5.4.12}
\end{eqnarray}

Let us consider the quotient algebra of all cosets $\g^a_b + \hat{M}'_1$ and $\ti{f}^{(a)}_{(b)} + \hat{M}'_1$.  
{\em In the rest of this section and in the accompanying appendices, we will write $\g^a_b + \hat{M}'_1$ and
$\ti{f}^{(a)}_{(b)} + \hat{M}'_1$ simply as $g^a_b$ and $\ti{f}^{(a)}_{(b)}$, respectively}.  The vector space 
spanned by all bosonic closed string states still provides a representation for this quotient algebra --- the 
actions of $g^a_b$ and $\ti{f}^{(a)}_{(b)}$ on this vector space are still described by Eqs.(\ref{5.4.5}) and 
(\ref{5.4.7}).  However, there is now a kernel in this representation:
\begin{equation}
   \ti{f}^{(a)}_{(b)} = g^a_b - g^{a+1}_{b+1}.
\label{5.4.13}
\end{equation}
Hence,
\begin{equation}
   g^a_b = \left\{ \begin{array}{ll}
           - \ti{f}^{(a)}_{(b)} - \ti{f}^{(a-1)}_{(b-1)} - \cdots - \ti{f}^{(a-b+1)}_{(1)} 
           + g^{a-b+1}_1 & \mbox{if $a \geq b$; and} \\
           - \ti{f}^{(a)}_{(b)} - \ti{f}^{(a-1)}_{(b-1)} - \cdots - \ti{f}^{(1)}_{(b-a+1)}
           + g^1_{b-a+1} & \mbox{if $a < b$.}
           \end{array} \right.
\label{5.4.14}
\end{equation}
A simplification brought about by setting $\Lambda$ to 1 is that we can write a basis for the quotient algebra 
easily.  Indeed, the set of all $\ti{f}^{(a)}_{(b)}$'s, $g^1_b$'s and $g^a_1$'s where $a$ and $b$ are arbitrary 
positive integers form a basis for the $\Lambda = 1$ quotient algebra.  The proof of this statement will be given 
in Appendix~\ref{sa5.2}.  The same proof also shows that not only does the vector space of bosonic closed string 
states provide a representation, but also it provides a {\em faithful} representation for this quotient algebra.  
From Eq.(\ref{5.4.10}), we deduce that the subspace $\tsaltone$ spanned by all the vectors of the form 
$\ti{f}^{(c)}_{(d)}$ form a proper ideal of this quotient algebra.

The set of all $\ti{f}^{(a)}_{(a)}$'s, where $a$ is arbitrary, and $g^1_1$ form a basis for a Cartan subalgebra.  
The proof of this statement will be seen in Appendix~\ref{sa5.3}.  In addition, since $\tsaltone$ is a proper ideal 
of the quotient algebra, the same proof reveals that any root vector with respect to this Cartan subalgebra must be 
a linear combination of a number of $\ti{f}^{(a)}_{(b)}$'s.  Therefore all root vectors are of the form 
$\ti{f}^{(c)}_{(d)}$ with $c \neq d$.  The corresponding eigenequations, which can be deduced from 
Eqs.(\ref{3.3.4}) and (\ref{3.3.5}), are
\begin{equation}
   \left[ g^1_1, \ti{f}^{(c)}_{(d)} \right] = \left( c - d \right) \ti{f}^{(c)}_{(d)}
\la{5.4.15}
\end{equation}
and
\begin{equation}
   \left[ \ti{f}^{(a)}_{(a)}, \ti{f}^{(c)}_{(d)} \right] = a \left( \delta^c_a - \delta^a_d \right) \ti{f}^{(c)}_{(d)}.
\la{5.4.16}
\end{equation}

As $\tsaltone$ is a proper ideal of this $\Lambda=1$ quotient algebra, we can form yet another quotient algebra of 
cosets of the form $v + \tsaltone$ where $v$ is an arbitrary vector of the previous quotient algebra.  This new
quotient algebra is spanned by the cosets $g^a_1 + \tsalt$ and $g^1_b + \tsaltone$, where $a$ and $b$ run over all 
positive integers.  It is a straightforward matter to show that the following Lie brackets are true:
\begin{eqnarray}
   \left[ g^a_1 + \tsaltone, g^c_1 + \tsaltone \right] & = &
   (c - a) \left( g^{a+c-1}_1 + \tsaltone \right) ; \nonumber \\
   \left[ g^a_1 + \tsaltone, g^1_d + \tsaltone \right] & = &
   \left\{ \begin{array}{ll}
   (2 - a - d) \left( g^{a-d+1}_1 + \tsaltone \right) & \mbox{if $d \leq a$, or} \\
   (2 - a - d) \left( g^1_{d-a+1} + \tsaltone \right) & \mbox{if $a \leq d$; and}
   \end{array} \right. \nonumber \\
   \left[ g^1_b + \tsaltone, g^1_d + \tsaltone \right] & = &
   (b - d) \left( g^1_{b+d-1} + \tsaltone \right).
\label{5.4.17}
\end{eqnarray}
Let us define
\begin{equation}
   L_a = \left\{ 
   \begin{array}{ll}
      - g^{a+1}_1 + \tsaltone & \mbox{if $a \geq 0$, and} \\
      - g^1_{1-a} + \tsaltone & \mbox{if $a \leq 0$.}
   \end{array} \right.
\label{5.4.18}
\end{equation}
Note that the $L$ here is {\em not} an integer sequence.  Then Eq.(\ref{5.4.17}) 
becomes
\begin{equation}
   \left[ L_a, L_b \right] = (a - b) L_{a+b}.
\la{5.4.19}
\end{equation}
This is the Witt algebra.  Consequently, the $\L = 1$ quotient algebra of all cosets $\g^a_b$ and 
$\ti{f}^{(a)}_{(b)}$ can be regarded as an extension of the Witt algebra by an algebra isomorphic to 
$gl_{+\infty}$.  The quotient algebras $\g^I_J + \hat{M}'_{\Lambda} + \tsalt$ for $\Lambda > 1$ can then be 
regarded as generalizations of the Witt algebra.

\section{The Ising Model Revisited}
\la{s5.5}

In this and the last two chapters, we have mostly been studying issues in mathematics.  Let us see how to use these
mathematical results in physics.  A good starting point should be a model which is simple enough that we know a
great deal of it so that we can see how the algebraic point of view we have been discussing fit in.  In 
Section~\ref{s2.7}, we introduced a number of exactly solvable matrix models.  The Ising model is the simplest, and
let us look into this model to see what can be learnt.

It is possible to understand the solvability of the Ising model in terms of the Dolan--Grady conditions \cite{dogr, 
davies91} and the Onsager algebra \cite{onsager, davies90}.  Suppose we have a system whose Hamiltonian can be 
written as 
\beq
	H = H_0 + V
\la{5.5.1}
\eeq
with the two terms in the Hamiltonian satisfying the Dolan--Grady conditions:
\beq
   \lbrack H_0, \lbrack H_0, \lbrack H_0, V \rbrack \rbrack \rbrack = 16 \lbrack H_0, V \rbrack,
\label{5.5.2}
\eeq
and 
\beq
   \lbrack V, \lbrack V, \lbrack V, H_0 \rbrack \rbrack \rbrack = 16 \lbrack V, H_0 \rbrack.
\label{5.5.3}
\eeq
Then we can construct operators satisfying an infinite-dimensional Lie algebra
\beq  
   \lbrack A_m, A_n \rbrack = 4 G_{m-n}, \; \lbrack G_m, A_n \rbrack = 2 A_{n+m} - 2 A_{n-m},
   \; \mbox{and} \; \lbrack G_m, G_n \rbrack = 0
\la{5.5.4}
\eeq
by the following recursion relations:
\beq
   & A_0 = H_0, \; A_1 = V, \; A_{n+1} - A_{n-1} = \frac{1}{2} \lbrack G_1, A_n \rbrack, & \nonumber \\
   & G_1 = \frac{1}{4} \lbrack A_1, A_0 \rbrack, \; \mbox{and} \; G_n = \frac{1}{4} \lbrack A_n, A_0 \rbrack. &
\la{5.5.5}
\eeq
This Lie algebra is called the {\em Onsager algebra}. It is known to be isomorphic to a fixed-point subalgebra of 
the $sl(2)$ loop algebra ${\cal L}(sl(2))$ with respect to the action of a certain involution \cite{roan}.  In 
particular, the system will admit an infinite number of conserved quantities \cite{honecker}:
\beq
   Q_m = -\frac{1}{2} \left( A_m + A_{-m} + \lambda A_{m+1} + \lambda A_{-m+1} \right).
\la{5.5.6}
\eeq
Hence any such system should be integrable. 

In the case of the Ising model we choose $H_0$ and $V$ as above. The Lie brackets of the $g^I_J$ then allow us to 
verify easily that the first of the Dolan--Grady conditions is satisfied.  This, together with the self-duality of 
the Ising model, guarantee that the other Dolan--Grady condition is satisfied also.  Eq.(\ref{5.5.3}) could also be 
verified directly.  The only caveat is that as the operators act on closed string states only, we have to treat all 
$\ti{f}$'s as linear combinations of $g$'s in order for Eq.(\ref{5.5.3}) to hold true.  Moreover we see that the 
Onsager algebra is a subalgebra of our algebra $\hat{C}_\Lambda$; all the conserved quantities are just linear 
combinations of our $g^I_J$ and $\ti{f}^I_J$.  This suggests that there may be other models that are integrable by 
this method; what we need to do is to identify pairs of elements in our algebra that satisfy the Dolan--Grady 
conditions.

Another deep relationship between integrable spin chain models and our algebras is that to each solution of the
Yang--Baxter equation (or Yang--Baxter--Sklyanin equation for the case of open chains), there is a maximal Abelian
subalgebra of the centrix or cyclix algebra.  This is just the subalgebra spanned by the conserved quantities of the
corresponding spin chain: each conserved quantity is an element of our centrix or cyclix algebra.  Thus it emerges
that the Lie algebras we have found are the underlying symmetries of many integrable models.  Perhaps we may even
be able to identify the relationship between the symmetries of integrable models and those of Yang--Mills theory
because of these Lie algebras.  We intend to develop these ideas in the future into a systematic theory.

\appendix

\chapter{Conventions Used in Chapter~\ref{c2} and Some Pertinent Proofs}
\label{ca1}

\section{Multi-Indices: Ordinary Sequences}
\label{sa1.1}

Much of our work involves manipulating tensors carrying multiple indices.  For the convenience of the reader, we 
give here a summary of the notation used in this article for multi-indices.

We will use lower case Latin letters such as $i, j, i_1, j_2$ to denote indices taking finite integer values. Often 
we will have to deal with a whole sequence of indices $i_1 i_2 i_3 \ld i_a$, which we will denote by the 
corresponding uppercase letter $I$ as the collective index of this sequence. The length of the  sequence $I$ will 
be denoted by $\#(I)$. 

Thus if  
\[ I = i_1 i_2 \ld i_a \]
and 
\[ J = j_1 j_2 \ld j_b \],
then we have  $\#(I) = a$ and $\#(J) = b$.  The composition of these two sequences will be denoted by 
\[ I J = i_1 i_2\ldots i_a j_1 j_2 \ldots j_b. \]
In particular, we have 
\beq
	Ij = i_1 i_2 \ld i_a j,
\eeq
when only a single index is added to the end. This operation of composing sequences is associative but not 
commutative:
\beq
	IJ \neq JI, \quad I(JK)=(IJ)K=IJK.
\eeq

Often we will have to allow the null sequence $\emptyset$ among the range of values of a collective index. A 
collective index that is allowed to take the null sequence as its value will have a dot over it. Thus the possible 
values of $\dot{I}$ are 
\beq
   \dot{I} = \emptyset, 1, 2, \ldots, \Lambda, 11, 12, \ldots 1\Lambda, 21, \ldots, \Lambda\Lambda, \ldots
\eeq
while those of $I$ do not include the empty set are
\beq
	I = 1, 2, \ldots, \Lambda, 11, 12, \ldots 1\Lambda, 21, \ldots, \Lambda\Lambda, \ldots
\eeq
Of course the length of the null sequence is zero. It is the identity element of the composition law above,
\beq
	\emptyset\dot{I} = \dot{I}\emptyset =\dot{I}.
\eeq
In fact the inclusion of the empty sequence in the set of all sequences turns it into a semi-group under the above
composition law.

The equation $I = J$ means that they have the same length $a$ (say) and
\[ i_1 = j_1; i_2 = j_2; \ldots i_a=j_a. \] 
In the same way we can define $\dot{I}=\dot{J}$ either if they are both the empty sequence, or if they are equal in 
the above sense.

The Kronecker delta function for integer sequences is defined as follows:
\[ \delta^I_J \equiv \left\{ \begin{array}{ll}
   				1 & \mbox{if $I = J$; or} \\
   				0 & \mbox{if $I \neq J$.} \\
   			     \end{array} \right\} \]
and similarly for dotted indices.  The summation sign in an expression such as 
\[ \sum_I X^I_J, \]  
means that all possible distinct sequences, {\em excluding the empty sequence}, are summed over.  (In all practical 
cases, it turns out that there is only a finite number of $I$'s such that $X^I_J \neq 0$ so there will be no 
convergence problems.)  On the other hand, the summation sign over a dotted index 
\[ \sum_{\dot{I}} X^{\dot{I}}_J \]
means that  all possible distinct sequences for $\dot{I}$, {\em including the empty sequence}, are summed over.

Often we will have to sum over all the ways of splitting a sequence into subsequences.  For example, the summation 
sign on the L.H.S. of the equation
\beq
   \sum_{I_1 I_2 = I} X^{I_1 }Y^{I_2}_K \equiv 
   \sum_{I_1I_2}\delta_I^{I_1I_2}X^{I_1}Y^{I_2}_K
\label{a1.0}
\eeq
denotes the sum over all the ways in which a given index $I$ can be split into two {\em nonempty} subsequences 
$I_1$ and $I_2$. If there is no way to split $I$ as required, then the sum simply yields 0.  For example, 
Eq.(\ref{a1.0}) yields 0 if $I$ has only one integer.  If the first subsequence is allowed to be empty in the sum, we 
would write instead,
\beq
   \sum_{\dot{I}_1 I_2 = I} X^{\dot{I}_1 }Y^{I_2}_K .
\eeq

There is a total ordering ({\em lexicographic ordering}) $>$ among integer sequences such that $\dot{I} >\dot{J}$ 
for the sequences $\dot{I}$ and $\dot{J}$ if either
\begin{enumerate}
\item $\#(\dot{I}) > \#(\dot{J})$; or
\item $\#(\dot{I}) = \#(\dot{J}) = a\neq 0$, and there exists an integer $r \leq a$ 
such that
      $i_1 = j_1$, $i_2 = j_2$, \ldots, $i_{r-1} = j_{r-1}$ and $i_r > j_r$.
\end{enumerate}
Explicitly, the total ordering can be presented as
\beq
   & & \emptyset<1<2<\cdots<\Lambda<11<12<\cdots<1\Lambda \\
   & & <21<\cdots<\Lambda1<\cdots \Lambda\Lambda<111\cdots.
\eeq
This also gives a one-one correspondence (counting rule) between the set of values of the indices and the set of 
natural numbers: $\emptyset \to 1, 1 \to 2, \ldots, \Lambda \to \Lambda + 1, 1\Lambda \to \Lambda+2,$ etc.. 

When we talk of an algebra spanned by operators such as $f^{\dot{I}}_{\dot{J}}$, $l^{\dot{I}}_{\dot{J}}$, 
$r^{\dot{I}}_{\dot{J}}$ or $\s^I_J$ \footnote{for definitions in Section~\ref{s2.2} or Chapter~\ref{c4}}, the underlying vector space is 
that of {\em finite} linear combinations. For example, a typical element of $\hatcentrix$ will be of the form 
$\sum_{IJ}c^J_I\sigma^I_J$. Although $I$ and $J$ can take an infinite number of values, only a finite number of the 
complex numbers $c^J_I$ can be non-zero. Thus there is never any issue of convergence in the sums of interest to 
us: they are all finite sums.  Viewed as operators, $f^{\dot{I}}_{\dot{J}}$'s are finite rank, 
$l^{\dot{I}}_{\dot{J}}$'s and $r^{\dot{I}}_{\dot{J}}$'s are bounded and $\sigma^I_J$ are unbounded. Hence their 
finite linear combinations are also finite rank, bounded and unbounded, respectively.

\section{Multi-Indices: Cyclic Sequences}
\la{sa1.2}

We define $(I)$ to be the equivalence class of all cyclic permutations of $I$. In fact $(I)$ can be viewed as a 
discrete model for a closed loop, or closed string.  The corresponding Kronecker delta function is defined by the 
following relation:
\begin{equation}
   \delta^I_{(J)} \equiv \delta^I_J + \sum_{J_1 J_2 = J} \delta^I_{J_2 J_1}. 
\label{a1.1}
\end{equation}
Eq.(\ref{a1.1}) means that the delta function returns the number of different cyclic permutations of $J$ such that 
each permuted sequence is identical with $I$. {\em Thus $\delta^I_{(J)}$ can take any non-negative integer as its 
value, not just 0 or 1.}  The reader can verify from this definition that
\begin{equation}
    \delta^I_{(J)} = \delta^J_{(I)} 
\label{a1.2}
\end{equation}
Next, the expression
\[ \sum_{(I)} X^I, \]
where $X^I$ is dependent on the equivalence class $(I)$, i.e., $X^I = X^J$ if
$(I) = (J)$, means that {\em all 
possible distinct equivalence classes $(I)$} are summed.  Note that each
equivalence class appears only {\em once} in the sum.  In all 
cases of interest to us, it turns out that there are  only a finite number of $(I)$'s such 
that $f(I) \neq 0$.

Now we can introduce the formula that defines the following summation:
\begin{equation}
   \sum_{I_1 I_2 \cdots I_n = (I)} X^{I_1, I_2, \ldots, I_n} \equiv
   \sum_{I_1, I_2, \ldots, I_n} \delta^{I_1 I_2 \cdots I_n}_{(I)}
   X^{I_1, I_2, \ldots, I_n}.
\label{a1.3}
\end{equation}   
In words, in Eq.(\ref{a1.3}) we sum over all distinct ways of cyclically permuting $I$, and then all {\em distinct 
sets} of $n$ non-empty sequences $I_1$, $I_2$, \ldots, $I_n$ (but note that within a particular set of $I_1$, 
$I_2$, \ldots, $I_n$, some of the sequences can be identical) such that $I_1 I_2 \ldots I_n$ is the same as this 
permuted sequence.  This equation then leads to
\begin{equation}
   \delta^{I_1 I_2 \cdots I_n}_{(I)} = \sum_{I'_1 I'_2 \cdots I'_n = (I)}
   \delta^{I_1}_{I'_1} \delta^{I_2}_{I'_2} \cdots \delta^{I_n}_{I'_n}.
\label{a1.5}
\end{equation}
A direct consequence of Eq.(\ref{a1.5}) is
\[ \delta^{I_1 I_2 \cdots I_n}_{(I)} = \delta^{I_2 I_3 \cdots I_n I_1}_{(I)}. \]   
In addition, the reader can verify from 
Eq.(\ref{a1.3}) together with Eqs.(\ref{a1.0}), (\ref{a1.1}) and (\ref{a1.2}) that
\begin{eqnarray}
   \sum_{I_1 I_2 \cdots I_n = (I)} & = & \sum_{I_1, I_2, \ldots, I_n}
   \left( \delta^{I_1 I_2 \cdots I_n}_I + \sum_{I_{11} I_{12} = I_1} 
   \delta^{I_{12} I_2 I_3 \cdots I_n I_{11}}_I + 
   \delta^{I_2 I_3 \cdots I_n I_1}_I \right. \nonumber \\
   & & + \sum_{I_{21} I_{22} = I_2}
   \delta^{I_{22} I_3 \cdots I_n I_1 I_{21}}_I + \cdots + 
   \delta^{I_n I_1 I_2 \cdots I_{n-1}}_I \nonumber \\
   & & \left. + \sum_{I_{n1} I_{n2} = I_n}
   \delta^{I_{n2} I_1 I_2 \cdots I_{n-1} I_{n1}}_I \right) .
\label{a1.4}
\end{eqnarray}
   
\section{Actions of Operators on Physical States in the Large-$N$ Limit}
\label{sa1.3}

We are going to illustrate (though not rigorously prove) why in the planar large-$N$ limit, an operator 
representing a term in a dynamical variable sends singlet states to singlet states.  We are going to confine 
ourselves to the action of operators of the second kind defined in Section~\ref{s2.2} only.  The reasoning is 
similar for operators of other kinds.

\begin{figure}
\epsfxsize=5in
\centerline{\epsfbox{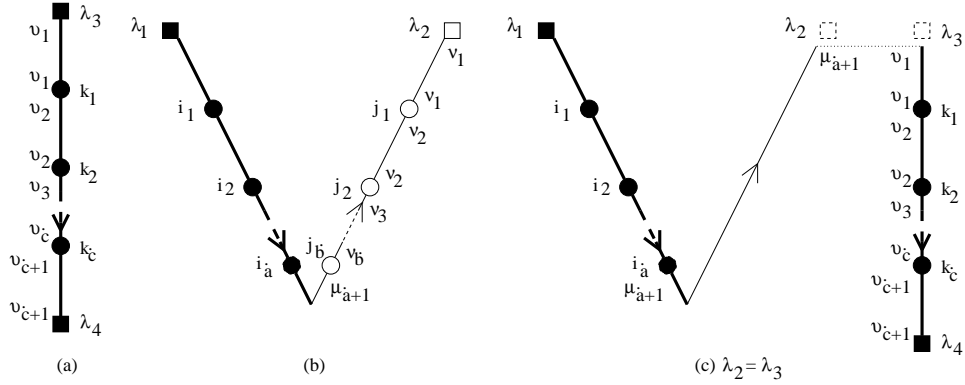}}
\caption{\em Detailed diagrammatic representations of singlet states, and the action of an operator of the second
kind with no annihilation operator of an adjoint parton on an open singlet state.}
\label{fa2.1}
\end{figure}

Assume that the operator is of the form Eq.(\ref{2.2.12}) and the open singlet state is of the form 
Eq.(\ref{2.2.5}).  Let $\dot{a} = \#(\dot{I})$, $\dot{b} = \#(\dot{J})$ and $\dot{c} = \#(\dot{K})$.  So there are 
$\dot{a}$ creation operators and $\dot{b}$ annihilation operators for adjoint partons in the operator of the second 
kind, and $\dot{c}$ creation operators for adjoint partons in the open singlet state.  There is a factor of 
$N^{-(\dot{a}+\dot{b})/2}$ in the operator of the second kind and a factor of $N^{-(\dot{c}+1)/2}$ in the open 
singlet state, so initially there is a total factor of $N^{-(\dot{a} + \dot{b} + \dot{c} + 1) / 2}$.  A term in the 
final state is either an open singlet state or a product of an open singlet and a number of closed singlets.  No 
matter how many closed singlets there are in this term, $N$ should be raised to the power of $-(\dot{a} - \dot{b} + 
\dot{c} + 1) / 2$ in order that the term survives the large-$N$ limit.  It is therefore that only the operations 
which produce a factor of $N^{\dot{b}}$ survive the large-$N$ limit.

To clarify the argument, we need to refine the diagrammatic representations of an operator of any kind and a 
singlet state more carefully.  Note that in Figs.~\ref{f2.1}(a) and (c), each square carries one color index 
whereas each circle carries two color indices.  We can put these indices at the ends of the thick or thin lines 
attaching to them.  Fig.\ref{fa2.1}(a) shows a typical open singlet state given by Eq.(\ref{2.2.5}).  Here the 
color indices are explicitly written out.  Note that each square carries 1 color index, whereas each circle carries 
2 color indices.  Moreover, the color indices at the two ends of a connecting solid line are the same.
Fig.\ref{fa2.1}(b) shows an operator of the second kind, given by Eq.(\ref{2.2.12}), with explicit color indices of 
the annihilation operators.

Consider the case when there are no annihilation operators for adjoint partons in the operator of the second kind, 
i.e., $\dot{b}=0$.  Fig.~\ref{fa2.1}(c) shows the action of such an operator on an open singlet state.  It destroys 
the solid square in Fig.~\ref{fa2.1}(a) and the hollow square in Fig.~\ref{fa2.1}(b).  The ends of the lines 
originally attached to the squares are now joined together by a dotted line with an appearance different from other 
dotted lines in the diagram.   Algebraically this dotted line is the Kronecker delta function 
$\delta^{\upsilon_1}_{\mu_{a+1}}$.  Clearly the final state is an open singlet state..  The factor involving 
$N$ in the final state is $N^{-(\dot{a} + \dot{c} + 1) / 2}$.  This is precisely the factor for an open singlet 
state with $\dot{a} + \dot{c}$ partons in the adjoint representation.  

\begin{figure}
\epsfxsize=5in
\centerline{\epsfbox{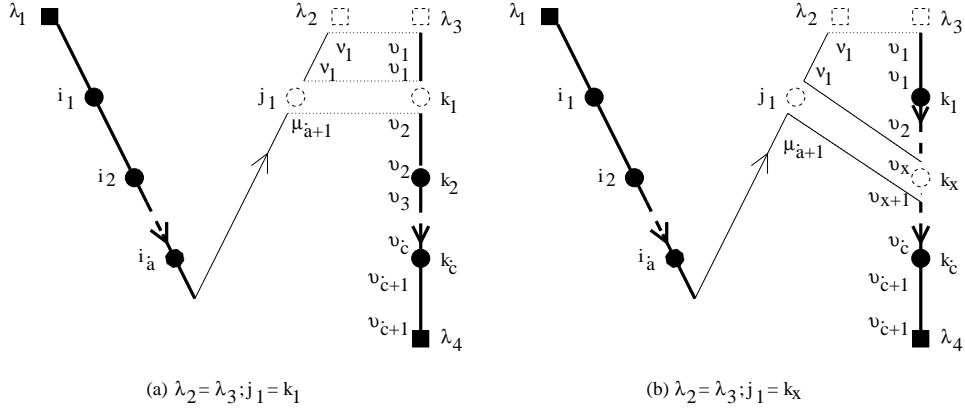}}
\caption{\em The action of an operator of the second kind with one annihilation operator of an adjiont parton on an 
open singlet state.}
\label{fa2.2}
\end{figure}

Now consider the case when there is one annihilation operator of an adjoint parton in the operator of the second
kind, i.e., $\dot{b} = 1$.  The action of this operator needs to produce a factor of $N$ in order for the final 
state to survive in the large-$N$ limit.  Consider the following two cases illustrated separately in 
Figs.~\ref{fa2.2}(a) and~(b).  In the former diagram, the annihilation operator of an adjoint parton in the 
operator of the second kind contracts with the creation operator of the first adjoint parton in the initial open
singlet state.  This results in a closed loop with no squares or circles but two dotted lines only.  
Algebraically this loop is the factor $\delta^{\upsilon_1}_{\nu_1} \delta^{\nu_1}_{\upsilon_1}$, which in turn is 
equal to $N$.  The remaining parts of this diagram form an open singlet state.  Thus the singlet 
state survives the large-$N$ limit.  In the latter diagram, the annihilation operator of the adjoint 
parton in the operator of the second kind contracts with the creation operator of a later adjoint parton in the 
adjoint parton sequence of the initial open singlet state.  This time the final state is a product of an open
singlet together with a closed singlet.  However, there is no closed loop with solid and dotted lines only. 
This implies that no extra factor of $N$ is produced and so this term can be neglected in the large-$N$ limit. 

\begin{figure}
\epsfxsize=5in
\centerline{\epsfbox{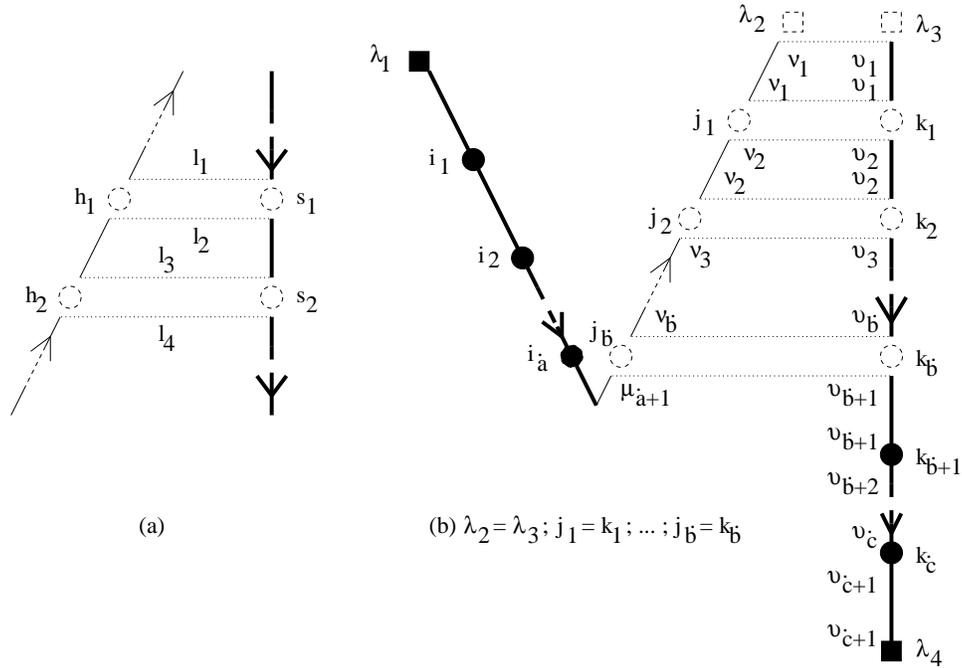}}
\caption{\em (a)  Contraction of circles.  (b)  The action of an operator of the second kind with $b$ annihilation 
operators of adjoint partons on an open singlet state.}
\label{fa2.3}
\end{figure}

Let us now turn to the case when $\dot{b}$ is an arbitrary positive integer.  The annihilation operator of a
conjugate parton in the operator of the second kind will contract with the creation operator of a conjugate parton
in the initial open singlet state.  This will produce 1 dotted line.  The $\dot{b}$ annihilation operators of 
adjoint partons in the operator of the second kind will contract with $\dot{b}$ creation operators of adjoint 
partons in the initial meson state.  This will further produce $2\dot{b}$ dotted lines.  Hence there are 
$2\dot{b}+1$ dotted lines.  One of these dotted lines has to be recruited to join the creation and annihilation 
operators in the final state.  In order for the final state to survive in the large-$N$ limit, we need a factor of 
$N^{\dot{b}}$, as explained above.  Since the minimum number of dotted lines to produce a closed loop without 
squares or circles is 2, the factor $N^{\dot{b}}$ can be obtained only if there are ${\dot{b}}$ closed loops, every 
closed loop has 2 dotted lines only, and there are no circles or squares in any closed loop.  This can be done 
only if the 2 dotted lines join 2 adjacent pairs of creation and annihilation operators.  Moreover, only 1 dotted 
line can be uninvolved in any closed loops.  Consider Fig~\ref{fa2.3}(a), where we contract 2 circles $h_1$ and 
$s_1$, producing 2 dotted lines $l_1$ and $l_2$.  Since one of these dotted lines has to be lie within a closed 
loop, a pair of circles adjacent to $h_1$ and $s_1$ has to be contracted.  In Fig~\ref{fa2.3}(a), $h_2$ and $s_2$ 
are contracted and hence we obtain $l_3$ and $l_4$.  Then either $l_1$ or $l_4$ (or both) has to lie within a 
closed loop.  If we continue this argument, we will obtain Fig.~\ref{fa2.3}(b).  As is clear from the figure, the 
final state is an open singlet state.  Thus we conclude that in the large-$N$ limit, an operator of the second kind
propagates open singlet states to open singlet states.  The actions of operators of other kinds can be understood 
similarly.

\chapter{Proofs of Some Results in Chapter~\ref{c4}}
\la{ca4}

\section{Associativity of the Algebra for the Operators of the Second Kind}
\label{sa4.1}

The reader can prove the associativity of this algebra by verifying the following identity:
\begin{eqnarray*}
   \lefteqn{l^{\dot{I}}_{\dot{J}} \left( l^{\dot{K}}_{\dot{L}} l^{\dot{M}}_{\dot{N}} \right) = 
   \left( l^{\dot{I}}_{\dot{J}} l^{\dot{K}}_{\dot{L}} \right) l^{\dot{M}}_{\dot{N}} = 
   \delta^{\dot{K}}_{\dot{J}} \delta^{\dot{M}}_{\dot{L}} l^{\dot{I}}_{\dot{N}} +
   \sum_{\dot{M_1} M_2 = \dot{M}} \delta^{\dot{K}}_{\dot{J}} \delta^{\dot{M_1}}_{\dot{L}} l^{\dot{I} M_2}_{\dot{N}}
   } \\
   & & + \sum_{\dot{L_1} L_2 = \dot{L}} \delta^{\dot{K}}_{\dot{J}} \delta^{\dot{M}}_{\dot{L_1}}
   l^{\dot{I}}_{\dot{N} L_2} +
   \sum_{\dot{K_1} K_2 = \dot{K}} \delta^{\dot{K_1}}_{\dot{J}} \delta^{\dot{M}}_{\dot{L}} l^{\dot{I} K_2}_{\dot{N}}
   + \sum_{\dot{J_1} J_2 = \dot{J}} \delta^{\dot{K}}_{\dot{J_1}} \delta^{\dot{M}}_{\dot{L}}
   l^{\dot{I}}_{\dot{N} J_2} \\
   & & + \sum_{\begin{array}{c}
   		  \dot{K_1} K_2 = \dot{K} \\
   		  \dot{M_1} M_2 = \dot{M}
   	       \end{array}}
   \delta^{\dot{K_1}}_{\dot{J}} \delta^{\dot{M_1}}_{\dot{L}} l^{\dot{I} K_2 M_2}_{\dot{N}} 
   + \sum_{\begin{array}{c}
   	      \dot{K_1} K_2 = \dot{K} \\
   	      \dot{L_1} L_2 = \dot{L}
	   \end{array}}
   \delta^{\dot{K_1}}_{\dot{J}} \delta^{\dot{M}}_{\dot{L_1}} l^{\dot{I} K_2}_{\dot{N} L_2} \\
   & & + \sum_{\begin{array}{c}
		  \dot{J_1} J_2 = \dot{J} \\
		  \dot{M_1} M_2 = \dot{M}
	       \end{array}}
   \delta^{\dot{K}}_{\dot{J_1}} \delta^{\dot{M_1}}_{\dot{L}} \delta^{M_2}_{L_2} l^{\dot{I}}_{\dot{N}}
   + \sum_{\begin{array}{c}
   	      \dot{J_1} J_2 = \dot{J} \\
   	      \dot{L_1} L_2 = \dot{L}
  	   \end{array}}
   \delta^{\dot{K}}_{\dot{J_1}} \delta^{\dot{M}}_{\dot{L_1}} l^{\dot{I}}_{\dot{N} L_2 J_2} \\
   & & + \sum_{\begin{array}{c}
   		  \dot{J_1} J_2 = \dot{J} \\
   		  \dot{M_1} M_2 M_3 = M
   	       \end{array}}
   \delta^{\dot{K}}_{\dot{J_1}} \delta^{\dot{M_1}}_{\dot{L}} \delta^{M_2}_{J_2} l^{\dot{I} M_3}_{\dot{N}} 
   + \sum_{\begin{array}{c}
   	      \dot{J_1} J_2 J_3 = J \\
   	      \dot{M_1} M_2 = M \\
	   \end{array}}
   \delta^{\dot{K}}_{\dot{J_1}} \delta^{\dot{M_1}}_{\dot{L}} \delta^{M_2}_{J_2} l^{\dot{I}}_{\dot{N} J_3} 	 
\end{eqnarray*}     	      
Q.E.D.

\section{Cartan Subalgebra of $\hatleftix$}
\label{sa4.2}

This can be seen by the following argument.  Let us call the subspace spanned by all the $l^{\dot{I}}_{\dot{I}}$'s 
${\cal M}$.  From Eq.(\ref{4.3.3}), $\lbrack l^{\dot{I}}_{\dot{I}}, l^{\dot{J}}_{\dot{J}} \rbrack = 0$ for any 
integer sequences $\dot{I}$ and $\dot{J}$.  Thus ${\cal M}$ is an Abelian subalgebra.  In particular, ${\cal M}$ is 
niltpotent.  To proceed on, we need the following two lemmas:
\begin{lemma}
Let 
\[ \lbrack l^{\dot{I}}_{\dot{I}}, l^{\dot{K}}_{\dot{L}} \rbrack = \sum_{k=1}^n \alpha^{\dot{N}_k}_{\dot{M}_k} 
   l^{\dot{M}_k}_{\dot{N}_k} \]
where $n$ is a finite positive integer, $\dot{M}_k$ and $\dot{N}_k$ are positive integer sequences such that 
$l^{\dot{M}_k}_{\dot{N}_k} \neq l^{\dot{M}_{k'}}_{\dot{N}_{k'}}$ for $k \neq k'$, and 
$\alpha^{\dot{N}_k}_{\dot{M}_k}$ are non-zero numerical coefficients.  Then
\[ \#(\dot{M}_k) - \#(\dot{N}_k) = \#(\dot{K}) - \#(\dot{L}) \]
for every $k = 1, 2, \ldots, n$.
\label{la4.2.1}
\end{lemma}
This lemma can be proved by using Eq.(\ref{4.3.3}) with $\dot{J} = \dot{I}$.  
\begin{lemma}
With the same assumptions as in the previous lemma, we have
\[ \#(\dot{M}_k) + \#(\dot{N}_k) \geq \#(\dot{K}) + \#(\dot{L}) \]
for every $k = 1, 2, \ldots, n$.
\label{la4.2.2}
\end{lemma}
This lemma can also be proved by using Eq.(\ref{4.3.3}) with $\dot{J} = \dot{I}$.  Let $m$ be a positive integer.  
Now we are ready to show for arbitrary non-zero complex numbers $\beta^{\dot{L}_i}_{\dot{K}_i}$, where $i = 1, 2, 
\ldots, m$ and arbitrary integer sequences $\dot{L}_i$'s and $\dot{K}_i$'s such that $l^{\dot{K}_i}_{\dot{L}_i} 
\neq l^{\dot{K}_{i'}}_{\dot{L}_{i'}}$ for $i \neq i'$, and $\dot{K}_i \neq \dot{L}_i$ for at least one $i$, that 
there exists a sequence $\dot{I}$ such that 
\[ \lbrack l^{\dot{I}}_{\dot{I}}, \sum_{i=1}^m \beta^{\dot{L}_i}_{\dot{K}_i} l^{\dot{K}_i}_{\dot{L}_i} \rbrack \]
does not belong to ${\cal M}$.  Indeed, let $j$ be an integer such that
\begin{enumerate}
   \item $\#(\dot{K}_j) - \#(\dot{L}_j) \geq \#(\dot{K}_i) - \#(\dot{L}_i)$ for all $i = 1, 2, \ldots$ and $m$; and
   \item $\#(\dot{K}_j) + \#(\dot{L}_j) \leq \#(\dot{K}_i) + \#(\dot{L}_i)$ for any $i = 1, 2, \ldots$ or $m$ such
         that $\#(\dot{K}_j) - \#(\dot{L}_j) = \#(\dot{K}_i) - \#(\dot{L}_i)$.
\end{enumerate}
If $\#(\dot{K}_j) \geq \#(\dot{L}_j)$, then consider
\begin{equation}
   \lbrack l^{\dot{K}_j}_{\dot{K}_j}, \sum_{i=1}^m \beta^{\dot{L}_i}_{\dot{K}_i}
   l^{\dot{K}_i}_{\dot{L}_i} \rbrack = \beta^{\dot{L}_j}_{\dot{K}_j} l^{\dot{K}_j}_{\dot{L}_j} -          
   \beta^{\dot{L}_j}_{\dot{K}_j} \sum_{\dot{K}_{j1} K_{j2} = \dot{K}_j} \delta^{\dot{K}_{j1}}_{\dot{L}_j} 
   l^{\dot{K}_j K_{j2}}_{\dot{K}_j} + \Gamma, 
\label{a4.2.1}
\end{equation}   
where
\begin{eqnarray}
   \Gamma & = & \lbrack l^{\dot{K}_j}_{\dot{K}_j}, \sum_{\begin{array}{c} i=1 \\ i \neq j \end{array}}^m 
   \beta^{\dot{L}_i}_{\dot{K}_i} l^{\dot{K}_i}_{\dot{L}_i} \rbrack 
   \nonumber \\
   & = & \sum_{\begin{array}{c} i=1 \\ i \neq j \end{array}}^m \sum_{k=1}^{n_i} 
   \beta^{\dot{L}_i}_{\dot{K}_i} \alpha^{\dot{N}_{ik}}_{\dot{M}_{ik}} l^{\dot{M}_{ik}}_{\dot{N}_{ik}}
\label{a4.2.2}   
\end{eqnarray}
where each $n_i$ for $i = 1, 2, \ldots, m$ but $i \neq j$ is dependent on $i$.  Let us assume that
\begin{equation}
   l^{\dot{M}_{ik}}_{\dot{N}_{ik}} = l^{\dot{K}_j}_{\dot{L}_j}
\label{a4.2.3}
\end{equation}
for some $i \in \{ 1, 2, \ldots, m \}$ but $i \neq j$ and $k \in \{ 1, 2, \ldots n_i \}$.  Then $\#(\dot{M}_{ik}) = 
\#(\dot{K}_j)$ and $\#(\dot{N}_{ik}) = \#(\dot{L}_j)$.  By Lemmas~\ref{la4.2.1} and~\ref{la4.2.2}, we get 
$\#(\dot{K}_i) = \#(\dot{K}_j)$ and $\#(\dot{L}_i) = \#(\dot{L}_j)$.  However, we also know that $\dot{K}_i \neq 
\dot{K}_j$ or $\dot{L}_i \neq \dot{L}_j$ and so there is no $k \in \{ 1, 2, \ldots, n_i \}$ such that 
Eq.(\ref{a4.2.3}) holds.  This leads to a contradiction and so we conclude that
\[ l^{\dot{M}_{ik}}_{\dot{N}_{ik}} \neq l^{\dot{K}_j}_{\dot{L}_j} \]
for {\em all} $i \in \{ 1, 2, \ldots, m \}$ but $i \neq j$ and $k \in \{ 1, 2, \ldots, n_i \}$.  From 
Eqs.(\ref{a4.2.1}) and (\ref{a4.2.2}), we deduce that $\lbrack l^{\dot{K}_j}_{\dot{K}_j}, \sum_{i=1}^m 
\beta^{\dot{L}_i}_{\dot{K}_i} l^{\dot{K}_i}_{\dot{L}_i} \rbrack$ does not belong to ${\cal M}$.  Similarly, if 
$\#(\dot{K}_j) \leq \#(\dot{L}_j)$, then $\lbrack l^{\dot{L}_j}_{\dot{L}_j}, \sum_{i=1}^m 
\beta^{\dot{L}_i}_{\dot{K}_i} l^{\dot{K}_i}_{\dot{L}_i} \rbrack$ does not belong to ${\cal M}$.  Hence, the 
normalizer of ${\cal M}$ is ${\cal M}$ itself.  We therefore conclude that ${\cal M}$ is a Cartan subalgebra of the 
algebra $\hatleftix$.  Q.E.D. 

\section{Root Vectors of $\hatleftix$}
\label{sa4.3}

All we need to do is to show that any root vector has to be of the form given by Eq.(\ref{4.3.7}).  Let $f \equiv 
\sum_{\dot{P}, \dot{Q}} a^{\dot{Q}}_{\dot{P}} l^{\dot{P}}_{\dot{Q}}$, where only a finite number of the numerical 
coefficients $a^{\dot{Q}}_{\dot{P}} \neq 0$, be a root vector.  In addition, we can assume without loss of 
generality that $\dot{P} \neq \dot{Q}$ if $a^{\dot{Q}}_{\dot{P}} = 0$.  Recall from Eq.(\ref{4.3.1}) that
\[ l^{\dot{I}}_{\dot{J}} s^{\dot{K}} = 
   \sum_{\dot{K}_1 \dot{K}_2 = \dot{K}} \delta^{\dot{K}_1}_{\dot{J}} s^{\dot{I} \dot{K}_2}. \]
Hence,
\[ l^{\dot{P}}_{\dot{Q}} s^{\dot{J}} = \sum_{\dot{I}} \sum_{\dot{J}_1 \dot{J}_2 = \dot{J}} 
   \delta^{\dot{J}_1}_{\dot{Q}} \delta^{\dot{P} \dot{J}_2}_{\dot{I}} s^{\dot{I}}. \]
Therefore,
\begin{equation}
   \left[ l^{\dot{M}}_{\dot{M}}, f \right] s^{\dot{K}} = 
   \sum_{\dot{I}, \dot{P}, \dot{Q}} a^{\dot{Q}}_{\dot{P}} \sum_{\begin{array}{c} \dot{I}_1 \dot{I}_2 = \dot{I} \\
   \dot{K}_1 \dot{K}_2 = \dot{K} \end{array}}
   \left( \delta^{\dot{M}}_{\dot{I}_1} \delta^{\dot{K}_1}_{\dot{Q}} \delta^{\dot{P} \dot{K}_2}_{\dot{I}} -
   \delta^{\dot{P}}_{\dot{I}_1} \delta^{\dot{K}_1}_{\dot{M}} \delta^{\dot{K}}_{\dot{Q} \dot{I}_2} \right) 
   s^{\dot{I}}.			   
\label{a4.3.1}
\end{equation}
Since $f$ is a root vector, we have
\begin{equation}
   \left[ l^{\dot{M}}_{\dot{M}}, f \right] = 
   \lambda_{\dot{M}} \sum_{\dot{P}, \dot{Q}} a^{\dot{Q}}_{\dot{P}} l^{\dot{P}}_{\dot{Q}}
\label{a4.3.2}
\end{equation}   
where $\lambda_{\dot{M}}$ is a root.  As a result, we can combine Eqs.(\ref{a4.3.1}) and (\ref{a4.3.2}) to obtain
\begin{eqnarray}
   \sum_{\dot{P}, \dot{Q}} a^{\dot{Q}}_{\dot{P}} \sum_{\begin{array}{c} \dot{I}_1 \dot{I}_2 = \dot{I} \\
   \dot{K}_1 \dot{K}_2 = \dot{K} \end{array}}
   \left( \delta^{\dot{M}}_{\dot{I}_1} \delta^{\dot{K}_1}_{\dot{Q}} \delta^{\dot{P} \dot{K}_2}_{\dot{I}} -
   \delta^{\dot{P}}_{\dot{I}_1} \delta^{\dot{K}_1}_{\dot{M}} \delta^{\dot{K}}_{\dot{Q} \dot{I}_2} \right) 
   & & \nonumber \\ 
   - \sum_{\dot{P}, \dot{Q}} a^{\dot{Q}}_{\dot{P}} \lambda_{\dot{M}} \sum_{\dot{K}_1 \dot{K}_2 = \dot{K}} 
   \delta^{\dot{K}_1}_{\dot{Q}} \delta^{\dot{P} \dot{K}_2}_{\dot{I}} & = & 0.
\label{a4.3.3}
\end{eqnarray}
This equation holds true for any arbitrarily chosen integer sequences $\dot{I}$, $\dot{K}$ and $\dot{M}$.

Let us find an $a^{\dot{S}}_{\dot{R}}$ in $f$ such that $\dot{R} \neq \dot{S}$, $a^{\dot{S}}_{\dot{R}} \neq 0$, and 
$a^{S_1}_{R_1} = 0$ for all $R_1$'s and $S_1$'s such that $R_1 \dot{R}_2 = \dot{R}$ and $S_1 \dot{S}_2 = \dot{S}$ 
for some $\dot{R}_2$ and $\dot{S}_2$.  The reader can easily convince himself or herself that such an 
$a^{\dot{S}}_{\dot{R}}$ always exists.  Let us choose $\dot{I} = \dot{R}$ and $\dot{K} = \dot{S}$ in 
Eq.(\ref{a4.3.3}).  Then we obtain from this equation that
\begin{equation}
   \lambda_{\dot{M}} = \sum_{\dot{R}_1 \dot{R}_2 = \dot{R}} \delta^{\dot{R}_1}_{\dot{M}} 
   - \sum_{\dot{S}_1 \dot{S}_2 = \dot{S}} \delta^{\dot{M}}_{\dot{S}_1}.
\label{a4.3.4}
\end{equation}
Therefore,
\[ \lambda_{\dot{M}} - \sum_{i=1}^{\Lambda} \lambda_{\dot{M}i} = \delta^{\dot{R}}_{\dot{M}} - 
   \delta^{\dot{M}}_{\dot{S}}. \]
Thence
\begin{eqnarray*}
   \left[ l^{\dot{R}}_{\dot{R}} - \sum_{i=1}^{\Lambda} l^{\dot{R}i}_{\dot{R}i}, f \right] 
   & \neq & 0 \mbox{; and} \\
   \left[ l^{\dot{S}}_{\dot{S}} - \sum_{i=1}^{\Lambda} l^{\dot{S}i}_{\dot{S}i}, f \right] 
   & \neq & 0.
\end{eqnarray*}
However, both $l^{\dot{R}}_{\dot{R}} - \sum_{i=1}^{\Lambda} l^{\dot{R}i}_{\dot{R}i}$ and $l^{\dot{S}}_{\dot{S}} - 
\sum_{i=1}^{\Lambda} l^{\dot{S}i}_{\dot{S}i} \in \salt$.  Thus $f \in \salt$.  Now Eq.(\ref{4.3.15}) shows clearly 
that $f = f^{\dot{R}}_{\dot{S}}$.  The same equation also shows that each root vector space must be 
one-dimensional.  Q.E.D. 
         
\section{Lie Bracket of $\hatcentrix$}
\label{sa4.4}

We are going to show that the commutator between two operators of the fourth kind, Eq.(\ref{4.5.4}), defines a Lie 
bracket between them.  This can be done by showing that Eq.(\ref{4.5.4}) satisfies Eq.(\ref{4.5.3}).  This involves 
a tedious compuatation involving the delta function defined in Appendix~\ref{sa1.1}.  The properties of this delta 
function will be extensively used.  Let us consider the action of the commutator of two $\s$'s  on $s^{\dot{P}}$.  
If $\dot{P}$ is empty, then Eq.(\ref{4.5.4}) certainly satisfies Eq.(\ref{4.5.3}).  Therefore we only need to 
consider the case when $\dot{P}$ is not empty.  In this case we can simply write $\dot{P}$ as $P$.  Then the action 
of the commutator on $s^P$ is
\begin{eqnarray}
   \lefteqn{ \left[ \s^I_J, \s^K_L \right] s^P = \sum_Q \left(
   \delta^I_Q \delta^K_J \delta^P_L + \sum_C \delta^I_Q \delta^{C K}_J
     \delta^P_{C L} + \sum_D \delta^I_Q \delta^{K D}_J \delta^P_{L D} \right. }
     \nonumber \\
   & & + \sum_{C, D} \delta^I_Q \delta^{C K D}_J \delta^P_{C L D} +
   \sum_A \delta^{A I}_Q \delta^K_{A J} \delta^P_L +
   \sum_{A, C} \delta^{A I}_Q \delta^{C K}_{A J} \delta^P_{C L} \nonumber \\
   & & + \sum_{A, D} \delta^{A I}_Q \delta^{K D}_{A J} \delta^P_{L D} +
   \sum_{A, C, D} \delta^{A I}_Q \delta^{C K D}_{A J} \delta^P_{C L D}
   + \sum_B \delta^{I B}_Q \delta^K_{J B} \delta^P_L  \nonumber \\
   & & + \sum_{B, C} \delta^{I B}_Q \delta^{C K}_{J B} \delta^P_{C L} + 
   \sum_{B, D} \delta^{I B}_Q \delta^{K D}_{J B} \delta^P_{L D} +
   \sum_{B, C, D} \delta^{I B}_Q \delta^{C K D}_{J B} \delta^P_{C L D}
   \nonumber \\
   & & + \sum_{A, B} \delta^{A I B}_Q \delta^K_{A J B} \delta^P_L +
   \sum_{A, B, C} \delta^{A I B}_Q \delta^{C K}_{A J B} \delta^P_{C L} +
   \sum_{A, B, D} \delta^{A I B}_Q \delta^{K D}_{A J B} \delta^P_{L D}
   \nonumber \\
   & & + \left. \sum_{A, B, C, D} \delta^{A I B}_Q \delta^{C K D}_{A J B} 
   \delta^P_{C L D} \right) s^Q - (I \leftrightarrow K, J \leftrightarrow L).
\label{a4.4.1}
\end{eqnarray}
Each of the terms on the R.H.S. of the above equations can be rewritten as follows;
\begin{eqnarray}
   \lefteqn{ \delta^I_Q \delta^K_J \delta^P_L = \delta^K_J \delta^I_Q \delta^P_L; } 
\nonumber \\
   \lefteqn{ \sum_C \delta^I_Q \delta^{C K}_J \delta^P_{C L} =  
   \sum_{J_1 J_2 = J} \delta^K_{J_2} \delta^I_Q \delta^P_{J_1 L}; } 
\nonumber \\
   \lefteqn{ \sum_D \delta^I_Q \delta^{K D}_J \delta^P_{L D} =  
   \sum_{J_1 J_2 = J} \delta^K_{J_1} \delta^I_Q \delta^P_{L J_2}; } 
\nonumber \\
   \lefteqn{ \sum_{C, D} \delta^I_Q \delta^{C K D}_J \delta^P_{C L D} =
   \sum_{J_1 J_2 J_3 = J} \delta^K_{J_2} \delta^I_Q \delta^P_{J_1 L J_3}; }
\nonumber \\
   \lefteqn{ \sum_A \delta^{A I}_Q \delta^K_{A J} \delta^P_L =
   \sum_{K_1 K_2 = K} \delta^{K_2}_J \delta^{K_1 I}_Q \delta^P_L; } 
\nonumber \\
   \lefteqn{ \sum_{A, C} \delta^{A I}_Q \delta^{C K}_{A J} \delta^P_{C L} =
   \sum_{K_1 K_2 = K} \sum_E \delta^{K_2}_J \delta^{E K_1 I}_Q \delta^P_{E L} +
   \sum_E \delta^K_J \delta^{E I}_Q \delta^P_{E L} } \nonumber \\
   & & + \sum_{J_1 J_2 = J} \sum_E \delta^K_{J_2} \delta^{E I}_Q \delta^P_{E J_1 L}; 
\nonumber \\  
   \lefteqn{ \sum_{A, D} \delta^{A I}_Q \delta^{K D}_{A J} \delta^P_{L D} =
   \sum_F \delta^{K F I}_Q \delta^P_{L F J} + \delta^{K I}_Q \delta^P_{L J}
   + \sum_{\begin{array}{l}
   	      J_1 J_2 = J \\
   	      K_1 K_2 = K
  	   \end{array}} \delta^{K_2}_{J_1} \delta^{K_1}_Q \delta^P_{L J_2}; } 
\nonumber \\ 
   \lefteqn{ \sum_{A, C, D} \delta^{A I}_Q \delta^{C K D}_{A J} \delta^P_{C L D} = 
   \sum_{J_1 J_2 J_3 = J} \sum_E \delta^K_{J_2} \delta^{E I}_Q \delta^P_{E J_1 L J_3}
   + \sum_{J_1 J_2 = J} \sum_E \delta^K_{J_1} \delta^{E I}_Q \delta^P_{E L J_2}} \nonumber \\ 
   & & + \sum_{\begin{array}{l}
   	      J_1 J_2 = J \\
   	      K_1 K_2 = K
   	   \end{array}} \delta^{K_2}_{J_1} \delta^{E K_1 I}_Q \delta^P_{E L J_2}
   + \sum_E \delta^{E K I}_Q \delta^P_{E L J} + \sum_{E, F} \delta^{E K F I}_Q \delta^P_{E L F J};
\nonumber \\ 
   \lefteqn{ \sum_B \delta^{I B}_Q \delta^K_{J B} \delta^P_L =
   \sum_{K_1 K_2 = K} \delta^{K_1}_J \delta^{I K_2}_Q \delta^P_L; } 
\nonumber \\
   \lefteqn{ \sum_{B, C} \delta^{I B}_Q \delta^{C K}_{J B} \delta^P_{C L} =
   \sum_{\begin{array}{l}
   	    J_1 J_2 = J \\
   	    K_1 K_2 = K
   	 \end{array}} \delta^{K_1}_{J_2} \delta^{I K_2}_Q \delta^P_{J_1 L} +
   \delta^{I K}_Q \delta^P_{J L} + \sum_F \delta^{I F K}_Q \delta^P_{J F L}; } 
\nonumber \\
   \lefteqn{ \sum_{B, D} \delta^{I B}_Q \delta^{K D}_{J B} \delta_{L D} =
   \sum_{J_1 J_2 = J} \sum_F \delta^K_{J_1} \delta^{I F}_Q \delta^P_{L J_2 F} +
   \sum_F \delta^K_J \delta^{I F}_Q \delta^P_{L F} } \nonumber \\
   & & + \sum_{K_1 K_2 = K} \sum_F \delta^{K_1}_J \delta^{I K_2 F}_Q \delta^P_{L F}; 
\nonumber \\ 
   \lefteqn{ \sum_{B, C, D} \delta^{I B}_Q \delta^{C K D}_{J B} \delta^P_{C L D} =
   \sum_{F, G} \delta^{I F K G}_Q \delta^P_{J F L G} + 
   \sum_G \delta^{I K G}_Q \delta^P_{J L G} } \nonumber \\
   & & + \sum_{\begin{array}{l}
   		  J_1 J_2 = J \\
   		  K_1 K_2 = K
  	       \end{array}} \delta^{K_1}_{J_2} \sum_F \delta^{I K_2 F}_Q \delta^P_{J_1 L F}  
   + \sum_{J_1 J_2 = J} \delta^K_{J_2} \delta^{I F}_Q \delta^P_{J_1 L F} \nonumber \\
   & & + \sum_{J_1 J_2 J_3 = J} \sum_F \delta^K_{J_2} \delta^{I F}_Q \delta^P_{J_1 L J_3 F};
\nonumber \\
   \lefteqn{ \sum_{A, B} \delta^{A I B}_Q \delta^K_{A J B} \delta^P_L =
   \sum_{K_1 K_2 K_3 = K} \delta^{K_2}_J \delta^{K_1 I K_3}_Q \delta^P_L; }
\nonumber \\
   \lefteqn{ \sum_{A, B, C} \delta^{A I B}_Q \delta^{C K}_{A J B} \delta^P_{C L} =
   \sum_{K_1 K_2 K_3 = K} \sum_E \delta^{K_2}_J \delta^{E K_1 I K_3}_Q \delta^P_{E L} } \nonumber \\
   & & + \sum_{K_1 K_2 = K} \sum_E \delta^{K_1}_J \delta^{E I K_2}_Q \delta^P_{E L}
   + \sum_{\begin{array}{l}
   	      J_1 J_2 = J \\
   	      K_1 K_2 = K
   	   \end{array}} \sum_E \delta^{K_1}_{J_2} \delta^{E I K_2}_Q \delta^P_{E J_1 L} \nonumber \\
   & & + \sum_E \delta^{E I K}_Q \delta^P_{E J L}
   + \sum_{E, F} \delta^{E I F K}_Q \delta^P_{E J F L}; 
\nonumber \\
   \lefteqn{ \sum_{A, B, D} \delta^{A I B}_Q \delta^{K D}_{A J B} \delta^P_{L D} =
   \sum_{F, G} \delta^{K F I G}_Q \delta^P_{L F J G} + 
   \sum_G \delta^{K I G}_Q \delta^P_{L J G} } \nonumber \\
   & & + \sum_{\begin{array}{l}
   		  J_1 J_2 = J \\
   		  K_1 K_2 = K
  	       \end{array}} \sum_F \delta^{K_2}_{J_1} \delta^{K_1 I F}_Q \delta^P_{L J_2 F}
   + \sum_{K_1 K_2 = K} \sum_F \delta^{K_2}_J \delta^{K_1 I F}_Q \delta^P_{L F} \nonumber \\
   & & + \sum_{K_1 K_2 K_3 = K} \sum_F \delta^{K_2}_J \delta^{K_1 I K_3 F}_Q \delta^P_{L F} \mbox{; and}
\nonumber \\
   \lefteqn{ \sum_{A, B, C, D} \delta^{A I B}_Q \delta^{C K D}_{A J B} \delta^P_{C L D} =
   \sum_{E, F, G} \delta^{E K F I G}_Q \delta^P_{E L F J G} + 
   \sum_{E, G} \delta^{E K I G}_Q \delta^P_{E L J G} } \nonumber \\
   & & + \sum_{\begin{array}{l}
   		  J_1 J_2 = J \\
   		  K_1 K_2 = K
  	       \end{array}} \sum_{E, F} \delta^{K_2}_{J_1} \delta^{E K_1 I F}_Q \delta^Q_{E L J_2 F}
   + \sum_{K_1 K_2 = K} \sum_{E, F} \delta^{K_2}_J \delta^{E K_1 I F}_Q \delta^P_{E L F} \nonumber \\
   & & + \sum_{K_1 K_2 K_3 = K} \sum_{E, F} \delta^{K_2}_J \delta^{E K_1 I K_3 F}_Q \delta^P_{E L F}
   + \sum_{J_1 J_2 = J} \sum_{E, F} \delta^K_{J_1} \delta^{E I F}_Q \delta^P_{E L J_2 F} \nonumber \\
   & & + \sum_{E, F} \delta^K_J \delta^{E I F}_Q \delta^P_{E L F} +
   \sum_{K_1 K_2 = K} \sum_{E, F} \delta^{K_1}_J \delta^{E I K_2 F}_Q \delta^P_{E L F} \nonumber \\
   & & + \sum_{J_1 J_2 J_3 = J} \sum_{E, F} \delta^K_{J_2} \delta^{E I F}_Q \delta^P_{E J_1 L J_3 F}
   + \sum_{J_1 J_2 = J} \sum_{E, F} \delta^K_{J_2} \delta^{E I F}_Q \delta^P_{E J_1 L F} \nonumber \\
   & & + \sum_{\begin{array}{l}
   		  J_1 J_2 = J \\
   		  K_1 K_2 = K
   	       \end{array}} \sum_{E, F} \delta^{K_1}_{J_2} \delta^{E I K_2 F}_Q \delta^P_{E J_1 L F}
   + \sum_{E, G} \delta^{E I K G}_Q \delta^P_{E J L F} \nonumber \\
   & & + \sum_{E, F, G} \delta^{E I F K G}_Q \delta^P_{E J F L G}. 
\label{a4.4.2}
\end{eqnarray}
We can now substitute the expressions in Eq.(\ref{a4.4.2}) into Eq.(\ref{a4.4.1}) to get
\begin{eqnarray}
   \lefteqn{ \left[ \s^I_J, \s^K_L \right] s^P = } \nonumber \\   
   & & \sum_Q \left\{ \delta^K_J ( \delta^I_Q \delta^P_L 
   + \sum_E \delta^{E I}_Q \delta^P_{E L}
   + \sum_F \delta^{I F}_Q \delta^P_{L F} + 
   \sum_{E, F} \delta^{E I F}_Q \delta^P_{E L F} ) \right. \nonumber \\
   & & + \sum_{J_1 J_2 = J} \delta^K_{J_2} (\delta^I_Q \delta^P_{J_1 L} 
   + \sum_E \delta^{E I}_Q \delta^P_{E J_1 L} + 
   \sum_F \delta^{I F}_Q \delta^P_{J_1 L F} \nonumber \\
   & & + \sum_{E, F} \delta^{E I F}_Q \delta^P_{E J_1 L F} )
   + \sum_{K_1 K_2 = K} \delta^{K_1}_J (\delta^{I K_2}_Q \delta^P_L
   + \sum_E \delta^{E I K_2}_Q \delta^P_{E L} \nonumber \\
   & & + \sum_F \delta^{I K_2 F}_Q \delta^P_{L F} +
   \sum_{E, F} \delta^{E I K_2 F}_Q \delta^P_{E L F} )
   + \sum_{J_1 J_2 = J} \delta^K_{J_1} (\delta^I_Q \delta^P_{L J_2} \nonumber \\
   & & + \sum_E \delta^{E I}_Q \delta^P_{E L J_2} +
   \sum_F \delta^{I F}_Q \delta^P_{L J_2 F} +
   \sum_{E, F} \delta^{E I F}_Q \delta^P_{E L J_2 F} ) \nonumber \\
   & & + \sum_{K_1 K_2 = K} \delta^{K_2}_J (\delta^{K_1 I}_Q \delta^P_L
   + \sum_E \delta^{E K_1 I}_Q \delta^P_{E L} +
   \sum_F \delta^{K_1 I F}_Q \delta^P_{L F} \nonumber \\
   & & + \sum_{E, F} \delta^{E K_1 I F}_Q \delta^P_{E L F} )
   + \sum_{\begin{array}{l}
   	      J_1 J_2 = J \\
   	      K_1 K_2 = K
   	   \end{array}} \delta^{K_1}_{J_2}
   (\delta^{I K_2}_Q \delta^P_{J_1 L} + 
   \sum_E \delta^{E I K_2}_Q \delta^P_{E J_1 L} \nonumber \\
   & & + \sum_F \delta^{I K_2 F}_Q \delta^P_{J_1 L F} +
   \sum_{E, F} \delta^{E I K_2 F}_Q \delta^P_{E J_1 L F} )
   + \sum_{\begin{array}{l}
   	      J_1 J_2 = J \\
   	      K_1 K_2 = K
   	   \end{array}} \delta^{K_2}_{J_1}
   (\delta^{K_1 I}_Q \delta^P_{L J_2} \nonumber \\
   & & + \sum_E \delta^{E K_1 I}_Q \delta^P_{E L J_2} +
   \sum_F \delta^{K_1 I F}_Q \delta^P_{L J_2 F} +
   \sum_{E, F} \delta^{E K_1 I F}_Q \delta^P_{E L J_2 F} ) \nonumber \\
   & & + \sum_{J_1 J_2 J_3 = J} \delta^K_{J_2}
   (\delta^I_Q \delta^P_{J_1 L J_3} +
   \sum_E \delta^{E I}_Q \delta^P_{E J_1 L J_3} \nonumber \\
   & & + \sum_F \delta^{I F}_Q \delta^P_{J_1 L J_3 F} +
   \sum_{E, F} \delta^{E, I, F} \delta^P_{E J_1 L J_3 F} )
   + \sum_{K_1 K_2 K_3 = K} \delta^{K_2}_J
   (\delta^{K_1 I K_3}_Q \delta^P_L \nonumber \\ 
   & & + \sum_E \delta^{E K_1 I K_3}_Q \delta^P_{E L} +
   \sum_F \delta^{K_1 I K_3 F}_Q \delta^P_{L F} +
   \sum_{E, F} \delta^{E K_1 I K_3 F}_Q \delta^P_{E L F} ) \nonumber \\
   & & \left. + \Gamma \right\} s^Q - (I \leftrightarrow K, J \leftrightarrow L)
\label{a4.4.3}
\end{eqnarray}    
where
\begin{eqnarray}
   \Gamma & = & \delta^{I K}_Q \delta^P_{J L} + 
   \sum_E \delta^{E I K}_Q \delta^P_{E J L} +
   \sum_F \delta^{I F K}_Q \delta^P_{J F L} +
   \sum_G \delta^{I K G}_Q \delta^P_{J L G} \nonumber \\
   & & + \sum_{E, F} \delta^{E I F K}_Q \delta^P_{E J F L} +
   \sum_{F, G} \delta^{I F K G}_Q \delta^P_{J F L G} +
   \sum_{E, G} \delta^{E I K G}_Q \delta^P_{E J L G} \nonumber \\
   & & + \sum_{E, F, G} \delta^{E I F K G}_Q \delta^P_{E J F L G}
   + (I \leftrightarrow K, J \leftrightarrow L)  
\label{a4.4.4}
\end{eqnarray}
If we substitute Eq.(\ref{a4.4.3}) without $\Gamma$ into Eq.(\ref{a4.4.1}), we will obtain exactly the action of 
operators on the R.H.S. of Eq.(\ref{4.5.4}) on $s^P$.  $\Gamma$ is reproduced when $I$ and $K$ are interchanged 
with $J$ and $L$, respectively, in Eq.(\ref{a4.4.3}) and so it is cancelled.  Consequently, Eq.(\ref{4.5.4}) is 
true.  Q.E.D.
   
\section{Root Vectors of $\hatcentrix$}
\label{sa4.5}

We need to show that any root vector has to be of the form given by Eq.(\ref{4.5.13}).  Let $f \equiv \sum_{P, Q} 
a^Q_P \s^P_{\bar{Q}}$, where only a finite number of the numerical coeffiicients $a^Q_P \neq 0$, be a root vector.  
In addition, we can assume without loss of generality that $P \neq Q$ if $a^Q_P = 0$.  Recall from Eq.(\ref{4.5.1}) 
that
\[ \s^I_J s^K = \delta^K_J s^I + \sum_{K_1 K_2 = K} \delta^{K_1}_J 
   s^{I K_2} + \sum_{K_1 K_2 = K} \delta^{K_2}_J s^{K_1 I} +
   \sum_{K_1 K_2 K_3 = K} \delta^{K_2}_J s^{K_1 I K_3} \]
Hence,
\begin{eqnarray*}
   \s^P_Q s^J & = & \sum_I \left( \delta^J_Q \delta^P_I + 
   \sum_{J_1 J_2 = J} \delta^{J_2}_Q \delta^{J_1 P}_I +
   \sum_{J_1 J_2 = J} \delta^{J_1}_Q \delta^{P J_2}_I \right. \\
   & & \left. \sum_{J_1 J_2 J_3 = J} \delta^{J_2}_Q \delta^{J_1 P J_3}_I
   \right) s^I.
\end{eqnarray*}
Therefore,
\begin{eqnarray}
   \lefteqn{ \left[ \s^M_M, f \right] s^K = } \nonumber \\ 
   & & \sum_I \left( \delta^M_I \delta^K_Q \delta^P_M +
   \sum_{K_1 K_2 = K} \delta^M_I \delta^{K_2}_Q \delta^{K_1 P}_M + 
   \sum_{K_1 K_2 = K} \delta^M_I \delta^{K_1}_Q \delta^{P K_2}_M \right.
     \nonumber \\
   & & + \sum_{K_1 K_2 K_3 = K} \delta^M_I \delta^{K_2}_Q \delta^{K_1 P K_3}_M
   + \sum_{I_1 I_2 = I} \delta^M_{I_2} \delta^K_Q \delta^P_I +
   \sum_{\begin{array}{l}
   	    I_1 I_2 = I \\
   	    K_1 K_2 = K
   	 \end{array}}
   \delta^M_{I_2} \delta^{K_2}_Q \delta^{K_1 P}_I \nonumber \\
   & & + \sum_{\begin{array}{l}
   	    I_1 I_2 = I \\
   	    K_1 K_2 = K
   	 \end{array}}
   \delta^M_{I_2} \delta^{K_1}_Q \delta^{P K_2}_I +
   \sum_{\begin{array}{l}
   	    I_1 I_2 = I \\
   	    K_1 K_2 K_3 = K
   	 \end{array}}
   \delta^M_{I_2} \delta^{K_2}_Q \delta^{K_1 P K_3}_I \nonumber \\
   & & + \sum_{I_1 I_2 = I} \delta^M_{I_1} \delta^K_Q \delta^P_I +
   \sum_{\begin{array}{l}
   	    I_1 I_2 = I \\
   	    K_1 K_2 = K
   	 \end{array}}
   \delta^M_{I_1} \delta^{K_2}_Q \delta^{K_1 P}_I +
   \sum_{\begin{array}{l}
   	    I_1 I_2 = I \\
   	    K_1 K_2 = K
   	 \end{array}}
   \delta^M_{I_1} \delta^{K_1}_Q \delta^{P K_2}_I \nonumber \\
   & & + \sum_{\begin{array}{l}
   		  I_1 I_2 = I \\
   		  K_1 K_2 K_3 = K
  	       \end{array}}
   \delta^M_{I_1} \delta^{K_2}_Q \delta^{K_1 P K_3}_I +
   \sum_{I_1 I_2 I_3 = I} \delta^M_{I_2} \delta^K_Q \delta^P_I \nonumber \\
   & & + \sum_{\begin{array}{l}
   		  I_1 I_2 I_3 = I \\
   		  K_1 K_2 = K
  	       \end{array}}
   \delta^M_{I_2} \delta^{K_2}_Q \delta^{K_1 P}_I
   + \sum_{\begin{array}{l}
   	      I_1 I_2 I_3 = I \\
   	      K_1 K_2 = K
  	   \end{array}}
   \delta^M_{I_2} \delta^{K_1}_Q \delta^{P K_2}_I \nonumber \\
   & & \left. + \sum_{\begin{array}{l}
   		         I_1 I_2 I_3 = I \\
   		         K_1 K_2 K_3 = K
   	              \end{array}}
   \delta^M_{I_2} \delta^{K_2}_Q \delta^{K_1 P K_3}_I \right) a^Q_P s^I
   \nonumber \\
   & & - (P \leftrightarrow M \;\mbox{in the superscripts},
   M \leftrightarrow Q \;\mbox{in the subscripts})
\label{a4.5.1}
\end{eqnarray}
Since $f$ is a root vector, we have
\begin{equation}
   \left[ \s^M_M, f \right] =
   \lambda_M \sum_{P, Q} a^Q_P \s^P_Q.
\label{a4.5.2}
\end{equation}
where $\lambda_M$ is a root.  As a result, we can combine Eqs.(\ref{a4.5.1}) and (\ref{a4.5.2}) together to obtain 
an equation which is too long to be written down here for {\em any} integer sequences $I$, $K$ and $M$.
   
Let us find an $a^S_R$ in $f$ such that $R \neq S$, $a^S_R \neq 0$, $a^{S_1}_{R_1} = a^{S_2}_{R_2} = 0$ for all 
$R_1$'s, $S_1$'s, $R_2$'s and $S_2$'s such that $R_1 R_2 = R$ and $S_1 S_2 = S$, and $a^{S_2}_{R_2} = 0$ for all 
$R_2$'s and $S_2$'s such that $R_1 R_2 R_3 = R$ and $S_1 S_2 S_3 = S$ for some $R_1$, $R_3$, $S_1$ and $S_3$.  The 
reader can easily convince himself or herself that such an $a^S_R$ always exists.  Let us choose $I = R$ and $K = S$
in Eq.(\ref{a4.5.1}).  Then when we combine Eqs.(\ref{a4.5.1}) and (\ref{a4.5.2}), we get
\begin{eqnarray}
   \lambda_M & = & \delta^R_M + \sum_{R_1 R_2 = R} \delta^{R_1}_M 
   + \sum_{R_1 R_2 = R} \delta^{R_2}_M + \sum_{R_1 R_2 R_3 = R}
\delta^{R_2}_M
   \nonumber \\
   & & - \delta^M_S - \sum_{S_1 S_2 = S} \delta^M_{S_1}
   - \sum_{S_1 S_2 = S} \delta^M_{S_2} - \sum_{S_1 S_2 S_3 = S}
\delta^M_{S_2}.
\label{a4.5.3}
\end{eqnarray}
Therefore, we obtain after some manipulation that
\[ \lambda_M - \sum_{j=1}^{{\Lambda}} \lambda_{Mj} - \sum_{i=1}^{{\Lambda}}
\lambda_{iM}        
   - \sum_{i, j = 1}^{{\Lambda}} \lambda_{iMj} = \delta^R_M - \delta^M_S. \]
This means
\begin{eqnarray*}
   \left[ f^R_R , f \right] & \neq & 0 \mbox{; and} \\
   \left[ f^S_S , f \right] & \neq & 0.
\end{eqnarray*}
Since $\salt$ is a proper ideal, so $f \in \salt$.  Now Eq.(\ref{4.5.13}) shows clearly that $f = f^R_S$.  The same 
equation also shows that each root vector space must be one-dimensional.  Q.E.D.   

\chapter{Proofs of Some Results in Chapter~\ref{c3}}
\la{ca3}

\section{Linear Independence of Color-Invariant Operators}
\label{sa3.1}

In this appendix, we are going to show by {\em ad absurdum} that the set of all $\ti{f}'^I_J$'s defined by 
Eq.(\ref{3.2.7}) and all $\g'^I_J$'s defined by Eq.(\ref{3.2.10}) is linearly independent.  Assume, on the 
contrary, that the set is linearly dependent.  Then there exists an equation
\beq
    \sum_{p=1}^r \a_p \ti{f}'^{I_p}_{J_p} + \sum_{p=r+1}^s \a_p \g^{I_p}_{J_p} = 0
\la{a3.1.1}
\eeq
such that $r$ and $s$ are positive integers with $r \leq s$ ($r=s$ means that the second sum vanishes), and all 
$\a_p$'s for $p = 1$, 2, \ldots, and $s$ are non-zero complex constants.  In addition, in the above equation, 
either $(I_p) \neq (I_q)$, or $(J_p) \neq (J_q)$, or both if $1 \leq p \leq r$, $1 \leq q \leq r$ and $p \neq q$, 
and either $I_p \neq I_q$, or $J_p \neq J_q$, or both if $r+1 \leq p \leq s$, $r+1 \leq q \leq s$ and $p \neq q$.  

Assume that $s > r$.  We can assume without loss of generality that 
\begin{enumerate}
\item $J_{r+1} = J_{r+2} = \cdots = J_{r+x}$;
\item $J_{r+1} \neq J_{r+x+1}$, $J_{r+1} \neq J_{r+x+2}$, \ldots, and $J_{r+1} \neq J_s$; and
\item $\#(J_{r+1}) \leq \#(J_p)$ for all $p = r+1, r+2$, \ldots, and $s$
\end{enumerate}
for some integer $x$ such that $r < r + x \leq s$.  Consider the action of the L.H.S. of Eq.(\ref{a3.1.1}) on 
$s'^{J_{r+1}}$.  We get
\beq
   \le( \sum_{p=1}^r \a_p \ti{f}'^{I_p}_{J_p} + \sum_{p=r+1}^s \a_p \g^{I_p}_{J_p} \ri) s'^{J_{r+1}} = 
   \sum_{p=r+1}^{r+x} \a_p s'^{I_p}.
\label{a3.1.2}
\eeq
Combining Eqs.(\ref{a3.1.1}) and (\ref{a3.1.2}) yields
$ \sum_{p=r+1}^{r+x} \alpha_p s'^{I_p} = 0, $
which is impossible because $s'^{I_r}$, $s'^{I_{r+1}}$, \ldots, $s'^{I_{r+x}}$ are linearly
independent.  Therefore, $s = r$ and Eq.(\ref{a3.1.1}) can be simplified to
\begin{equation}
   \sum_{p=1}^r \alpha_p \ti{f}'^{I_p}_{J_p} = 0.
\label{a3.1.3}
\end{equation}
Again we can assume without loss of generality that
\begin{enumerate}
\item $(J_1) = (J_2) = \cdots = (J_y)$; and
\item $(J_1) \neq (J_{y+1})$, $(J_1) \neq (J_{y+2}), \ldots,$ and $(J_1) \neq (J_r)$
\end{enumerate}
for some integer $y$ such that $1 \leq y \leq r$.  Consider the action of the L.H.S. of Eq.(\ref{a3.1.3}) on 
$\Ps'^{J_1}$:
\[ \le( \sum_{p=1}^r \alpha_p \ti{f}'^{I_p}_{J_p} \ri) \Ps'^{J_1} = \sum_{p=1}^y \a_p 
   \le( \d^{J_1}_{J_p} \Ps'^{I_p} + 
   \sum_{J_{p1} J_{p2} = J} (-1)^{\ep(J_{p1}) \ep(J_{p2})} \d^{J_1}_{J_{p2} J_{p1}} \Ps'^{I_p} \ri). \]
However, the R.H.S. of this equation is impossible to vanish because $\Ps'^{I_1}$, $\Ps'^{I_2}$, \ld, $\Ps^{I_y}$ 
are linearly independent.  Thus the set of all $\ti{f}'^I_J$'s together with all $\g^I_J$'s is linearly 
independent.  Q.E.D.

\section{Product of Two Color-Invariant Operators}
\label{sa3.2}

We will show by contradiction that the product of two color-invariant operators is in general not well defined.
Consider the case when $\L = 1$, and assume that the operators $\a^{\m}_{\n}(1)$ and $\a^{\da\m}_{\n}(1)$ are 
bosonic.  Let $\g'^a_b = \g'^{11\ld 1}_{11\ld 1}$ and $\ti{f}'^{(a)}_{(b)} = \ti{f}'^{11\ld 1}_{11\ld 1}$, where 
the number 1 shows up $a$ times in the superscript and $b$ times in the subscript of $\g'$.  Moreover, let 
$\Ps'^{(c)} = \Ps'^{(11\ldots 1)}$ and $s'^c = s^{11\ldots 1}$, where the number 1 shows up $c$ times.  

Assume that $\g'^1_1 \g'^1_1 = \sum_{p=1}^r \a_p \g'^p_p + \sum_{q=1}^s \b_q \ti{f}'^{(q)}_{(q)}$, where $\a_1$, 
$\a_2$, \ld, $\a_r$, $\b_1$, $\b_2$, \ld, and $\b_s$ are non-zero complex numbers for some positive integers $r$ 
and $s$.  Then from the equations $\g'^1_1 \g'^1_1 (s'^1) = \g'^1_1 (\g'^1_1 s'^1) = 1^2 s'^1$, $\g'^1_1 \g'^1_1 
(s'^2) = \g'^1_1 (\g'^1_1 s'^1) = 2^2 s'^1$, \ld, and $\g'^1_1 \g'^1_1 (s'^r) = \g'^1_1 (\g'^1_1 s'^1) = r^2 s'^1$, 
we deduce that $\a_1 = 1$ and $\a_2 = \a_3 = \cd = \a_r = 2$.  Hence $\g'^1_1 \g'^1_1 = \g'^1_1 + 2 \sum_{p=2}^r 
\g'^p_p + \sum_{q=1}^s \b_q \ti{f}^{(q)}_{(q)}$.  However, $\g'^1_1 \g'^1_1 (s'^{r+1}) = (r + 1)^2 s'^{r+1}$ and 
$\g'^1_1 + 2 \sum_{p=2}^r \g'^p_p + \sum_{q=1}^s \b_q \ti{f}'^{(q)}_{(q)} (s'^{r+1}) = (r^2 + 2 r - 1) s'^{r+1}$, 
leading to a contradiction.  (This proves our assertion at the beginning of Section~\ref{s4.5}, namely that
the product of two operatros of the fourth kind cannot be written as a finite linear combination of operators of
this kind.)  

Thus we assume instead that $\g'^1_1 \g'^1_1 = \sum_{q=1}^s \b_q \ti{f}'^{(q)}_{(q)}$, where the $\b_q$'s are
non-zero complex numbers.  However, $\g'^1_1 \g'^1_1 (\Ps'^{(s+1)}) = (s+1)^2 \Ps'^{(s+1)}$ whereas 
$\sum_{q=1}^s \b_q \ti{f}'^{(q)}_{(q)} \Ps'^{(s+1)} = 0$, leading to a contradiction, too.  Consequently, it is 
impossible to write $\g'^1_1 \g'^1_1$ as a finite linear combination of $\g'$'s and $\tilde{f}$'s.

This proof can be easily modified to more general cases.  Q.E.D. 

\section{Lie Brackets of a Precursor Algebra}
\label{sa3.3}
    			   	   
What we need to do is to show that Eqs.(\ref{3.3.1}), (\ref{3.3.2}) and (\ref{3.3.3}) satisfy
\beq
   \lbrack \g^I_J, \g^K_L \rbrack \Ps^{(P)} & = & \g^I_J (\g^K_L \Ps^{(P)}) - \g^K_L (\g^I_J \Ps^{(P)});
\label{a3.3.1} \\
   \lbrack \g^I_J, \ti{f}^{(K)}_{(L)} \rbrack \Ps^{(P)} & = & \g^I_J (\ti{f}^{(K)}_{(L)} \Ps^{(P)}) - 
   \ti{f}^{(K)}_{(L)} (\g^I_J \Ps^{(P)});
\label{a3.3.2} \\
   \lbrack \ti{f}^{(I)}_{(J)}, \ti{f}^{(K)}_{(L)} \rbrack \Ps^{(P)} & = & 
   \ti{f}^{(I)}_{(J)} (\ti{f}^{(K)}_{(L)} \Ps^{(P)}) - \ti{f}^{(K)}_{(L)} (\ti{f}^{(I)}_{(J)} \Ps^{(P)});
\label{a3.3.3} \\
   \lbrack \g^I_J, \g^K_L \rbrack s^M & = & \g^I_J (\g^K_L s^M) - \g^K_L (\g^I_J s^M);
\label{a3.3.4} \\
   \lbrack \g^I_J, \ti{f}^{(K)}_{(L)} \rbrack s^M & = & 
   \g^I_J (\ti{f}^{(K)}_{(L)} s^M) - \ti{f}^{(K)}_{(L)} (\g^I_J s^M); \; \mbox{and}
\label{a3.3.5} \\
   \lbrack \ti{f}^{(I)}_{(J)}, \ti{f}^{(K)}_{(L)} \rbrack s^M & = & \ti{f}^{(I)}_{(J)} (\ti{f}^{(K)}_{(L)} s^M) - 
   \ti{f}^{(K)}_{(L)} (\ti{f}^{(I)}_{(J)} s^M).
\label{a3.3.6}
\eeq
for any integer sequences $I$, $J$, $K$, $L$, $M$ and $P$.  Eqs.(\ref{a3.3.5}) and (\ref{a3.3.6}) are trivially
true.  That Eq.(\ref{3.3.3}) satisfies Eq.(\ref{a3.3.3}) is also straightforward.  We have proved the validity of
Eq.(\ref{a3.3.4}) in Appendix~\ref{sa4.4}.  The remaining questions are whether Eq.(\ref{3.3.2}) satisfies 
Eq.(\ref{a3.3.2}), and whether Eq.(\ref{3.3.1}) satisfies Eq.(\ref{a3.3.1}).

Consider Eq.(\ref{3.3.2}).  The action of the Lie bracket operator on the L.H.S. of this equation on $\Ps^{(P)}$, 
where $P$ is arbitrary, can be obtained using Eqs.(\ref{3.2.9}) and (\ref{3.2.12}), and we get
\begin{eqnarray}
   \lbrack \g^I_J, \ti{f}^{(K)}_{(L)} \rbrack \Ps^{(P)} & = & \sum_{(Q)}      
   \left( \delta^I_{(Q)} \delta^K_{(J)} \delta^P_{(L)} +
   \sum_A \delta^{I A}_{(Q)} \delta^K_{(J A)} \delta^P_{(L)} \right. \nonumber \\ 
   & & - \left. \delta^K_{(Q)} \delta^I_{(L)} \delta^P_{(J)}
   - \sum_{A'} \delta^K_{(Q)} \delta^{I A'}_{(L)} \delta^P_{(J A')} \right)       
   \Ps^{(Q)}.
\label{a3.3.7}
\end{eqnarray}
On the other hand, the action of the operators on the R.H.S. of Eq.(\ref{3.3.2}) (let us call this linear 
combination of operators $g \ti{f} ^{I, K}_{J, L}$) on $\Ps^{(P)}$ is 
\begin{eqnarray}
   g\ti{f} ^{I, K}_{J, L} \Ps^{(P)} & = & \sum_{(Q)} \left( \delta^K_{(J)} \delta^I_{(Q)} \delta^P_{(L)} 
   + \sum_{K_1 K_2 = (K)} \delta^{K_1}_J \delta^{I K_2}_{(Q)} \delta^P_{(L)} \right. \nonumber \\
   & & \left. - \delta^I_{(L)} \delta^K_{(Q)} \delta^P_{(J)} - \sum_{L_1 L_2 = (L)}   
   \delta^I_{L_2} \delta^K_{(Q)} \delta^P_{(L_1 J)} \right) \Ps^{(Q)}.  
\label{a3.3.8}
\end{eqnarray}
The R.H.S of Eqs.(\ref{a3.3.7}) and (\ref{a3.3.8}) can be seen to the same by using the delta function defined in 
Eq.(\ref{a1.1}).  Thus Eq.(\ref{a3.3.2}) holds true.

Now let us verify Eq.(\ref{a3.3.1}).  Consider the Lie bracket operator on the L.H.S. of Eq.(\ref{3.3.1}) on  
$\Ps^{(P)}$, where $P$ is again arbitrary.  We then obtain
\begin{eqnarray}
   \left[ \g^I_J, \g^K_L \right] \Ps^{(P)} & = & \sum_{(Q)} \left(
   \delta^I_{(Q)} \delta^K_{(J)} \delta^P_{(L)} + \sum_A \delta^{I A}_{(Q)}
   \delta^K_{(J A)} \delta^P_{(L)} + \sum_B \delta^I_{(Q)} \delta^{K B}_{(J)}
   \delta^P_{(L B)} \right. \nonumber \\
   & & \left. + \sum_{A, B} \delta^{I A}_{(Q)} \delta^{K B}_{(J A)} 
   \delta^P_{(L B)} \right) \Ps^{(Q)} - 
   (I \leftrightarrow K, J \leftrightarrow L).
\label{a3.3.9}
\end{eqnarray}
The first three summations on the R.H.S. of this equation can be turned into the following expressions:
\begin{eqnarray}
   \delta^I_{(Q)} \delta^K_{(J)} \delta^P_{(L)} & = & 
   \delta^K_J \delta^I_{(Q)} \delta^P_{(L)} +
   \sum_{\begin{array}{l}
   	    J_1 J_2 = J \\
   	    K_1 K_2 = K
   	 \end{array}}
   \delta^{K_1}_{J_2} \delta^{K_2}_{J_1} \delta^I_{(Q)} \delta^P_{(L)};
\label{a3.3.10} \\
   \sum_A \delta^{I A}_{(Q)} \delta^K_{(J A)} \delta^P_{(L)} & = &
   \sum_{K_1 K_2 = K} \delta^{K_1}_J \delta^{I K_2}_{(Q)} \delta^P_{(L)} +
   \sum_{K_1 K_2 = K} \delta^{K_2}_J \delta^{I K_1}_{(Q)} \delta^P_{(L)}
   \nonumber \\
   & & + \sum_{K_1 K_2 K_3 = K} \delta^{K_2}_J \delta^{K_1 I K_3}_{(Q)}
   \delta^P_{(L)} \nonumber \\
   & & + \sum_{\begin{array}{l}
   		  J_1 J_2 = J \\
   		  K_1 K_2 K_3 = K
   	       \end{array}}
   \delta^{K_1}_{J_2} \delta^{K_3}_{J_1} \delta^{I K_2}_{(Q)} \delta^P_{(L)}
   \mbox{; and}	    
\label{a3.3.11} \\
   \sum_B \delta^I_{(Q)} \delta^{K B}_{(J)} \delta^P_{(L B)} & = &
   \sum_{J_1 J_2 = J} \delta^K_{J_2} \delta^I_{(Q)} \delta^P_{(J_1 L)} +
   \sum_{J_1 J_2 = J} \delta^K_{J_1} \delta^I_{(Q)} \delta^P_{(J_2 L)} 
   \nonumber \\
   & & + \sum_{J_1 J_2 J_3 = J} \delta^K_{J_2} \delta^I_{(Q)}
   \delta^P_{(J_1 L J_3)} \nonumber \\
   & & + \sum_{\begin{array}{l}
   		  J_1 J_2 J_3 = J \\
   		  K_1 K_2 = K
   	       \end{array}}
   \delta^{K_1}_{J_3} \delta^{K_2}_{J_1} \delta^I_{(Q)} \delta^P_{(J_2 L)}.
\label{a3.3.12}
\end{eqnarray}
The fourth summation is more complicated.  This can be manipulated to be:
\begin{eqnarray}
   \lefteqn{\sum_{A, B} \delta^{I A}_{(Q)} \delta^{K B}_{(J A)} \delta^P_{(L B)} 
   = \delta^K_J \sum_C \delta^{I C}_{(Q)} \delta^P_{(L C)} +
   \sum_{J_1 J_2 = J} \sum_C \delta^K_{J_1} \delta^{I C}_{(Q)}
   \delta^P_{(L J_2 C)} } \nonumber \\
   & & + \sum_{K_1 K_2 = K} \sum_C \delta^{K_1}_J \delta^{I K_2 C}_{(Q)} 
   \delta^P_{(L C)} +
   \sum_{J_1 J_2 = J} \sum_C \delta^K_{J_2} \delta^{I C}_{(Q)} 
   \delta^P_{(J_1 L C)} \nonumber \\
   & & + \sum_{K_1 K_2 = K} \sum_C \delta^{K_2}_J \delta^{K_1 I C}_{(Q)}
   \delta^P_{(L C)} +
   \sum_{\begin{array}{l}
   		J_1 J_2 = J \\
   		K_1 K_2 = K
   	     \end{array}}
   \delta^{K_2}_{J_1} \delta^{I K_1}_{(Q)} \delta^P_{(J_2 L)} \nonumber \\	     	      
   & & + \sum_{\begin{array}{l}
   		  J_1 J_2 = J \\
  		  K_1 K_2 = K
  	       \end{array}}
   \delta^{K_1}_{J_2} \delta^{I K_2}_{(Q)} \delta^P_{(J_1 L)} +
   \sum_{\begin{array}{l}
	    J_1 J_2 = J \\
	    K_1 K_2 = K
	 \end{array}} \sum_C
   \delta^{K_1}_{J_2} \delta^{I K_2 C}_{(Q)} \delta^P_{(J_1 L C)} \nonumber \\
   & & + \sum_{\begin{array}{l}
		  J_1 J_2 = J \\
		  K_1 K_2 = K
	       \end{array}} \sum_C
   \delta^{K_2}_{J_1} \delta^{K_1 I C}_{(Q)} \delta^P_{(L J_2 C)} \nonumber \\
   & & + \sum_{J_1 J_2 J_3 = J} \sum_C
   \delta^K_{J_2} \delta^{I C}_{(Q)} \delta^P_{(J_1 L J_3 C)} \nonumber \\
   & & + \sum_{K_1 K_2 K_3 = K} \sum_C
   \delta^{K_2}_J \delta^{K_1 I K_3 C}_{(Q)} \delta^P_{(L C)} \nonumber \\
   & & + \sum_{(Q)} \sum_{\begin{array}{l}
			     J_1 J_2 J_3 = J \\
			     K_1 K_2 K_3 = K
			  \end{array}}
   \delta^{K_1}_{J_3} \delta^{K_3}_{J_1} \delta^{I K_2}_{(Q)} \delta^P_{(J_2 L)}  
   + \Gamma_2
\label{a3.3.13}
\end{eqnarray}
where
\begin{eqnarray}
   \Gamma_2 & = & \sum_{\begin{array}{l}
   			 P_1 P_2 = (P) \\
   			 Q_1 Q_2 = (Q)
   		      \end{array}}
   \delta^I_{Q_1} \delta^K_{Q_2} \delta^{P_1}_L \delta^{P_2}_J +
   \sum_{\begin{array}{l}
   	    P_1 P_2 P_3 = (P) \\
   	    Q_1 Q_2 Q_3 = (Q)
   	 \end{array}}
   \delta^K_{Q_1} \delta^I_{Q_2} \delta^{P_1}_{Q_3} \delta^{P_2}_L 
   \delta^{P_3}_J \nonumber \\
   & & + \sum_{\begin{array}{l}
   		  P_1 P_2 P_3 = (P) \\
   		  Q_1 Q_2 Q_3 = (Q)
  	       \end{array}}		      	 
   \delta^I_{Q_1} \delta^K_{Q_2} \delta^{P_1}_{Q_3} \delta^{P_2}_J 
   \delta^{P_3}_L \nonumber \\
   & & + \sum_{\begin{array}{l}
   		  P_1 P_2 P_3 P_4 = (P) \\
   		  Q_1 Q_2 Q_3 Q_4 = (Q)
  	       \end{array}}
   \delta^I_{Q_1} \delta^K_{Q_3} \delta^{P_1}_J \delta^{P_3}_L 
   \delta^{P_2}_{Q_2} \delta^{P_4}_{Q_4}.
\label{a3.3.14}
\end{eqnarray}

If we substitute Eqs.(\ref{a3.3.10}), (\ref{a3.3.11}), (\ref{a3.3.12}) and (\ref{a3.3.13}) without $\Gamma_2$ into 
Eq.(\ref{a3.3.9}), we will obtain exactly the action of the operators on the R.H.S. of Eq.(\ref{3.3.1}) on 
$\Ps^{(P)}$.  $\Gamma_2$ is reproduced when $I$ and $K$ are interchanged with $J$ and $L$, respectively, in 
Eq.(\ref{a3.3.14}) and so it is cancelled.  Hence Eq.(\ref{a3.3.1}) is also satisfied.  Consequently, 
Eq.(\ref{3.3.1}) is true.  Q.E.D. 

\chapter{Proofs of Results in Chapter~{\ref{c5}}}
\la{ca5}

\section{Grand String Superalgebra}
\la{sa5.1}

We would like to show that the binary operations given in Section~\ref{s5.2} are Lie superbrackets.  Thus they 
constitute a Lie superalgebra.

Define the following operators:
\beq
   l'^I_J & \equiv & \g'^I_J - \sum_{i=1}^{2 \L + 4 \L_F} (-1)^{\ep(i) \lb \ep(I) + \ep(J) \rb} \g'^{iI}_{iJ} 
		     - \ti{f}'^I_J; \nn \\
   r'^I_J & \equiv & \g'^I_J - \sum_{j=1}^{2 \L + 4 \L_F} \g'^{Ij}_{Jj} - \ti{f}'^I_J; \; \mbox{and} \nn \\
   f'^I_J & \equiv & l'^I_J - \sum_{j=1}^{2 \L + 4 \L_F} l'^{Ij}_{Jj} \nn \\
	  & = & r'^I_J - \sum_{i=1}^{2 \L + 4 \L_F} (-1)^{\ep(i) \lb \ep(I) + \ep(J) \rb} r'^{iI}_{iJ} \nn \\
	  & = & \g'^I_J - \sum_{i=1}^{2 \L + 4 \L_F} (-1)^{\ep(i) \lb \ep(I) + \ep(J) \rb} \g'^{iI}_{iJ}
		- \sum_{j=1}^{2 \L + 4 \L_F} \g'^{Ij}_{Jj} \nn \\
          & & + \sum_{i,j=1}^{2 \L + 4 \L_F} (-1)^{\ep(i) \lb \ep(I) + \ep(J) \rb} \g'^{iIj}_{iIj}
	      - \ti{f}'^I_J + \sum_{j=1}^{2 \L + 4 \L_F} \ti{f}'^{Ij}_{Jj}.
\la{a5.1.1}
\eeq
Eqs.(\ref{3.3.6}), (\ref{3.3.7}), (\ref{3.3.8}) and (\ref{3.3.9}) are now generalized to
\beq
   l'^I_J s'^K & = & \sum_{K_1 \dot{K}_2 = K} \d^{K_1}_J s'^{I \dot{K}_2}; \nn \\
   r'^I_J s'^K & = & \sum_{\dot{K}_1 K_2 = K} (-1)^{\ep(\dot{K}_1) \lb \ep(I) + \ep(J) \rb} 
		     \d^{K_2}_J s'^{\dot{K}_1 I}; \nn \\
   f'^I_J s'^K & = & \d^K_J s'^I;
\la{a5.1.2}
\eeq
and
\beq
   l'^I_J \Ps'^K = r'^I_J \Ps'^K = f'^I_J \Ps'^k = 0.
\la{a5.1.3}
\eeq

Consider the subspace of the singlet states spanned by all states of the form
\beq
   \bar{\ph}^{\r_1} \otimes s^{\dot{K}} \otimes \ph^{\r_2} \equiv s'^{\r_1 + 2 \L, \dot{K}, \r_2 + 2 \L + 2 \L_F}
   \; \mbox{and} \; \Ps^K \equiv \Ps'^K,
\la{a5.1.4}
\eeq
where any integer in $K$ and $\dot{K}$ is between 1 and $2 \L$ inclusive, and $1 \leq \r_1, \r_2 \leq 2 \L_F$.
(The justification of the use of the direct products $\otimes$ will be obvious shortly.)  From Eqs.(\ref{3.2.6})
tells us that this definition of $\Ps^K$ satisfies Eq.(\ref{2.2.7})
\[ \Ps^K = (-1)^{\ep(K_1) \ep(K_2)} \Ps^{K_2 K_1}. \]
Next, consider a subset of color-invariant operators in the heterix superalgebra consisting of all finite linear 
combinations of
\beq
   \g^I_J & \equiv & \g'^I_J; \nn \\
   \ti{f}^I_J & \equiv & \ti{f}'^I_J; \nn \\
   \bar{\X}^{\l_1}_{\l_2} \otimes l^{\dot{I}}_{\dot{J}} & \equiv &  
   l'^{\l_1 + 2 \L, \dot{I}}_{\l_2 + 2 \L, \dot{J}}; \nn \\
   r^{\dot{I}}_{\dot{J}} \otimes \X^{\l_3}_{\l_4} & \equiv &
   r'^{\dot{I}, \l_3 + 2 \L + 2 \L_F}_{\dot{J}, \l_4 + 2 \L + 2 \L_F} \mbox{; and} \nn \\
   \bar{\X}^{\l_1}_{\l_2} \otimes f^{\dot{I}}_{\dot{J}} \otimes \X^{\l_3}_{\l_4} & \equiv &
   f'^{\l_1 + 2 \L, \dot{I}, \l_3 + 2 \L + 2 \L_F}_{\l_2 + 2 \L, \dot{J}, \l_4 + 2 \L + 2 \L_F}
\la{a5.1.6}
\eeq
where any integer in $I$, $J$, $\dot{I}$ or $\dot{J}$ is between 1 and $2 \L$ inclusive, and 
$1 \leq \l_1, \l_2, \l_3 \; \mbox{and} \; \l_4 \leq 2 \L$.  (Again it will be obvious shortly why the direct 
products are appropriate.)  It can be shown that this subset of color-invariant operators form a subalgebra of the 
heterix superalgebra.  {\em A fortiori}, this subset forms a Lie superalgebra.  Moreover, the subspace of the 
states defined above is a representation space for this Lie superalgebra, albeit a {\em reducible} one according to
Eq.(\ref{5.3.1}).

\section{A Basis for a $\L = 1$ Quotient Lie Algebra}
\label{sa5.2}

We are going to prove that the set of all $\ti{f}^{(a)}_{(b)}$, $g^1_b$ and $g^a_1$ where $a$ and $b$ are arbitrary 
positive integers form a basis for the quotient algebra of all {\em cosets} $\g^a_b$ and $\ti{f}^{(a)}_{(b)}$.  
Indeed, consider the equation
\begin{equation}
   \sum_{a,b=1}^{\infty} \alpha^{(b)}_{(a)} \ti{f}^{(a)}_{(b)} +
   \sum_{d=1}^{\infty} \alpha^d g^1_d + \sum_{c=2}^{\infty} \alpha_c g^c_1 = 0
\label{a5.2.1}
\end{equation}
where only a finite number of the $\a$'s are non-zero.  Let $n$ be an integer such that for all $a$'s and $b$'s 
such that $\a^{(b)}_{(a)} \neq 0$, we have $n > b$ and for all $d$'s such that $\a^d \neq 0$, we have $n > d$ also. 
Then
\begin{eqnarray*}
   \lefteqn{ \left( \sum_{a,b=1}^{\infty} \alpha^{(b)}_{(a)} \ti{f}^{(a)}_{(b)} +
   \sum_{d=1}^{\infty} \alpha^d g^1_d + \sum_{c=2}^{\infty} \alpha_c g^c_1
   \right) \Ps^{(n)} = } \\
   & & \sum_{d=1}^{\infty} n \alpha^d \Ps^{(n-d+1)} + \sum_{c=2}^{\infty}
   n \alpha_c \Ps^{(n+c-1)}.
\end{eqnarray*}
For the R.H.S. of this equation to vanish, we need $\alpha^d = 0$ and $\alpha_c = 0$ for all $c$'s and $d$'s.  
Hence Eq.(\ref{a5.2.1}) becomes
\[ \sum_{a,b=1}^{\infty} \alpha^{(b)}_{(a)} \ti{f}^{(a)}_{(b)} = 0. \]
Then
\[ \sum_{a,b=1}^{\infty} \alpha^{(b)}_{(a)} \ti{f}^{(a)}_{(b)} \Ps^{(e)}
   = \sum_{a=1}^{\infty} e \alpha^{(e)}_{(a)} \Ps^{(a)} \]
for any positive integer $e$.  Thus $\alpha^{(e)}_{(a)} = 0$ also for any positive $a$ and $e$.  Consequently, the 
set of all $\ti{f}^{(a)}_{(b)}$'s, $g^1_b$'s and $g^a_1$'s where $a$ and $b$ are arbitrary integers is linearly 
independent.  Q.E.D.             

\section{Cartan Subalgebra of a $\Lambda = 1$ Quotient Lie Algebra}
\label{sa5.3}

This can be seen as follows.  From the results of Section~\ref{s3.3}, we know that the subspace spanned by all 
$\ti{f}^{(a)}_{(a)}$'s and $g^1_1$ forms an Abelian subalgebra.  Moreover, consider a vector $v$ of the form
\[ v = \sum_{a, b = 1}^{\infty} \alpha^{(b)}_{(a)} \ti{f}^{(a)}_{(b)} +               
   \sum_{d=2}^{\infty} \alpha^d g^1_d + \sum_{c=2}^{\infty} \alpha_c g^c_1 \]
where only a finite number of the $\alpha$'s not equal to 0, and where $\alpha^{(a)}_{(a)} = 0$ for all $a = 1, 2, 
\cdots, \infty$.  If all the $\alpha^d$'s and $\alpha_c$'s vanish, then choose a particular $a_0$ such that there 
exists a $b$ with $\alpha^{(b)}_{(a_0)} \neq 0$.  Then 
\begin{eqnarray*}
   \left[ \ti{f}^{(a_0)}_{(a_0)}, v \right] & = & 
   \sum_{b = 1}^{\infty} a_0 \alpha^{(b)}_{(a_0)} \ti{f}^{(a_0)}_{(b)} -
   \sum_{a = 1}^{\infty} a_0 \alpha^{(a_0)}_{(a)} \ti{f}^{(a)}_{(a_0)} \\
   & \neq & 0.
\end{eqnarray*}
Hence $v$ does not commute with the subspace spanned by all $\ti{f}^{(a)}_{(a)}$'s and $g^1_1$.  If there exists at 
least one non-zero $\alpha^d$ or $\alpha_c$, set $m$ to be the maximum of all $a$'s and $b$'s such that 
$\alpha^{(b)}_{(a)} \neq 0$.  Then $\alpha^{(b')}_{(a')} = 0$ if either $b'$ or $a' > m$.  Use Eq.(\ref{5.4.13}) to 
rewrite each $g^1_d$ and $g^c_1$ such that $\alpha^d \neq 0$ and $\alpha_c \neq 0$ as
\[ g^1_d = \ti{f}^{(1)}_{(d)} + \ti{f}^{(2)}_{(d+1)} + \cdots + \ti{f}^{(m)}_{(m+d-1)} +
           g^{m+1}_{d+m} \]
and
\[ g^c_1 = \ti{f}^{(c)}_{(1)} + \ti{f}^{(c+1)}_{(2)} + \cdots + \ti{f}^{(m+c-1)}_{(m)} +
	   g^{c+m}_{m+1}. \]
Then
\[ v = \sum_{a, b = 1}^{\infty} \alpha'^{(b)}_{(a)} \ti{f}^{(a)}_{(b)} +
   \sum_{d=2}^{\infty} \alpha^d g^{m+1}_{d+m} + 
   \sum_{c=2}^{\infty} \alpha_c	g^{c+m}_{m+1} \]   
where 
\[ \alpha'^{(b)}_{(a)} = \left\{ \begin{array}{ll}
   \alpha^{(b)}_{(a)} + \alpha_{a-b+1} & \mbox{if $a>b$; and} \\
   \alpha^{(b)}_{(a)} + \alpha^{b-a+1} & \mbox{if $b>a$.}
   \end{array} \right. \]
It is possible for $\alpha'^{(b)}_{(a)} \neq 0$ only if $b \leq m$ or $a \leq m$.  If $\alpha'^{(b_0)}_{(a_0)} \neq 
0$ for a particular pair of numbers $b_0$ and $a_0$, then
\begin{eqnarray*}
   \left[ \ti{f}^{(b_0)}_{(b_0)}, v \right] & \neq & 0 \;\mbox{if $b_0 \leq m$; or} \\
   \left[ \ti{f}^{(a_0)}_{(a_0)}, v \right] & \neq & 0 \;\mbox{if $a_0 \leq m$}
\end{eqnarray*}
because
\begin{eqnarray*}
   \left[ \ti{f}^{(b_0)}_{(b_0)}, \sum_{d=2}^{\infty} \alpha^d g^{m+1}_{d+m} \right] & = & 0 \mbox{; and} \\
   \left[ \ti{f}^{(b_0)}_{(b_0)}, \sum_{c=2}^{\infty} \alpha_c g^{c+m}_{m+1} \right] & = & 0
\end{eqnarray*}
if $b_0 \leq m$, and
\begin{eqnarray*}
   \left[ \ti{f}^{(a_0)}_{(a_0)}, \sum_{d=2}^{\infty} \alpha^d g^{m+1}_{d+m} \right] & = & 0 \mbox{; and} \\
   \left[ \ti{f}^{(a_0)}_{(a_0)}, \sum_{c=2}^{\infty} \alpha_c g^{c+m}_{m+1} \right] & = & 0
\end{eqnarray*}
if $a_0 \leq m$.  If all $\alpha'^{(b)}_{(a)}$'s vanish, then
\[ v = \sum_{d=2}^{\infty} \alpha^d g^{m+1}_{d+m} + 
   \sum_{c=2}^{\infty} \alpha_c	g^{c+m}_{m+1} \]
and so
\[ \left[ \ti{f}^{(m+1)}_{(m+1)}, v \right] =  \sum_{d=2}^{\infty} (m + 1) \alpha^d   
   \ti{f}^{(m+1)}_{(d+m)} - \sum_{c=2}^{\infty} (m + 1) \alpha_c \ti{f}^{(c+m)}_{(m+1)} \neq 0.
\]       
Q.E.D.

\end{document}